\documentclass[11pt,a4paper]{article}
\usepackage{jcappub}

\usepackage{epsfig,psfrag,graphics,verbatim}
\usepackage{graphicx}% Include figure files
\usepackage{multirow}

\usepackage{dcolumn}% Align table columns on decimal point
\usepackage{bm}% bold math
\usepackage{color}
\usepackage{ulem}
\newcommand{\neut}{{\tilde{\chi}^0_1}}
\newcommand{\params}{{\mathbf \Theta}}

\newcommand{\loc}{\mathrm{loc}}
\newcommand{\lsr}{\mathrm{lsr}}
\newcommand{\esc}{\mathrm{esc}}

\newcommand{\be}{\begin{equation}}
\newcommand{\ee}{\end{equation}}
\newcommand{\sigmaSI}{\sigma_{\neut-p}^\text{SI}}
\newcommand{\sigmaSD}{\sigma_{\neut-p}^\text{SD}}
\newcommand{\fiveinvfb}{5.8 fb$^{-1}$}

\newcommand{\BR}{BR}
\newcommand\RBtaunu{\frac{\BR(B_u \to \tau \nu)}{\BR(B_u \to \tau \nu)_{SM}}}
\newcommand\DeltaO{\Delta_{0-}}
\newcommand\RBDtaunuBDenu{\frac{\BR(B \to D \tau \nu)}{\BR(B \to D e \nu)}}
\newcommand\Rl{R_{l23}}

\newcommand\Dstaunu{\BR(D_s \to \tau \nu)}
\newcommand\Dsmunu{\BR(D_s\to \mu \nu)} 
\newcommand\Dmunu{\BR(D \to \mu \nu)} 
\newcommand\brbsmumu{\BR(\overline{B}_s\to\mu^+\mu^-)}
\newcommand{\amusm}{a_{\mu}^{\text{SM}}}
\newcommand{\gmtwo}{(g-2)_{\mu}}

\newcommand\mhl{m_h}

\newcommand{\brbsgamma}{BR(\bar{B} \rightarrow X_s\gamma) }

\newcommand{\cl}{\text{CL}}
\newcommand{\gmt}{$g-2$}
\newcommand{\like}{\mathcal{L}}

\newcommand{\rowofthree}{trim = 8mm 15mm 5mm 0mm, clip, width=0.32\linewidth}
\newcommand{\rowoffour}{trim = 8mm 15mm 5mm 0mm, clip, width=0.244\linewidth}

\begin{document}

%\preprint{APS/123-QED}

\title{Global Fits of the cMSSM and NUHM including the LHC Higgs discovery and new XENON100 constraints}

\author[a]{C. Strege}
\author[b]{G. Bertone}
\author[c]{F. Feroz}
\author[d]{M. Fornasa}
\author[e]{R. Ruiz de Austri}
\author[a]{R. Trotta}

\affiliation[a]{Astrophysics Group, \& Imperial Centre for Inference and Cosmology, Imperial College London, Blackett Laboratory, Prince Consort Road, London SW7 2AZ, UK}
\affiliation[b]{GRAPPA Institute, University of Amsterdam, Science Park 904, 1090 GL Amsterdam, Netherlands}
\affiliation[c]{Cavendish Laboratory, University of Cambridge, J.J. Thomson Avenue, Cambridge CB3 0HE, UK}
\affiliation[d]{School of Physics and Astronomy, University of Nottingham, University Park NG7 2RD, UK}
\affiliation[e]{Instituto de F\'isica Corpuscular, IFIC-UV/CSIC, Valencia, Spain} 
%\affiliation[f]{African Institute for Mathematical Sciences, 6 Melrose Rd, Muizenberg, 7945, Cape Town, South Africa}

\abstract{
We present global fits of the constrained Minimal Supersymmetric Standard Model (cMSSM) and the Non-Universal Higgs Model (NUHM), including the most recent CMS constraint on the Higgs boson mass, 5.8 fb$^{-1}$ integrated luminosity null Supersymmetry searches by ATLAS, the new LHCb measurement of $\brbsmumu$ and the 7-year WMAP dark matter relic abundance determination. We include the latest dark matter constraints from the XENON100 experiment, marginalising over astrophysical and particle physics uncertainties. We present  Bayesian posterior and profile likelihood maps of the highest resolution available today, obtained from up to 350M points. We find that the new constraint on the Higgs boson mass has a dramatic impact, ruling out large regions of previously favoured cMSSM and NUHM parameter space. In the cMSSM, light sparticles and predominantly gaugino-like dark matter with a mass of a few hundred GeV are favoured. The NUHM exhibits a strong preference for heavier sparticle masses and a Higgsino-like neutralino with a mass of 1 TeV. The future ton-scale XENON1T direct detection experiment will probe large portions of the currently favoured cMSSM and NUHM parameter space. The LHC operating at 14 TeV collision energy will explore the favoured regions in the cMSSM, while most of the regions favoured in the NUHM will remain inaccessible. Our best-fit points achieve a satisfactory quality-of-fit, with p-values ranging from 0.21 to 0.35, so that none of the two models studied can be presently excluded at any meaningful significance level.}

%\pacs{}% PACS, the Physics and Astronomy
%                             % Classification Scheme.


%\arxivnumber{1107.1715}

\maketitle
%%%%%%%%%%%%%%%%%%%%%%%%%%%%%%%%%%%%%%%%%%%%%%%%%%%%%%%%%%%%%%%%%%%%%%%
%
% INTRODUCTION
%
%%%%%%%%%%%%%%%%%%%%%%%%%%%%%%%%%%%%%%%%%%%%%%%%%%%%%%%%%%%%%%%%%%%%%%%

\section{Introduction}
\label{secintro}

The Large Hadron Collider (LHC) is providing valuable information on possible extensions of the Standard Model of particle physics. In particular, LHC data are probing some of the most popular models of Supersymmetry (SUSY). At the same time, dark matter experiments searching for direct evidence of dark matter particles scattering off atomic nuclei are providing stringent constraints on the parameter space of supersymmetric models, that are highly complementary to the LHC results (see e.g. Refs. \cite{reviews} and references therein).

We have recently presented global fits of one of the most popular SUSY models, the so-called constrained Minimal Supersymmetric Standard Model (cMSSM, see e.g.~\cite{cMSSM1,cMSSM2}), including LHC data with 1 fb$^{-1}$ of integrated luminosity and XENON100 direct detection data presented in 2011 \cite{Bertone:2011nj,Strege:2011pk}. Other global fits studies of the cMSSM can for example be found in Ref.~\cite{arXiv:1104.3572,arXiv:1111.6098,arXiv:1105.5162,arXiv:1110.3568,Buchmueller:2011ab,Buchmueller:2012hv,Fowlie:2012im,Bechtle:2012zk}. In this paper, we extend our analysis to the Non-Universal Higgs Model (NUHM, see e.g.~\cite{NUHM1,NUHM2,NUHM3}), a less constrained model of SUSY with less restrictive boundary conditions applied at the Grand Unification scale than in the cMSSM. 

In the following we discuss the impact of new LHC data, including the discovery of the Higgs boson, and the latest XENON100 data on both the cMSSM and the NUHM. More specifically, we implement in our cMSSM global fits the latest exclusion limits in the ($m_0$,$m_{1/2}$) plane obtained by the ATLAS collaboration for squarks and gluinos in final states that contain missing $E_T$, jets and 0 leptons in 5.8 fb $^{-1}$ integrated luminosity of data at $\sqrt{s} = 8$ TeV collision energy  \cite{LHCSUSY}. For the NUHM we additionally include the exclusion limit in the ($m_A$,$\tan \beta$) plane from a CMS search for the decay of neutral Higgs bosons into final states containing two muons and missing $E_T$, based on 4.5 fb$^{-1}$ integrated luminosity of data collected at $\sqrt{s} = 7$ TeV collision energy \cite{LHCSUSYNUHM}. 

Furthermore, we include the most recent experimental constraint from the CMS collaboration on the mass of the lightest Higgs boson $m_h = 125.8 \pm 0.6$ GeV. This constraint is derived from a combination of 5.1 fb$^{-1}$ $\sqrt{s} = 7$ TeV data and 12.2 fb$^{-1}$ $\sqrt{s} = 8$ TeV data \cite{Higgs}. For the first time, we include the new measurement of  $\brbsmumu = (3.2^{+1.5}_{-1.2}) \times 10^{-9}$ from the LHCb collaboration, derived from 1 fb$^{-1}$ of data at $\sqrt{s} = 7$ TeV collision energy and 1.1 fb$^{-1}$ of data at $\sqrt{s} = 8$ TeV collision energy \cite{:2012ct}. Finally, we also include the most recent dark matter results from the XENON100 direct detection experiment, obtained from $224.6$ live days of data and $34$ kg fiducial volume, collected between February 2011 and March 2012~\cite{Aprile:2012nq}.

The main aim of this paper is to study the impact of new experimental results from LHC SUSY and Higgs searches, from the LHCb collaboration, and from direct detection searches for dark matter with 225 live days of XENON100 data on the cMSSM and the NUHM. We determine the most favoured regions of the cMSSM and NUHM parameter space in light of these experimental constraints, from both the Bayesian and the profile likelihood statistical perspective. We discuss the viability of the best-fit points for both models, and evaluate the extent to which these models are disfavoured by present-day experimental constraints.

The paper is organized as follows. In section \ref{sec:theory} we present the theoretical and statistical framework, and provide details about the supersymmetric models we study, the priors used for the model and nuisance parameters, and the implementation of the experimental constraints. In section \ref{sec:cMSSM} we present the favoured regions and best-fit point for the cMSSM and discuss the impact of different experimental constraints, in particular of the \gmt\ measurement, as well as the prospects for discovery with future direct detection experiments and the LHC operating at 14 TeV collision energy. In Section \ref{sec:NUHM} we repeat this analysis for the NUHM, and discuss the phenomenological differences between the cMSSM and the NUHM. We present our conclusions in section \ref{secconclusion}.

%%%%%%%%%%%%%%%%%%%%%%%%%%%%%%%%%%%%%%%%%%%%%%%%%%%%%%%%%%%%%%%%%%%%%%%
%
% THEORY AND STATS
%
%%%%%%%%%%%%%%%%%%%%%%%%%%%%%%%%%%%%%%%%%%%%%%%%%%%%%%%%%%%%%%%%%%%%%%%

\section{Theoretical and statistical framework}
\label{sec:theory}
In this paper we consider two different SUSY models, the cMSSM and the NUHM. The cMSSM is a model that has been widely studied in the past. It employs strong universality conditions at the Grand Unification scale, as a result it can be described by only five free parameters: the universal scalar and gaugino mass parameters $m_0$ and $m_{1/2}$, the universal trilinear coupling $A_0$, the ratio of the two Higgs vacuum expectation values $\tan \beta$, and the sign of the Higgs/Higgsino mass parameter sgn$(\mu)$. In the following analysis we fix sgn$(\mu) = +1$, which is favoured by the measured value of the magnetic moment of the muon~\cite{Feroz:2008wr}. The universality assumptions in the cMSSM are motivated by the natural link between SUSY and Grand Unification Theories; due to its low number of parameters the cMSSM has been an extremely popular model to study in the past.

A more general model with less restrictive boundary conditions applied at the Grand Unification scale is the NUHM. Specifically, in the NUHM the soft masses of the two Higgs doublets $m_{H_u}$ and $m_{H_d}$ are free parameters, independent of the sfermion masses. This is a reasonable assumption to make, since the Higgs and the matter fields are described by different supermultiplets, so that there is no strong motivation to assume unified Higgs and sfermion masses. The NUHM is described by six free parameters: the four cMSSM parameters $m_0$, $m_{1/2}$, $A_0$, $\tan \beta$, the (continuous) Higgs/Higgsino mass parameter $\mu$ and the mass of the pseudoscalar Higgs, $m_A$ (instead of the last two parameters one could use $m_{H_u}$ and $m_{H_d}$, as has been done in previous works, e.g.~\cite{Roszkowski:2009sm}). The larger number of free parameters in the NUHM compared to the cMSSM leads to a richer phenomenology.

Although we use highly efficient Bayesian methods to explore the favoured regions of the cMSSM and NUHM parameter spaces given all available constraints, we present our results both in Bayesian and in frequentist terms. Our approach starts from Bayes' theorem~\cite{Trotta:2008qt} 
\begin{equation}
p(\params|\mathbf{D})=\frac{p(\mathbf{D}|\params) p(\params)}{p(\mathbf{D})},
\label{eqn:Bayes}
\end{equation}
where $\mathbf{D}$ are the data and $\params$ are the model parameters of interest. Bayes' theorem states that the posterior probability distribution function (pdf) $p(\params|\mathbf{D})$ for the parameters is obtained from the likelihood function $p(\mathbf{D}|\params) \equiv \like(\params)$ and the prior pdf (or ``prior'' for short) $p(\params)$. In this paper we are primarily interested in parameter inference, therefore the Bayesian evidence $p(\mathbf{D})$ merely acts as a normalisation constant, and will not be considered further in the following analysis. 

In order to study the constraints on a single parameter of interest $\theta_i$, one can consider either the one-dimensional marginal posterior pdf (Bayesian), or the one-dimensional profile likelihood (frequentist). The marginal pdf is obtained from the full posterior distribution by integrating (marginalising) over the unwanted parameters in the $n$-dimensional parameter space:
\begin{equation}
p(\theta_i|\mathbf{D})=\int p(\params|\mathbf{D}) d\theta_1 ... d\theta_{i-1} d\theta_{i+1} ... d\theta_{n}.
\end{equation}

The frequentist profile likelihood function for $\theta_i$, instead, is found by maximising over the parameters that are not of interest:
\begin{equation}
{\mathcal L}(\theta_i)= \max_{\theta_1,...,\theta_{i-1},\theta_{i+1},...,\theta_{n}}\mathcal{L}(\params).
\end{equation}
The extension of these concepts to more than one parameter is straightforward. The profile likelihood function and the marginal posterior pdf are two different statistical quantities that may lead to different conclusions about the parameter space of interest. The marginal posterior pdf integrates over hidden parameter directions and therefore correctly accounts for volume effects; it peaks at the region of highest posterior mass. The profile likelihood function peaks at the region of highest likelihood. It is oblivious to volume effects, but is an excellent quantity to find small regions of high likelihood in parameter space. Especially when studying complicated parameter spaces of high dimensionality, such as SUSY parameter spaces, these two quantities will usually not lead to the same conclusions, and the maximum of information about the model parameter space is obtained by studying both of these quantities~\cite{Feroz:2011bj,Scott:2009jn}. Therefore, in the following we present results for both the marginalised Bayesian posterior pdf and the profile likelihood function.

\subsection{Model and nuisance parameters}

As can be seen from Eq.~\eqref{eqn:Bayes}, the posterior pdf is dominated by the likelihood whenever the prior is flat in the parameter space region where the likelihood has support. However, this is not generally the case, and therefore the choice of prior distribution can have a significant impact on the resulting posterior.  One approach is the adoption of an informative prior, that for example encapsulates in a natural way the theoretical prejudice that finely-tuned regions ought to be penalised (in an Occam's razor sense, see e.g.~\cite{Trotta:2008qt}). In this work, we compare the posterior distributions for two different choices of priors, in order to evaluate the prior dependence of the posterior. Such a prior-dependence has been found to be commonplace in high-dimensional parameter spaces with complex, multi-modal likelihoods which are typical of SUSY phenomenology (see e.g.~\cite{Bertone:2011nj,Strege:2011pk,Trotta:2008bp}). If the posterior distribution displays some residual dependence on the choice of prior distribution, the resulting constraints on the observables have to be interpreted with care~\cite{Trotta:2008bp,Scott:2009jn}. 

In order to assess to what extent the posterior distribution is influenced by the choice of prior, we repeat each of our scans for two different choices of (non-informative) prior distributions. ``Flat" priors are uniform on the mass parameters of the cMSSM ($m_0$, $m_{1/2}$) and NUHM ($m_0$, $m_{1/2}$, $m_A$), ``log" priors are uniform in the log of the mass parameters. Both choices of priors are uniform on $A_0$, $\tan \beta$ (cMSSM and NUHM) and $\mu$ (NUHM). Prior ranges for the cMSSM and the NUHM are summarised in Table~\ref{tab:params}.
\begin{table}
\begin{center}
\begin{tabular}{l l l}
\hline
\hline
\multicolumn{3}{c}{cMSSM Parameters} \\
% & \multicolumn{2}{c}{Range covered} \\
 & Flat priors & Log priors \\
\hline
\multicolumn{3}{c}{cMSSM parameters} \\ \hline 
$m_0$ [GeV] & (50.0, 4000.0) & ($10^{1.7}$, $10^{3.6}$) \\
$m_{1/2} $ [GeV] & (50.0, 4000.0) & ($10^{1.7}$, $10^{3.6}$) \\
$A_0$ [GeV] & \multicolumn{2}{c}{(-4000.0, 4000.0)} \\
$\tan\beta$ & \multicolumn{2}{c}{(2.0, 65.0)} \\ \hline
\multicolumn{3}{c}{NUHM parameters as above, and additionally:}\\ 
\hline 
$\mu$ [GeV] & \multicolumn{2}{c}{(-2000.0, 2000.0)} \\
$m_A$ [GeV] & (50.0, 4000.0) & ($10^{1.7}$, $10^{3.6}$)\\ 
\hline
\end{tabular}
\end{center}
\caption{\fontsize{9}{9}\selectfont cMSSM and NUHM parameters and the range of their values explored by the scan. Flat priors are uniform in the masses; log priors are uniform in the logarithm of the masses.}\label{tab:params}
\end{table}

%\begin{table}
%\begin{center}
%\begin{tabular}{l l l}
%\hline
%\hline
%\multicolumn{3}{c}{NUHM Parameters} \\
%% & \multicolumn{2}{c}{Range covered} \\
% & Flat priors & Log priors \\
%\hline
%$m_0$ [GeV] & (50.0, 4000.0) & ($10^{1.7}$, $10^{3.6}$) \\
%$m_{1/2} $ [GeV] & (50.0, 4000.0) & ($10^{1.7}$, $10^{3.6}$) \\
%$A_0$ [GeV] & \multicolumn{2}{c}{(-4000.0, 4000.0)} \\
%$\tan\beta$ & \multicolumn{2}{c}{(2.0, 65.0)} \\
%\hline
%\end{tabular}
%\end{center}
%\caption{\fontsize{9}{9}\selectfont NUHM parameters and the range of their values covered by the scan. Flat priors are uniform in the masses; log priors are uniform in the logarithm of the masses.}\label{tab:NUHM_params}
%\end{table}

In addition to the above model parameters, we include several nuisance parameters in the scans. Residual uncertainties on measurements of certain Standard Model parameters have been shown to have an important impact on the results of SUSY studies~\cite{Roszkowski:2007fd}. To correctly account for this effect, we include four SM parameters as nuisance parameters in our scan (the top mass, the bottom mass, the electroweak coupling constant and the strong coupling constant). When including direct detection constraints in the analysis one additionally needs to take into account uncertainties in astrophysics and the nuclear physics quantities. Here, we adopt the same strategy as presented in Ref.~\cite{Bertone:2011nj}, and include three additional hadronic nuisance parameters (the hadronic matrix elements $f_{Tu}$, $f_{Td}$ and $f_{Ts}$ that parameterise the contribution of the light quarks to the proton composition) and four astrophysical nuisance parameters (the local dark matter density $\rho_{\loc}$, and three quantities parameterising the WIMP velocity distribution: the local circular velocity $v_{\lsr}$, the escape velocity $v_{\esc}$ and the velocity dispersion $v_d$) in the analysis, see~\cite{Bertone:2011nj} for full details. Nuisance parameters are well constrained by experimental data, and therefore we adopt informative Gaussian priors for these quantities. The mean and standard deviation of the Gaussian priors are chosen to reflect up-to-date experimental constraints, and are given in Table~\ref{tab:nuis_params}.

\begin{table}
\begin{center}
\begin{tabular}{l l l l }
\hline
\hline
 & Gaussian prior  & Range scanned & Ref. \\
\hline
\multicolumn{4}{c}{SM nuisance parameters} \\
\hline
$M_t$ [GeV] & $173.2 \pm 0.9$  & (170.5, 175.9) &  \cite{topmass:1} \\
$m_b(m_b)^{\bar{MS}}$ [GeV] & $4.20\pm 0.07$ & (3.99, 4.41) &  \cite{pdg07}\\
$[\alpha_{em}(M_Z)^{\bar{MS}}]^{-1}$ & $127.955 \pm 0.030$ & (127.865, 128.045) &  \cite{pdg07}\\
$\alpha_s(M_Z)^{\bar{MS}}$ & $0.1176 \pm  0.0020$ &  (0.1116, 0.1236) &  \cite{Hagiwara:2006jt}\\
\hline
\multicolumn{4}{c}{Astrophysical nuisance parameters} \\
\hline
$\rho_{\loc}$ [GeV/cm$^3$] & $0.4\pm 0.1$ & (0.1, 0.7) & \cite{Pato:2010zk}\\
$v_{\lsr}$ [km/s] & $230.0 \pm 30.0$ &  (140.0, 320.0) & \cite{Pato:2010zk}\\
$v_{\esc}$ [km/s] & $544.0 \pm 33.0$ & (445.0, 643.0) & \cite{Pato:2010zk}\\
$v_d$ [km/s] & $282.0 \pm 37.0$ & (171.0, 393.0) & \cite{Pato:2010zk}\\
\hline
\multicolumn{4}{c}{Hadronic nuisance parameters} \\
\hline
$f_{Tu}$ & $0.02698 \pm 0.00395$ & (0.015, 0.039) &  \cite{Ellis}\\
$f_{Td}$ & $0.03906 \pm 0.00513$ &  (0.023, 0.055) &  \cite{Ellis}\\
$f_{Ts}$ & $0.363 \pm 0.119$ & (0.0006, 0.72) &  \cite{Ellis}\\
\hline
\hline
\end{tabular}
\end{center}
\caption{\fontsize{9}{9}\selectfont List of nuisance parameters included in the scans of the cMSSM and NUHM parameter spaces. The mean value and standard deviation adopted for the Gaussian prior on each parameter is shown, as well as the range of values explored by the scan.}\label{tab:nuis_params} 
\end{table}

\subsection{Experimental constraints and the likelihood function}

The likelihood function entering in Eq.~\eqref{eqn:Bayes} is composed of several different parts, corresponding to the different experimental constraints that are applied in our global fits analysis:
\begin{equation}
\ln \like = \ln \like_\text{LHC} + \ln \like_\text{WMAP}  + \ln \like_\text{EW}+ \ln \like_\text{B(D)} + \ln \like_{g-2}+ \ln \like_\text{Xe100}.      
\end{equation}

The full list of experimental constraints included in the likelihood function is given in Table~\ref{tab:exp_constraints}.
\begin{itemize}
\item $ \ln \like_\text{LHC}$. 
The LHC likelihood implements recent results from SUSY null searches from ATLAS and CMS. Exclusion limits in the ($m_0$,$m_{1/2}$) plane are based on a search by the ATLAS collaboration for squarks and gluinos in final states that contain missing $E_T$, jets and 0 leptons in 5.8 fb $^{-1}$ integrated luminosity of data at $\sqrt{s} = 8$ TeV collision energy \cite{LHCSUSY}. While this exclusion limits was obtained in the cMSSM framework for fixed values of $\tan \beta = 10$ and $A_0 = 0$, the result is fairly insensitive to $\tan \beta$ and $A_0$, so that we can use this limit even when varying these quantities. Furthermore, we can also apply this limit to the NUHM, since the signal is dominated by the strong cross-sections of the two first generation of squarks and gluino production, which do not depend significantly on other parameters than $m_0$ and $m_{1/2}$\footnote{Additionally, as we will show below, the LHC exclusion limit has essentially no impact on the NUHM results, so that small variations of this exclusion limit will not change our conclusions.}. The LHC exclusion limit is included in the likelihood function by defining the likelihood of samples corresponding to masses below the limit to be zero. For the NUHM we additionally include the exclusion limit in the ($m_A$,$\tan \beta$) plane from a CMS search for the decay of neutral Higgs bosons into final states containing two muons and missing $E_T$, based on 4.5 fb$^{-1}$ integrated luminosity of data collected at $\sqrt{s} = 7$ TeV collision energy \cite{LHCSUSYNUHM}. 

We furthermore include the most recent experimental constraint from the CMS collaboration on the mass of the lightest Higgs boson $m_h = 125.8 \pm 0.6$ GeV. This constraint is derived from a combination of 5.1 fb$^{-1}$ $\sqrt{s} = 7$ TeV data and 12.2 fb$^{-1}$ $\sqrt{s} = 8$ TeV data \cite{Higgs}. The statistical significance of the signal is $6.8 \sigma$. We use a Gaussian likelihood and we add in quadrature a theoretical error of 2 GeV to the experimental error. We do not impose the experimental constraint on the Higgs production cross-section in this analysis, since all of our samples fall within a very narrow range $\sigma/\sigma_{SM} = [0.95,1.00]$, which is in good agreement with the CMS constraint $\sigma/\sigma_{SM} = 0.88 \pm 0.21$ reported in \cite{Higgs}. 

Finally, we include the new LHCb constraint on $\brbsmumu = (3.2^{+1.5}_{-1.2}) \times 10^{-9}$, derived from a combined analysis of 1 fb$^{-1}$ data at $\sqrt{s} = 7$ TeV collision energy and 1.1 fb$^{-1}$ data at $\sqrt{s} = 8$ TeV collision energy \cite{:2012ct}. We implement this constraint as a Gaussian distribution with a conservative experimental error of $\sigma = 1.5 \times 10^{-9}$, and a $10\%$ theoretical error. 

\item $\ln \like_\text{WMAP}$. The WMAP measurement of the dark matter relic abundance is included as a Gaussian in $\ln \like_\text{WMAP}$. We use the WMAP 7-year value $\Omega_\chi h^2 = 0.1109 \pm 0.0056$ and we add a fixed 10\% theoretical uncertainty in quadrature. We assume that neutralinos make up all of the dark matter in the universe. 

\item $\ln \like_\text{EW}$ implements precision tests of the electroweak sector. The electroweak precision observables
$M_W$ and $\sin^2\theta_{eff}$ are included with a Gaussian likelihood. 

\item $\ln \like_\text{B(D)}$. The $B$ and $D$ physics observables are included as a Gaussian likelihood. The full list of $B$ and $D$ physics observables included in our analysis is shown in Table~\ref{tab:exp_constraints}.

\item $\ln \like_{g-2}$. The measured anomalous magnetic moment of the muon, included as a Gaussian datum, provides important information about supersymmetric parameter spaces, since it can be experimentally measured to very good precision. By comparing the theoretical value of this quantity favoured in the Standard Model with the experimental result \cite{Davier:2010nc} the supersymmetric contribution $\delta a_{\mu}^{SUSY}$ to this quantity can be constrained. The experimental measurement of the muon anomalous magnetic moment is discrepant with the SM prediction by $\delta a_{\mu}^{SUSY} = (28.7 \pm 8.0) \times 10^{-9}$ \cite{Davier:2010nc}. This analysis, based on e$^+$e$^-$ data, leads to a $3.6\sigma$ discrepancy between the experimental result and the theoretical SM value. However, the corresponding analysis based on $\tau$ data leads to a smaller discrepancy of $2.4\sigma$ \cite{Davier:2010nc}. The significance of this discrepancy has to be interpreted with care, since the calculation of the theoretical value of the muon anomalous magnetic moment in the Standard Model is subject to important theoretical uncertainties, arising mostly in the computation of the hadronic loop contributions. In past studies of the cMSSM \cite{Strege:2011pk} and NUHM \cite{Roszkowski:2009sm} it was found that the $\delta a_{\mu}^{\mathrm{SUSY}}$ constraint plays a dominant role in driving the global fits results for both models. Therefore, in the following we will present results for both an analysis including and excluding the experimental constraint on $\delta a_{\mu}^{\mathrm{SUSY}}$, in order to evaluate the dependence of our conclusions on this constraint.

\item $\ln \like_\text{Xe100}$. We also include the most recent dark matter results from the XENON100 direct detection experiment, obtained from $224.6$ live days of data and $34$ kg fiducial volume, collected between February 2011 and March 2012~\cite{Aprile:2012nq}. XENON100 currently places the tightest direct detection constraints on the WIMP properties. The collaboration reported the detection of two candidate WIMP scattering events in the pre-defined signal region, with an expected background of $b = 1.0 \pm 0.2$ events. The detected events are compatible with the background, so that new exclusion limits could be derived in the $(m_{\neut},\sigmaSI)$ plane. 

We use an updated version of the approximate likelihood function from our previous analyses~\cite{Bertone:2011nj,Strege:2011pk}, for a detailed description we refer the reader to Ref.  \cite{Bertone:2011nj}. In addition to changes in the fiducial mass and exposure time,  and the number of detected and background events, updates to the likelihood function include the reduction of the lower energy threshold for the analysis to 3 photoelectron events (PE) and an update to the response to 122 keV gamma-rays from calibration measurements to $L_y = 2.28$ PE/keVee, in accordance with the values reported in Ref.~\cite{Aprile:2012nq}. We make the simplifying assumption of an energy-independent acceptance of data quality cuts, and adjust the acceptance-corrected exposure to accurately reproduce the exclusion limit in the $(m_{\neut},\sigmaSI)$ plane  reported in Ref.~\cite{Aprile:2012nq} in the mass range of interest.

\end{itemize}

\begin{table*}
\begin{center}
\begin{tabular}{|l | l l l | l|}
\hline
\hline
Observable & Mean value & \multicolumn{2}{c|}{Uncertainties} & Ref. \\
 &   $\mu$      & ${\sigma}$ (exper.)  & $\tau$ (theor.) & \\\hline
$M_W$ [GeV] & 80.399 & 0.023 & 0.015 & \cite{lepwwg} \\
$\sin^2\theta_{eff}$ & 0.23153 & 0.00016 & 0.00015 & \cite{lepwwg} \\
$\delta a_\mu^{\mathrm{SUSY}} \times 10^{10}$ & 28.7 & 8.0 & 2.0 & \cite{Davier:2010nc} \\
$\brbsgamma \times 10^4$ & 3.55 & 0.26 & 0.30 & \cite{hfag}\\
$R_{\Delta M_{B_s}}$ & 1.04 & 0.11 & - & \cite{deltambs} \\
$\RBtaunu$   &  1.63  & 0.54  & - & \cite{hfag}  \\
$\DeltaO  \times 10^{2}$   &  3.1 & 2.3  & - & \cite{delta0}  \\
$\RBDtaunuBDenu \times 10^{2}$ & 41.6 & 12.8 & 3.5  & \cite{Aubert:2007dsa}  \\
$\Rl$ & 0.999 & 0.007 & -  &  \cite{Antonelli:2008jg}  \\
$\Dstaunu \times 10^{2}$ & 5.38 & 0.32 & 0.2  & \cite{hfag}  \\
$\Dsmunu  \times 10^{3}$ & 5.81 & 0.43 & 0.2  & \cite{hfag}  \\
$\Dmunu \times 10^{4}$  & 3.82  & 0.33 & 0.2  & \cite{hfag} \\
$\Omega_\chi h^2$ & 0.1109 & 0.0056 & 0.012 & \cite{wmap} \\
$\mhl$ [GeV] & 125.8  & 0.6  & 2.0 & \cite{Higgs} \\
$\brbsmumu$ &  $3.2 \times 10^{-9}$ & $1.5  \times 10^{-9}$ & 10\% & \cite{:2012ct}\\
\hline\hline
   &  Limit (95\%~$\cl$)  & \multicolumn{2}{r|}{$\tau$ (theor.)} & Ref. \\ \hline
%$m_{\tilde{q}}$ & $>375$ GeV  & & 5\% & \cite{pdg07}\\
%$m_{\tilde{g}}$ & $>289$ GeV  & & 5\% & \cite{pdg07}\\
Sparticle masses  &  \multicolumn{3}{c|}{As in table~4 of
  Ref.~\cite{deAustri:2006pe}.}  & \\
$m_0, m_{1/2}$ & \multicolumn{3}{l|}{ATLAS, $\sqrt{s} = 8$ TeV, \fiveinvfb\ 2012 limits} & \cite{LHCSUSY} \\
$m_A, \tan \beta$ & \multicolumn{3}{l|}{CMS, $\sqrt{s} = 7$ TeV, $4.7$ fb$^{-1}$ 2012 limits} & \cite{LHCSUSYNUHM} \\
$m_\chi - \sigmaSI$ & \multicolumn{3}{l|}{XENON100 2012 limits ($224.6 \times 34$ kg days)} & \cite{Aprile:2012nq} \\
\hline
\end{tabular}
\end{center}
\caption{\fontsize{9}{9} \selectfont Summary of experimental constraints that enter in the computation of the likelihood function. The upper part lists the observables for which a positive measurement exists. For these quantities mean values, experimental ($\sigma$) and theoretical ($\tau$) uncertainties are given, which are added in quadrature in the Gaussian likelihood. $\delta a_\mu^\text{SUSY}= a_\mu^{\rm exp}-\amusm$ corresponds to the discrepancy between the experimental value and the SM prediction of the anomalous magnetic moment of the muon $\gmtwo$; $\mhl$ stands for the mass of the lightest Higgs boson, for which we use the latest CMS constraint~\cite{Higgs}.  The lower part shows observables for which only experimental limits currently exist, including recent limits from LHC SUSY searches~\cite{LHCSUSY,LHCSUSYNUHM}, and constraints on the dark matter mass and spin-independent cross-section from the XENON100 direct detection experiment \cite{Aprile:2012nq}. \label{tab:exp_constraints}}
\end{table*}

\subsection{Scanning methodology}

To map out the posterior pdf and the profile likelihood we use the \texttt{SuperBayeS-v2.0} package, an evolution of the publicly available \texttt{SuperBayeS-v1.5}~\cite{deAustri:2006pe,Roszkowski:2007fd,Trotta:2008bp,Bertone:2011nj}, which has been developed for this work. \texttt{SuperBayeS-v2.0} will shortly be released to the public.  
 
This latest version of \texttt{SuperBayeS} is interfaced with SoftSUSY 3.1.7 as SUSY spectrum calculator, 
MicrOMEGAs 2.4 \cite{MicrOMEGAs,Belanger:2006is} to compute the abundance of DM,
DarkSUSY 5.0.5~\cite{DarkSUSY,Gondolo:2004sc} for the computation of $\sigmaSI$ and $\sigmaSD$, SuperIso 3.0 \cite{SuperIso,Mahmoudi:2008tp} to compute $\delta a_\mu^{\mathrm{SUSY}}$ and B(D) physics 
observables, SusyBSG 1.5 for the determination of $\brbsgamma$ 
\cite{SusyBSG,Degrassi:2007kj} and FeynHiggs 1.9 \cite{feynhiggs} to compute the Higgs production cross-sections. The infrastructure of the code has also been revised, to allow for a larger choice of supersymmetric models at compilation time, including (but not limited to) the cMSSM and the NUHM studied here. 

The  \texttt{SuperBayeS-v2.0} package uses the publicly available MultiNest v2.18~\cite{Feroz:2007kg,Feroz:2008xx} nested sampling algorithm to explore the cMSSM and NUHM model parameter space. MultiNest  is an extremely efficient scanning algorithm that can reduce the number of likelihood evaluations required for an accurate mapping of the posterior pdf by up to two orders of magnitude with respect to conventional MCMC methods. This Bayesian algorithm, originally designed to compute the model likelihood and to accurately map out the posterior,  is also able to reliably evaluate the profile likelihood, given appropriate MultiNest settings, as demonstrated in~\cite{Feroz:2011bj}. We use the settings recommended in Ref. \cite{Feroz:2011bj} (number of live points nlive = 20,000, tolerance tol = $10^{-4}$), tuned to obtain an accurate map of the profile likelihood function. 

Our cMSSM posterior results are based on approximately 220M (128M) likelihood evaluations for log (flat) priors. cMSSM posterior results excluding the constraint on $\delta a_\mu^{\mathrm{SUSY}}$ are derived from 199M (124M) likelihood evaluations for log (flat) priors. For the NUHM, the chains for log (flat) priors were generated from 132M (73M) likelihood evaluations. NUHM results excluding the $\delta a_\mu^{\mathrm{SUSY}}$ constraint are based on 95M (52M) likelihood evaluations. The profile likelihood function, which is in principle prior-independent, is derived from combined chains of the log and flat prior scans. In order to achieve a higher resolution even in the tail of the profile likelihood, we save the value and coordinates of {\em all} likelihood evaluations for the profile likelihood analysis. This procedure thus includes a large number of samples that would normally not have been saved in the posterior chains (as they belong to rejected steps in the sampling). This results in a combined total of 348M (205M) samples for the cMSSM (NUHM) scans including the $\delta a_\mu^{\mathrm{SUSY}}$ constraint, and 323M (147M) samples for scans excluding the $\delta a_\mu^{\mathrm{SUSY}}$ constraint, out of which the profile likelihood results are obtained. For the cMSSM, this is a factor of $\sim100$ more than our previous works \cite{Bertone:2011nj,Strege:2011pk}, and a factor of $> 3$ more than the frequentist global fits analysis presented in Ref. \cite{Buchmueller:2012hv} (for the NUHM our resolution is comparable).

As another check of the robustness of our scanning procedure, we run 10 scans in parallel (for both flat and log priors) for each case we consider and we compare the resulting profile likelihood (and best-fit points) across the different scans, and between each scan and the merged samples obtained from all the scans together. We have found that while each scan is more noisy than the combined samples (as expected), our results are consistent across all the scans.  

The total computational effort for the various cases considered is approximately 22 (13) CPU years for the cMSSM scans, and 72 (61) CPU years for the NUHM scans including (excluding) the $\delta a_\mu^{\mathrm{SUSY}}$ constraint, for a total computational time of approximately 168 CPU years. 

%%%%%%%%%%%%%%%%%%%%%%%%%%%%%%%%%%%%%%%%%%%%%%%%%%%%%%%%%%%%%%%%%%%%%%%
%
% RESULTS
%
%%%%%%%%%%%%%%%%%%%%%%%%%%%%%%%%%%%%%%%%%%%%%%%%%%%%%%%%%%%%%%%%%%%%%%%

%\section{Results}
%\label{secresults}

%%%%%%%%%%%%%%%%%%%%%%%%%%%%%%%%%%%%%%%%%%%%%%%%%%%%%%%%%%%%%%%%%%%%%%%
%
% CMSSM 
%
%%%%%%%%%%%%%%%%%%%%%%%%%%%%%%%%%%%%%%%%%%%%%%%%%%%%%%%%%%%%%%%%%%%%%%%

\section{Results for the cMSSM}
\label{sec:cMSSM}

\begin{figure*}[htp]
\centering
\expandafter\includegraphics\expandafter[\rowofthree]{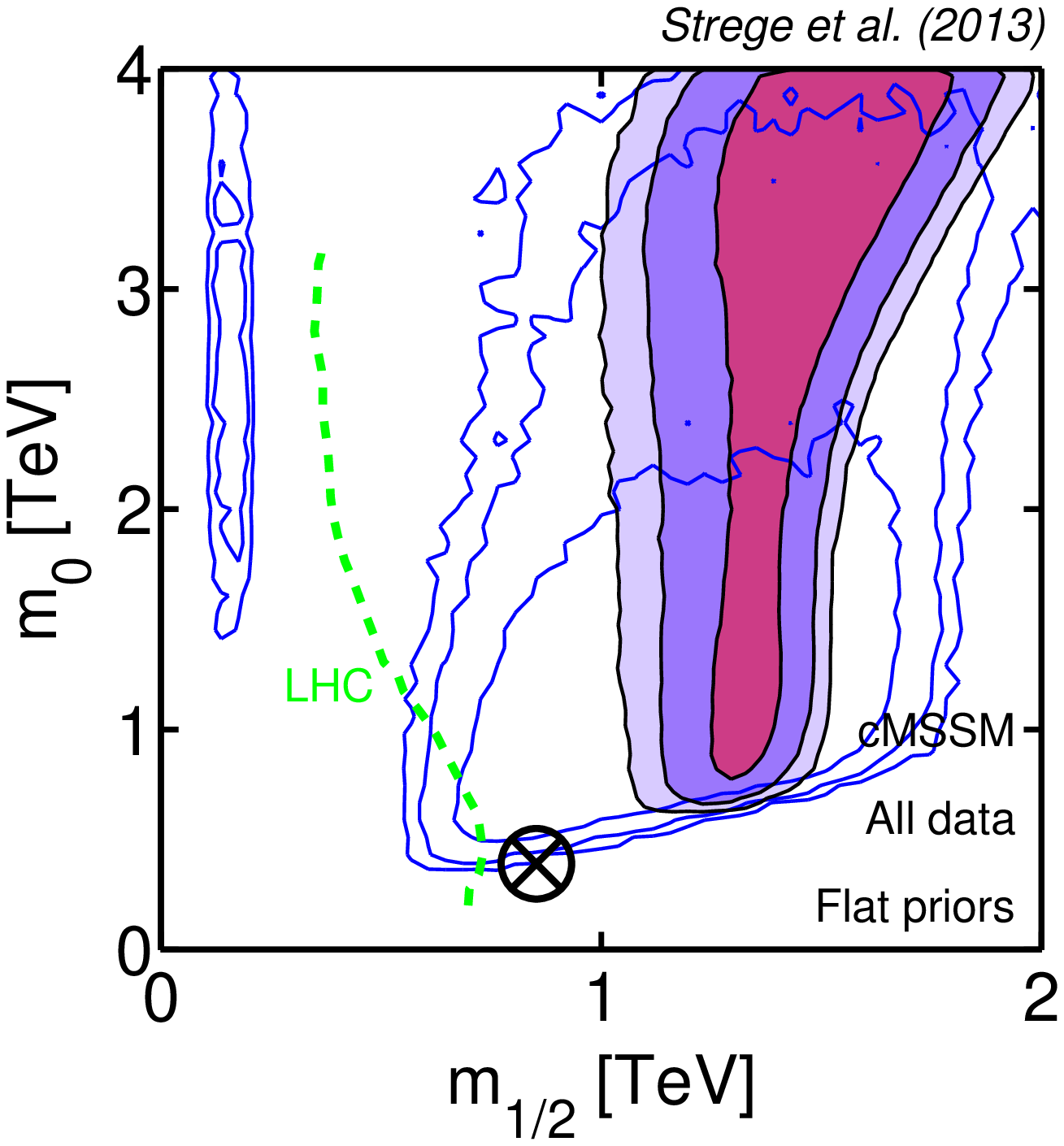}
\expandafter\includegraphics\expandafter[\rowofthree]{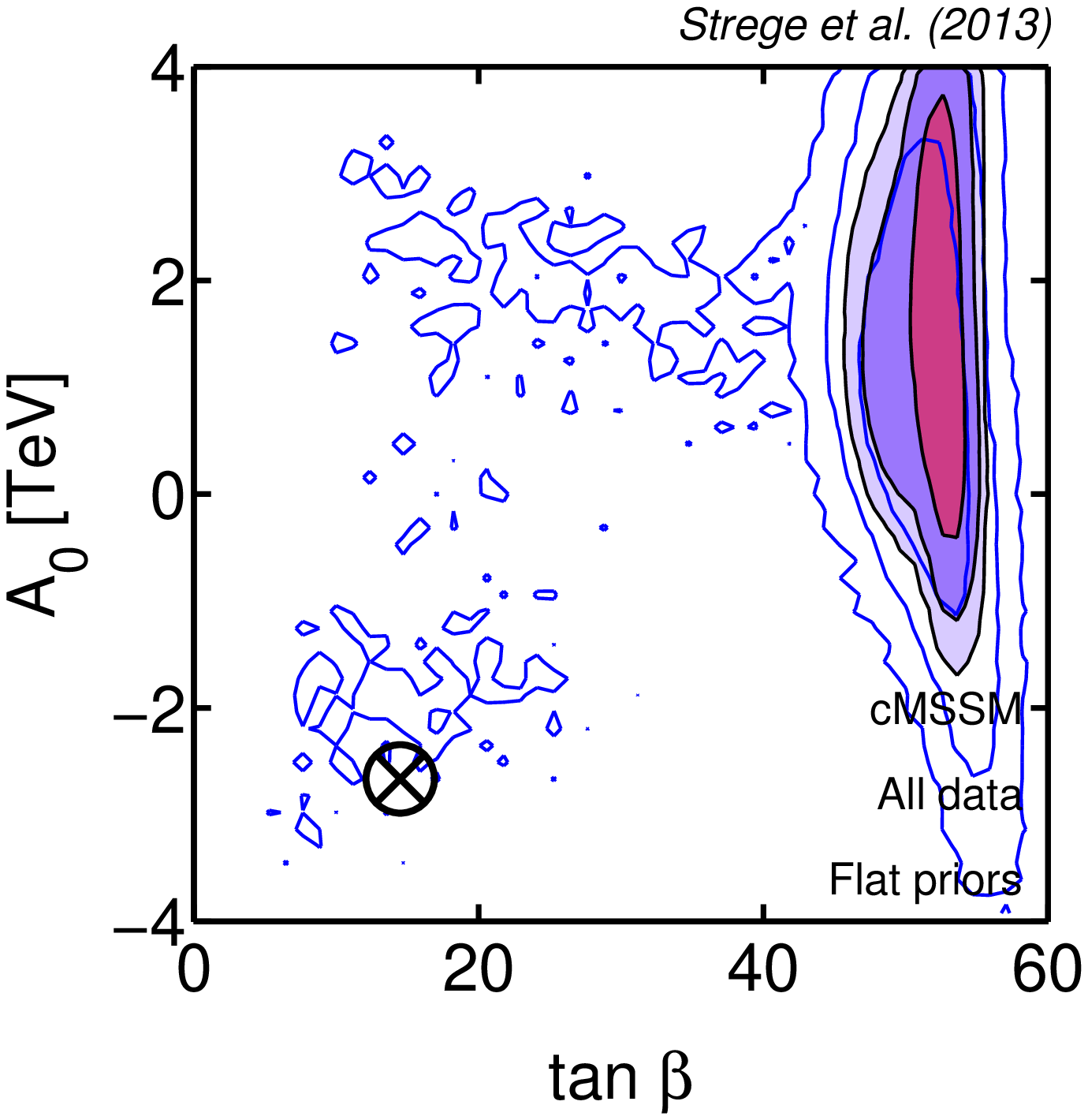}
\expandafter\includegraphics\expandafter[\rowofthree]{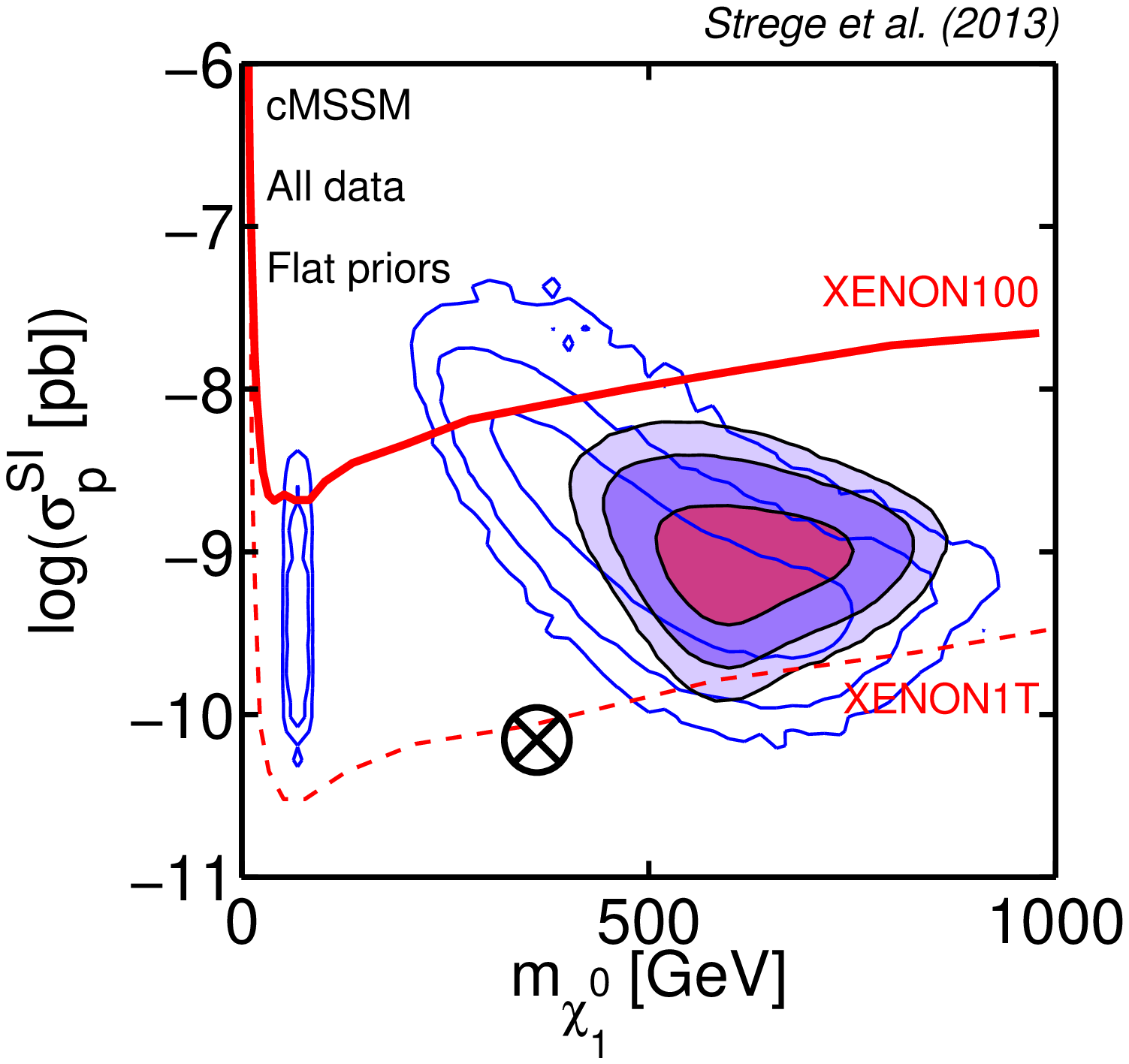} \\
\expandafter\includegraphics\expandafter[\rowofthree]{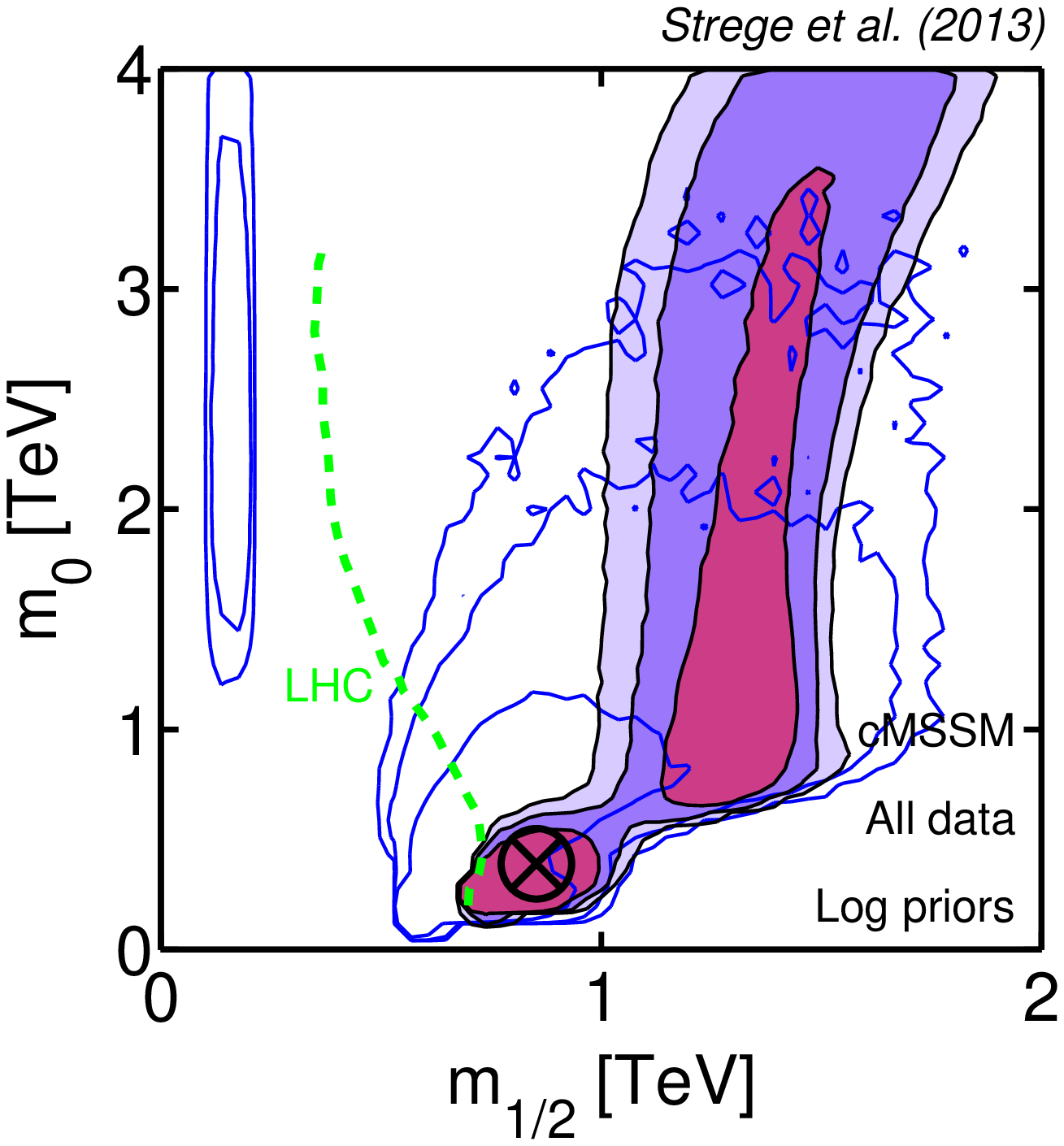}
\expandafter\includegraphics\expandafter[\rowofthree]{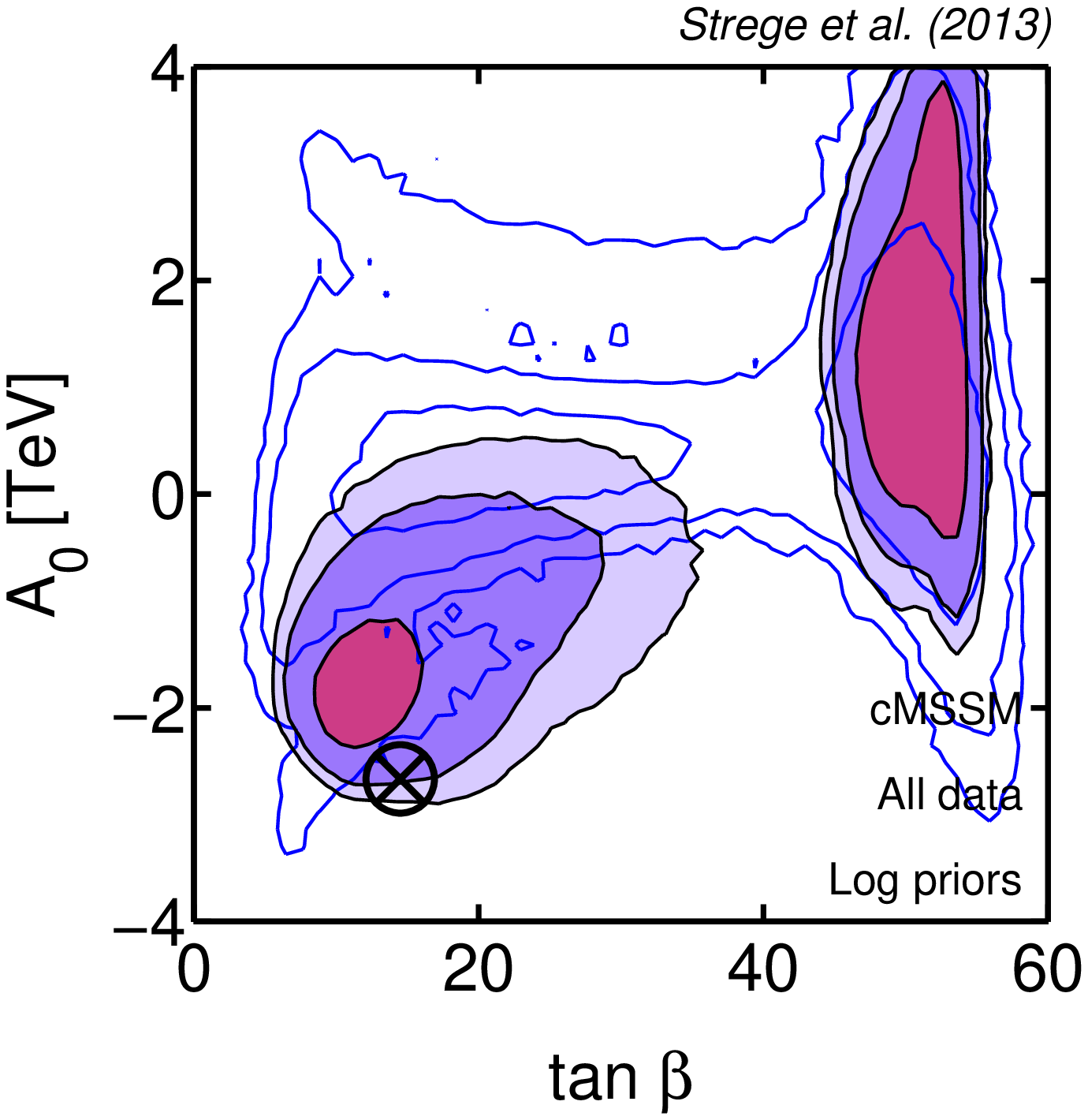}
\expandafter\includegraphics\expandafter[\rowofthree]{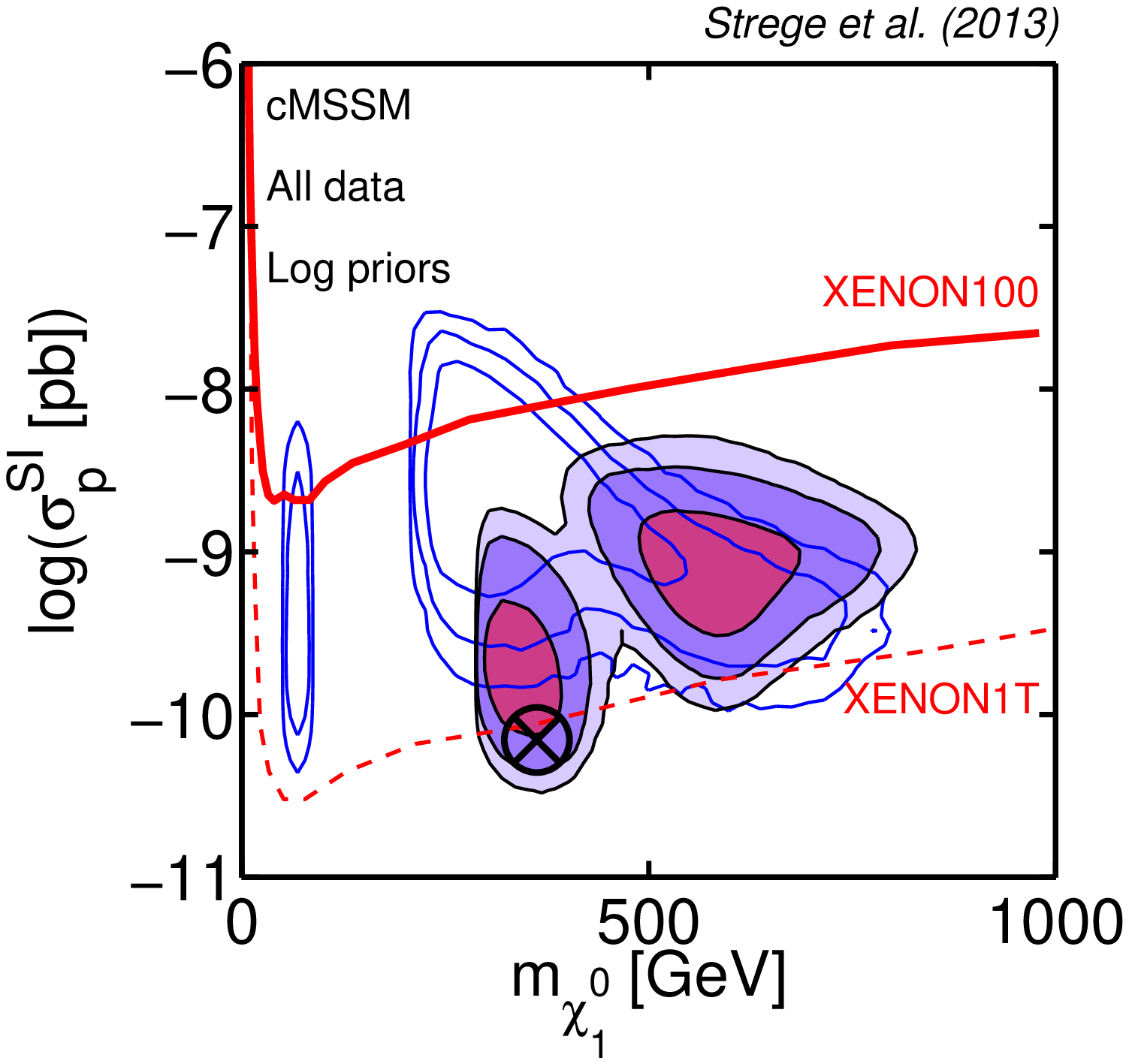} \\
\expandafter\includegraphics\expandafter[\rowofthree]{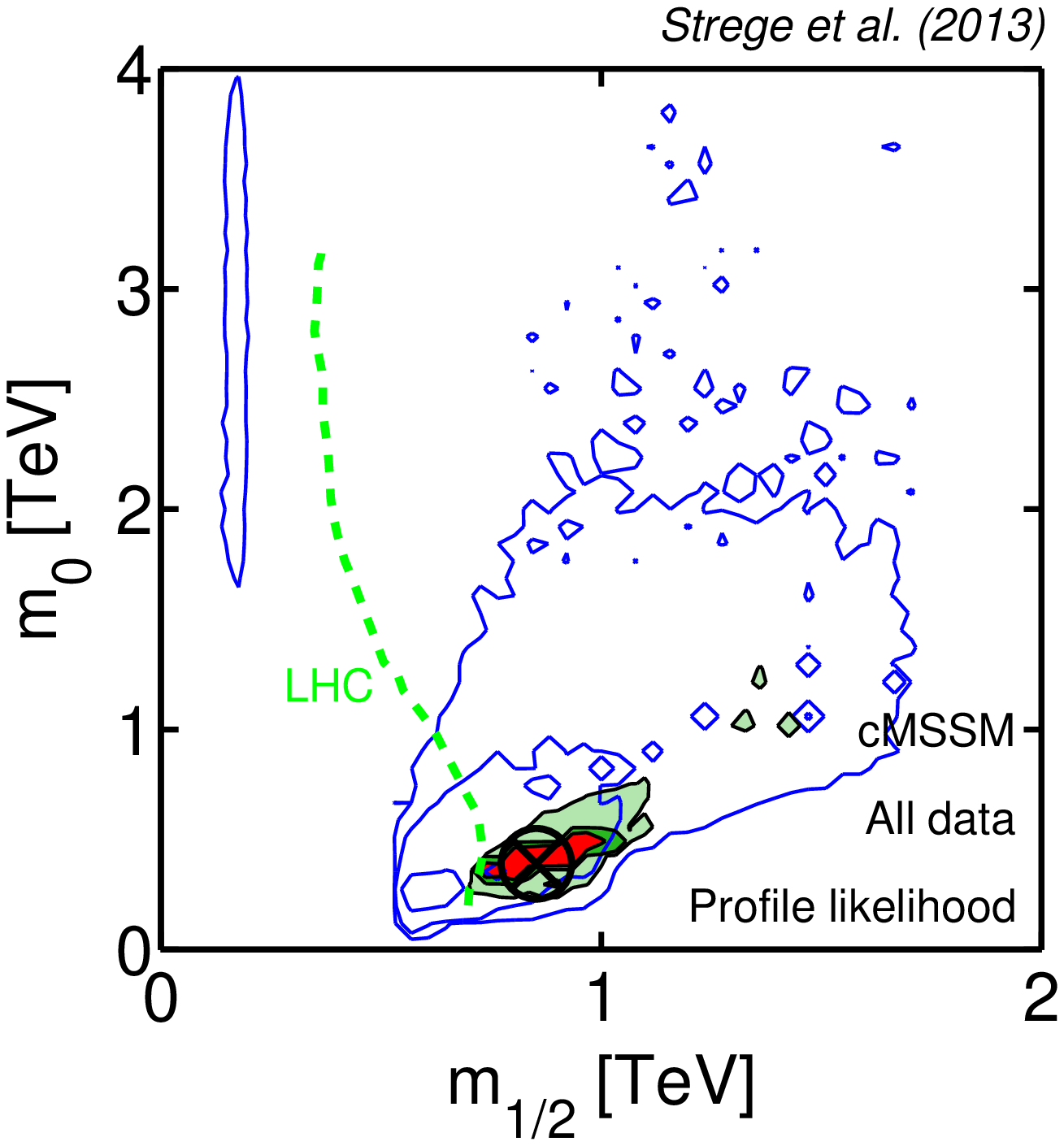}
\expandafter\includegraphics\expandafter[\rowofthree]{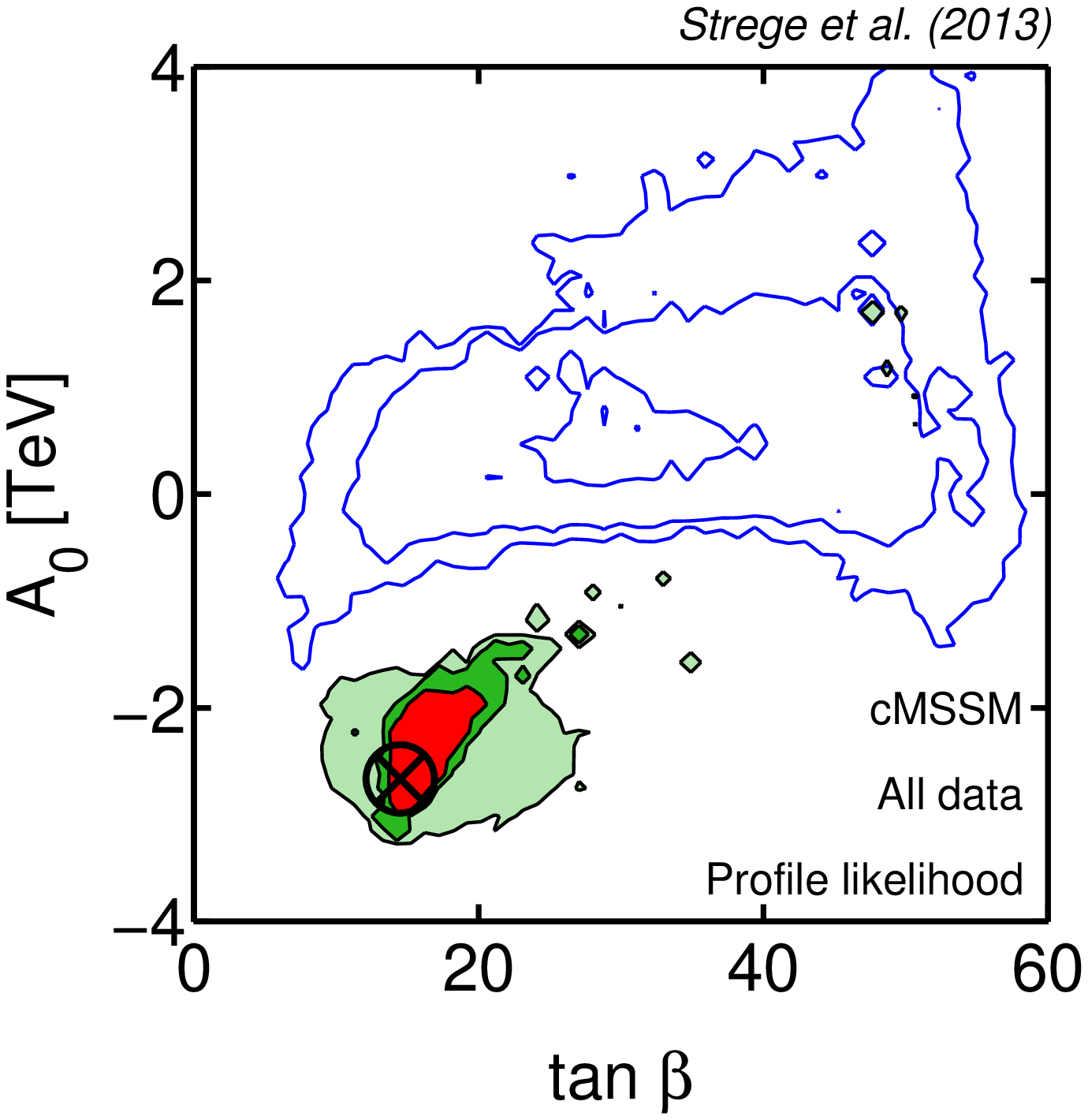}
\expandafter\includegraphics\expandafter[\rowofthree]{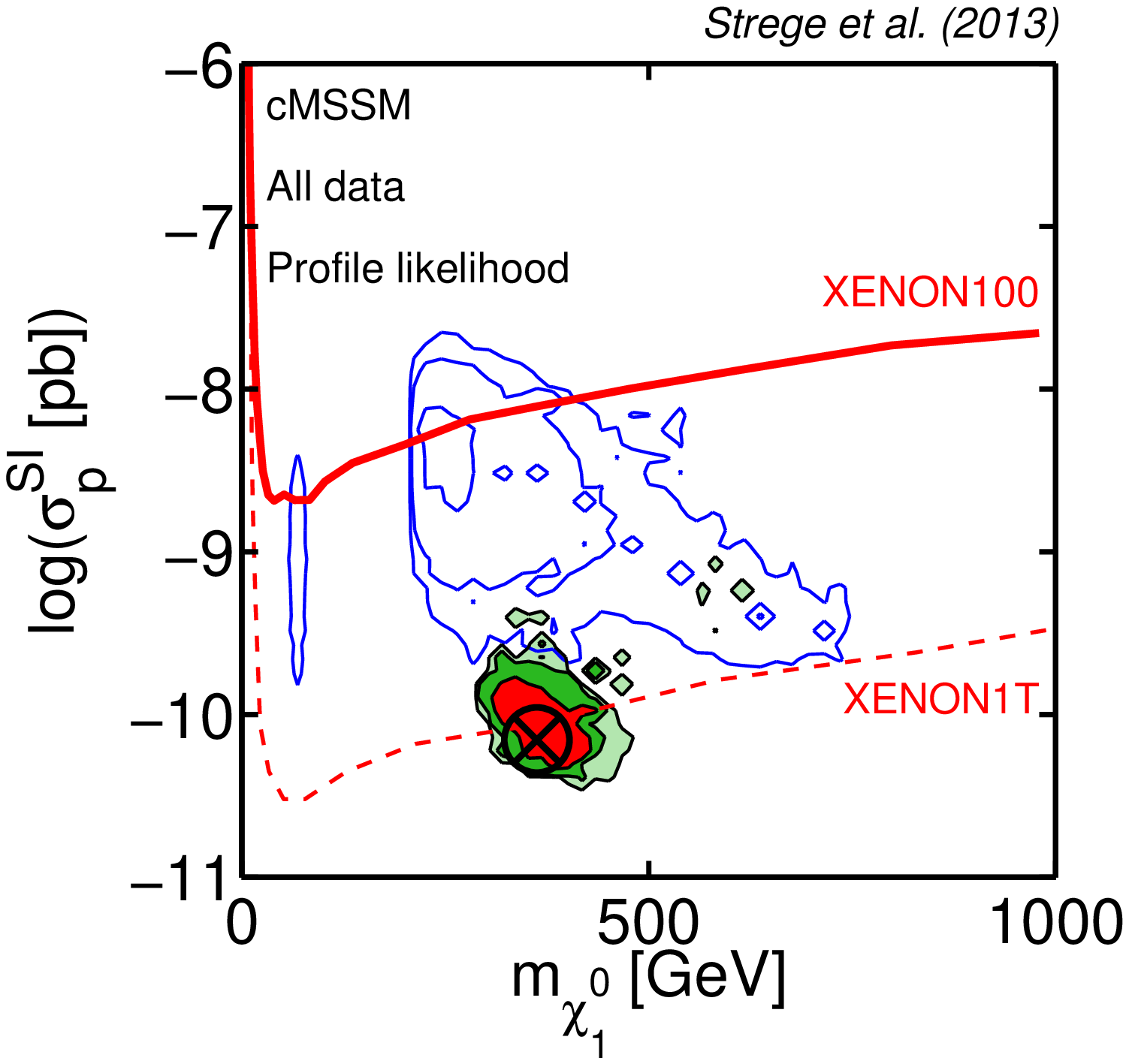} \\
\caption{\fontsize{9}{9} \selectfont Constraints on the cMSSM including all available present-day data (WMAP 7-year, ATLAS \fiveinvfb\ SUSY null search and CMS Higgs mass value, XENON100 2012 direct detection limits, and others -- see Table~\ref{tab:exp_constraints}). Black, filled contours depict the marginalised posterior pdf (top row: flat priors; middle row: log priors) and the profile likelihood (bottom row), showing 68\%, 95\% and 99\% credible/confidence regions. The encircled black cross is the overall best-fit point, obtained from about 350M likelihood evaluations. Blue/empty contours show constraints as of Dec 2011, before the latest LHC and XENON100 results, for comparison (from \cite{Strege:2011pk}). In the left-hand plots, the dashed/green line shows the current LHC 95\% exclusion limit, while in the plots on the right the red/solid line represents the 90\% XENON100 exclusion limit (from Ref.~\cite{Aprile:2012nq}) rescaled to our fiducial astrophysical dark matter distribution. We also show the expected reach of XENON1T as a red/dashed line. \label{cMSSM_2D_alldata}}  
\end{figure*}

We begin by showing in Fig.~\ref{cMSSM_2D_alldata} the combined impact of all present-day experimental constraints, including the ATLAS exclusion limit on the cMSSM mass parameters derived from 5.8 fb$^{-1}$ total integrated luminosity, the latest measurement of the mass of the lightest Higgs boson from CMS and the most recent XENON100 limit on the dark matter parameters, on the cMSSM parameter space. Results are shown in the $(m_{1/2},m_0)$ plane (left), the $(\tan \beta,A_0)$ plane (centre) and the $(m_{\neut},\sigmaSI)$ plane (right). The top row depicts the posterior pdf for flat priors,  the central row the posterior pdf for log priors, and the bottom row the profile likelihood. In each panel, the 68\%, 95\% and 99\% credible/confidence intervals are shown. For comparison, blue/empty contours show the favoured regions from Ref.~\cite{Strege:2011pk}, which included all experimental constraints available in December 2011, previous to the discovery of the Higgs boson and the latest XENON100 results.

\begin{table*}
\begin{center}
%\begin{tabular}{|l|ll| l l}
\begin{tabular}{| l | c | c |}
\hline
\multirow{3}{*}{cMSSM} & \multirow{2}{*}{LHC 2012} & LHC 2012 \\
&  \multirow{2}{*}{+ XENON100} & + XENON100 \\
& & w/o $\delta a_\mu^{SUSY}$ \\\hline 
\multicolumn{3}{|c|}{Best-fit cMSSM parameters}\\\hline
$ m_0$ [GeV] & 389.51 & 321.08\\
$ m_{1/2}$ [GeV] & 853.03 & 839.84\\
$A_0$ [GeV] & -2664.79 & -2163.28\\
$ \tan\beta$ & 14.50 & 13.48\\
\hline 
\multicolumn{3}{|c|}{Best-fit nuisance parameters}\\\hline
$M_t$ [GeV] & 173.539 & 173.779\\
$m_b(m_b)^{\bar{MS}}$ [GeV] & 4.234 & 4.202\\
$[\alpha_{em}(M_Z)^{\bar{MS}}]^{-1}$ & 127.956 & 127.970\\
$\alpha_s(M_Z)^{\bar{MS}}$ & 0.118 & 0.119\\ \hline
$\rho_{\loc}$ [GeV/cm$^3$] & 0.417 & 0.373 \\
$v_{\lsr}$ [km/s] & 224.6 & 228.8 \\
$v_{\esc}$ [km/s] & 549.0 & 518.2 \\
$v_d$ [km/s] & 276.7 & 274.7 \\ \hline
$f_{Tu} \times 10^{2}$ & 2.708 & 2.693 \\
$f_{Td} \times 10^{2}$ & 3.814 & 3.886 \\
$f_{Ts}$ & 0.372 & 0.421 \\ \hline
\multicolumn{3}{|c|}{Best-fit observables}\\\hline
$m_h$ [GeV] & 123.8 & 123.3\\
$m_\neut$ [GeV] & 362.7 & 355.9\\
$\delta a_\mu^{SUSY} \times 10^{10}$ & 25.47 & 3.68 \\ 
$BR(\bar{B} \rightarrow X_s\gamma) \times 10^4$ & 3.11 & 3.17\\
$\DeltaO  \times 10^{2}$ & 8.35 & 5.34\\ 
$\brbsmumu \times 10^{9}$ & 2.94 & 2.94\\
$\Dstaunu \times 10^{2}$ & 5.10 & 5.10\\
$\Dsmunu \times 10^{3}$ & 5.24 & 5.24\\
$\Dmunu \times 10^{4}$ & 3.85 & 3.85\\
$\sigmaSI$ [pb] & $ 7.0 \times 10^{-11}$ & $1.1 \times 10^{-10}$ \\
$\sigmaSD$ [pb] & $ 2.1 \times 10^{-9}$ & $5.5 \times 10^{-9}$ \\
$\Omega_\neut h^2$ & 0.1203 & 0.1105\\ 
\hline
\end{tabular}
\end{center}
\caption{Best-fit model parameters (top section), nuisance parameters (central section) and derived observables (bottom section) in the cMSSM. The column ``LHC 2012+XENON100'' denotes the case where all data, including LHC and XENON100 data, are applied in the analysis; astrophysical and hadronic nuisance parameters have been included in the scan and profiled over. The column ``LHC 2012 + XENON100 w/o $\delta a_\mu^{SUSY}$" is for the case where all constraints except for the $\delta a_\mu^{SUSY}$ constraint are applied. \label{cMSSMbf1}}
\end{table*}

\begin{table*}
\begin{center}
\begin{tabular}{| l | c | c |}
\hline
\multirow{3}{*}{cMSSM} & \multirow{2}{*}{LHC 2012} & LHC 2012 \\
&  \multirow{2}{*}{+ XENON100} & + XENON100 \\
& & w/o $\delta a_\mu^{SUSY}$ \\\hline 
& \multicolumn{2}{c|}{Gaussian constraints} \\
\hline
SM nuisance (4 parameters) & 0.437 & 1.047\\
Astro nuisance (4 parameters)& 0.106 & 0.725\\
Hadronic nuisance (3 parameters)& 0.039 & 0.237\\
$M_W$ & 1.434 & 1.424\\
$\sin^2\theta_{eff}$ & 0.039 & 0.039\\
$\delta a_\mu^{SUSY} $ & 0.153 & N/A \\
$BR(\bar{B} \rightarrow X_s\gamma) $ & 1.251 & 0.933\\
$\Delta M_{B_s}$ & 0.133 & 0.132\\
$\RBtaunu$  & 1.376 & 1.377\\
$\DeltaO$  & 5.219 & 0.946\\
$\RBDtaunuBDenu $ & 0.801 & 0.801\\
$\Rl$ & 0.020 & 0.020\\
$\Dstaunu $ & 0.540 & 0.541\\
$\Dsmunu  $ & 1.466 & 1.466\\
$\Dmunu $  & 0.008 & 0.008\\
$\Omega_\neut h^2$ & 0.500 & 0.001 \\
$m_h$ & 0.925 & 1.413\\
$\brbsmumu$ & 0.028 & 0.030\\
\hline
& \multicolumn{2}{c|}{Exclusion limits} \\\hline
XENON100 & 0.633 & 0.572\\
LHC & 0.0 & 0.0 \\ 
Sparticles (LEP) & 0.0 & 0.0 \\ \hline
Total $\chi^2$ & 15.11 & 11.71\\
Total Gaussian $\chi^2$ (dof) & 14.48 (11) & 11.14 (10) \\
Gaussian $\chi^2/$dof & 1.32 & 1.11 \\
p-value (Gaussian constraints only) & 0.21 & 0.35  \\
\hline
\end{tabular}
\end{center}
\caption{Breakdown of the total best-fit $\chi^2$ by observable for the cMSSM, for both the case where all data, including LHC 2012 constraints and XENON100 data, were applied, and nuisance parameters were profiled over (left column), and when all data except the $\delta a_\mu^{SUSY}$ constraint were applied (right column).  \label{cMSSMbf2}}
\end{table*}

\subsection{Impact of ATLAS exclusion limit}

The latest  ATLAS exclusion limit cuts further into the low-mass regions of the cMSSM, now excluding the entire $h$-pole region which was previously viable and is visible as a vertical region of empty/blue contours at small $m_{1/2}$ in the $(m_{1/2}, m_0)$ plane.  Additionally, contours are further pushed towards higher values of $m_{1/2}$. Focusing on the $(m_{1/2}, m_0)$ plane, we observe that the posterior pdf for the log prior exhibits a bimodal nature, with two connected favoured regions, one corresponding to the A-funnel (AF) region at high masses, and one to the stau-coannihilation (SC) region, corresponding to $m_{1/2} \sim 800-1000$ GeV and $m_0 \sim 300-400$. 
 In the SC region the lightest stau is only slightly heavier than the neutralino LSP. Therefore,  in this region the relic density of the neutralino is reduced by neutralino-stau coannihilations in the early universe, in agreement with the WMAP constraint. The AF region is characterised by a relatively light pseudoscalar Higgs, with $m_{\neut} \approx 2m_{A^0}$, which can mediate resonant annihilations of the neutralino LSP, making it easier to satisfy the WMAP relic density constraint.

The SC region is also where our overall best-fit point is located (see below for further discussion). In contrast, the SC region is not favoured in the flat prior scan. The flat prior gives a much larger statistical {\it a priori} weight to regions at large values of the mass parameters, so that the corresponding posterior pdf is strongly affected by volume effects at high masses, and therefore peaks there. The log prior scan explores the low mass regions in much more detail (see~\cite{Trotta:2008bp} for a detailed discussion), so that the posterior distribution for log priors also favours the SC region at small values of $m_0$ and $m_{1/2}$, still allowed by the LHC exclusion limit.

The profile likelihood function (bottom left panel) favours the SC region around the best-fit point, and is much more strongly localised than the Bayesian pdf. Small scalar and gaugino masses are strongly favoured, values of $m_0 > 1$ TeV are disfavoured at $99\%$ confidence level. Compared to the results before inclusion of the Higgs mass measurement (blue/empty contours), the profile likelihood contours are confined to a much smaller region. This is the result of two effects: pre-Higgs constraints favoured relatively low Higgs masses, with a best-fit value $m_h = 115.6$ GeV (see~\cite{Strege:2011pk}). In the region now favoured by the profile likelihood the constraint on $m_h$ can be fulfilled due to the maximal mixing scenario (see below). Additionally, small masses are strongly favoured by the constraint on the anomalous magnetic moment of the muon, which can only be satisfied in the SC region, since in the AF region the value of \gmt\ tends to 0, producing a $\sim 3\sigma$ discrepancy with the data. Largely due to these two constraints, the profile likelihood function favours a small region at low $m_0$, $m_{1/2}$ that achieves high likelihood values. The importance of the \gmt\ constraint in confining the profile likelihood function to small masses will be discussed in detail in section~\ref{sec:cMSSM_wogm2}.

\subsection{Impact of latest Higgs mass measurement} \label{sec:cMSSM_Higgs}

The new measurement of the Higgs mass has a significant impact on the cMSSM parameter space. For both choices of priors, large regions of parameter space previously favoured at $68\%$ marginal posterior probability are ruled out by this constraint. Contours are pushed towards larger values of $m_{1/2}$, the posterior pdf with flat priors excludes values of $m_{1/2} < 1$ TeV at 99$\%$ level. Both posterior distributions now favour much larger values of $m_0$, with the 68\% (95\%) contour touching the prior boundary for flat (log) priors.  The strong impact of the LHC Higgs constraint on this parameter space is expected, since in the cMSSM smaller values of $m_h$ are otherwise favoured~\cite{Strege:2011pk}. Larger values of  $m_h$ can be achieved by radiative corrections. At one-loop level $m_h$ is most sensitive to the stop mass, which is mainly determined by the value of $m_{1/2}$, so that large $m_{1/2}$ can lead to large values of $m_h$, satisfying the experimental constraint. The value of $m_0$ is much less important for $m_{\tilde{t}_{1,2}}$, so that relatively low values of $m_0$ are still allowed by the new Higgs constraint. 

\begin{figure*}%[htp]
%\expandafter\includegraphics\expandafter
%[\rowofthree]{Plots/cMSSM/all/cMSSM_flat_pp_2D_4_blue_sq.eps}
%\expandafter\includegraphics\expandafter
%[\rowofthree]{Plots/cMSSM/all/cMSSM_log_pp_2D_4_blue_sq.eps}
%\expandafter\includegraphics\expandafter
%[\rowofthree]{Plots/cMSSM/all/cMSSM_PL_pp_2D_4_blue_sq.eps} \\
\expandafter\includegraphics\expandafter
[\rowofthree]{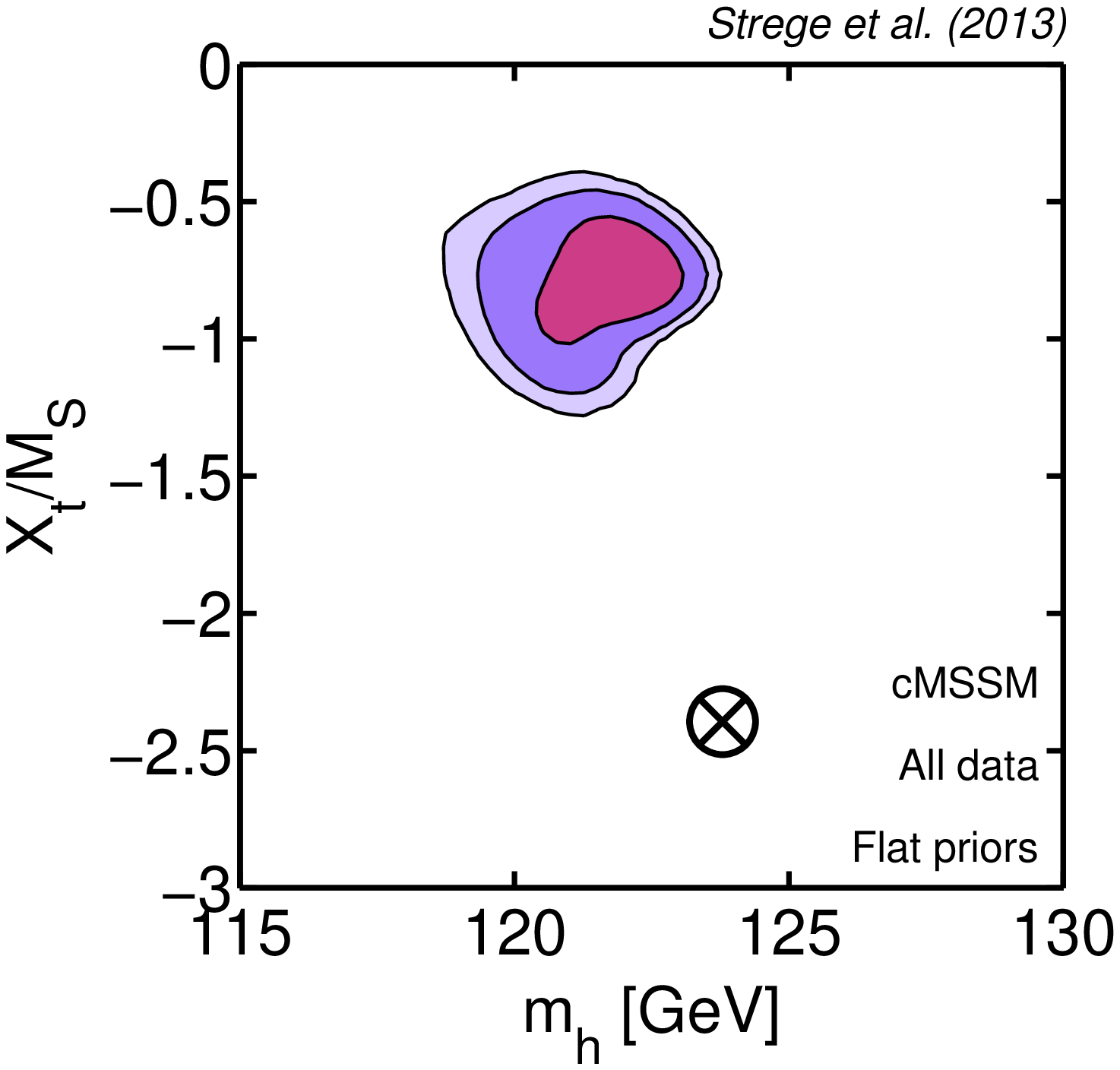}
\expandafter\includegraphics\expandafter
[\rowofthree]{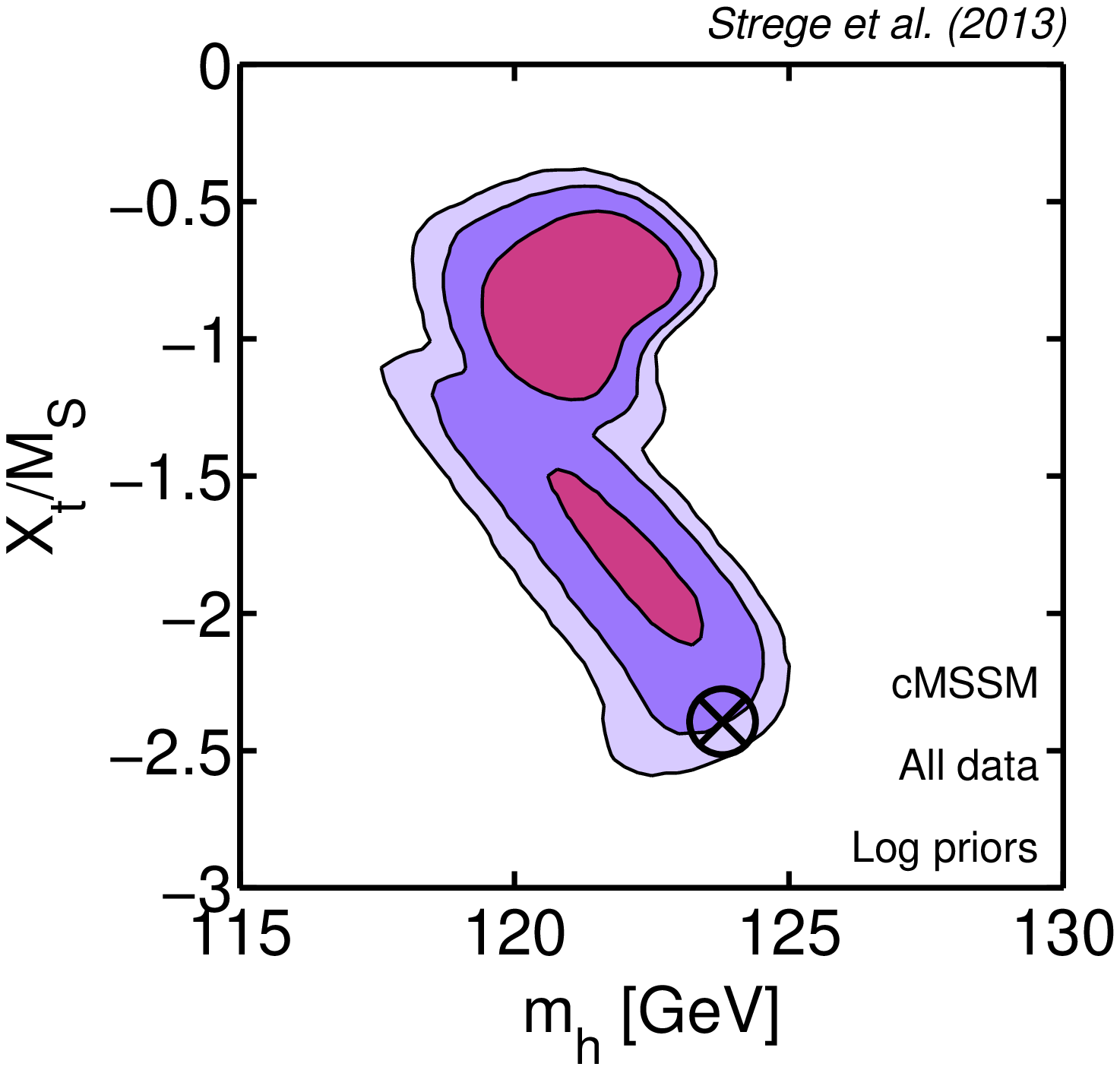}
\expandafter\includegraphics\expandafter
[\rowofthree]{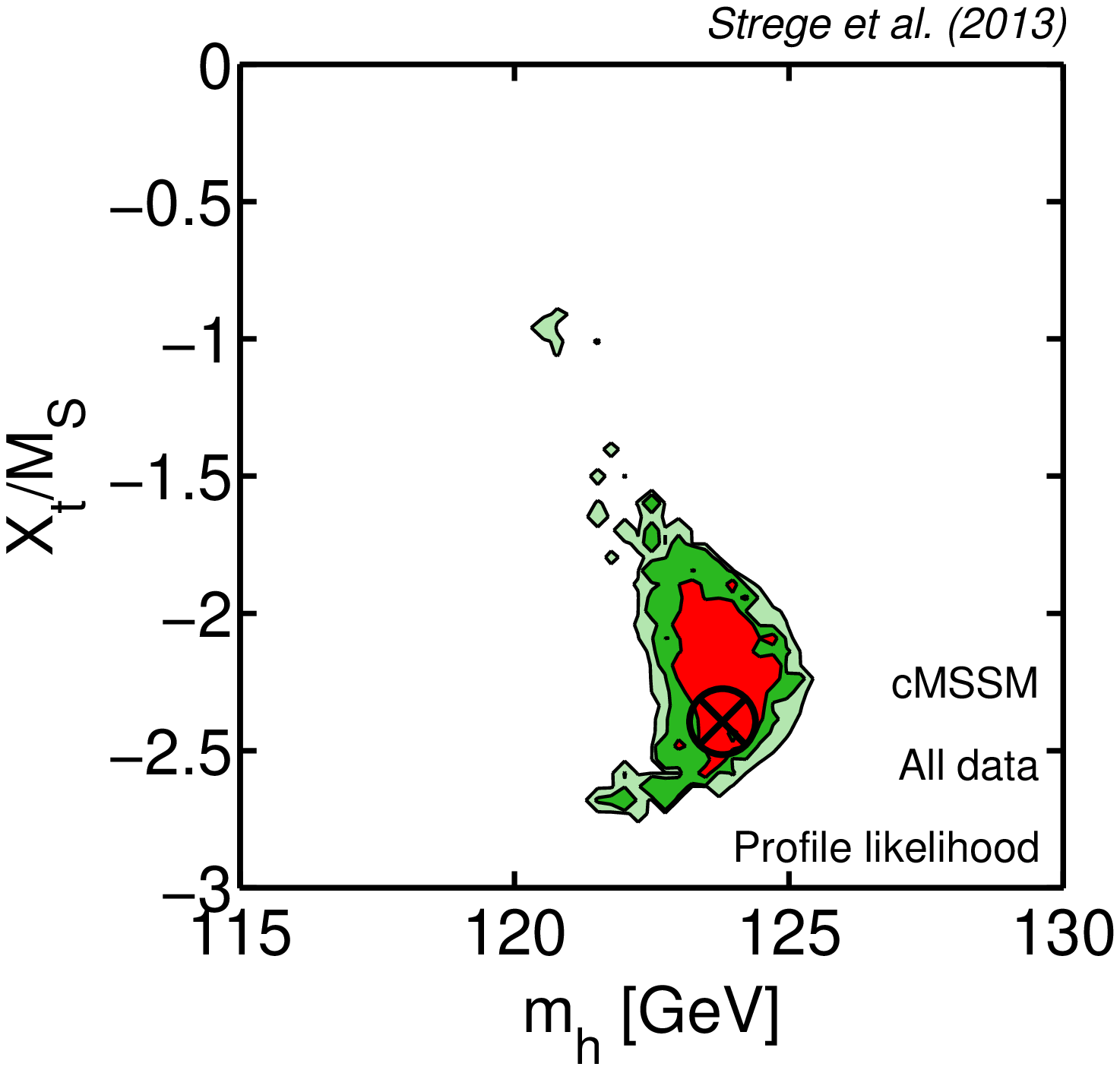} 
\caption{\fontsize{9}{9} \selectfont Favoured regions in the cMSSM in the $X_t/M_S$ vs $m_h$ plane (all data included; from left to right: posterior pdf with flat and log priors, and profile likelihood).  The maximal mixing scenario ($X_t/M_S \approx 2.44$) is realised in the stau co-annihilation region (where the best-fit point is located, encircled black cross), while it can not be achieved in the A-funnel region (where $M_S$ is larger and $|X_t/M_S|$ is reduced).\label{cMSSM_2D_Xt}}  
\end{figure*}

A second possibility to achieve $m_h \approx 125$ GeV is the so-called maximal mixing scenario. If the stop mixing parameter $X_t$ approaches a value $\sqrt{6} M_S$, with $M_S^2 = 0.5(m_{\tilde{t}_{1}}^2 + m_{\tilde{t}_{2}}^2)$, the contribution of the stop to $m_h$ is maximised, and hence the Higgs mass increases (see for instance \cite{Djouadi:2005gj}). This is illustrated in Fig.~\ref{cMSSM_2D_Xt}, where we plot the ratio $X_t/M_S$ vs $m_h$, showing that indeed the highest Higgs mass values are found in the maximal mixing region, where $|X_t/M_S| \approx \sqrt{6} \approx 2.45.$ 
In the cMSSM, this effect can be achieved in the low-mass SC region, but is very difficult to achieve for large $m_{1/2}$~\cite{Brummer:2012ns}. Therefore, the posterior pdf with flat priors (left-hand panel of Fig.~\ref{cMSSM_2D_Xt}), which favours the AF region at larger values of $m_{1/2}$, is concentrated at moderate values of $X_t/M_S$. A relatively large Higgs mass of $m_h \sim 122$ GeV can still be achieved in this region, due to the large stop masses. Larger values of $m_h$, in agreement with the experimental constraint, would require very large values of the stop masses. However, the AF region can only be achieved at intermediate values of $m_{1/2}$ that are not large enough to result in very high stop masses leading to $m_h \approx 126$ GeV. 

In contrast, the posterior with log priors (central panel) and the profile likelihood function (right-hand panel) favour much larger values of $X_t/M_S$, corresponding to the mode at small $m_{1/2}$, and a small number of fine-tuned points exist for which the maximal mixing scenario is realised. Nevertheless, without an additional contribution from large stop masses, it is difficult to achieve the measured value of the Higgs mass, and our best-fit value is $m_h = 123.8$ GeV,  which is compatible with the experimental constraint ($m_h = 125.8 \pm 0.6$ GeV) at the $\sim 1 \sigma$ level only due to the inclusion of a theoretical error of 2 GeV in the likelihood. 

Larger values of $m_h$ could in principle be achieved in the Focus Point region, however, inside our prior range for $m_0$ this region is strongly disfavoured by the XENON100 constraint. Therefore, a value of $m_h \approx 126$ GeV cannot be achieved in the cMSSM within the prior ranges adopted for this work\footnote{For an analysis of the impact of the measurement of the Higgs mass on the cMSSM at masses outside our prior ranges see e.g.\ Ref.~\cite{Akula:2012kk}.}.

%\rr{As I mentioned above the DM constraint drives the bulk of the pdf to the AF region enterely for flat priors and partially to log priors therefore large tanb values are favoured.}
A similar pattern as in the  $(m_{1/2},m_0)$ plane is observed in the $(\tan \beta,A_0)$ plane (central panels in Fig.~\ref{cMSSM_2D_alldata}). Previously favoured regions shrink significantly due to inclusion of the new Higgs constraint. The posterior pdf with flat priors spans a large range of $A_0$ values, with a preference for positive $A_0$. Large values of $\tan \beta$ are favoured, as required for the AF region. As mentioned above, this region however corresponds to slightly lower Higgs masses ($m_h \sim 120-122$ GeV), and is hence disfavoured by the Higgs constraint. The posterior with log priors shows a bimodal shape, with the mode at low $\tan \beta$ corresponding to the region at small $m_{1/2}$ in the left-hand panel. Compared to our previous profile likelihood results~\cite{Strege:2011pk}, we observe a strong shift of the favoured region in the  $(\tan \beta,A_0)$ plane, with negative $A_0$ now favoured. This is a consequence of the new Higgs constraint, forcing the best-fit point to a region of maximal mixing. This is despite the constraint on the isospin asymmetry $\DeltaO$ disfavouring negative $A_0$ values~\cite{Ahmady:2006yr}.

\subsection{Best-fit point}

The coordinates of the best-fit point are given in Table~\ref{cMSSMbf1} for the input cMSSM and nuisance parameters, as well as for some notable derived quantities. Compared to our previous best-fit (in Ref. \cite{Strege:2011pk}) we observe an upward shift of $\sim 100-200$ GeV in the mass parameters and a strong shift to negative $A_0$ (as explained above), while low $\tan\beta$ remains favoured.

The overall best-fit $\chi^2$, broken down in terms of the contribution of each observable, is given in Table~\ref{cMSSMbf2}. As can be seen, by far the largest contribution to the best-fit $\chi^2$ results from the isospin asymmetry $\DeltaO$. The SM prediction of this quantity is already in tension with the experimental measurement at $\sim 2\sigma$~\cite{Ahmady:2006yr}. Any positive SUSY contribution to $\Delta_{0-}$ will therefore further worsen the fit. The SUSY contribution is minimised at positive values of $A_0$, small $\tan \beta$ and large $m_{1/2}$~\cite{Ahmady:2006yr}. This preference is in tension with other constraints, most importantly the Higgs mass measurement, which favours strongly negative $A_0$. This leads to an additional SUSY contribution to $\Delta_{0-}$, which further increases the contribution of this observable to the total $\chi^2$. Other contributions to the overall best-fit $\chi^2$ are much smaller; in particular, the best-fit point simultaneously satisfies the constraint on the Higgs mass, the exclusion limit from the XENON100 direct detection experiment, the relic density constraint and the constraint on the anomalous magnetic moment of the muon. The value of $\brbsmumu$ is also in very good agreement with the new LHCb constraint on this quantity. 

When evaluating the p-value for the best-fit point, we only consider contributions to the $\chi^2$ from Gaussian-distributed observables in the likelihood. This allows us to compute the (approximate) p-value analytically from the corresponding chi-squared distribution with the number of degrees of freedom (dof) given by the number of Gaussian data points minus the number of free parameters (4 cMSSM model parameter, 4 SM nuisance parameters, 4 astrophysical nuisance parameters and 3 hadronic nuisance parameters, giving a total of 15 parameters). We find that the p-value for all data sets combined is 0.21. Therefore, even a strongly constrained model such as the cMSSM is not ruled out at any meaningful confidence level by the latest experimental data sets. 

\subsection{Implications for direct detection and future SUSY searches} 

The implications for the spin-independent scattering cross-section and neutralino mass are displayed in the right-most column of Fig.~\ref{cMSSM_2D_alldata}.  As a consequence of the new Higgs measurement, the favoured region in the $(m_{\neut},\sigmaSI)$ plane is shifted towards larger neutralino masses and much lower spin-independent cross-sections, especially from the profile likelihood statistical perspective. The best-fit point corresponds to a very small SI scattering cross-section of $\sigmaSI = 7\times 10^{-11}$ pb, which is challenging to explore even with future ton-scale direct detection experiments.  For comparison, the expected 90$\%$ exclusion limit from the future XENON1T direct detection experiment is indicated on the plots in the right-hand panels. Therefore, the discovery of a Higgs boson with $m_h \approx 126$ GeV renders direct detection of the cMSSM more difficult.  However, the posterior pdf for both choices of priors displays a large island of probability density around $m_{\neut} \sim 600$ GeV and for SI scattering cross-sections $\sigmaSI \sim 10^{-9} - 10^{-8}$, corresponding to the A-funnel region in the $(m_{1/2}, m_0)$ plane. This region can be fully probed by future ton-scale direct detection experiments. Therefore, from a Bayesian statistical perspective, significant regions of the cMSSM parameter space currently favoured at the 95\% level will be within reach of the next generation of direct detection experiments. This result is obtained also from a frequentist perspective if the \gmt\ constraint is excluded from the analysis, see section~\ref{sec:cMSSM_wogm2} below. 

The favoured values of the spin-dependent scattering cross-section in the cMSSM are confined to the range $\sigmaSD \in [10^{-9} , 10^{-6}]$ pb, with the best-fit point located at $\sigmaSD \approx 10^{-9}$ pb, and hence outside the reach of even future multion-scale direct detection experiments such as DARWIN~\cite{Baudis:2012bc}. Detection prospects for neutrino telescopes are similarly pessimistic. 

We do not study in detail the impact of indirect detection experiments, such as the Fermi Large Area Telescope. Current limits on annihilating dark matter from observations of dwarf spheroidals only significantly constrain low-mass dark matter $m_{\neut} < 30$ GeV~\cite{Fermidwarf}, that is not realised in the cMSSM. This may change as more data become available: 10 years of Fermi observations of dwarf spheroidals have the capability to exclude WIMPs with masses $m_{\neut} < 700$ GeV that have a thermal annihilation cross-section$ <\sigma v> = 3 \times 10^{-26}$ cm$^3$/s (assuming 30 dwarf spheroidals~\cite{Fermidwarf})\footnote{While a thermal self-annihilation cross-section can be realised in several regions of cMSSM parameter space, in some regions much lower values are favoured. For example, in the SC region the neutralino annihilation cross-section can be much lower than the thermal value, since the relic density is reduced by co-annihilations in the early universe.}. Therefore, future Fermi data will have a powerful impact on simple SUSY parameter spaces such as the cMSSM.

% \rt{more comments on SD plane once figures finished. Do we want to comment on ID, as well? Gf?}. \rr{We could plot the IC with 79 strings exclusion in the SD-mchi plane using the best 
%scenario which relies on the W+W- annihilation channel. Then one can conclude that IC is far 
%away from future DD experiments in terms of proving/rejecting the model since the 79 strings result 
%is going to be similar to the final DC configuration}

\begin{figure*}%[htp]
%\centering
\expandafter\includegraphics\expandafter[\rowoffour]{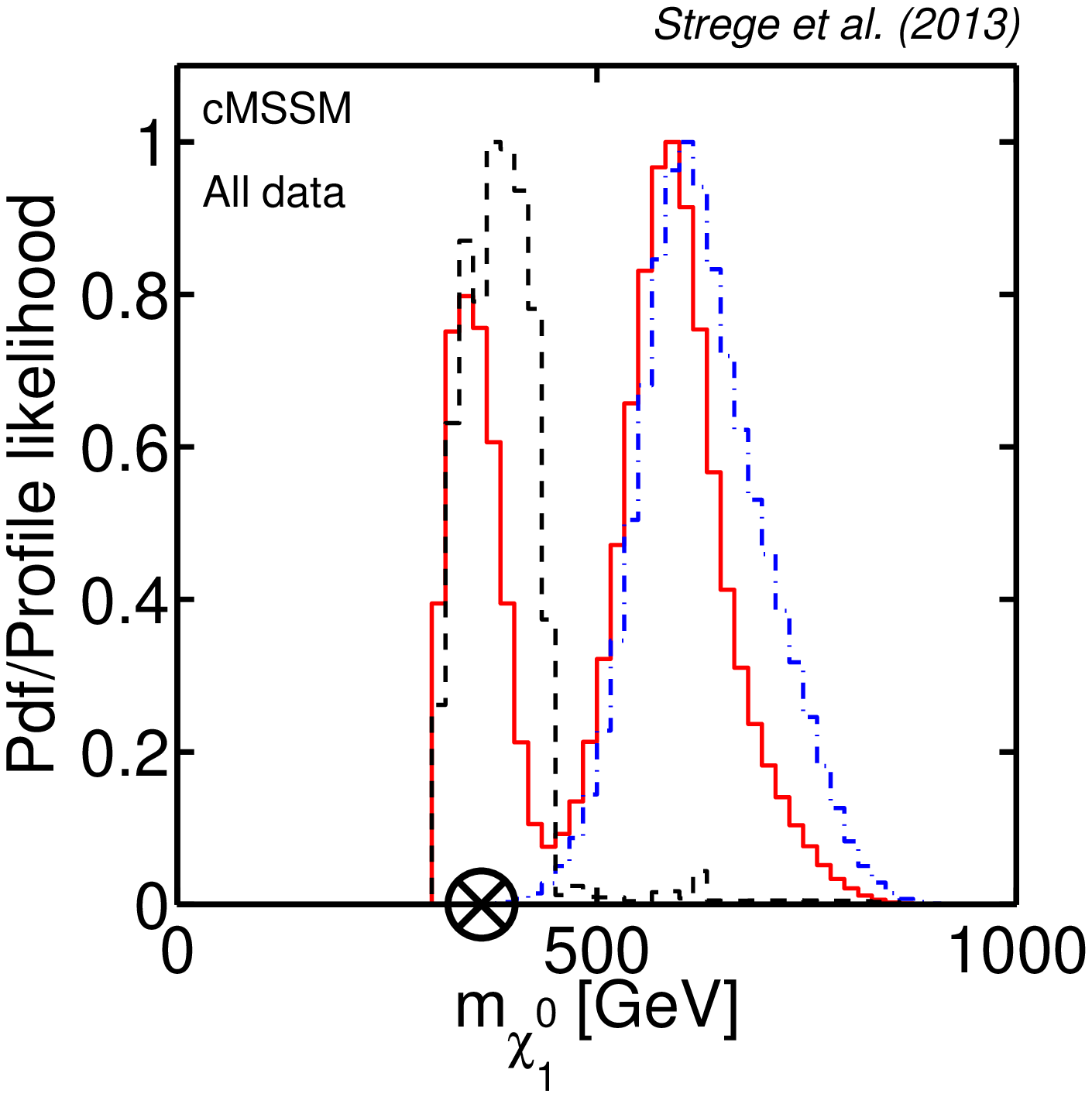}
\expandafter\includegraphics\expandafter[\rowoffour]{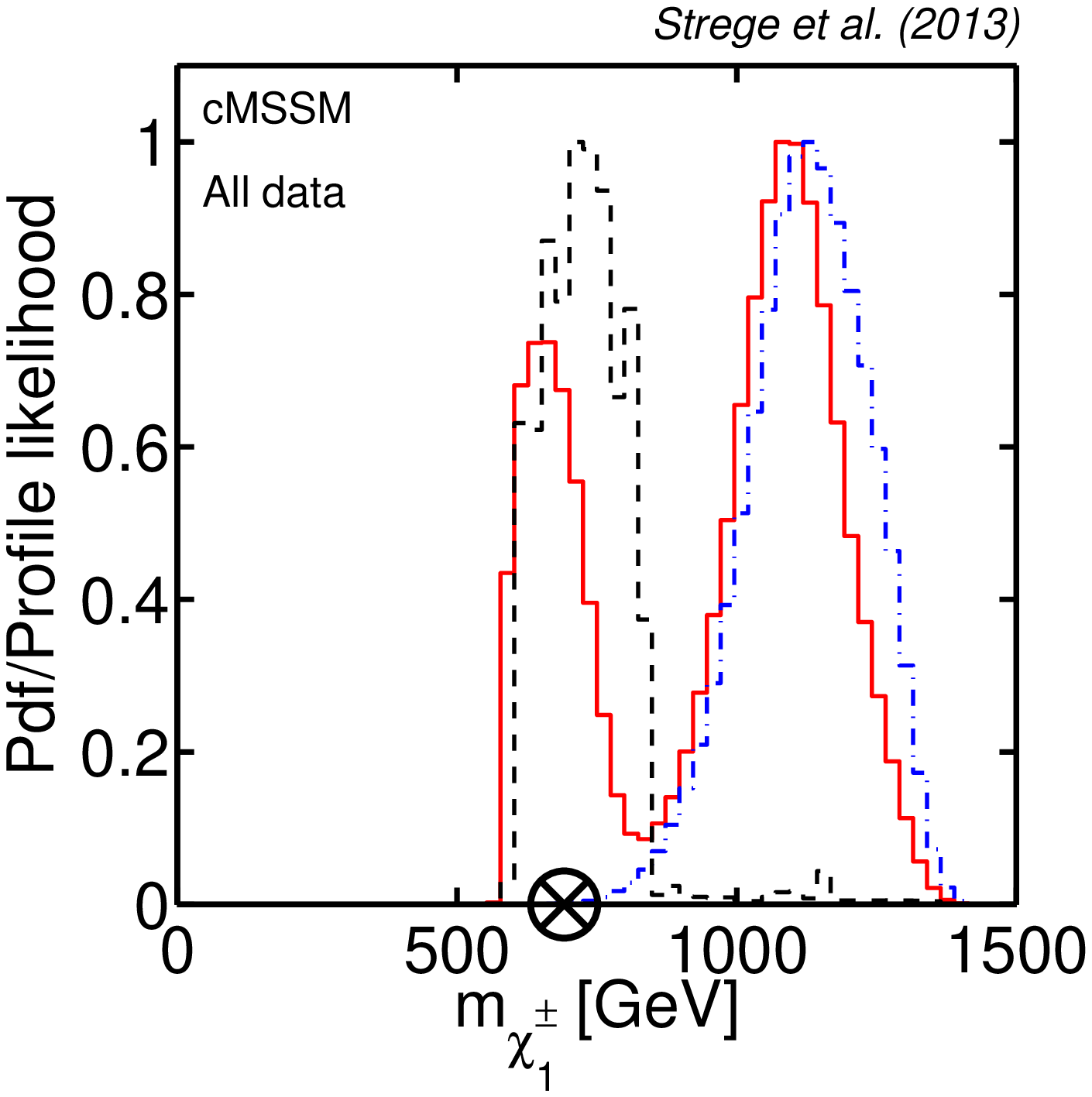} 
\expandafter\includegraphics\expandafter[\rowoffour]{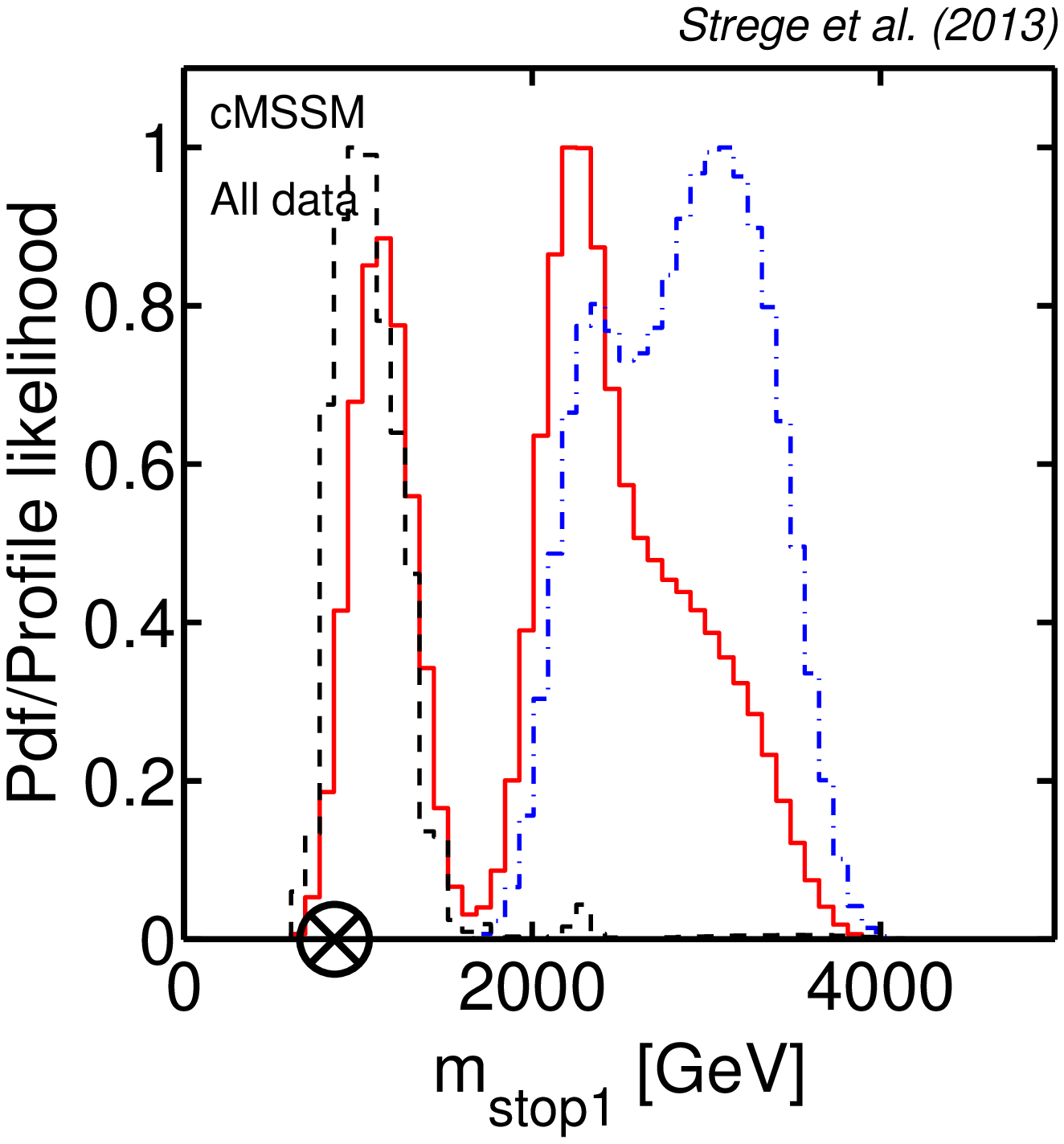}
\expandafter\includegraphics\expandafter[\rowoffour]{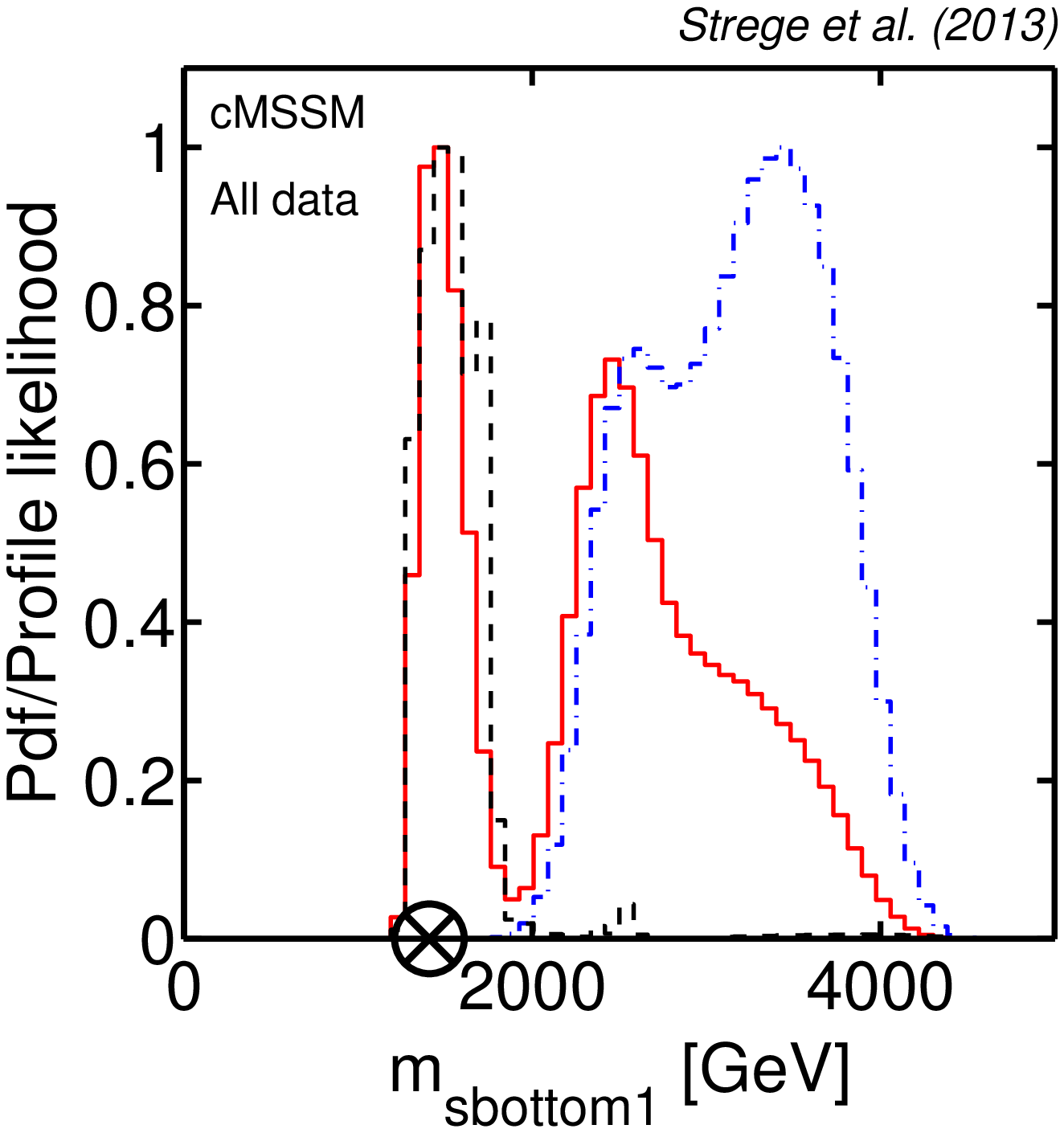}\\
\expandafter\includegraphics\expandafter[\rowoffour]{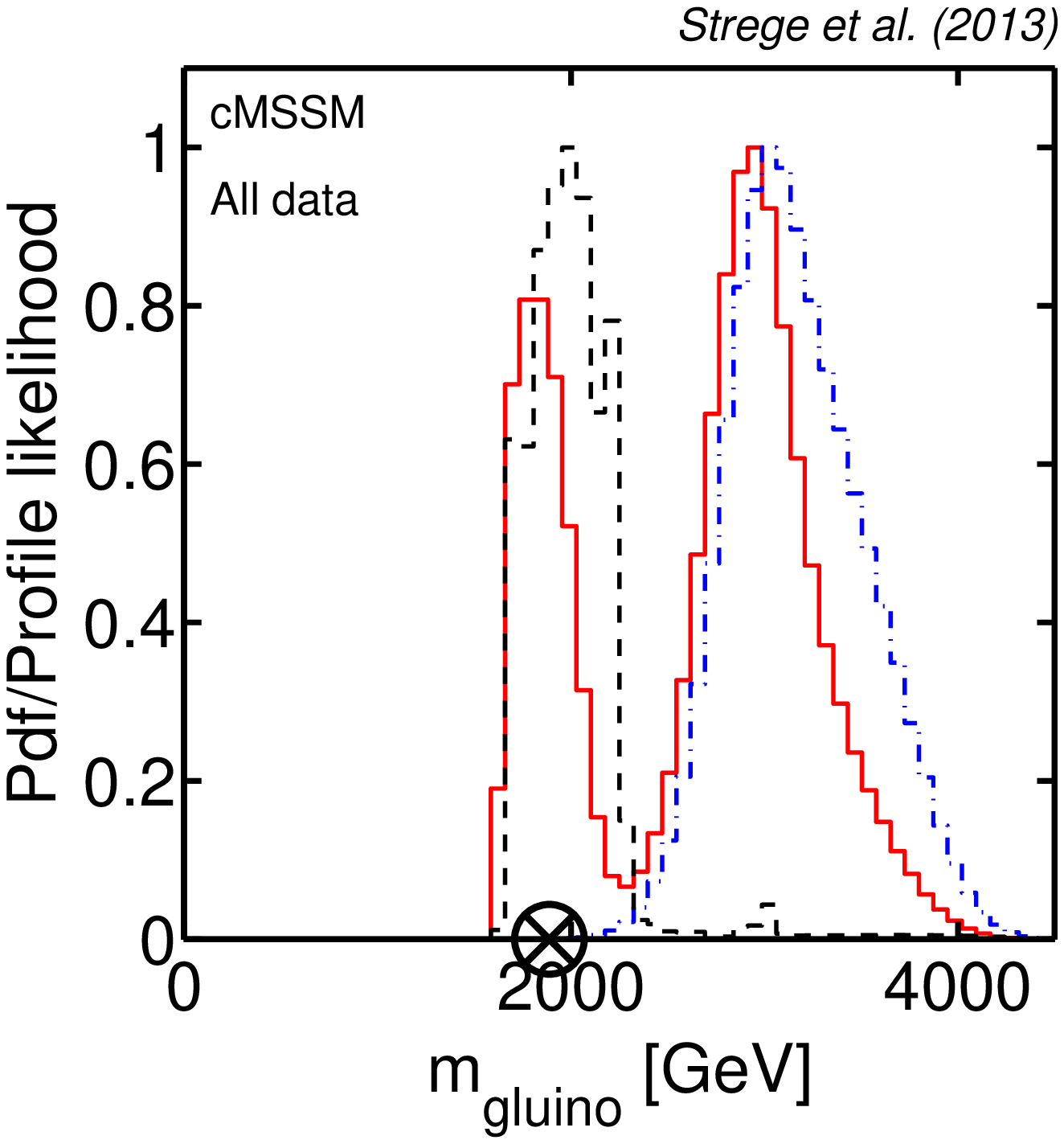} 
\expandafter\includegraphics\expandafter[\rowoffour]{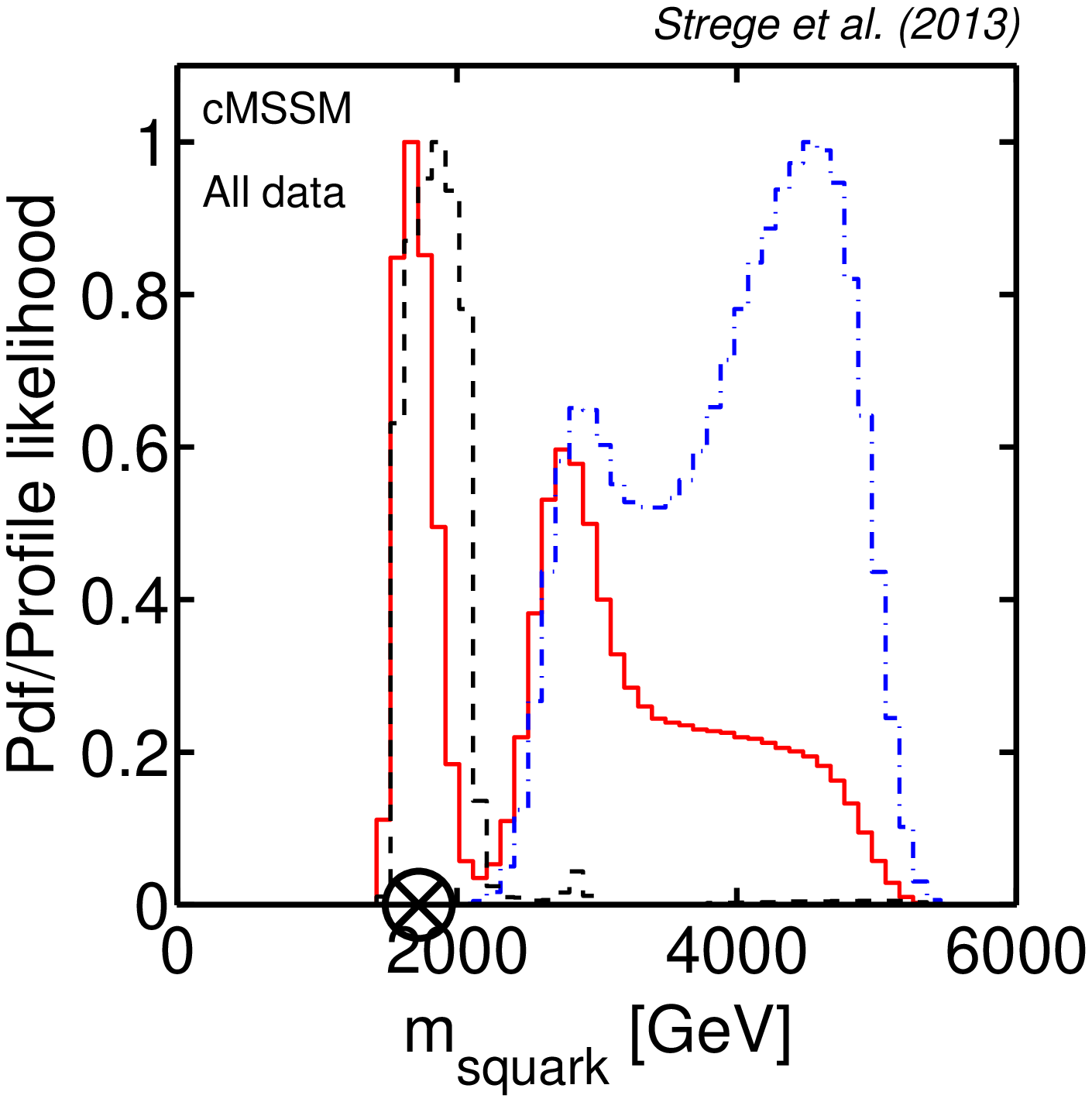}
\expandafter\includegraphics\expandafter[\rowoffour]{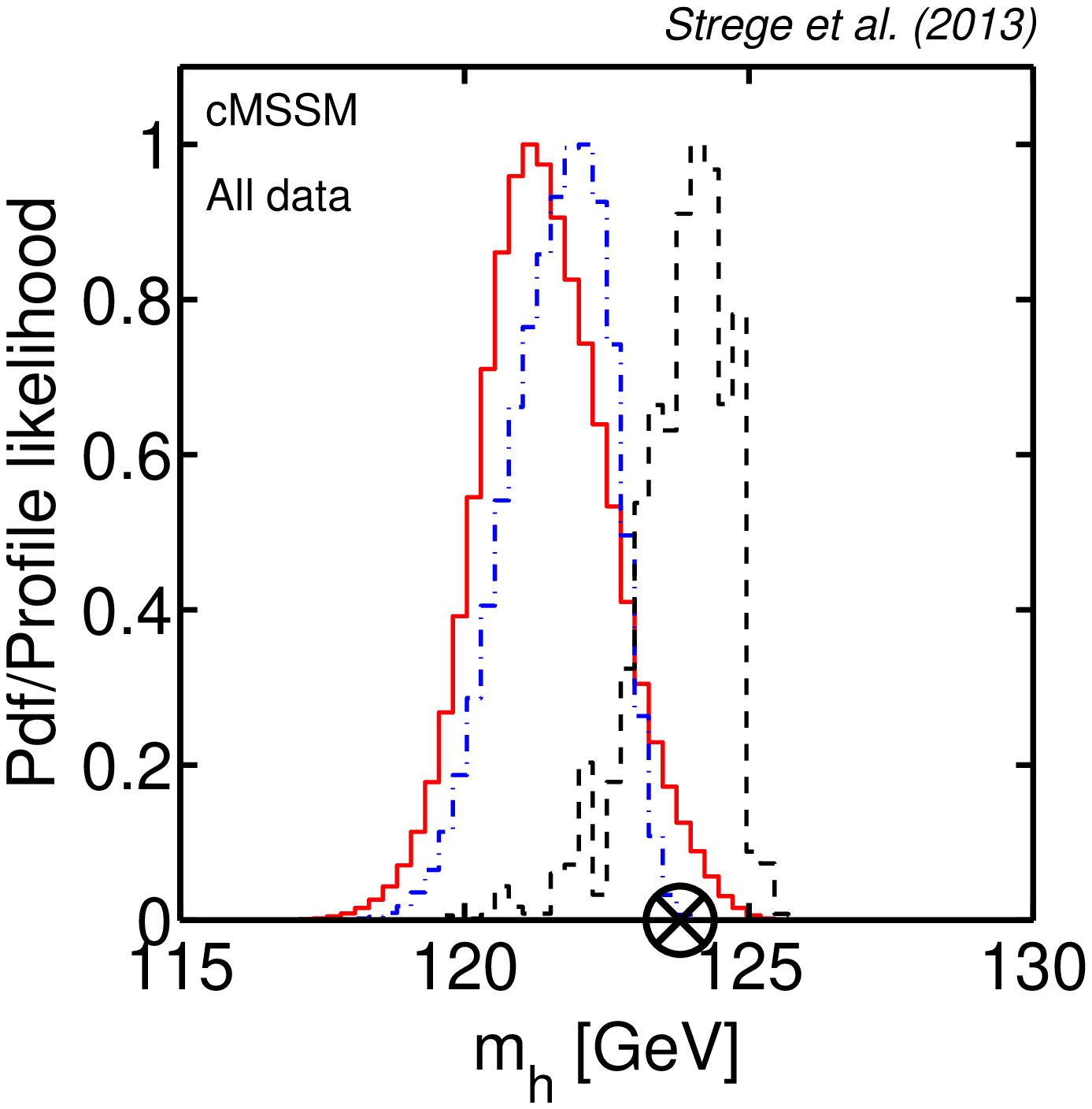} 
\expandafter\includegraphics\expandafter[\rowoffour]{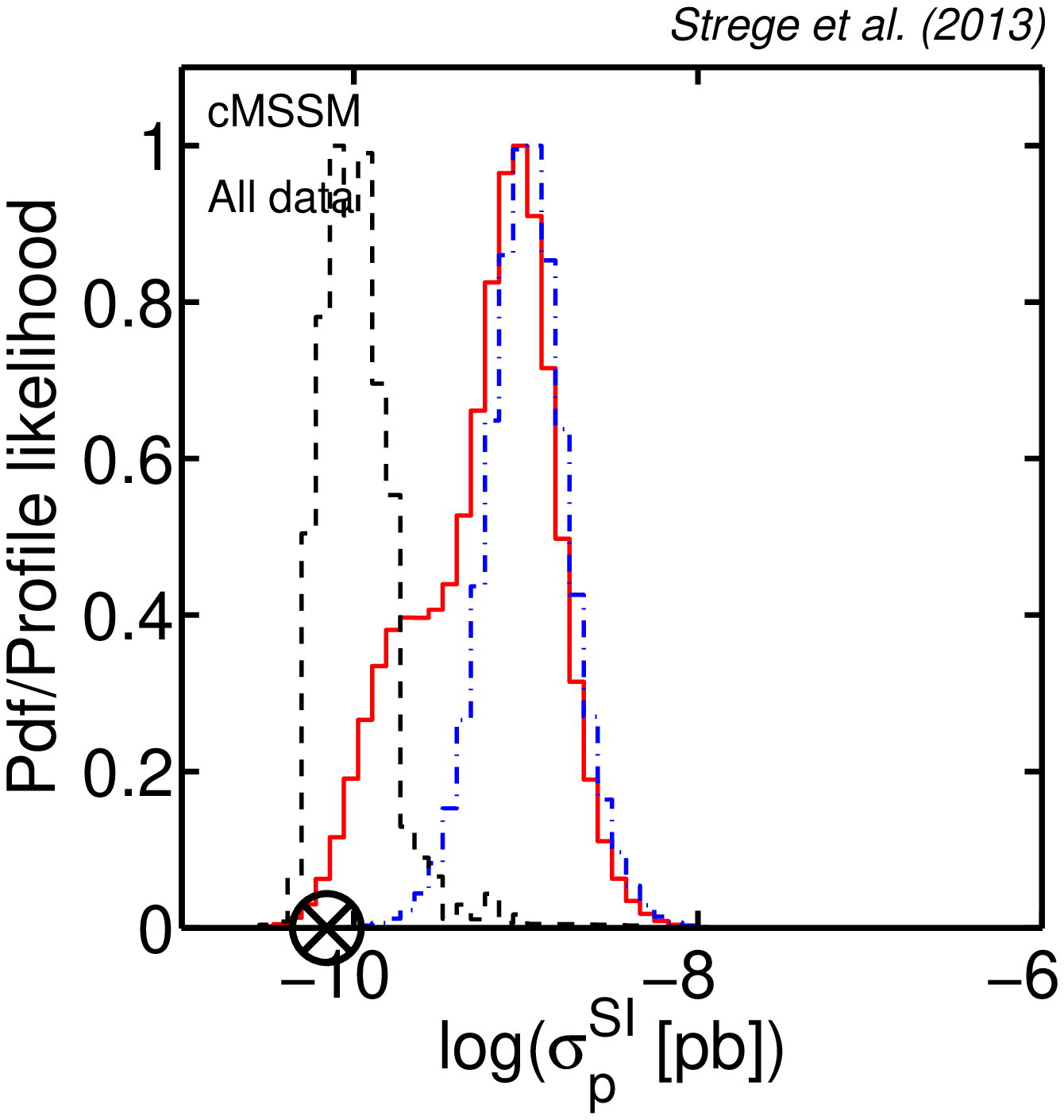}\\  
\expandafter\includegraphics\expandafter[\rowoffour]{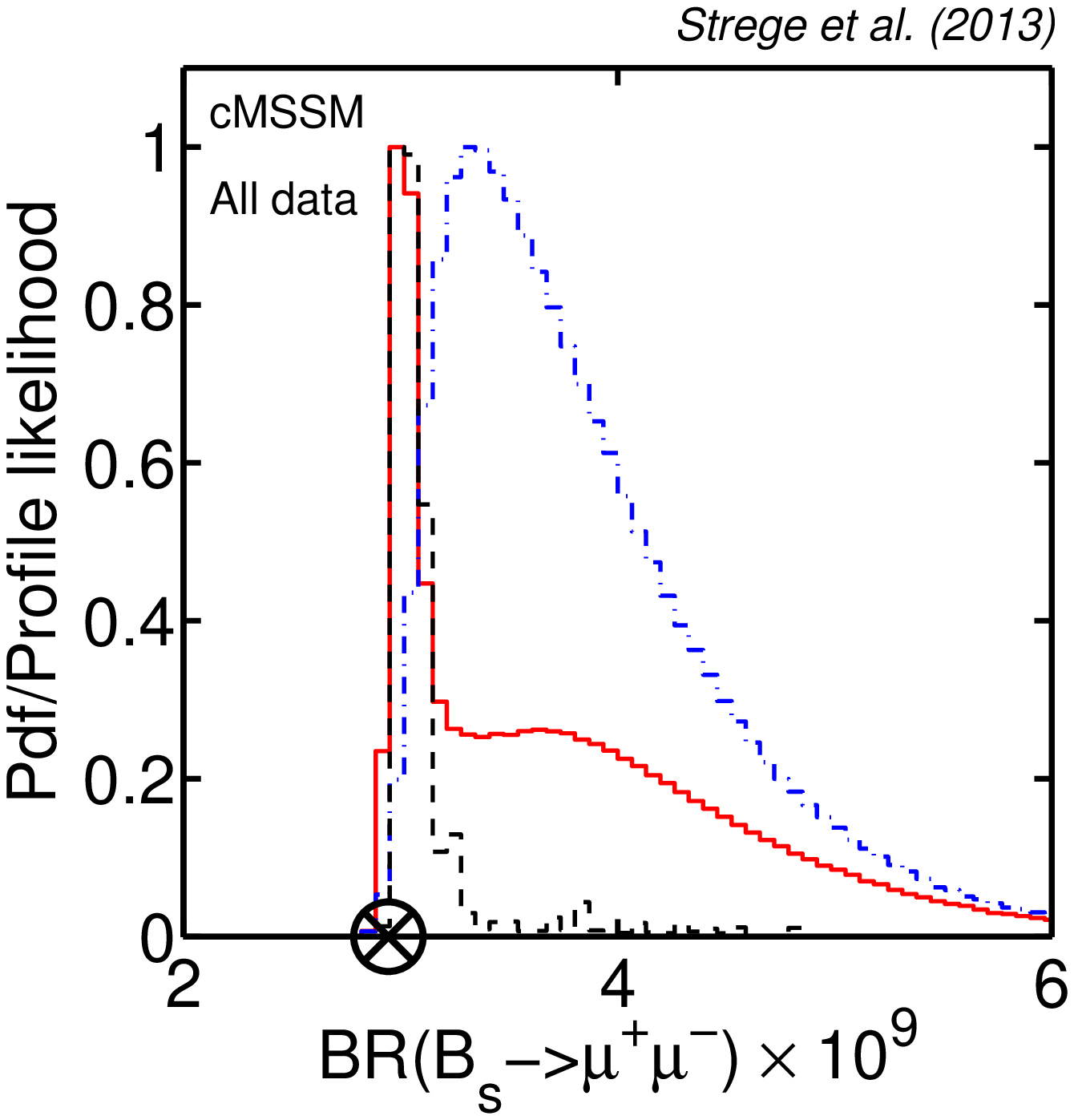}
\expandafter\includegraphics\expandafter[\rowoffour]{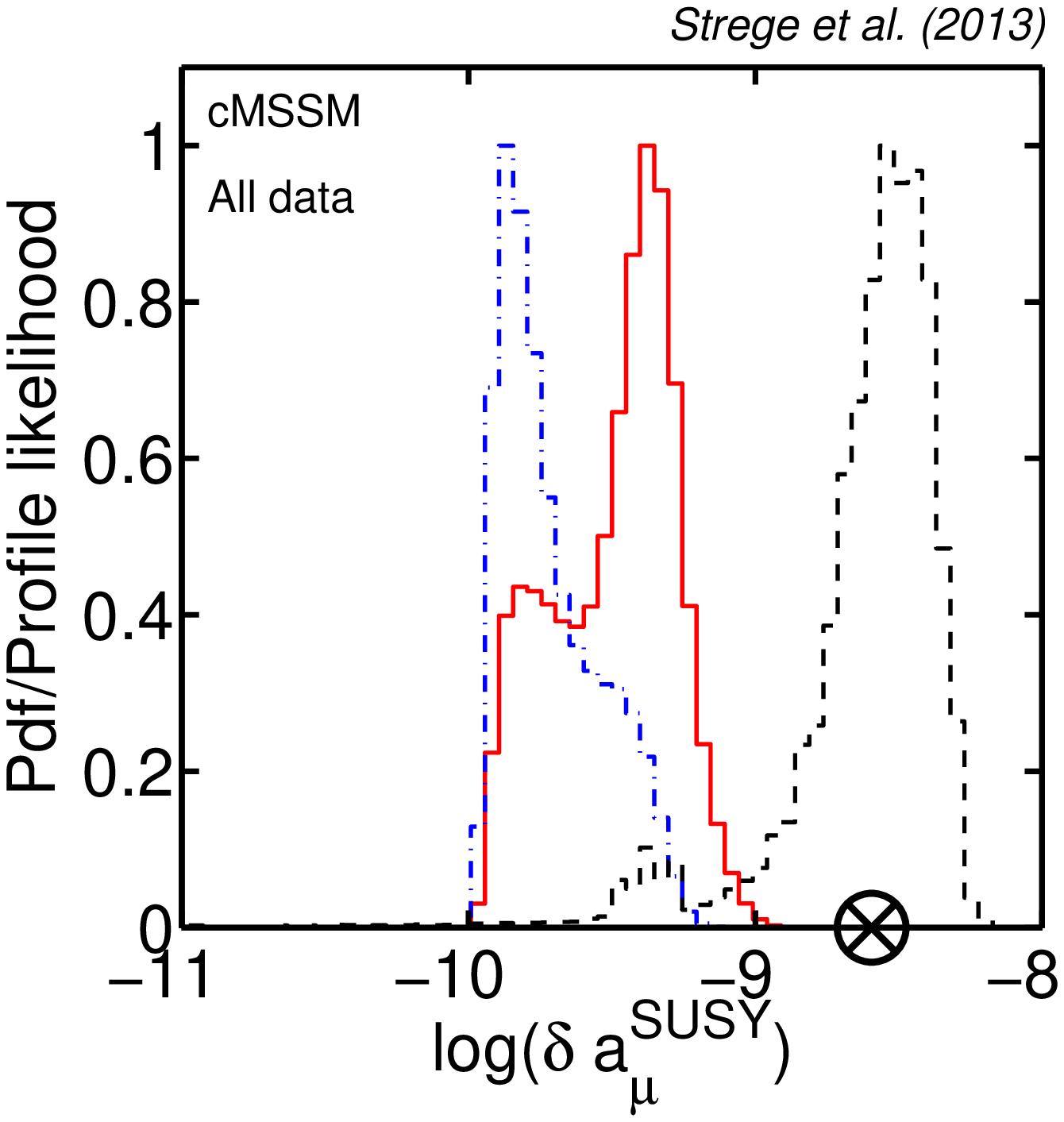}
\expandafter\includegraphics\expandafter[\rowoffour]{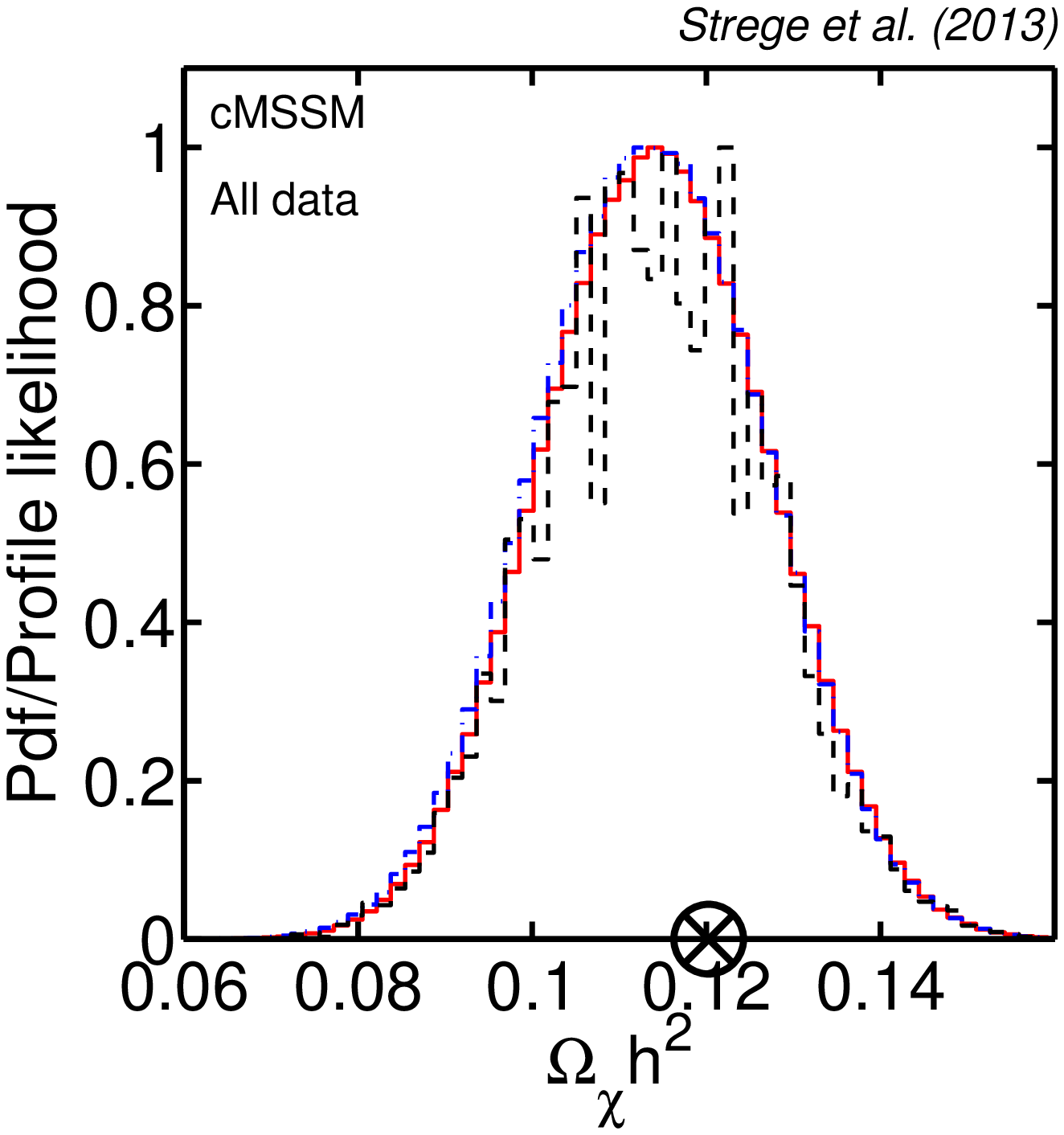}
\expandafter\includegraphics\expandafter[\rowoffour]{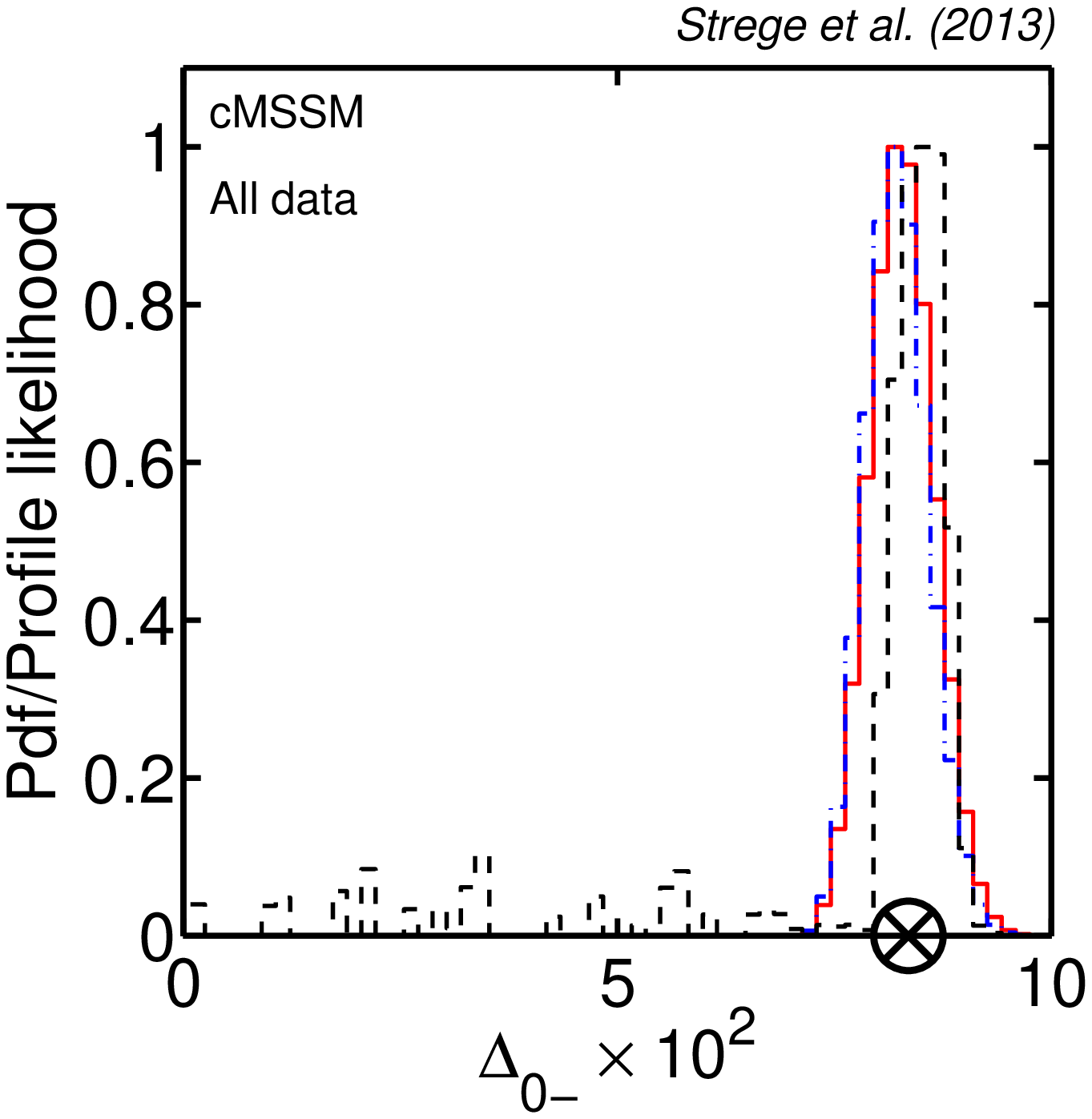}
\caption{\fontsize{9}{9} \selectfont 1D marginal pdf for flat priors (dash-dot/blue), log priors (thick solid/red) and 1D profile likelihood (dashed/black) in the cMSSM, including all up-to-date experimental constraints.  Top row, from left to right: neutralino mass, lightest chargino mass, lightest stop and sbottom masses. Central row: gluino mass, average squark mass, lightest Higgs boson mass, spin-independent scattering cross-section. Bottom row:  $\brbsmumu$ branching ratio, anomalous magnetic moment of the muon, dark matter relic abundance and isospin asymmetry. The best-fit point is indicated by the encircled black cross. \label{fig:1D_CMSSM}} 
\end{figure*}

The 1D marginalised pdfs and profile likelihood for some derived quantities are shown in Fig.~\ref{fig:1D_CMSSM}. The 1D marginal posterior distributions for log (flat) priors are shown in red/solid (dash-dot/blue), the 1D profile likelihood function is shown in black/dashed.

The LHC exclusion limit pushes the sparticle masses to larger values. The difference in the regions of parameter space favoured by the profile likelihood and the posterior pdf with flat priors, as well as the bimodal behaviour of the posterior pdf with log priors, are clearly visible for all of the sparticle masses. Due to its confinement to small $m_{1/2}$, the profile likelihood function shows a strong preference for the smallest allowed values of the lightest stop and sbottom masses, with $m_{\textnormal{stop1}},m_{\textnormal{sbottom1}} < 2$ TeV. The posterior pdf stretches to much larger values of these quantities, up to $m_{\textnormal{stop1}},m_{\textnormal{sbottom1}} \approx 4$ TeV. Similarly, the profile likelihood for the lightest chargino mass favours relatively low values of $m_{\chi_1^{\pm}} \sim [500, 1000]$ GeV, while the posterior pdfs reach larger values of $m_{\chi_1^{\pm}} < 1500$ GeV.
The range of gluino masses favoured by the posterior pdfs is $m_{\textnormal{gluino}} \sim [2,4]$ TeV, while the profile likelihood is confined to a relatively narrow region around $m_{\textnormal{gluino}} \approx 2$ TeV. The profile likelihood also favours a very small range of $m_{\textnormal{squark}} \approx 2$ TeV. The posterior pdf allows a much larger range of average squark masses, extending to $m_{\textnormal{squark}} < 5$ TeV.

With the upgrade of the LHC to 14 TeV collision energy, the sensitivity to heavy SUSY particles will be increased significantly. With 
a total integrated luminosity of 300 fb$^{-1}$, gluinos and squarks of the first two generations with masses around $3-4$ TeV, and the third generation of squarks and charginos with masses around $800$ GeV will be accessible. With the increase of the integrated luminosity to 3000 fb$^{-1}$, as planned with the High-Luminosity (HL-LHC) upgrade, the sensitivity will further improve by a few hundred GeVs \cite{HL_LHC}. Therefore, detection prospects of the cMSSM at the LHC remain very promising. The majority of the parameter space favoured by the Bayesian posterior will be accessible to the LHC with 14 TeV collision energy. From the profile likelihood statistical perspective an important fraction of the favoured gluino and squark masses will already be explored with 20 fb$^{-1}$ of data collected at a collision energy of 8 TeV, the remaining currently favoured region will be explored by the LHC operating at 14 TeV collision energy.

The 1D posterior distributions for $m_h$  for both choices of priors peak at relatively low $m_h \sim 121 - 122$ GeV, while the profile likelihood function favours slightly larger values $m_h \sim 124$ GeV. The discrepancy between the posterior pdfs and the profile likelihood illustrates the difficulty of satisfying the experimental constraint on $m_h$ in the cMSSM. The posterior pdf takes into account volume effects, and therefore peaks at lower $m_h$ that are easier to achieve. The profile likelihood is dominated by a relatively small number of points of high likelihood that achieve a value of $m_h$ closer to the experimental constraint due to maximal stop mixing. Both distributions are offset from the measured mass $m_h = 125.8$ GeV, this value is basically not achieved in the cMSSM.

The 1D profile likelihood for $\brbsmumu$ strongly favours a small range of values around $\brbsmumu \sim 3.0 \times 10^{-9}$. The posterior pdf spreads over a much larger range of $\brbsmumu \geq 3.0 \times 10^{-9}$, but also peaks at relatively small values and falls of at larger $\brbsmumu$. As can be seen, the new LHCb constraint on $\brbsmumu$ has a very limited impact on our results. The experimental value $\brbsmumu = 3.2 \times 10^{-9}$ agrees well with the peak of both the 1D posterior pdf and profile likelihood function for this quantity. Significantly smaller values, that would be discrepant with the experimental measurement, are not realised in the cMSSM. Larger values of $\brbsmumu$ can be achieved, and are disfavoured by this constraint. However, the previous upper limit was even slightly more constraining at large $\brbsmumu$ than the current constraint. A more precise measurement of $\brbsmumu$ is needed for this constraint to have a strong impact on the cMSSM parameter space.

The 1D profile likelihood and posterior pdf for $\delta a_{\mu}^{SUSY}$ are also shown in Fig.~\ref{fig:1D_CMSSM}. Results for the two statistical perspectives differ strongly. The profile likelihood function peaks at relatively large values of $\delta a_{\mu}^{SUSY}$, in good agreement with the experimental constraint. In contrast, the posterior pdf for both choices of priors favours a SM-like value of the anomalous magnetic moment of the muon, and therefore peaks at significantly smaller values. While these values are in strong disagreement with the experimental constraint, they are much easier to achieve in the cMSSM, especially for high values of the mass parameters. The posterior pdf takes into account these volume effects, while the profile likelihood function, which peaks at the region of highest likelihood, favours values that reproduce the experimental measurement.

From this discrepancy one can see that the constraint on $\delta a_{\mu}^{SUSY}$ may have a very strong impact on our conclusions, especially on the profile likelihood results. In the following section we will discuss the impact of this constraint in more detail.

The 1D profile likelihood and posterior pdfs for the dark matter relic density are in good agreement with each other. Both quantities peak at the experimentally favoured value. In contrast, the distributions for the isospin asymmetry $\Delta_{0-}$ are discrepant with the experimental measurement $\Delta_{0-} = 3.1 \times 10^{-2}$ by more than $2\sigma$. Smaller values of $\Delta_{0-}$ are difficult to achieve in the cMSSM, since the SM-like value is already strongly discrepant with the experimental measurement, and the vast majority of points in cMSSM parameter space lead to a positive contribution to $\Delta_{0-}$. As can be seen, several points leading to smaller $\Delta_{0-}$ are found, but are in conflict with other constraints, which leads to a low likelihood value. As a result, the best-fit point is also located at large values $\Delta_{0-} = 8.35 \times 10^{-2}$, which explains the large contribution to the best-fit $\chi^2$ from the $\Delta_{0-}$ constraint, observed in the previous section.

\subsection{Impact of the  $\delta a_\mu^{SUSY}$ constraint}
\label{sec:cMSSM_wogm2}

The magnetic anomaly of the muon, $a_\mu= \frac{1}{2}(g-2)_\mu$ is an exciting and powerful test for new physics. At present, the experimental measurement and theoretical determinations are very precise, enough to either strongly constrain, or even give a positive signal of, new physics. However, the situation is still somewhat uncertain, essentially due to inconsistencies between alternative determinations of the SM hadronic contribution, specifically the contribution from the hadronic vacuum polarization diagram $\delta_{\rm had}^{\rm SM}a_\mu$. This contribution can be expressed in terms of the total hadronic cross-section $e^+ e^-\rightarrow$ hadrons. When experimentally measuring this cross-section, one obtains a value for $a_\mu$ which is discrepant from current experimental measurements by more than 3$\sigma$. In the past this discrepancy has widely been interpreted as a signal of new physics. In this case, the discrepancy should be cured by contributions from new physics, in our case MSSM contributions. The immediate implication is that supersymmetric particles should have relatively small masses, in order to produce a large enough contribution, $\delta^{\rm SUSY}a_\mu $, to reconcile theory and experiment. Hence, SUSY should be found at low energies (accessible to LHC), mainly because of the experimental measurement of $a_\mu$.

The lack of a SUSY signal at the LHC strongly challenges this interpretation. In previous works we have investigated the impact of the \gmt\ constraint on the cMSSM~\cite{Roszkowski:2007fd,Trotta:2008bp,Strege:2011pk}. We found that this constraint plays a dominant role in determining the favoured regions of parameter space in terms of the profile likelihood, and that it was the single most important datum disfavouring large values of $m_0$ and $m_{1/2}$.  We thus repeat the analysis presented in the previous section, but exclude the experimental constraint on \gmt, in order to assess the robustness of our conclusions with respect to omission of this contraint. 
%\cs{Buchmueller et al say for without \gmt mh = 125 GeV is easy to realise, but from our previous paper this is not so true..}
\begin{figure*}%[htp]
\centering
\expandafter\includegraphics\expandafter
[\rowofthree]{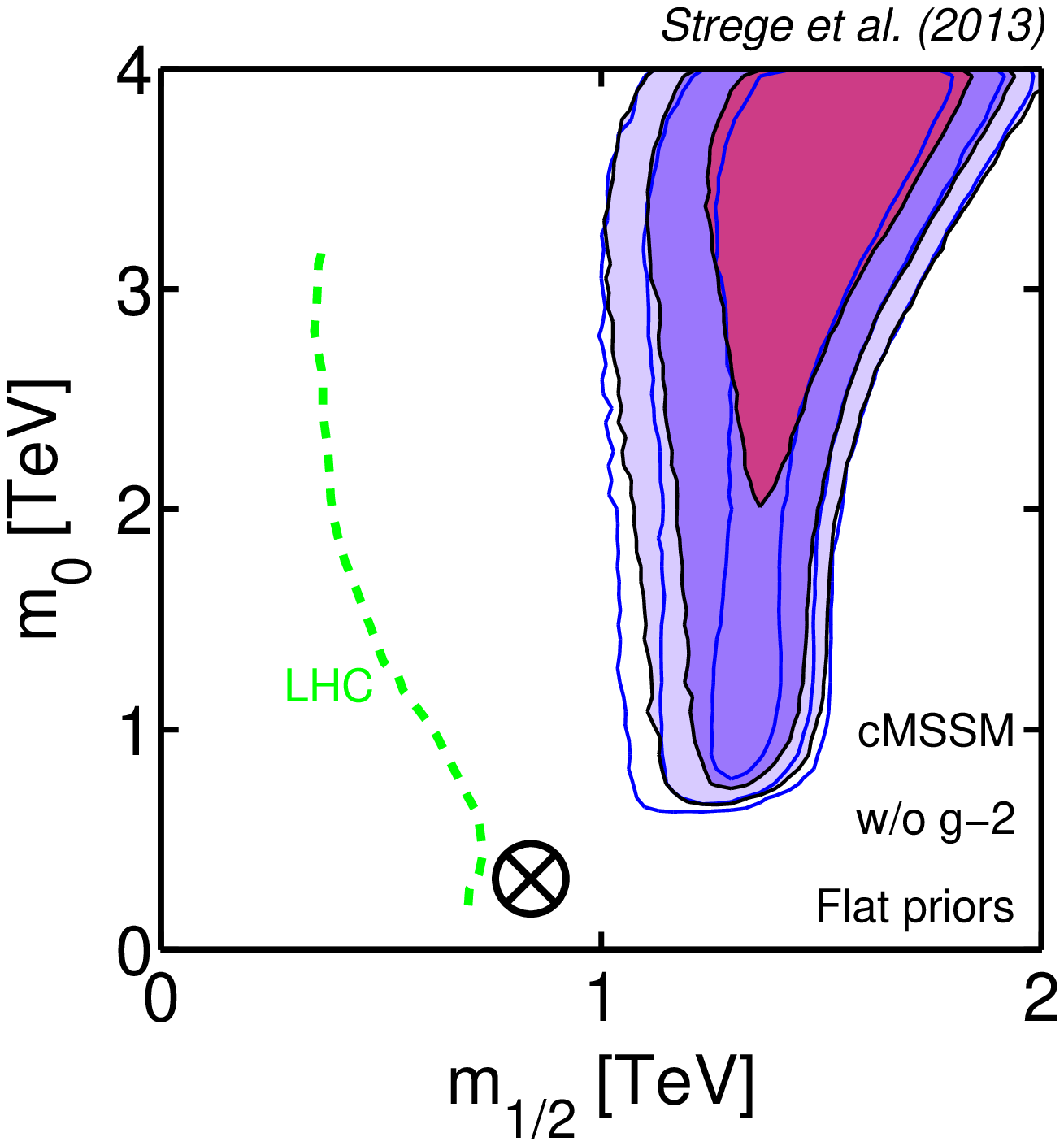}
\expandafter\includegraphics\expandafter
[\rowofthree]{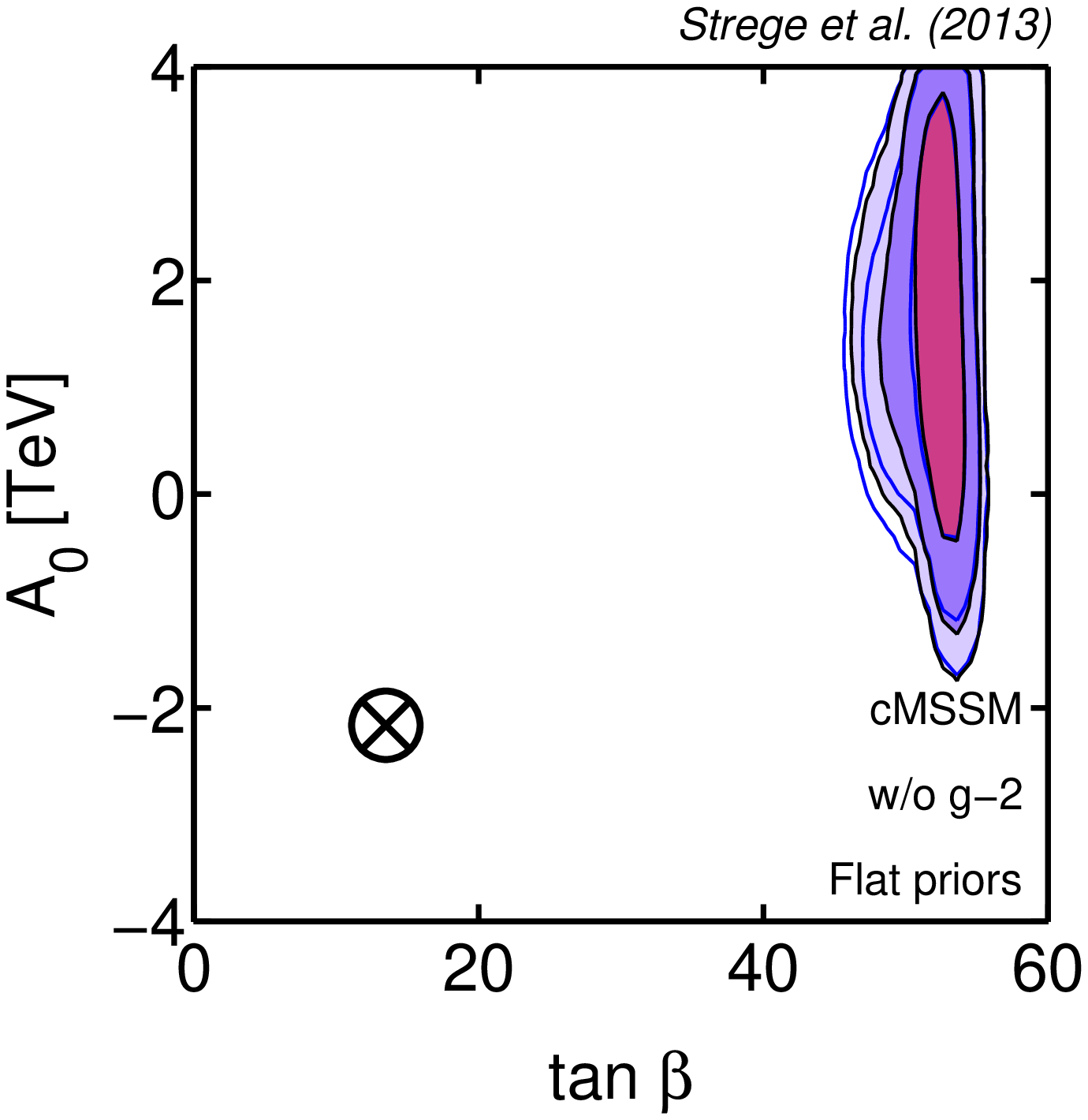}
\expandafter\includegraphics\expandafter
[\rowofthree]{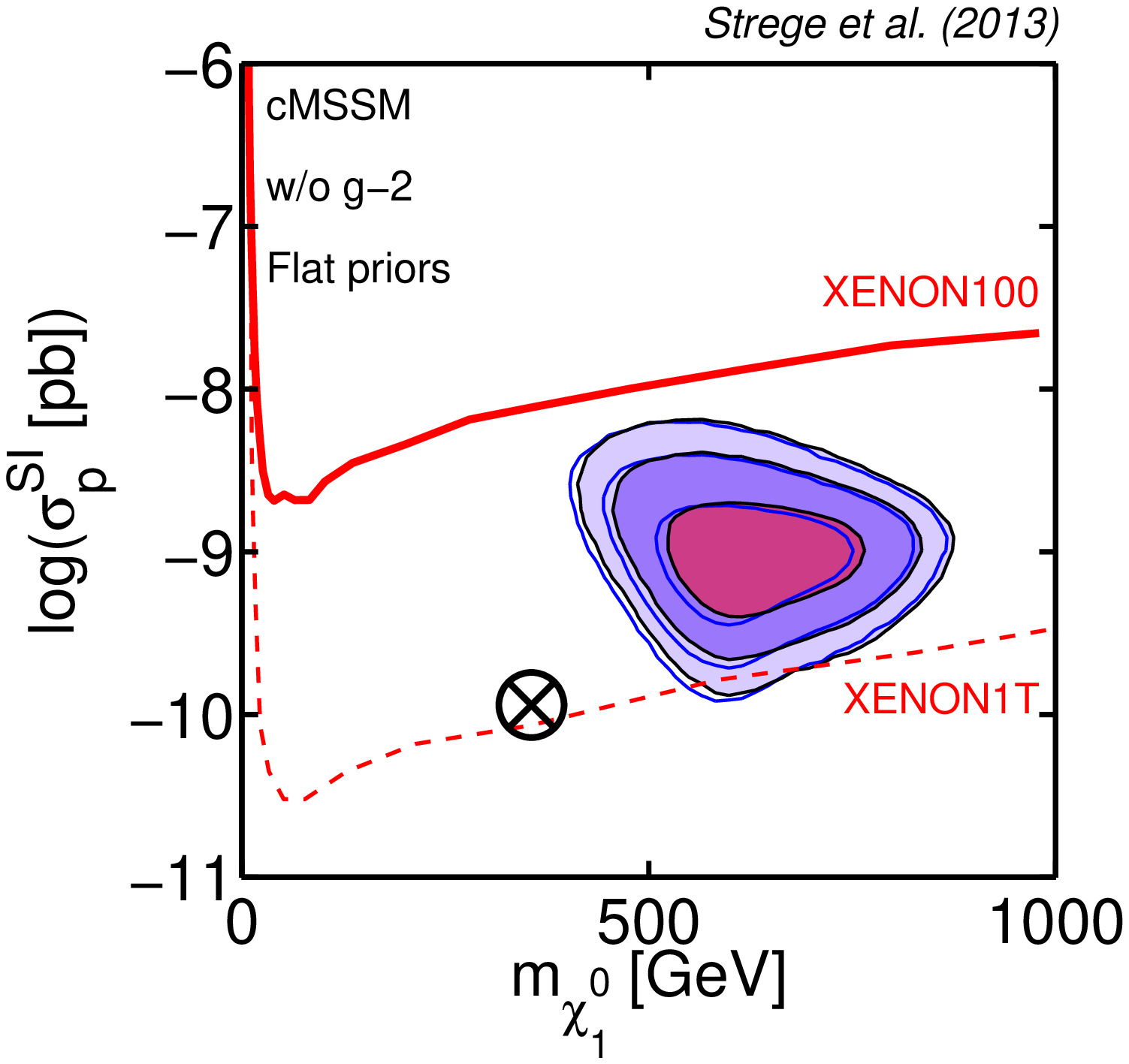} \\
\expandafter\includegraphics\expandafter
[\rowofthree]{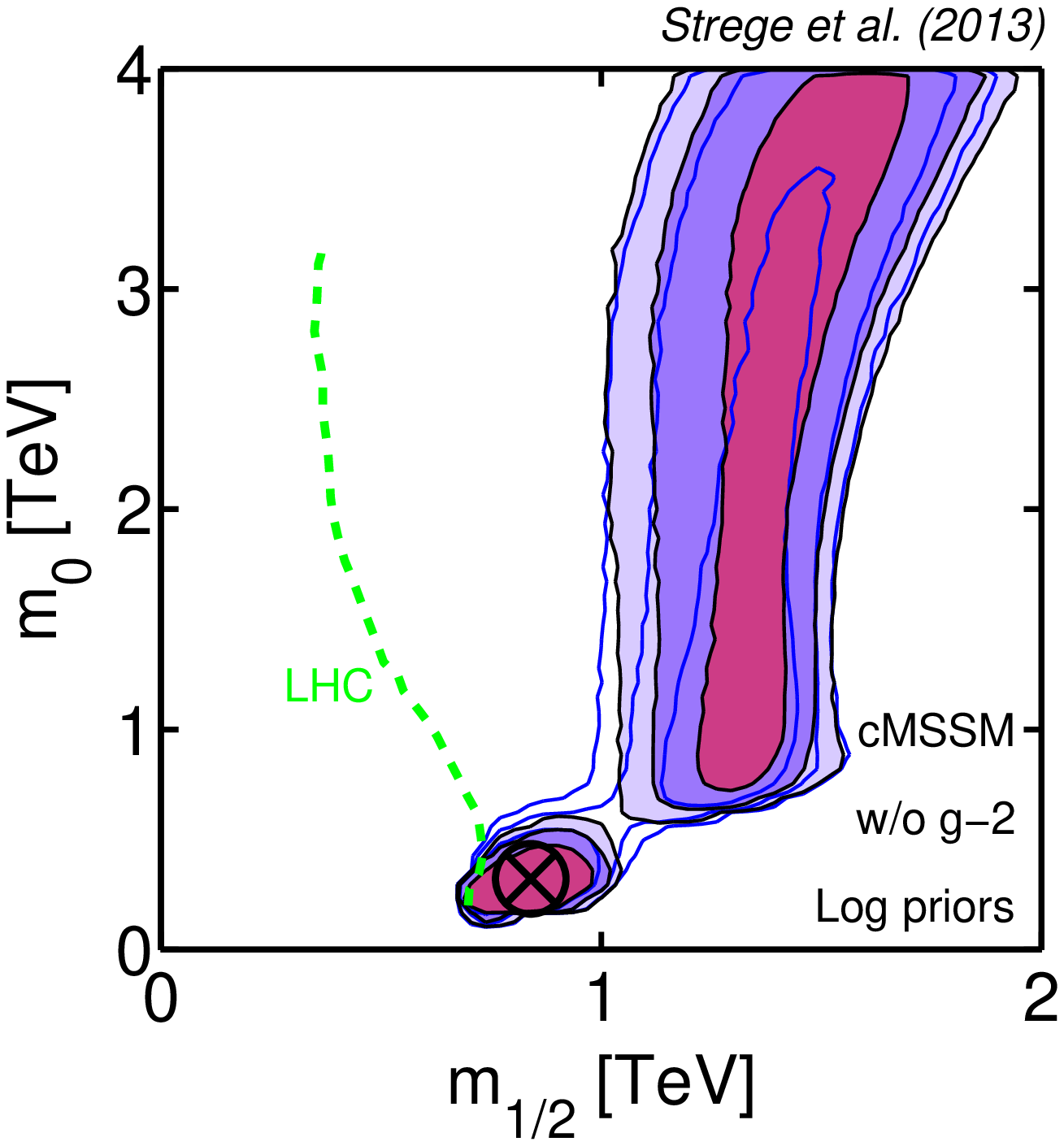}
\expandafter\includegraphics\expandafter
[\rowofthree]{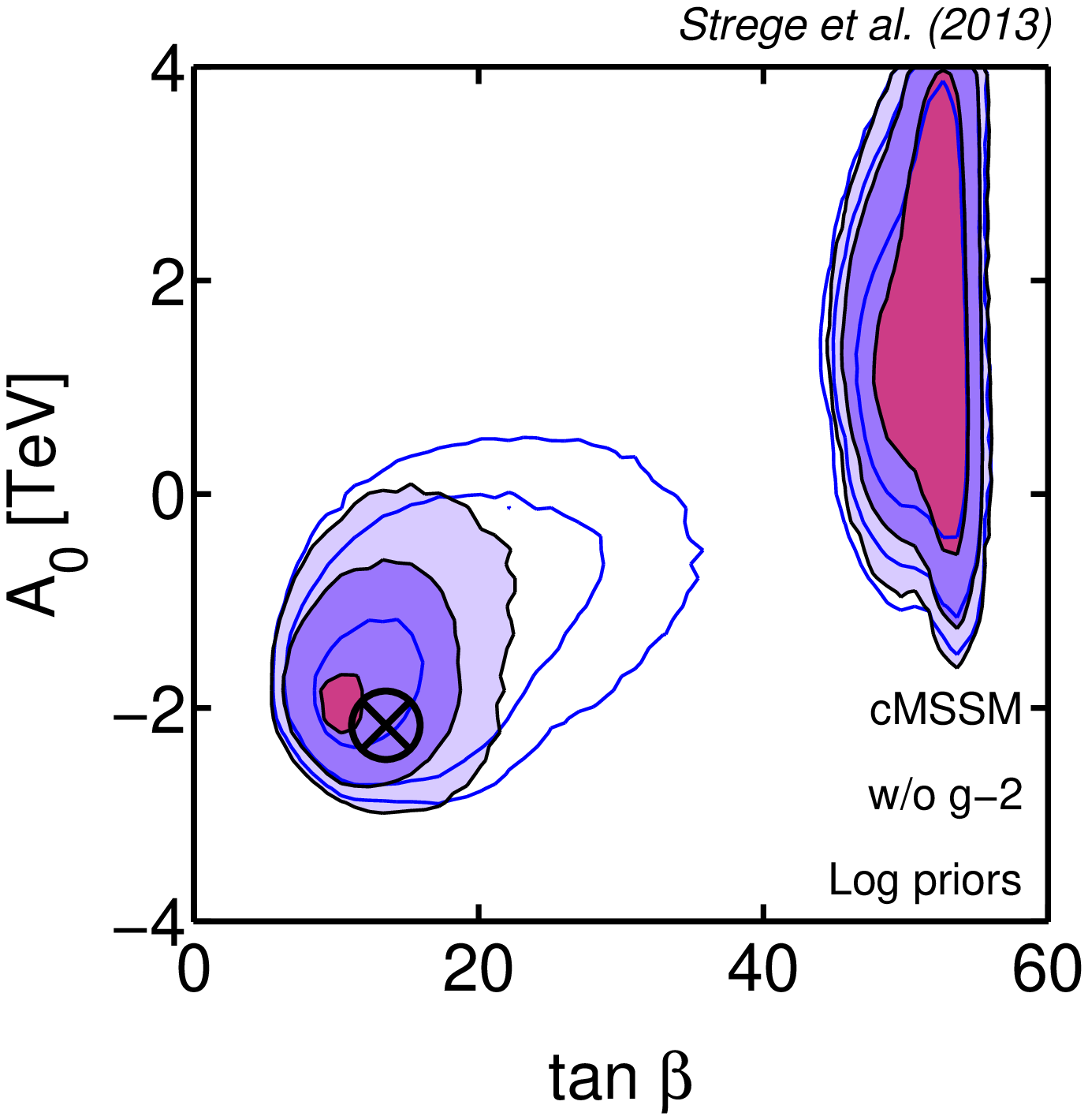}
\expandafter\includegraphics\expandafter
[\rowofthree]{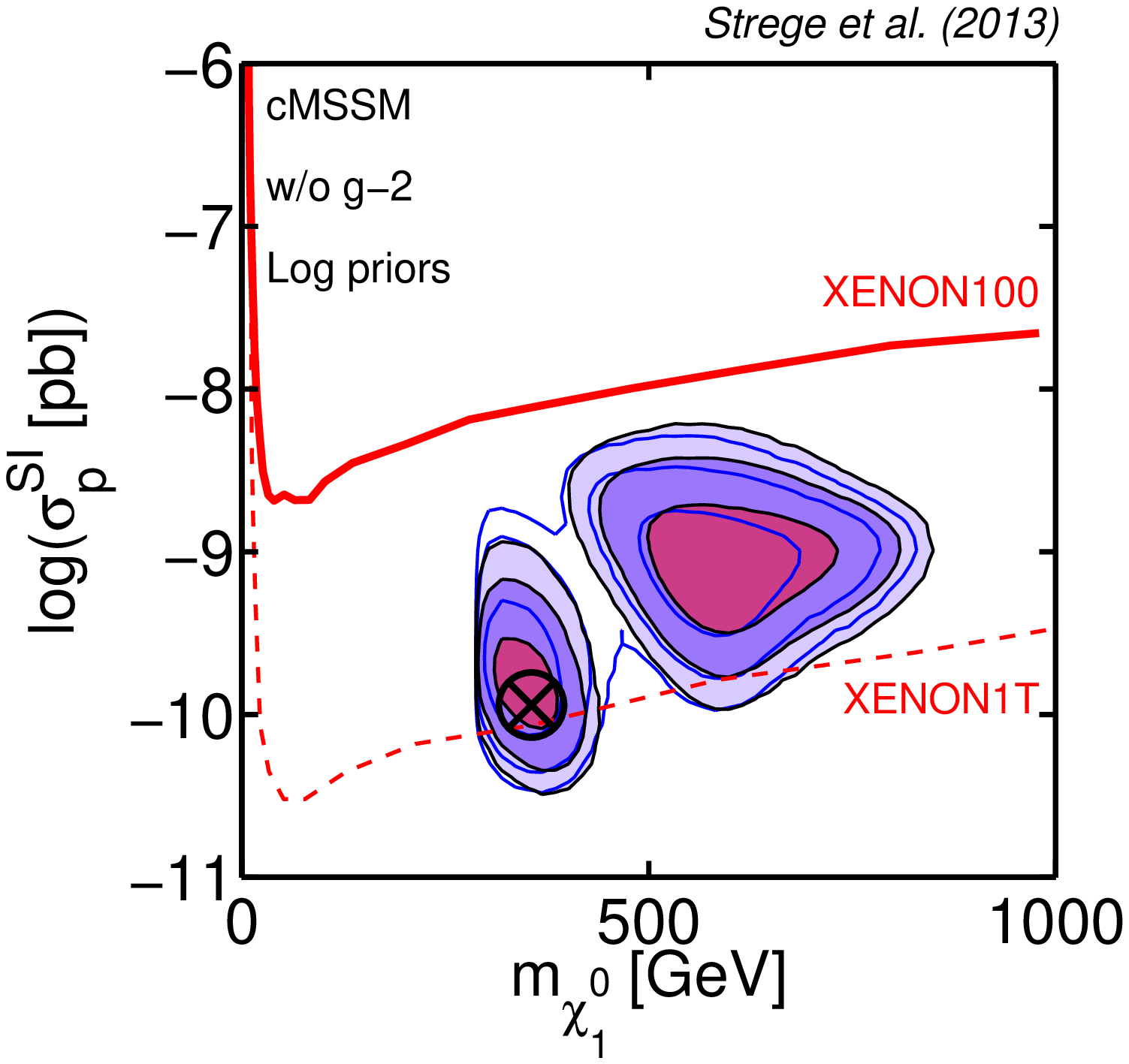} \\
\expandafter\includegraphics\expandafter
[\rowofthree]{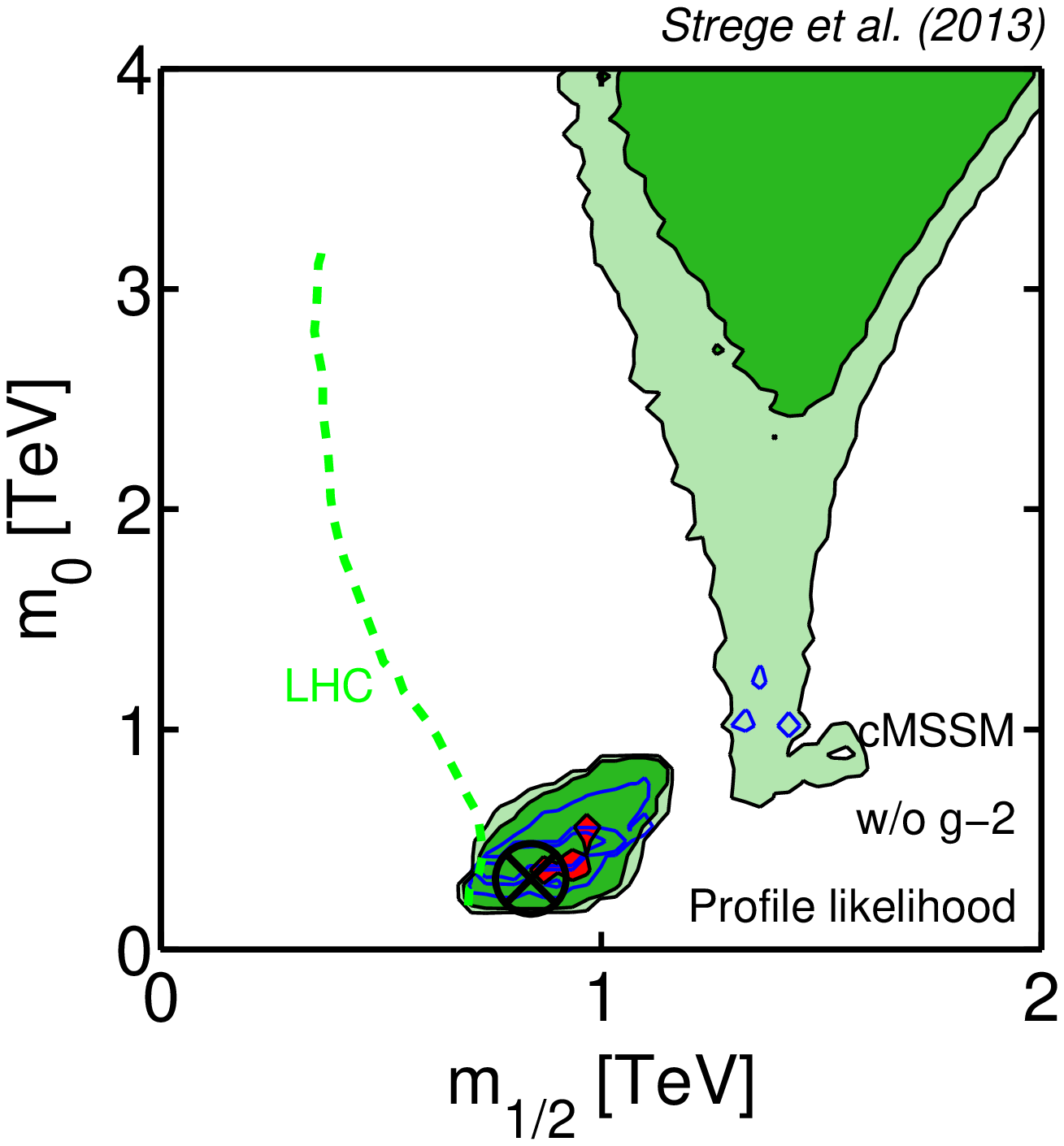}
\expandafter\includegraphics\expandafter
[\rowofthree]{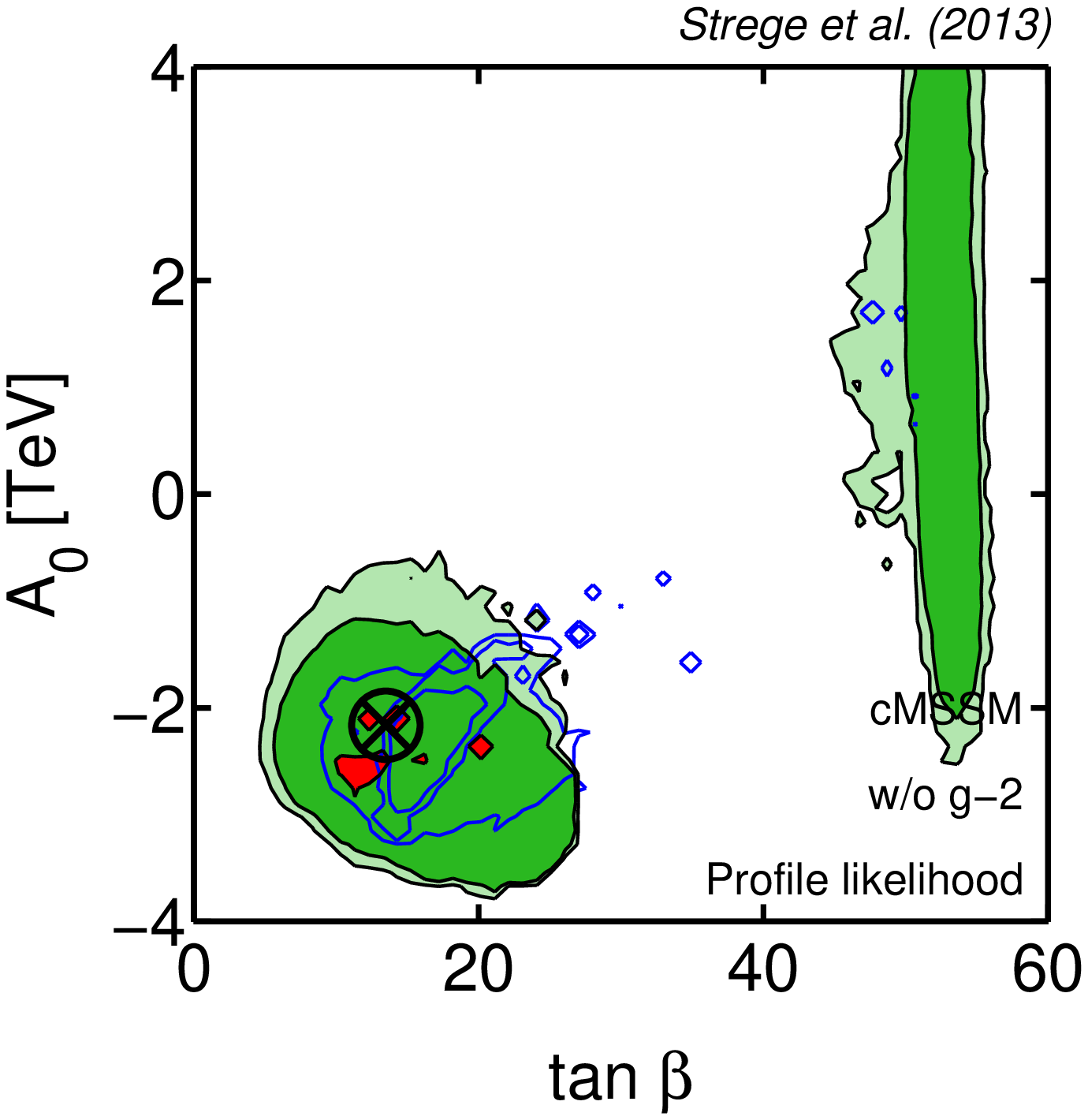}
\expandafter\includegraphics\expandafter
[\rowofthree]{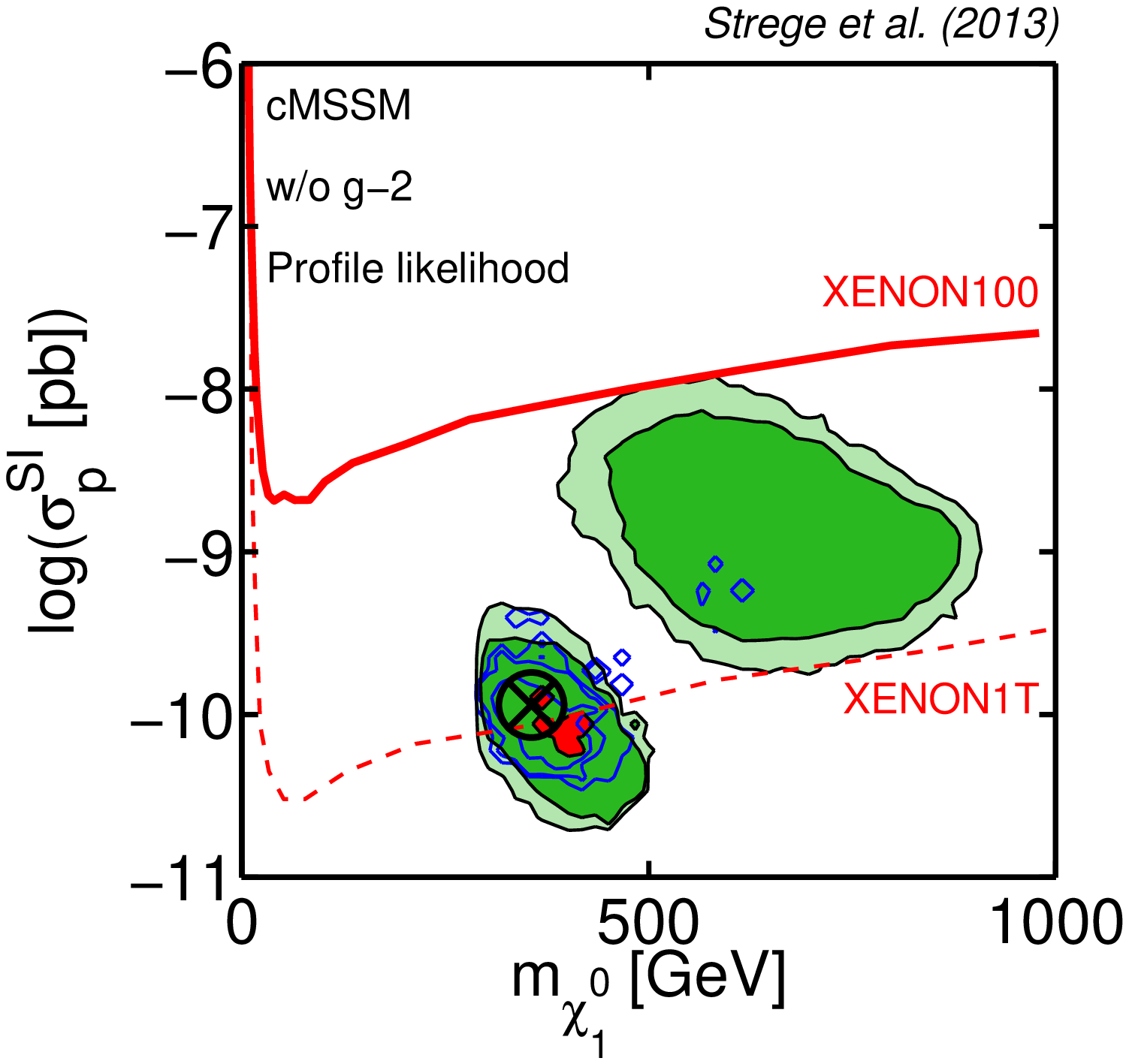} \\
\caption{\fontsize{9}{9} \selectfont   Constraints on the cMSSM as in Fig. \ref{cMSSM_2D_alldata}, but now excluding the \gmt\ constraint (black/filled contours). From top to bottom: posterior pdf with flat priors, posterior pdf with log priors and profile likelihood. The encircled black cross is the overall best-fit point. For comparison, blue/empty contours do include the \gmt\ constraint (as the filled contours in Fig.~\ref{cMSSM_2D_alldata}). Dropping the constraint on the anomalous magnetic moment of the muon hardly changes the best-fit point, but does open up the A-funnel region in the profile likelihood analysis.\label{cMSSM_2D_wogm2}}
\end{figure*}

Results of this analysis are shown in Fig.~\ref{cMSSM_2D_wogm2} (black/filled contours), where they are compared with the findings presented above (blue/empty contours, corresponding to the filled contours in Fig.~\ref{cMSSM_2D_alldata}). 
The posterior distributions in the $(m_{1/2},m_0)$ plane are very similar to the results for the Bayesian analysis including the \gmt\ constraint, for both log and flat priors. Since a large value of \gmt\ can only be obtained for low gaugino and scalar masses, removing this constraint from the analysis leads to a slight shift of the favoured regions towards larger values of $m_0$.  Nevertheless, even when excluding the \gmt\ constraint from the analysis, the posterior pdf with log priors still favours the SC region at the 68\% level, although the posterior probability mass associated with this mode is now reduced with respect to Fig.~\ref{cMSSM_2D_alldata}.  The smaller size of this mode is also reflected in the $(\tan \beta,A_0)$ plane. The posterior pdf with flat priors does not find the mode at low masses and instead shows a strong preference for very large values of $m_0$ and $m_{1/2}$. This prior dependence of the posterior distributions was already observed in section \ref{sec:cMSSM} and is unrelated to the \gmt\ constraint. Instead, this is a result of volume effects at high masses influencing the results for the flat prior posterior distribution.

The impact of dropping the \gmt\ constraint on the profile likelihood is more significant (bottom row in Fig.~\ref{cMSSM_2D_wogm2}). Without \gmt\, the A-funnel region (which is excluded at 99\% level when the \gmt\ constraint is taken into account) is now viable at the $95\%$ confidence level.  Large values of $m_0$ remain viable, and the 95\% confidence region extends all the way to the 4 TeV prior boundary. However, the best-fit point and thus the region most favoured from the profile likelihood statistical perspective, is still found in the SC region. Indeed, the coordinates of the best-fit (as displayed in Table~\ref{cMSSMbf2}) are only slightly different from before.  We also notice that the profile likelihood results agree fairly well with the posterior pdf obtained with the log prior. From the hypothesis testing perspective, the best-fit point has $\chi^2/\text{dof} = 1.11$, corresponding to a p-value of 0.35. Therefore, when dropping the \gmt\ constraint, the cMSSM best-fit point remains perfectly viable in light of the latest data. 

The bimodal behaviour of the profile likelihood function is also apparent in the $(m_{\neut},\sigmaSI)$ plane (right-most panels in Fig.~\ref{cMSSM_2D_wogm2}) . In contrast to the results including \gmt,  a large region corresponding to large $\tan \beta$ values and larger neutralino masses of $m_{\neut} > 500$ GeV is now allowed at 95\% confidence level. Compared to the case with \gmt, values of $\sigma_p^{SI}\sim 10^{-8}$ pb for a neutralino mass $m_{\neut} \approx 650$ GeV are now within the 95\% CL.  In this case, the latest XENON100 results actively constrain this region, which is bounded from above by the direct detection limit. This entire AF region will be explored by the XENON1T experiment. The spin-dependent cross-section remains, as before, out of reach of proposed future experiments.

\begin{figure*}%[htp]
%\centering
\expandafter\includegraphics\expandafter[\rowoffour]{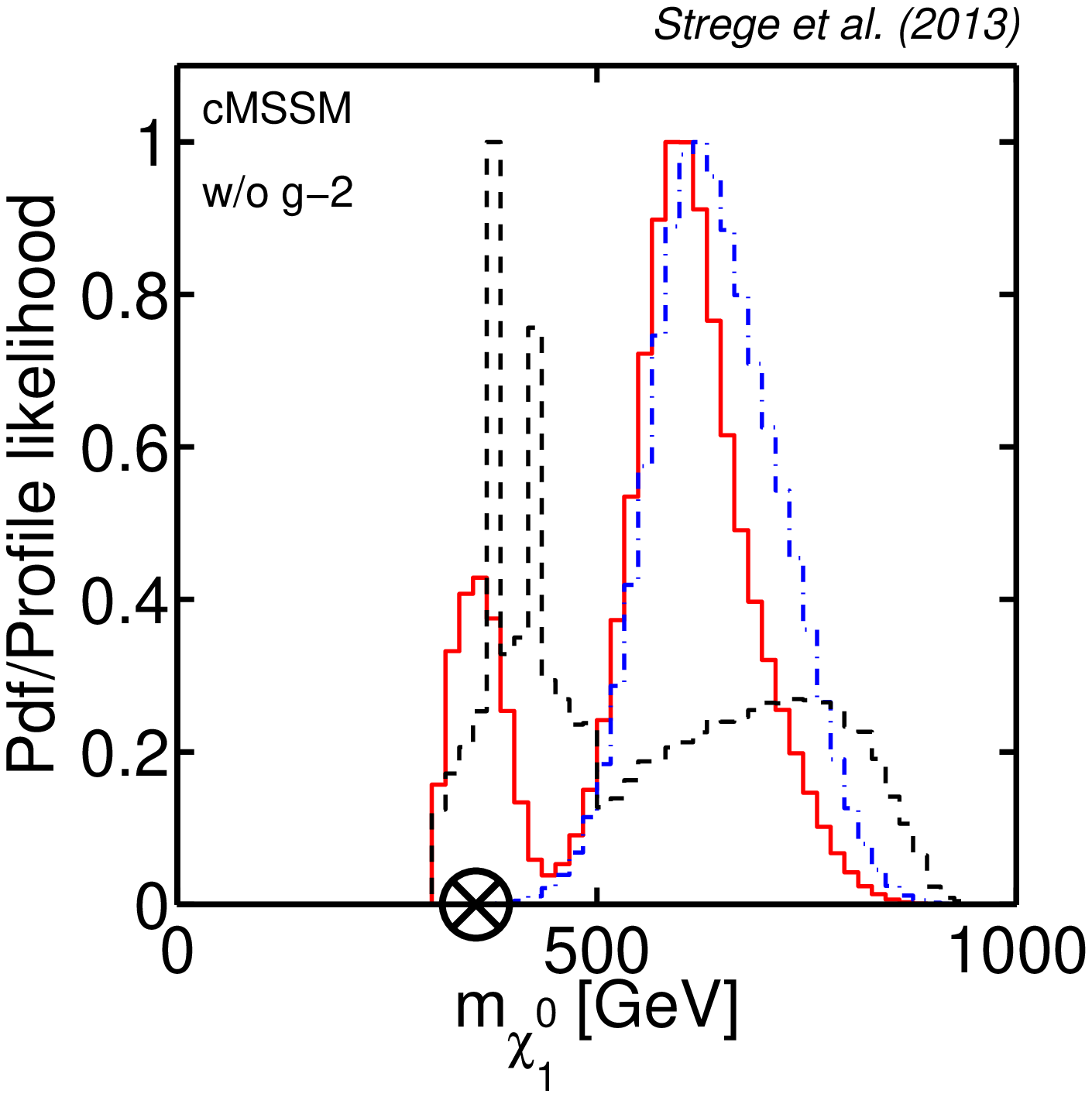}
\expandafter\includegraphics\expandafter[\rowoffour]{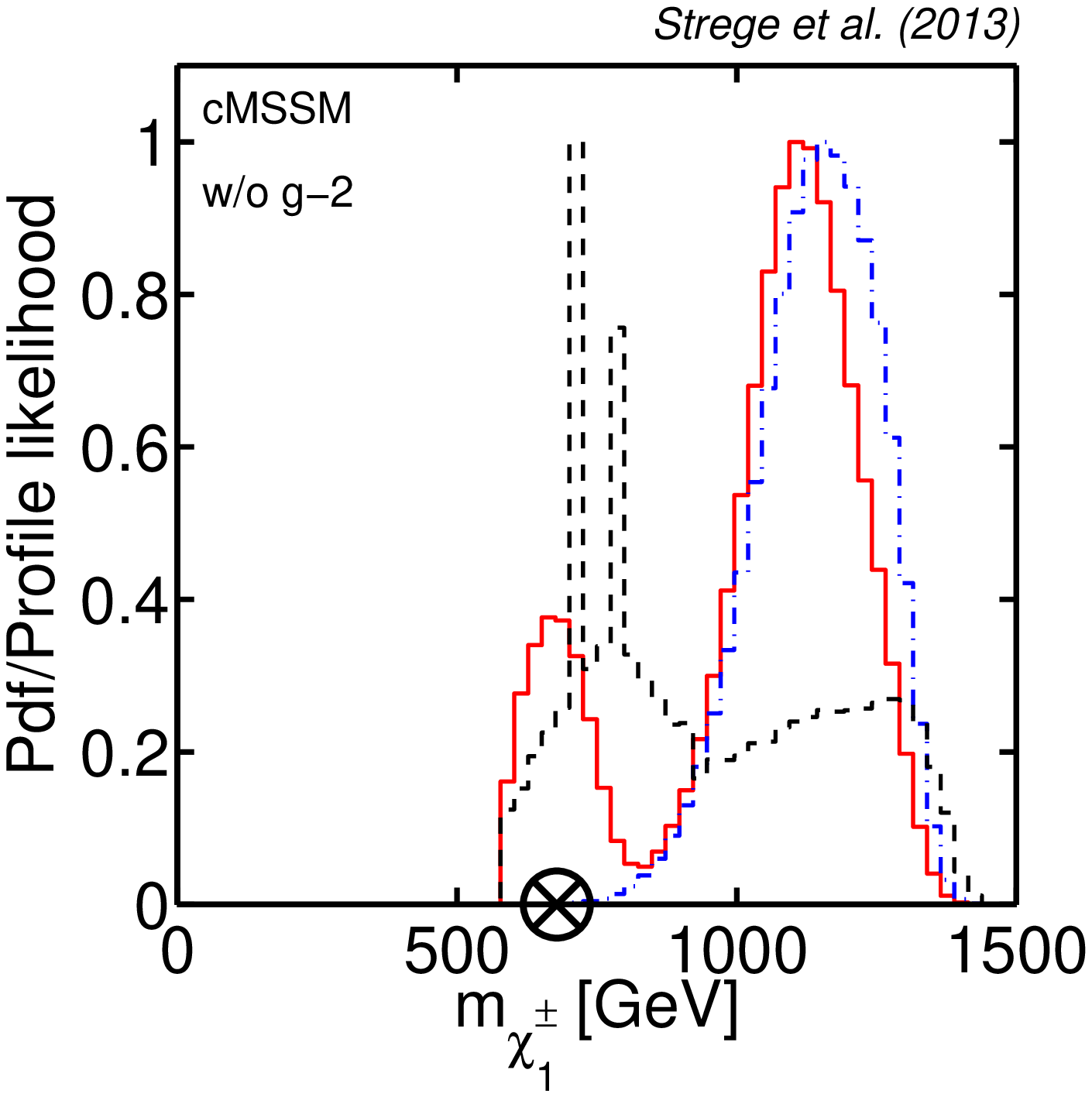}
\expandafter\includegraphics\expandafter[\rowoffour]{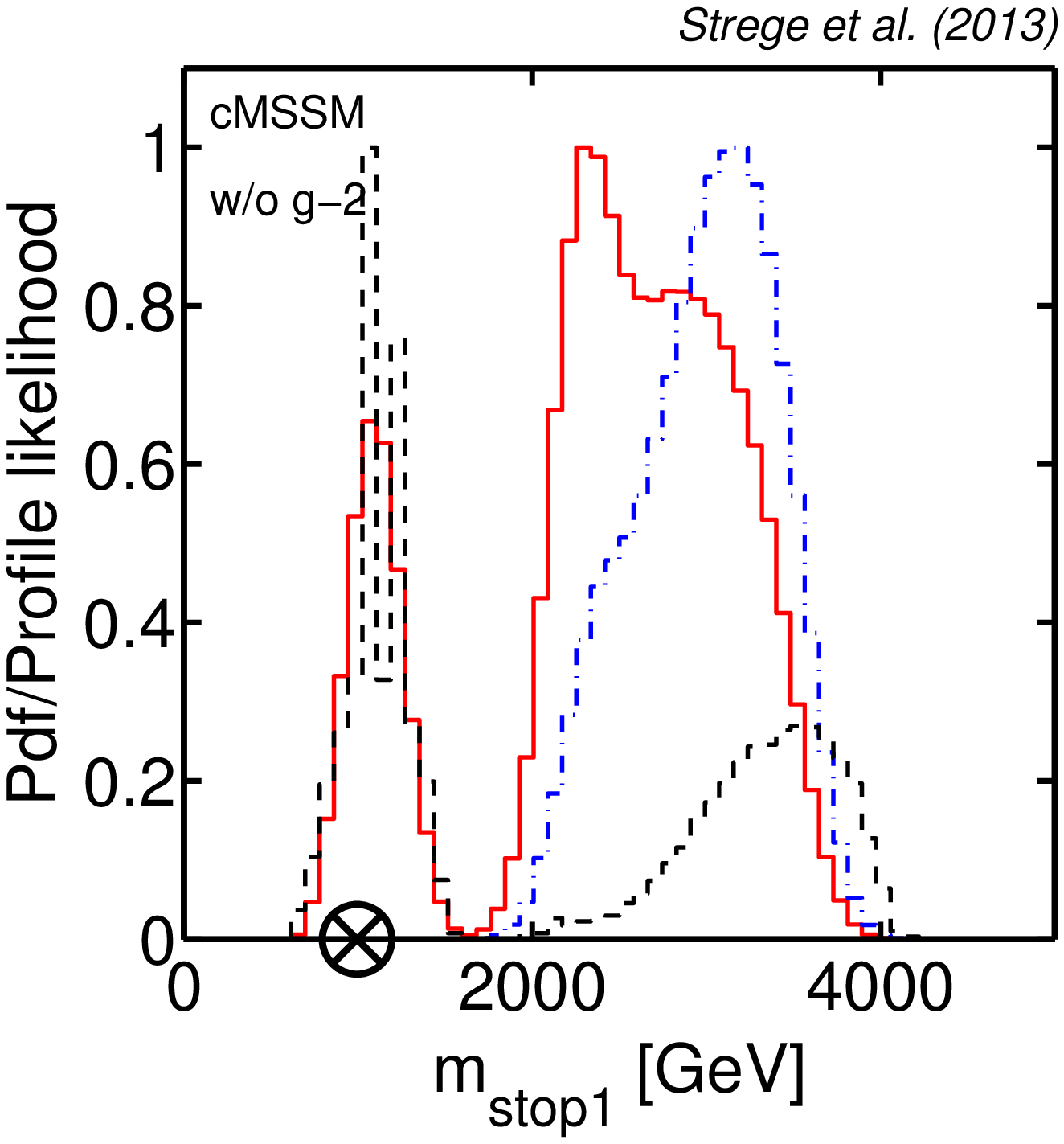}
\expandafter\includegraphics\expandafter[\rowoffour]{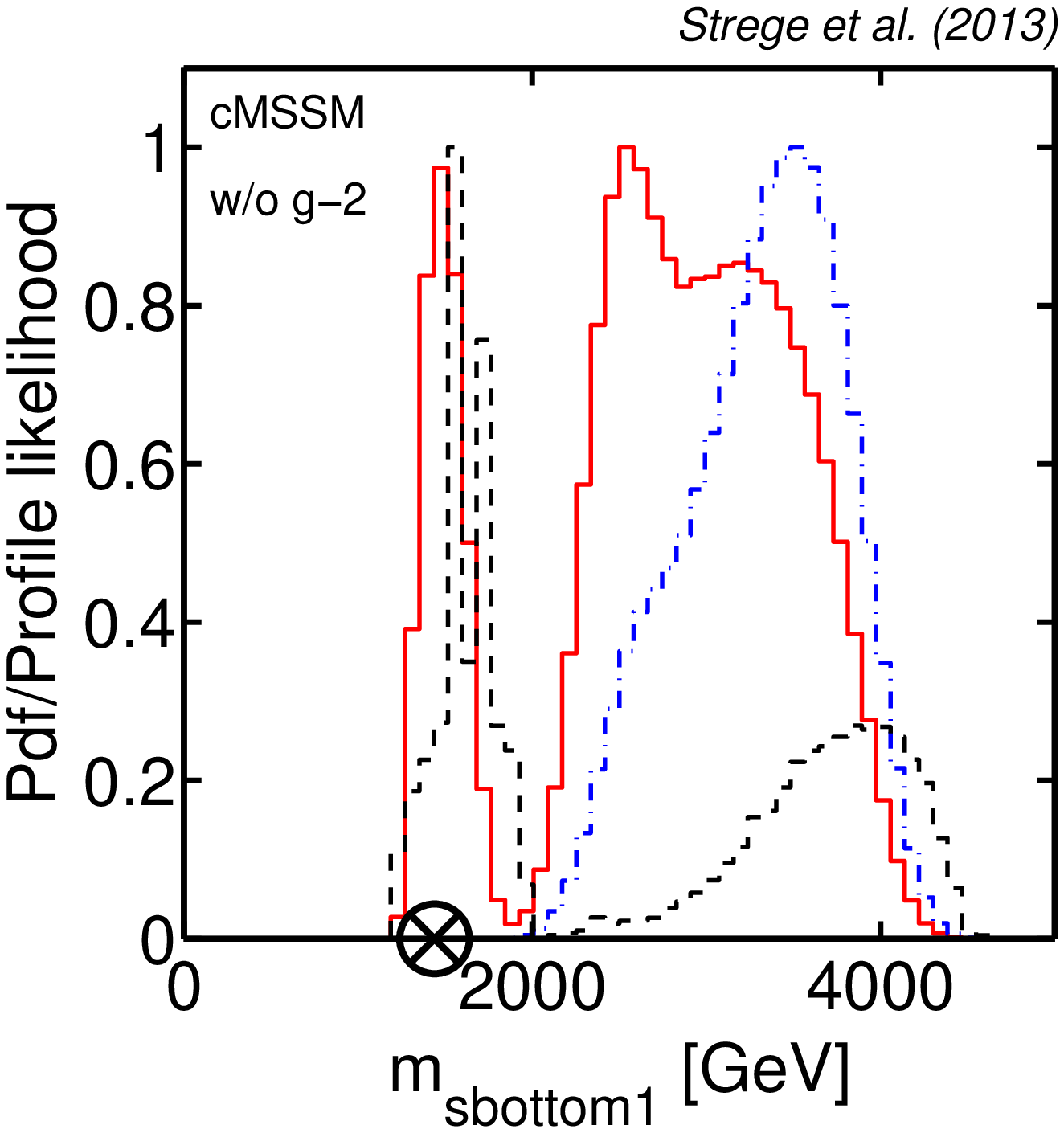}\\
\expandafter\includegraphics\expandafter[\rowoffour]{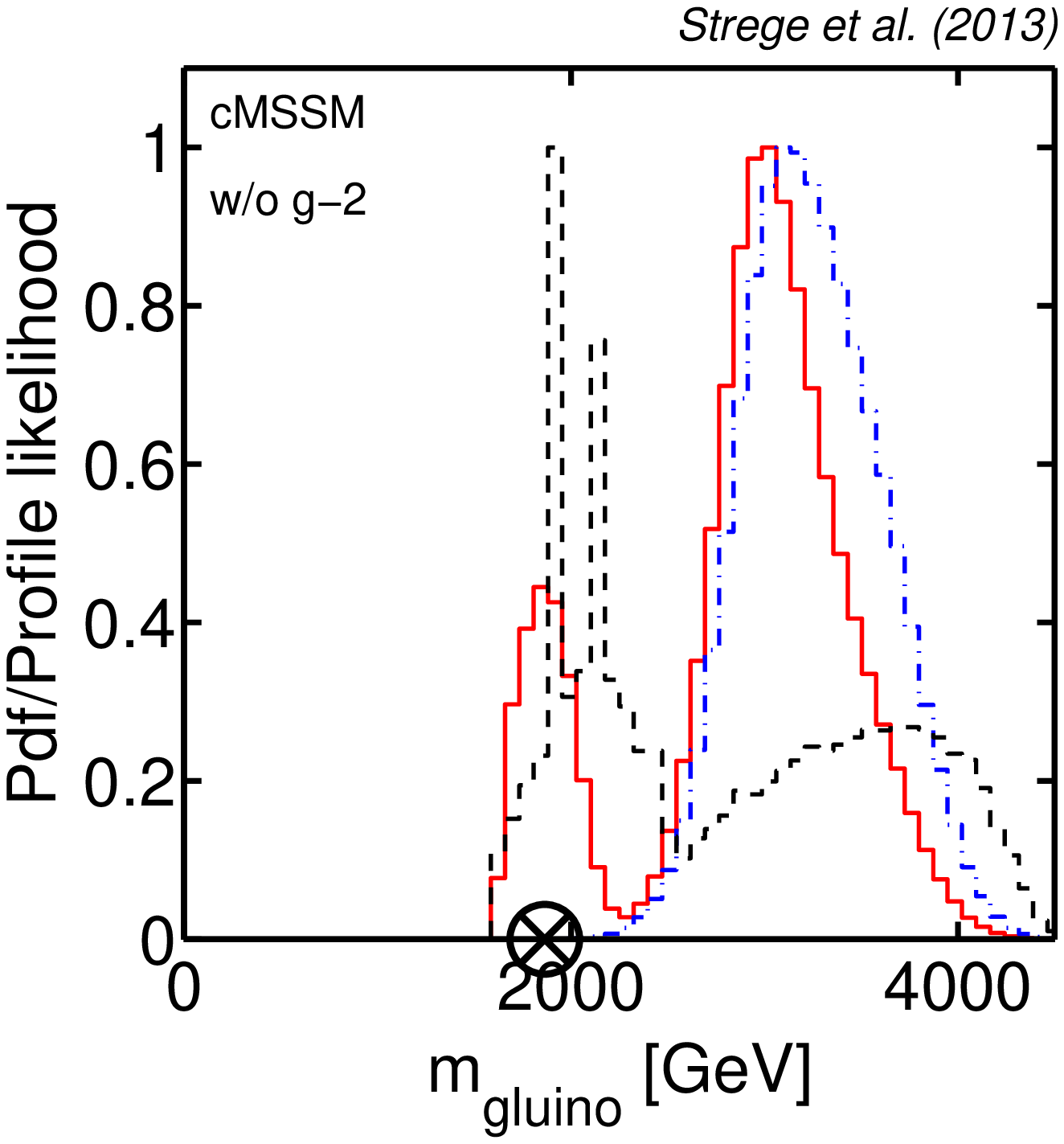}  
\expandafter\includegraphics\expandafter[\rowoffour]{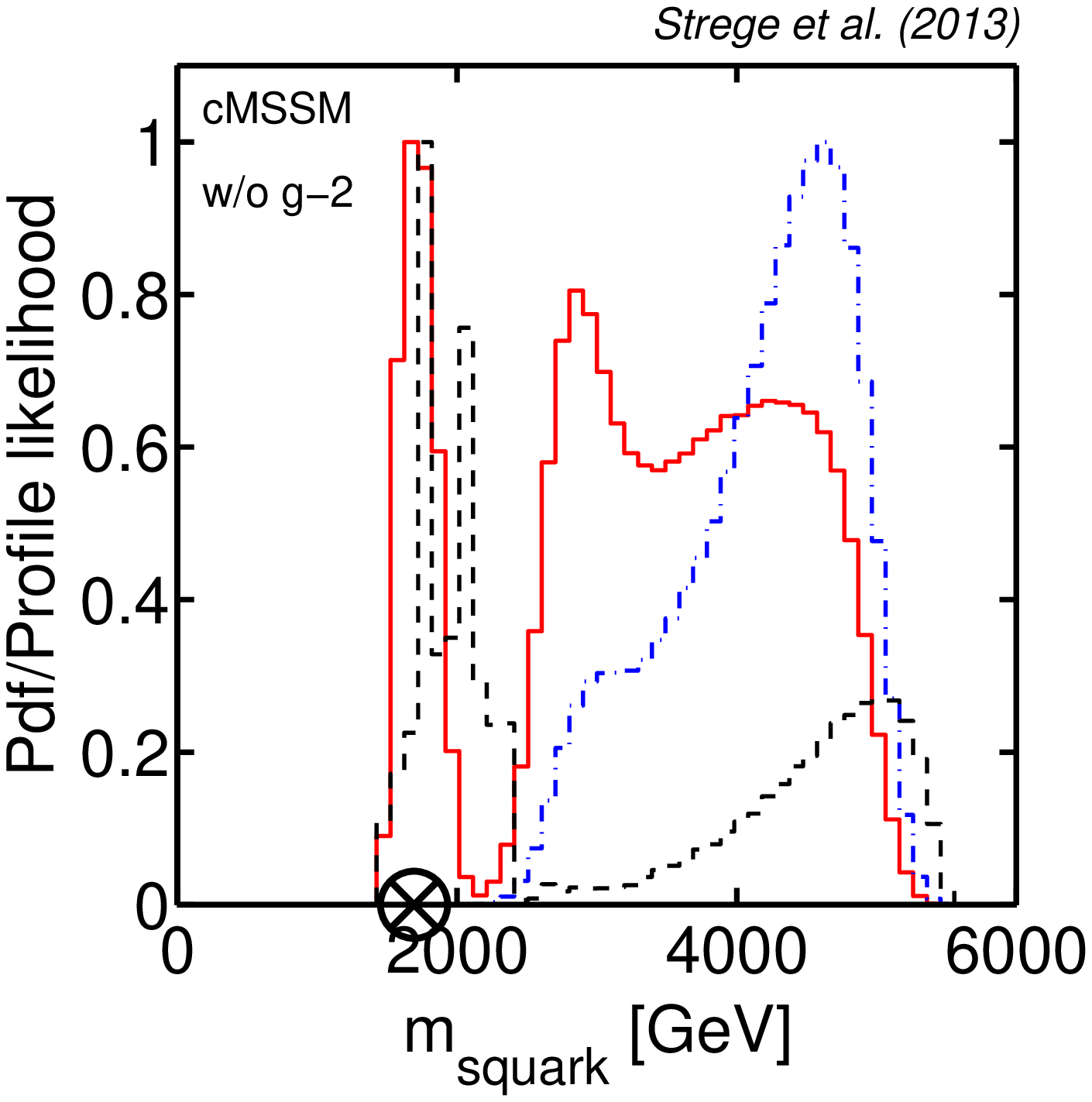}
\expandafter\includegraphics\expandafter[\rowoffour]{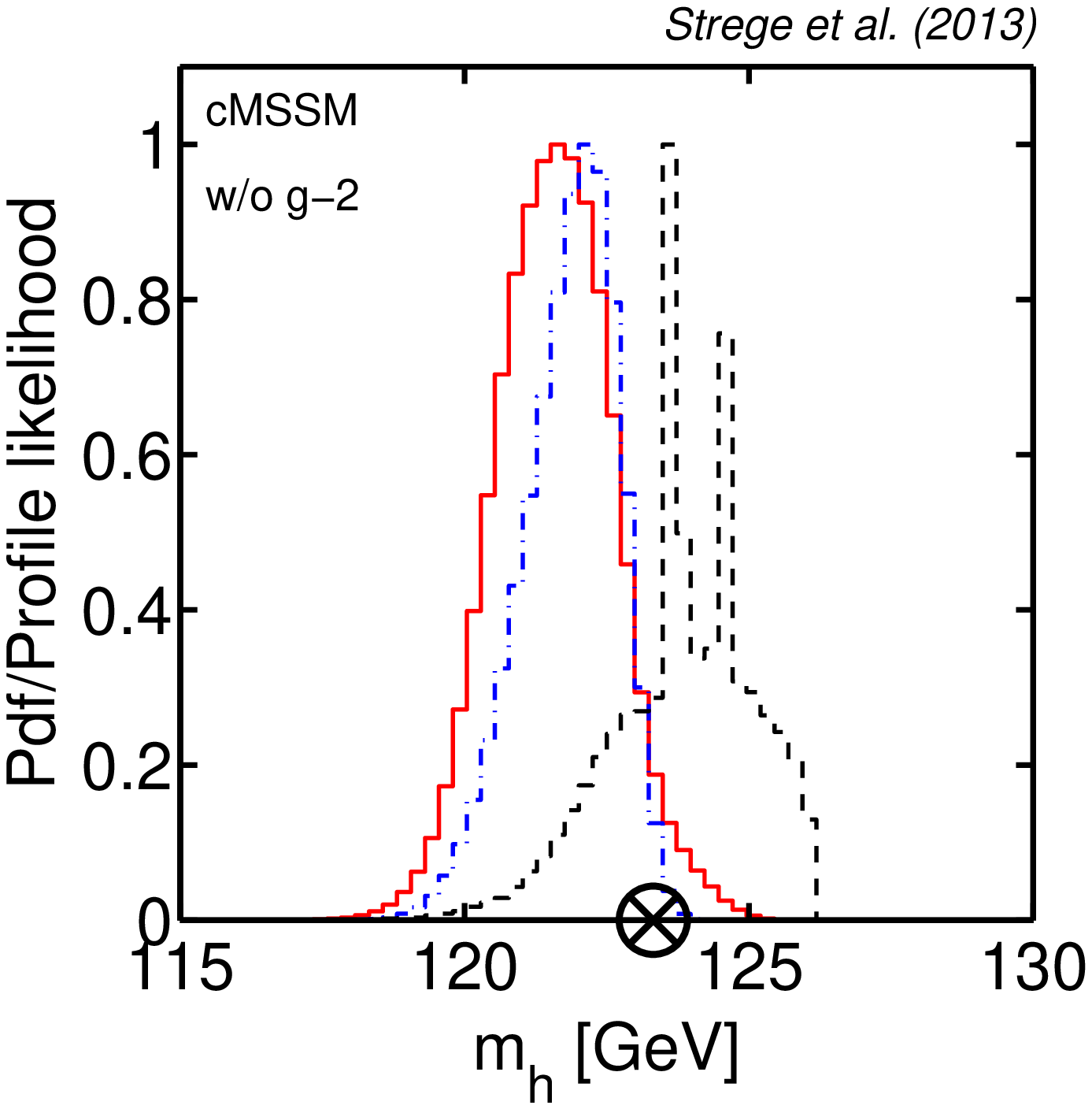}
\expandafter\includegraphics\expandafter[\rowoffour]{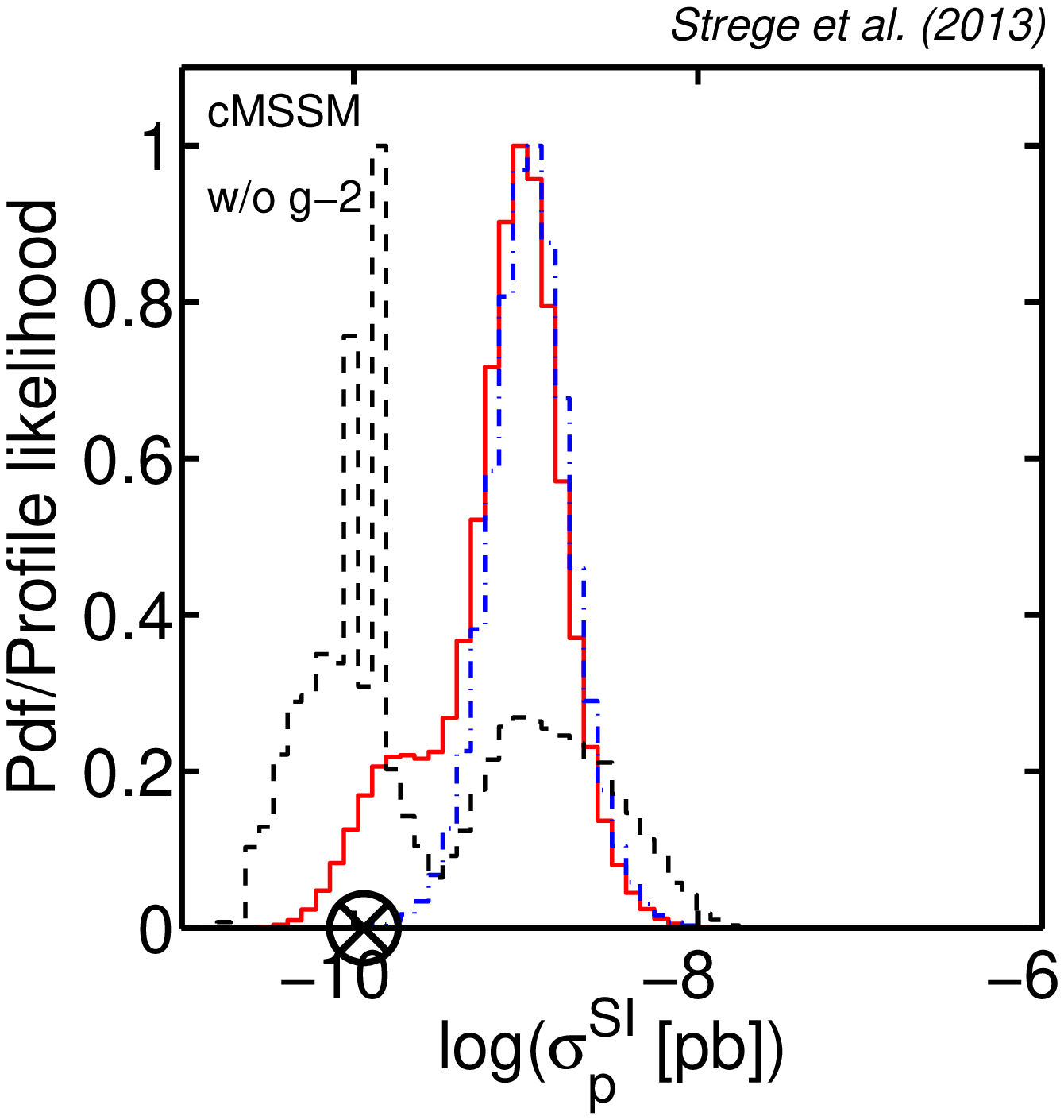} \\
\expandafter\includegraphics\expandafter[\rowoffour]{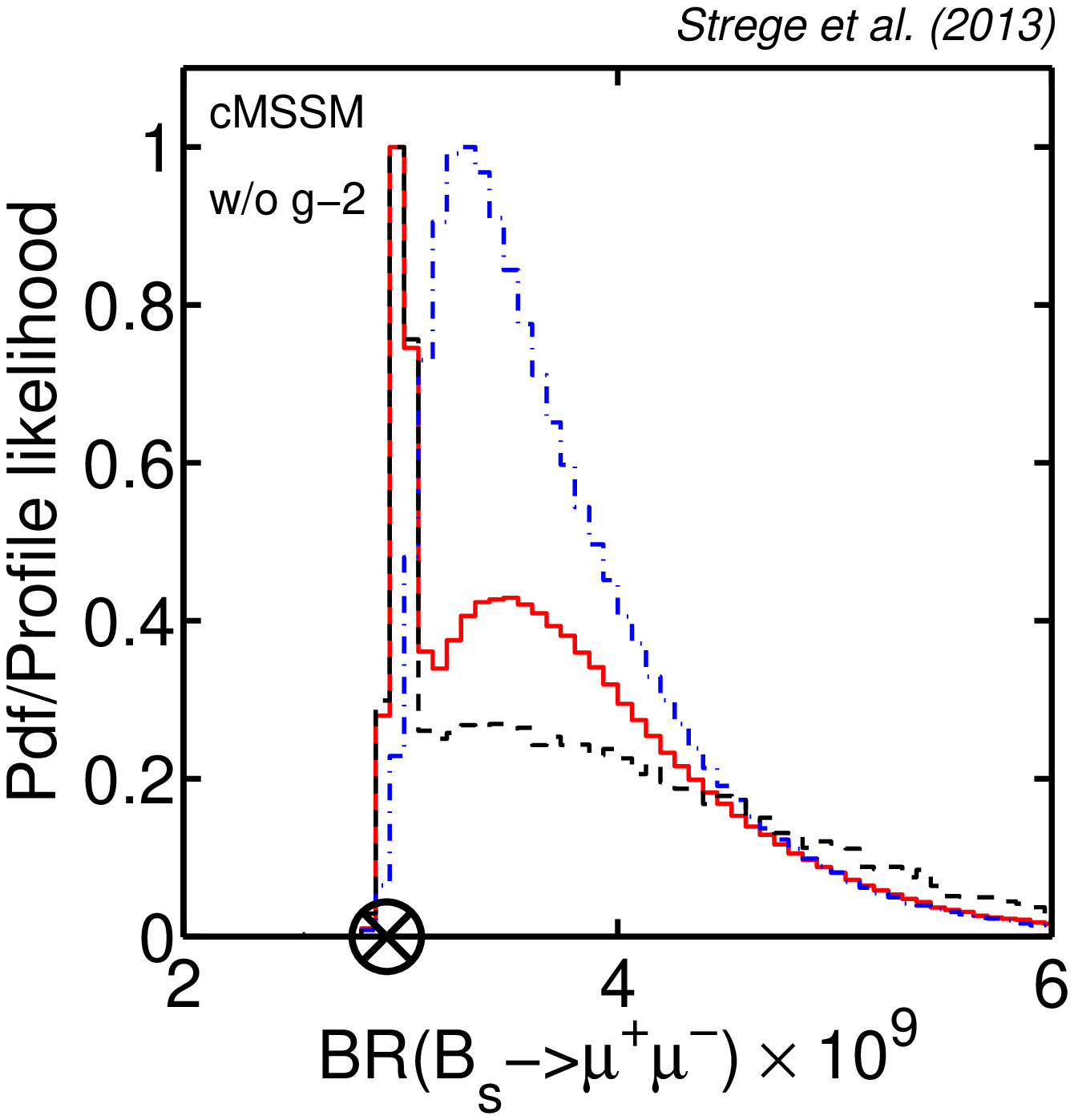} 
\expandafter\includegraphics\expandafter[\rowoffour]{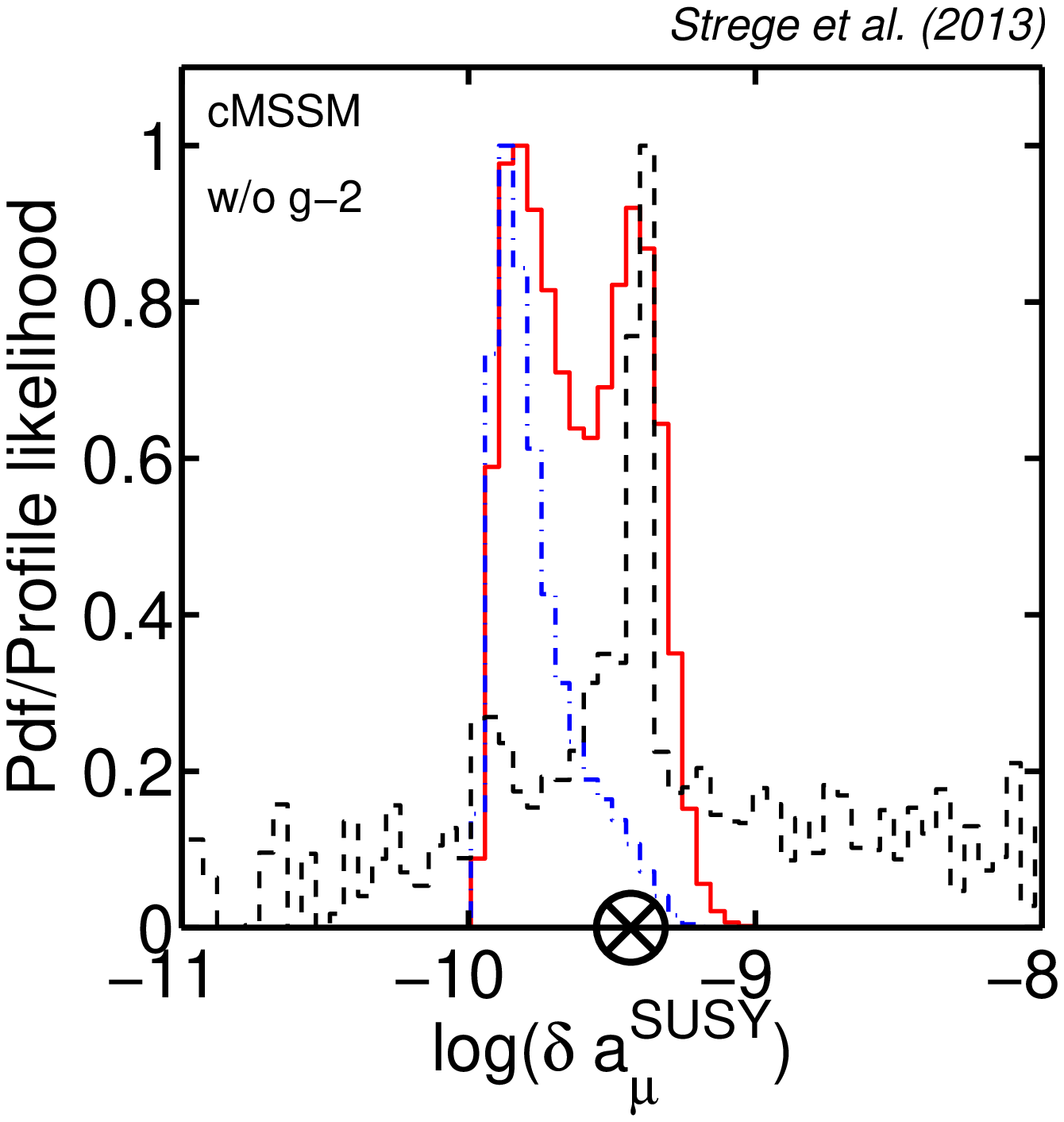} 
\expandafter\includegraphics\expandafter[\rowoffour]{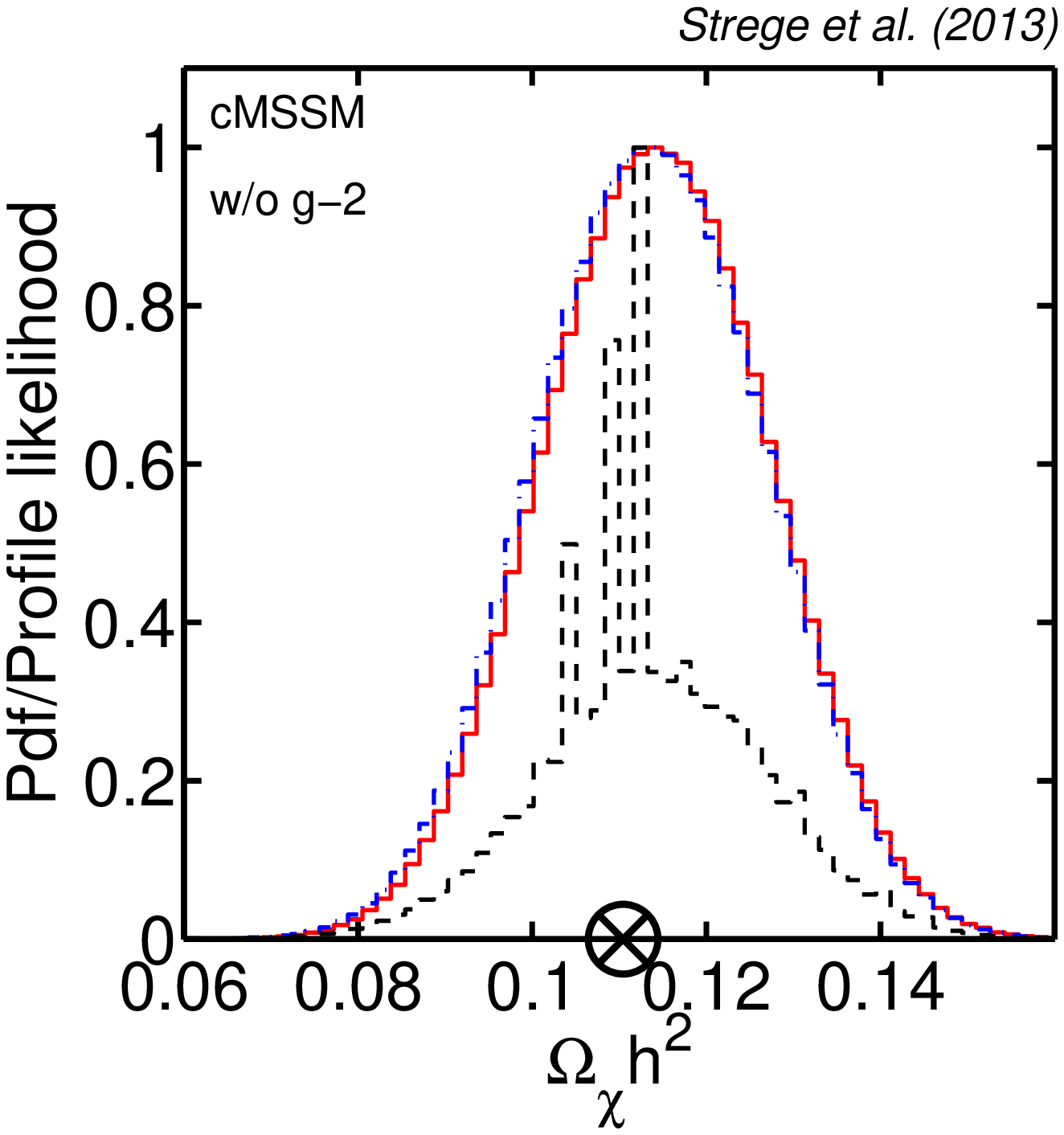} 
\expandafter\includegraphics\expandafter[\rowoffour]{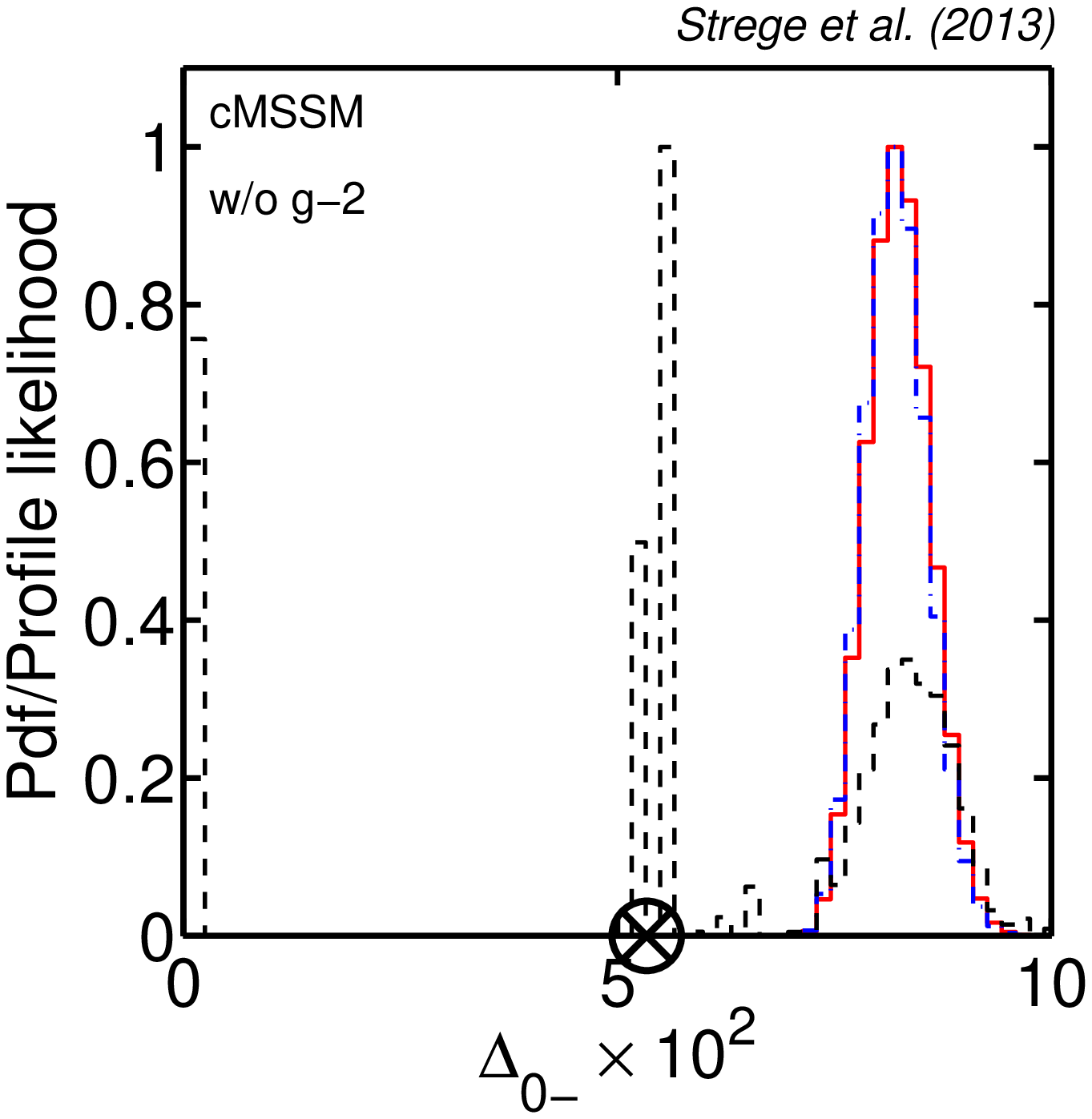} 
\caption{\fontsize{9}{9} \selectfont  As in Fig.~\ref{fig:1D_CMSSM}, 1D marginal pdf for flat priors (dash-dot/blue), log priors (thick solid/red) and 1D profile likelihood (dashed/black) in the cMSSM including all current experimental constraints {\em except the \gmt\ constraint}, for some derived quantities of interest. The best-fit point is indicated by the encircled black cross. \label{fig:1D_CMSSM_no_gmt}} 
\end{figure*}

The 1D marginalised pdfs and profile likelihood for the same derived quantities as in Fig.~\ref{fig:1D_CMSSM} are shown in Fig.~\ref{fig:1D_CMSSM_no_gmt}.  The posterior pdf with flat priors is qualitatively very similar to before. This is to be expected, since large $m_0$ and $m_{1/2}$ which lead to values discrepant with the experimental constraint on \gmt\ were already favoured in the previous analysis. Results for the posterior pdf with log priors also qualitatively agree well with previous results, although a significant shift of posterior probability from the low-mass (in the SC region) to the high-mass (in the AF region) mode can be observed for all sparticle masses. Results for the 1D profile likelihood functions are, instead, very different from before. While the \gmt\ constraint confined the profile likelihood for the sparticle masses to narrow regions around the best-fit point at small masses, after dropping this constraint the profile likelihood function is much more spread out (although the best-fit point hardly changes). While the increased preference for larger sparticle masses for both the posterior distributions and the profile likelihood function slightly worsens detection prospects in this case, the majority of the favoured regions are still accessible for the LHC operating at 14 TeV collision energy.

The posterior pdfs for $m_h$ are almost identical to the distributions shown in Fig.~\ref{fig:1D_CMSSM} (up to numerical noise). The profile likelihood function is shifted to slightly larger values of $m_h$, favouring values of $m_h = [123, 126]$ GeV. This shows that when dropping the \gmt\ constraint from the analysis it is much easier to achieve large $X_t$, fine-tuned to lead to values of $m_h$ that are in reasonably good agreement with the experimental constraint.

The posterior pdfs for $\brbsmumu$ are qualitatively very similar to the results in the previous section. The profile likelihood function however now spreads over a much larger range of values and closely resembles the shape of the posterior pdf with log priors. 

The posterior distributions for $\delta a_{\mu}^{SUSY}$ are almost identical to the distributions shown in Fig.~\ref{fig:1D_CMSSM}. The posterior pdf with log priors still displays a bimodal shape, however now the majority of the posterior mass is found in the mode at lower values of $\delta a_{\mu}^{SUSY}$. Results for the 1D profile likelihood function change significantly. In the absence of the experimental constraint on $\delta a_{\mu}^{SUSY}$ the profile likelihood is no longer pushed to large values. Instead, it now spreads over a large range of \gmt\ values and is in much better agreement with the values favoured by the posterior pdfs (although it still peaks at slightly larger values of $\delta a_{\mu}^{SUSY}$).

Both the posterior pdfs and the profile likelihood for the dark matter relic density are in good agreement with the experimental constraint, similarly to the results of the analysis including the \gmt\ constraint, although now the profile likelihood appears less Gaussian. Likewise, the posterior pdfs for the isospin asymmetry, shown in the left-hand panel of the bottom row, are in good agreement with the previous results. However, the 1D profile likelihood for $\Delta_{0-}$ shows a very different behaviour. This distribution is dominated by a small number of strongly fine-tuned points in parameter space that achieve a negative SUSY contribution to $\Delta_{0-}$, so that this quantity is in better agreement with the experimental constraint, while also reproducing other measurements. These points show up as `spikes' of high likelihood in all panels of Fig.~\ref{fig:1D_CMSSM_no_gmt}. In the cMSSM, a large amount of fine-tuning is required to satisfy the \gmt\ constraint. In the absence of this constraint there is significantly more freedom to find regions in parameter space that are strongly fine-tuned to satisfy other experimental constraints, such as $\Delta_{0-}$. The presence of a small number of fine-tuned points achieving a very high likelihood value also explains the small size of the 2D 68\% confidence level in Fig.~\ref{cMSSM_2D_wogm2}. The 95\% and 99\% C.L. also receive contributions from points at higher $\Delta_{0-}$, and are therefore much more spread out.

\subsection{Comparison with other analyses}

This is the first study that includes the new LHCb measurement of $\brbsmumu$, the most up-to-date CMS constraint on the mass of the Higgs boson (derived from a combination of 5.1 fb$^{-1}$ $\sqrt{s} = 7$ TeV data and 12.2 fb$^{-1}$ $\sqrt{s} = 8$ TeV data), and the 5.8 fb $^{-1}$ integrated luminosity exclusions limit from ATLAS SUSY searches.

Other recent global fits analyses of the cMSSM can be found in Ref.~\cite{Buchmueller:2012hv,Fowlie:2012im,Bechtle:2012zk}. In Ref.~\cite{Buchmueller:2012hv} a frequentist  analysis of the cMSSM is presented that can be compared to our profile likelihood analysis. In Ref.~\cite{Buchmueller:2012hv} the best-fit point is found in the SC region, in agreement with our findings. However, the p-value is significantly lower than ours (0.085 compared to 0.21), perhaps as a consequence of the lower resolution of their scan, which could not find a better best-fit. As a consequence, the profile likelihood contours in Ref.~\cite{Buchmueller:2012hv} encompass the AF region at $\sim 1\sigma$ level even when including the \gmt\ constraint. 
A Bayesian analysis of the cMSSM is found in Ref.~\cite{Fowlie:2012im}. The Bayesian posterior results agree qualitatively with ours, although there are important quantitative differences due to our more constraining data sets. Also, we found the Focus Point region to be disfavoured at $>99\%$ level by XENON100 data, which are not included in Ref.~\cite{Fowlie:2012im}. In contrast to this work, Ref.~\cite{Fowlie:2012im} presents results for only one choice of priors (log priors), and does not discuss the prior dependence of the results. Their best-fit point is found in the AF region, which is excluded at 99\% C.L in our profile likelihood analysis. We notice that the MultiNest settings used in Ref.~\cite{Fowlie:2012im} are inadequate to achieve a reliable exploration of the profile likelihood, as demonstrated by Ref.~\cite{Feroz:2011bj}, which means that the best-fit point found in \cite{Fowlie:2012im} is unlikely to be reliable.
Global fits of the cMSSM from the Fittino group can be found in Ref.\cite{Bechtle:2012zk}, for both the Bayesian and the profile likelihood statistical perspective. This analysis does not include the experimental constraint on the mass of the Higgs boson from the CMS or ATLAS collaborations. A discussion of a potential Higgs discovery at $m_h \approx 126$ GeV is provided, that qualitatively agrees with our results. However, since the experimental and theoretical errors on $m_h$ assumed in Ref.~\cite{Bechtle:2012zk} are significantly larger than in this work, and the most recent limit from the XENON100 experiment is not included in the analysis, our results are not directly comparable.

%
%\begin{figure*}%[htp]
%\expandafter\includegraphics\expandafter
%[\rowofthree]{Plots/cMSSM_wog2/all/cMSSM_flat_wog2_pp_2D_4_blue_sq.eps}
%\expandafter\includegraphics\expandafter
%[\rowofthree]{Plots/cMSSM_wog2/all/cMSSM_log_wog2_pp_2D_4_blue_sq.eps}
%\expandafter\includegraphics\expandafter
%[\rowofthree]{Plots/cMSSM_wog2/all/cMSSM_PL_wog2_pp_2D_4_blue_sq.eps}
%\expandafter\includegraphics\expandafter
%[\rowofthree]{Plots/cMSSM_wog2/all/cMSSM_flat_wog2_pp_2D_5_blue_sq.eps}
%\expandafter\includegraphics\expandafter
%[\rowofthree]{Plots/cMSSM_wog2/all/cMSSM_log_wog2_pp_2D_5_blue_sq.eps}
%\expandafter\includegraphics\expandafter
%[\rowofthree]{Plots/cMSSM_wog2/all/cMSSM_PL_wog2_pp_2D_5_blue_sq.eps}
%\caption{\fontsize{9}{9} \selectfont Implications for the SD scattering cross-section vs neutralino mass in the cMSSM, for a scan without including the \gmt\ constraint. \rt{Probably to be dropped, just say in words that even in this case the SD is out of reach of future detectors. \cs{New plots added in bottom row.}}\label{cMSSM_2D_SD_wgmt}}  
%\end{figure*}

%%%%%%%%%%%%%%%%%%%%%%%%%%%%%%%%%%%%%%%%%%%%%%%%%%%%%%%%%%%%%%%%%%%%%%%
%
% NUHM
%
%%%%%%%%%%%%%%%%%%%%%%%%%%%%%%%%%%%%%%%%%%%%%%%%%%%%%%%%%%%%%%%%%%%%%%%

\section{Results for the NUHM}
\label{sec:NUHM}

\subsection{Impact of all present-day experimental constraints}

\begin{figure*}%[htp]
\centering
\expandafter\includegraphics\expandafter
[\rowofthree]{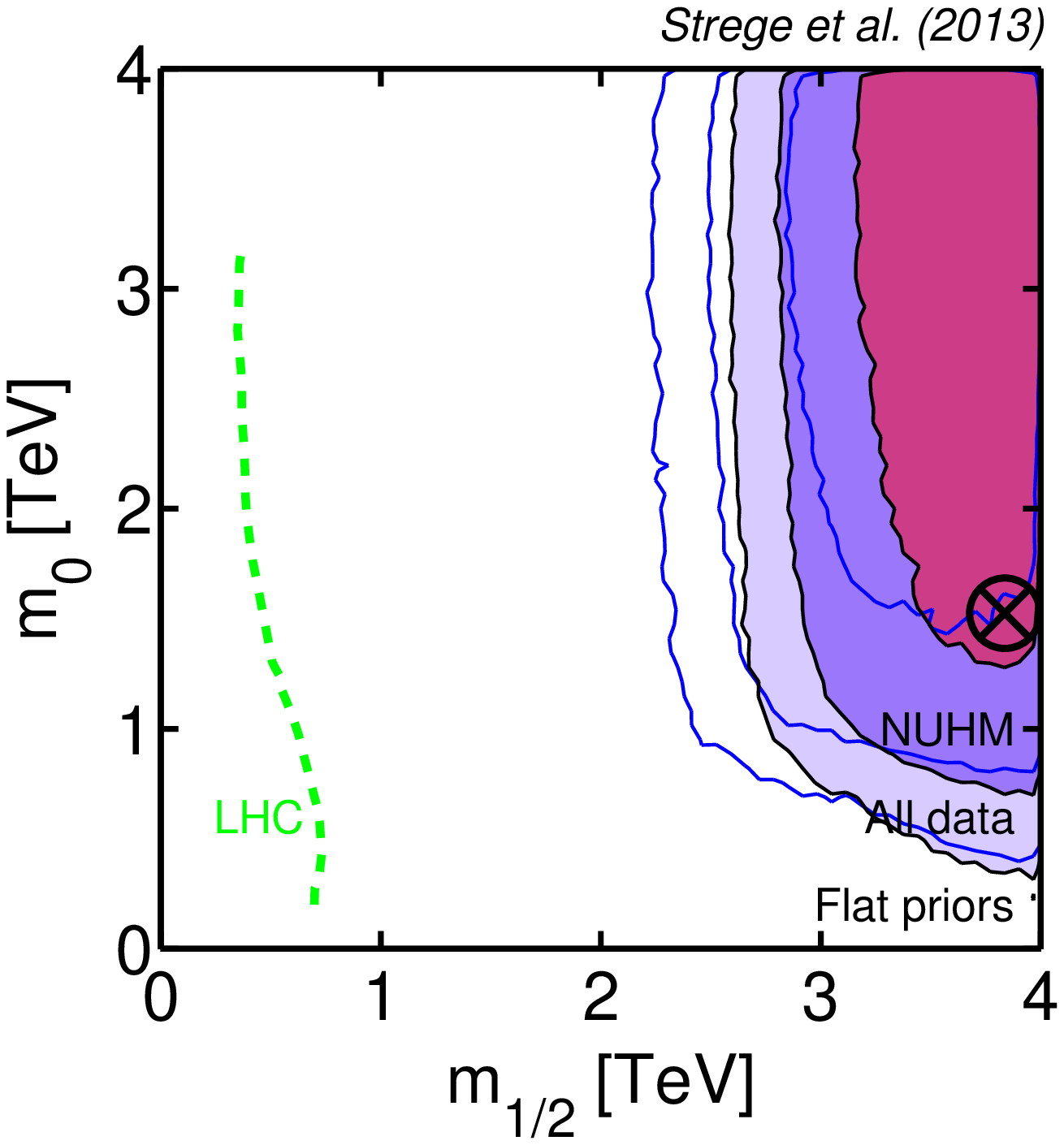}
\expandafter\includegraphics\expandafter
[\rowofthree]{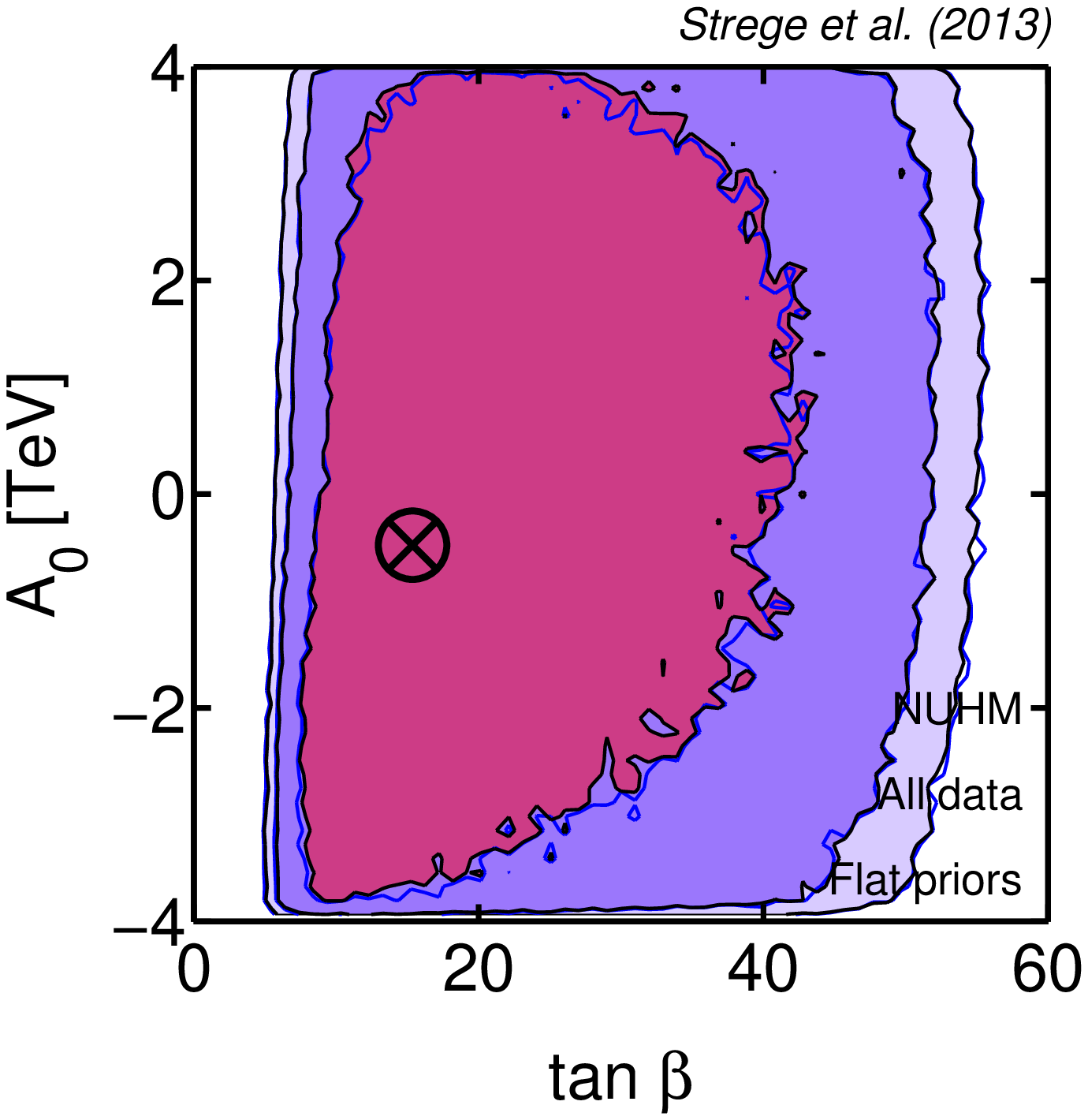}
\expandafter\includegraphics\expandafter
[\rowofthree]{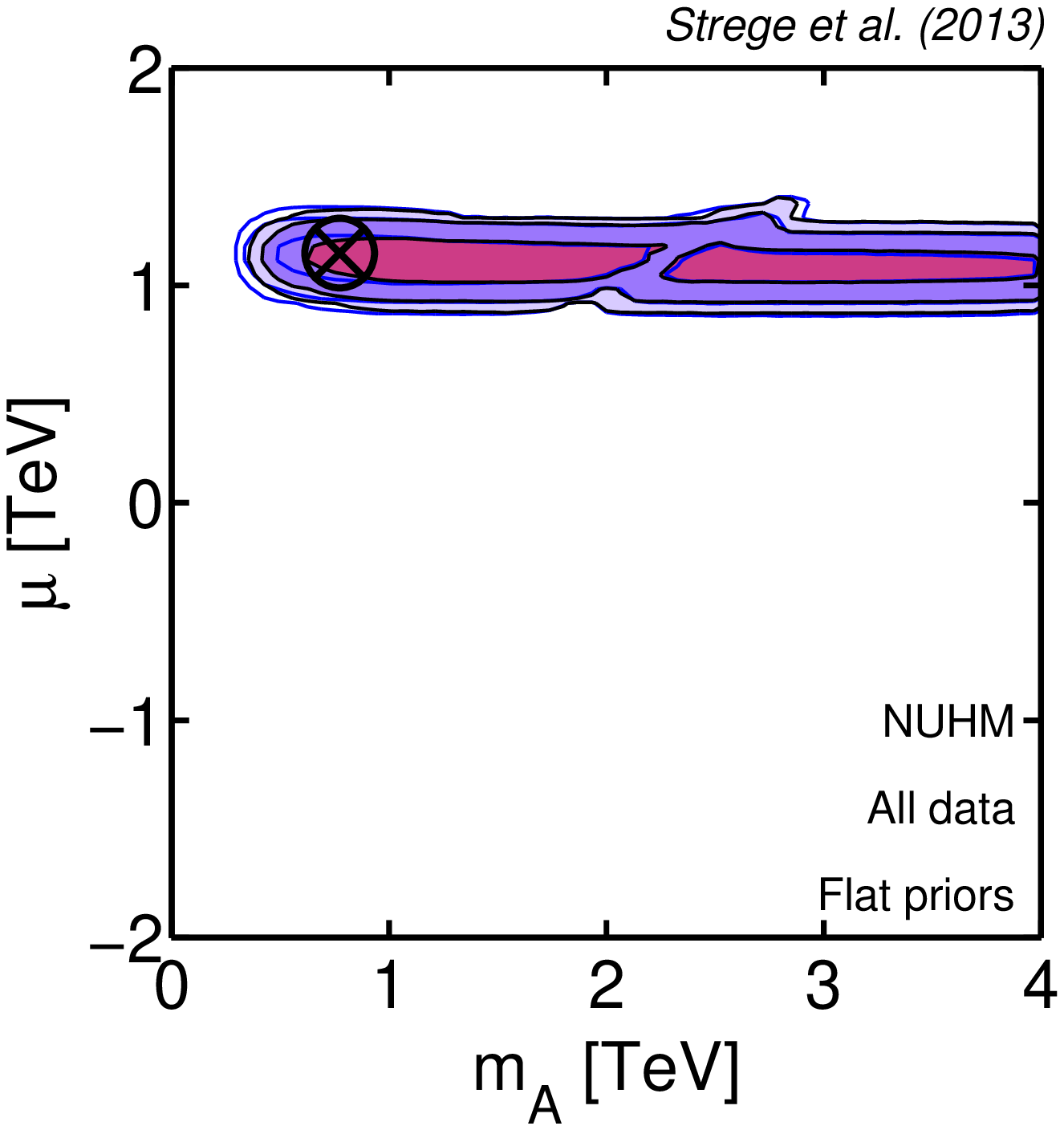} \\
\expandafter\includegraphics\expandafter
[\rowofthree]{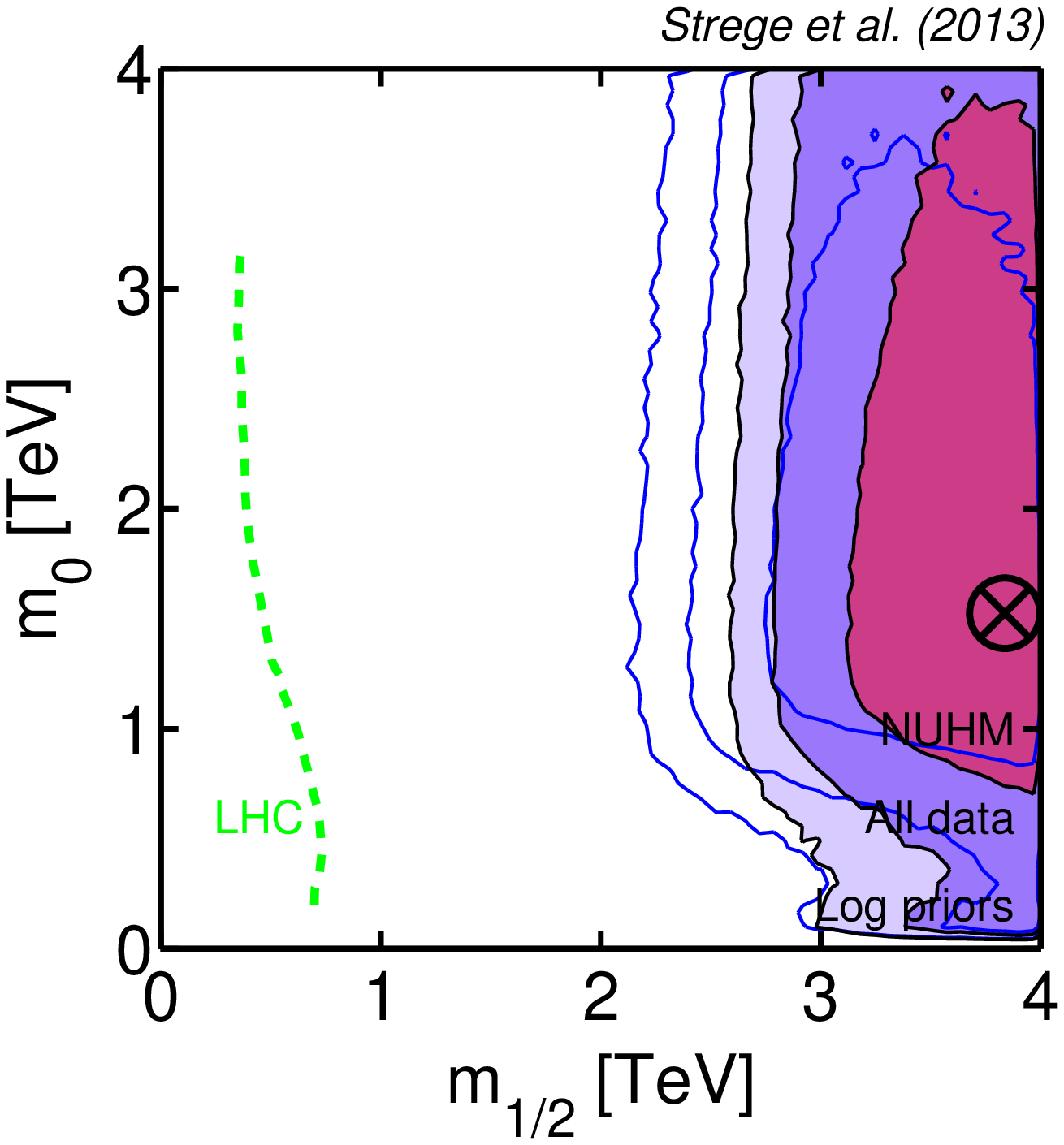}
\expandafter\includegraphics\expandafter
[\rowofthree]{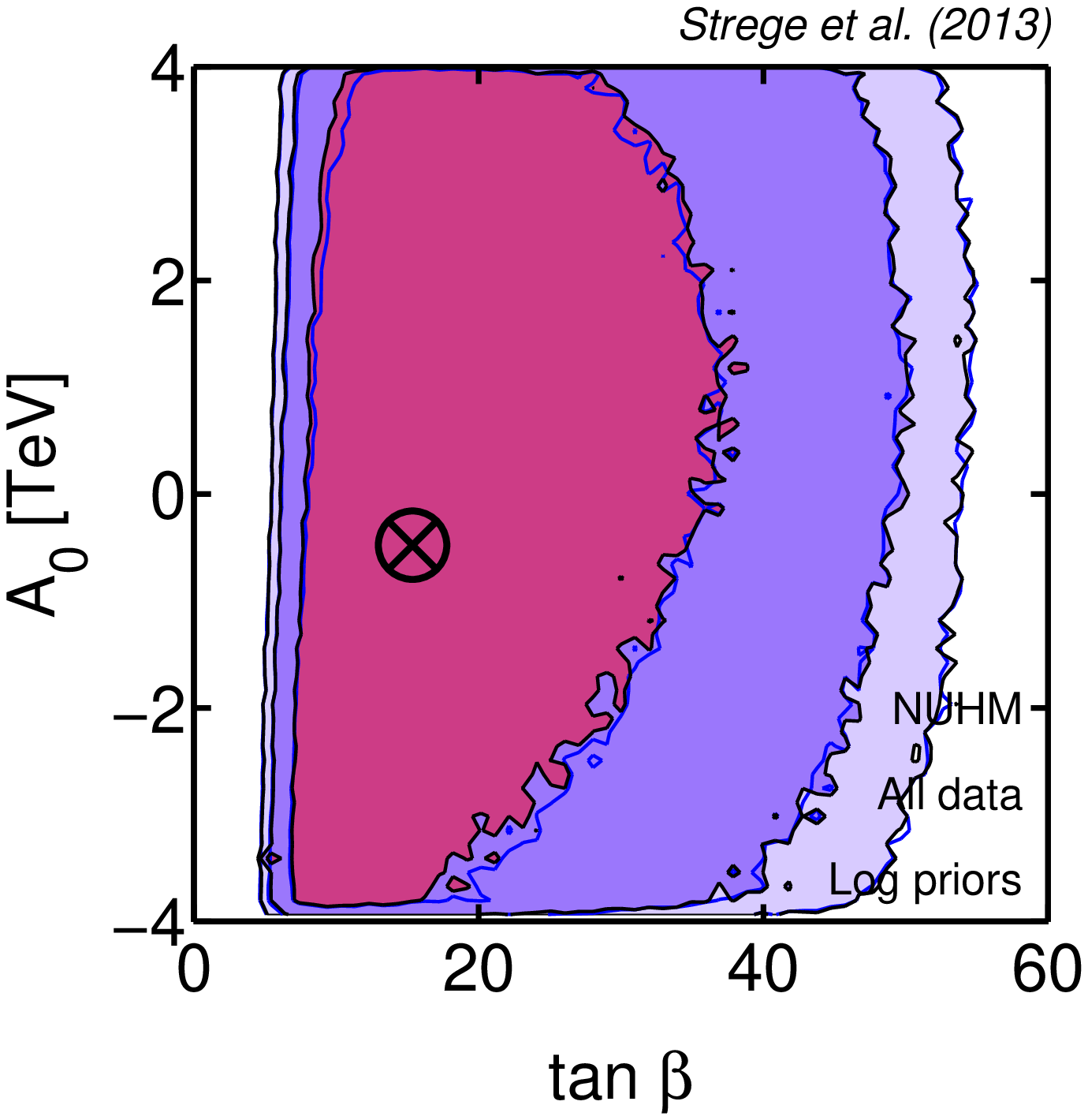}
\expandafter\includegraphics\expandafter
[\rowofthree]{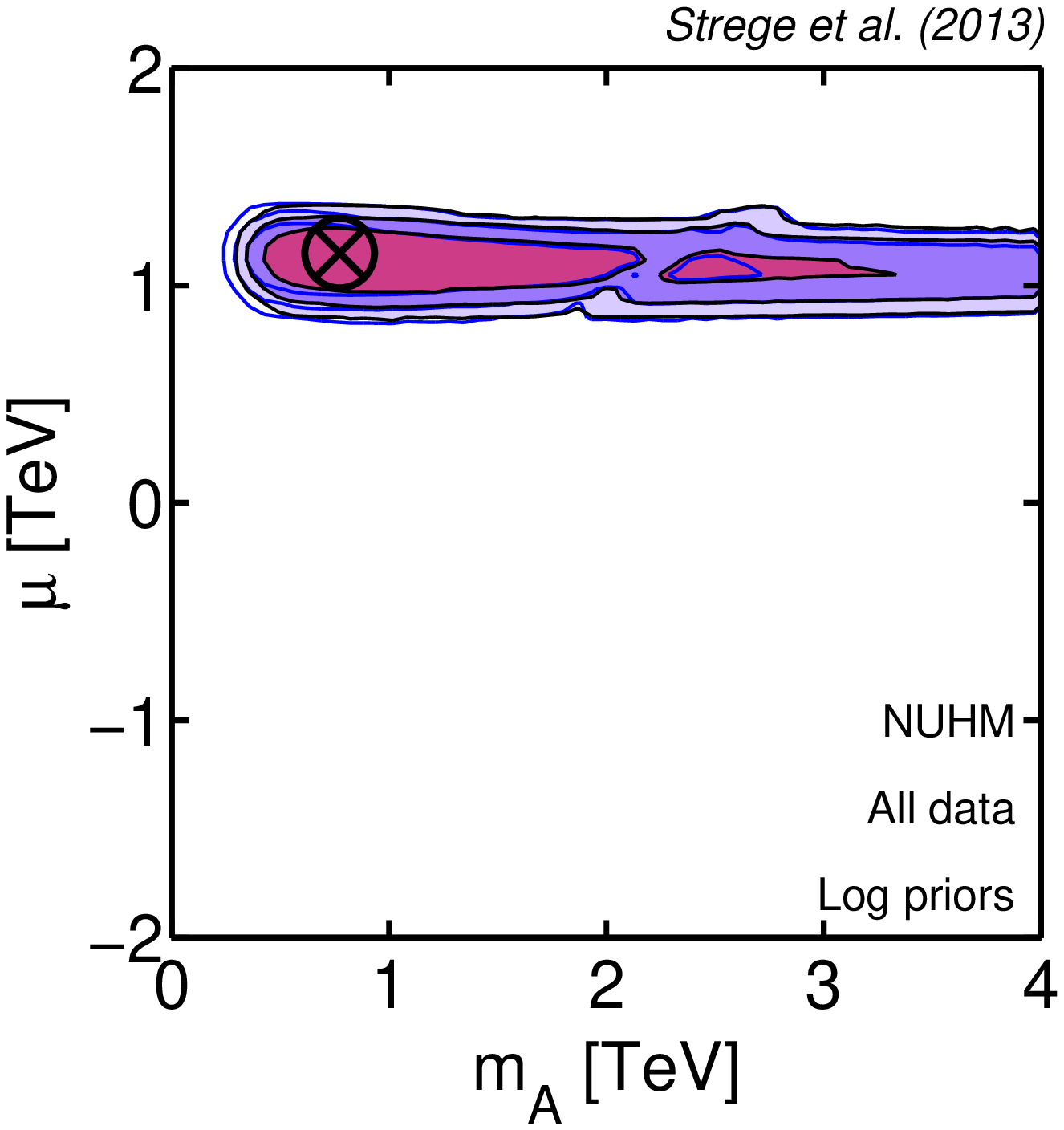}\\
\expandafter\includegraphics\expandafter
[\rowofthree]{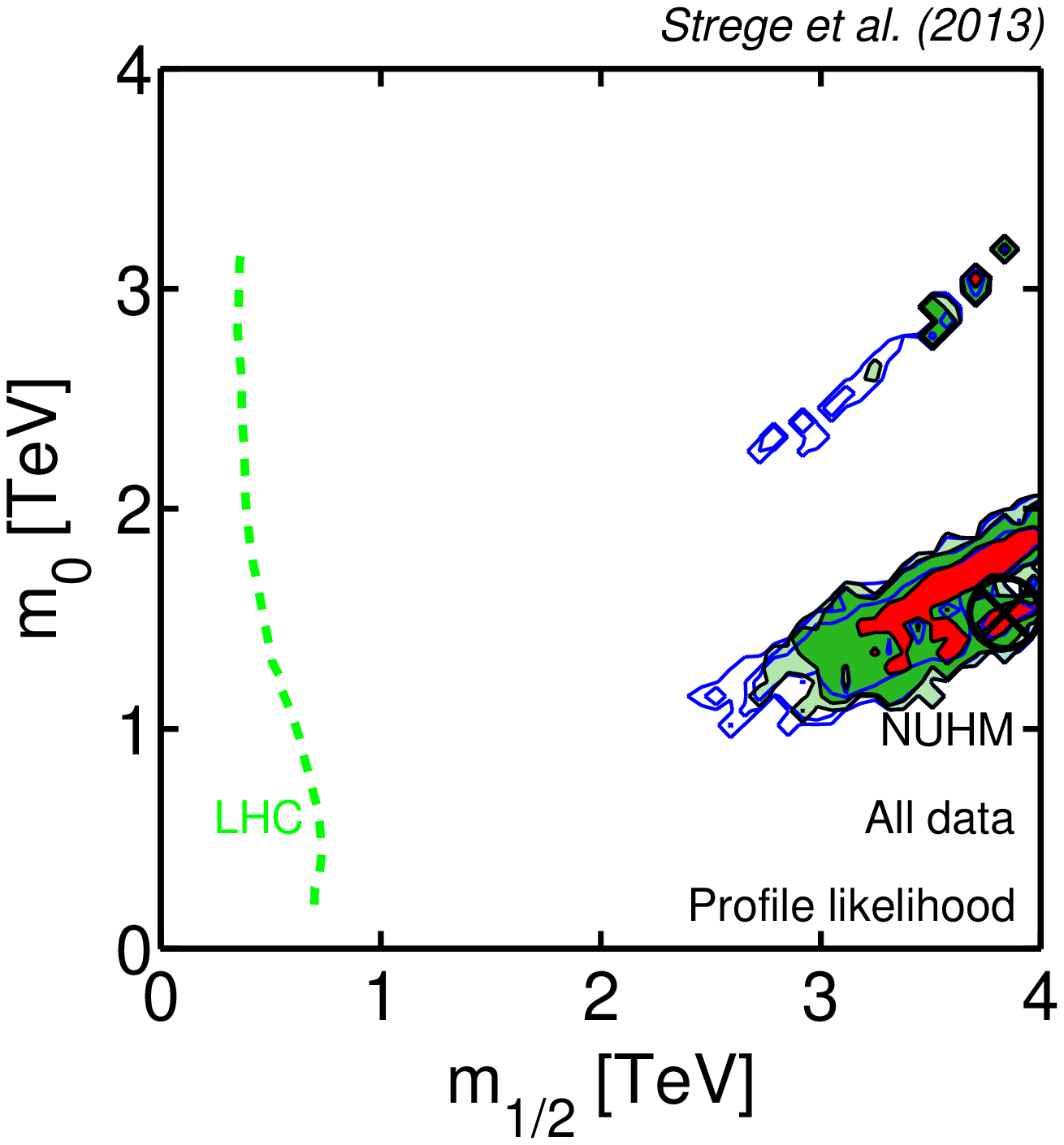} 
\expandafter\includegraphics\expandafter
[\rowofthree]{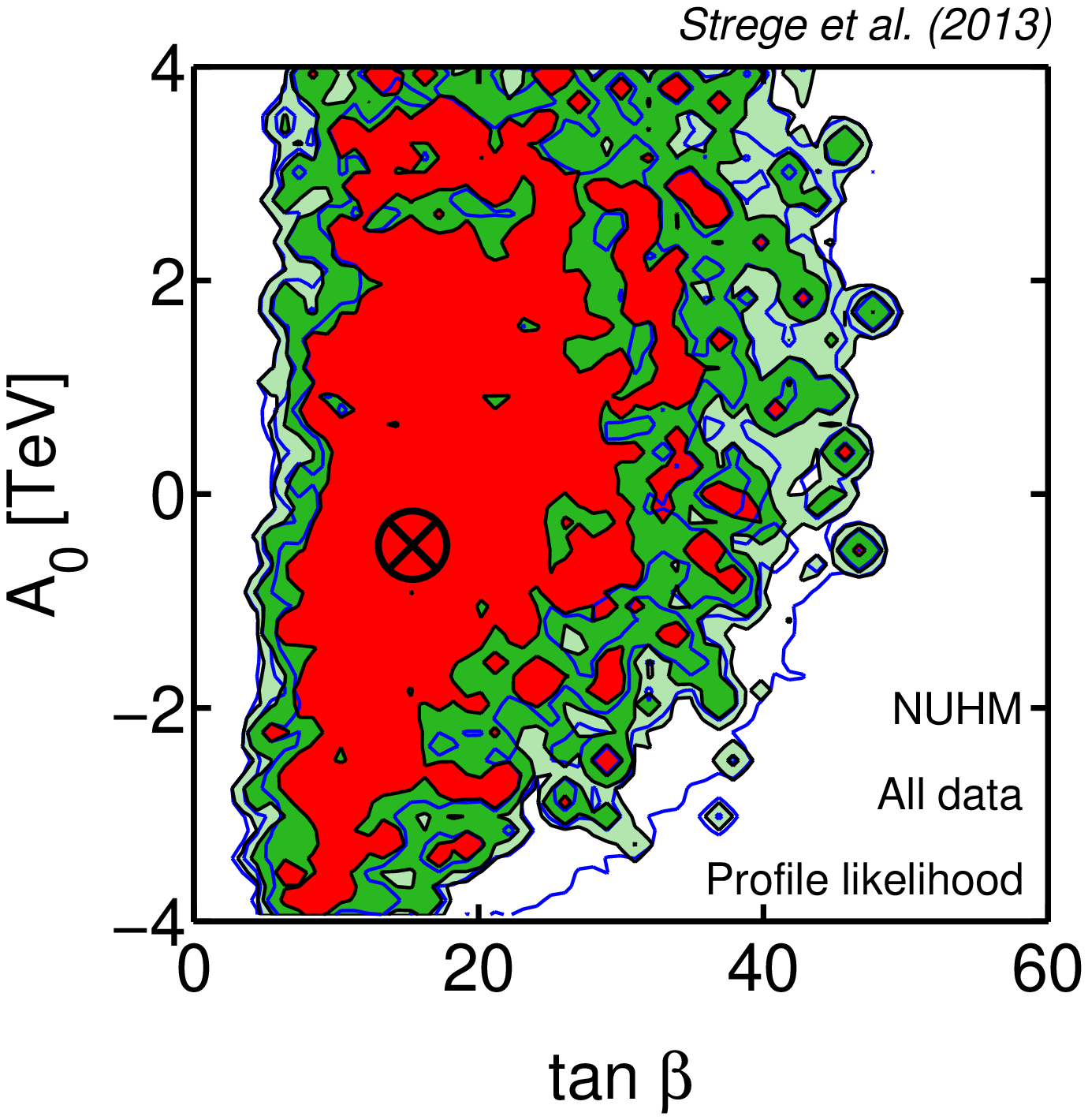} 
\expandafter\includegraphics\expandafter
[\rowofthree]{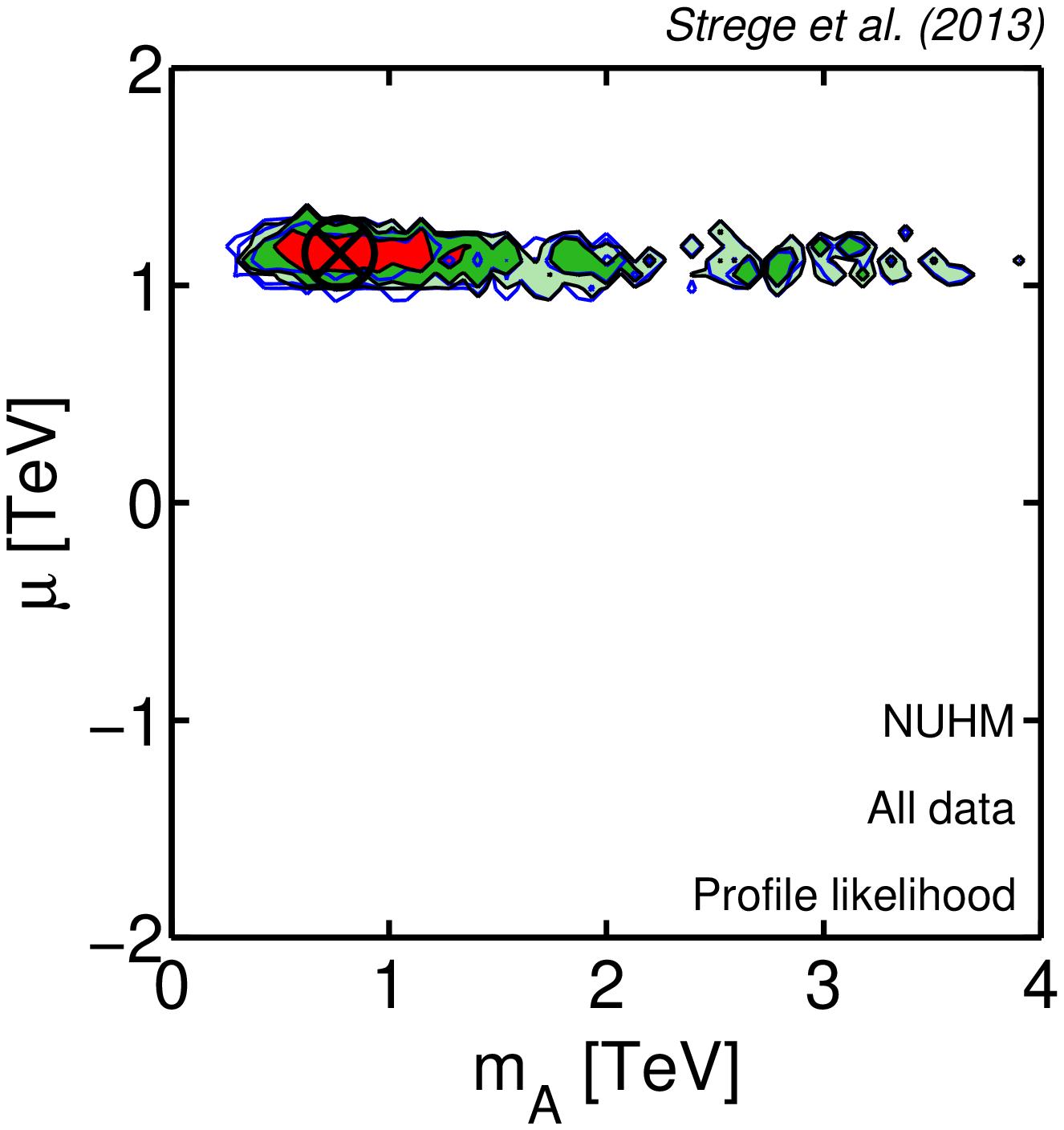} 
\caption{\fontsize{9}{9} \selectfont  
\selectfont Constraints on the NUHM parameters including all available present-day data (WMAP 7-year, LHC \fiveinvfb\ SUSY null search and Higgs detection, XENON100 2012 direct detection limits). Black, filled contours depict the marginalised posterior pdf (top row: flat priors; middle row: log priors) and the profile likelihood (bottom row), showing 68\%, 95\% and 99\% credible/confidence regions. The encircled black cross is the overall best-fit point, obtained from approximately 200M likelihood evaluations. Blue/empty contours show constraints without the latest XENON100 results, for comparison. In the left-hand plots, the dashed/green line shows the current LHC 95\% exclusion limit. \label{NUHM2D}}  
\end{figure*}

The constraints on the NUHM parameters obtained from all present-day data sets, including the LHC 5.8 fb$^{-1}$ exclusion limit and the constraint on the mass of the lightest Higgs boson, are shown in Fig. \ref{NUHM2D}. This figure also compares the results to the constraints obtained without inclusion of the XENON100 2012 results. Results are shown in the $(m_{1/2},m_0)$ plane, the $(\tan \beta,A_0)$ plane and the $(m_A,\mu)$ plane, in terms of the posterior pdf for flat (log) priors (top and central row) and of the profile likelihood (bottom row).

\begin{figure*}%[htp]
\centering
\expandafter\includegraphics\expandafter[\rowofthree]{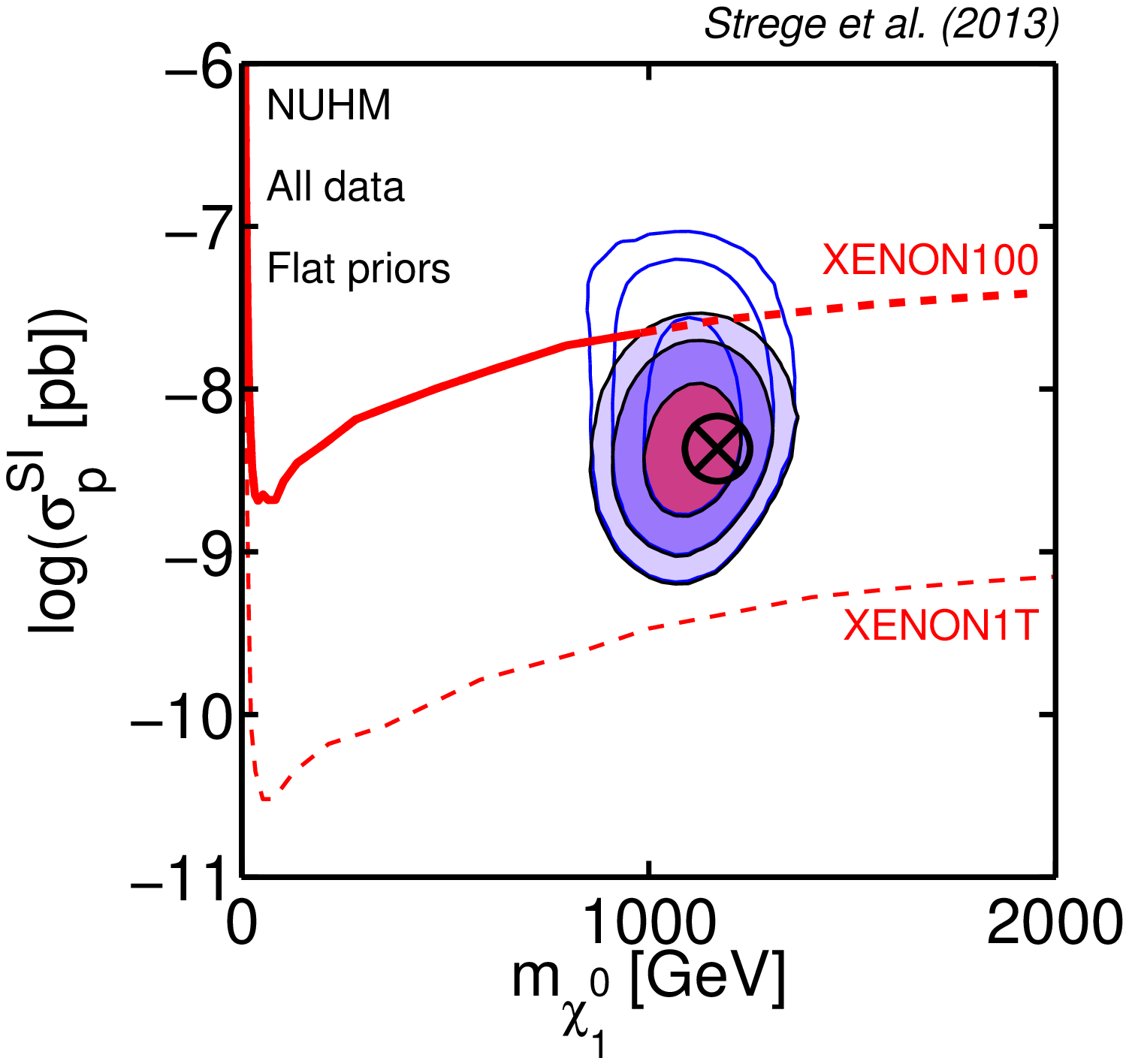}
%\expandafter\includegraphics\expandafter
%[\rowofthree]{Plots/NUHM/all/NUHM_flat_pp_2D_5_sq.eps}
%\expandafter\includegraphics\expandafter
%[\rowofthree]{Plots/NUHM/all/NUHM_flat_pp_2D_6_sq.eps}\\
\expandafter\includegraphics\expandafter[\rowofthree]{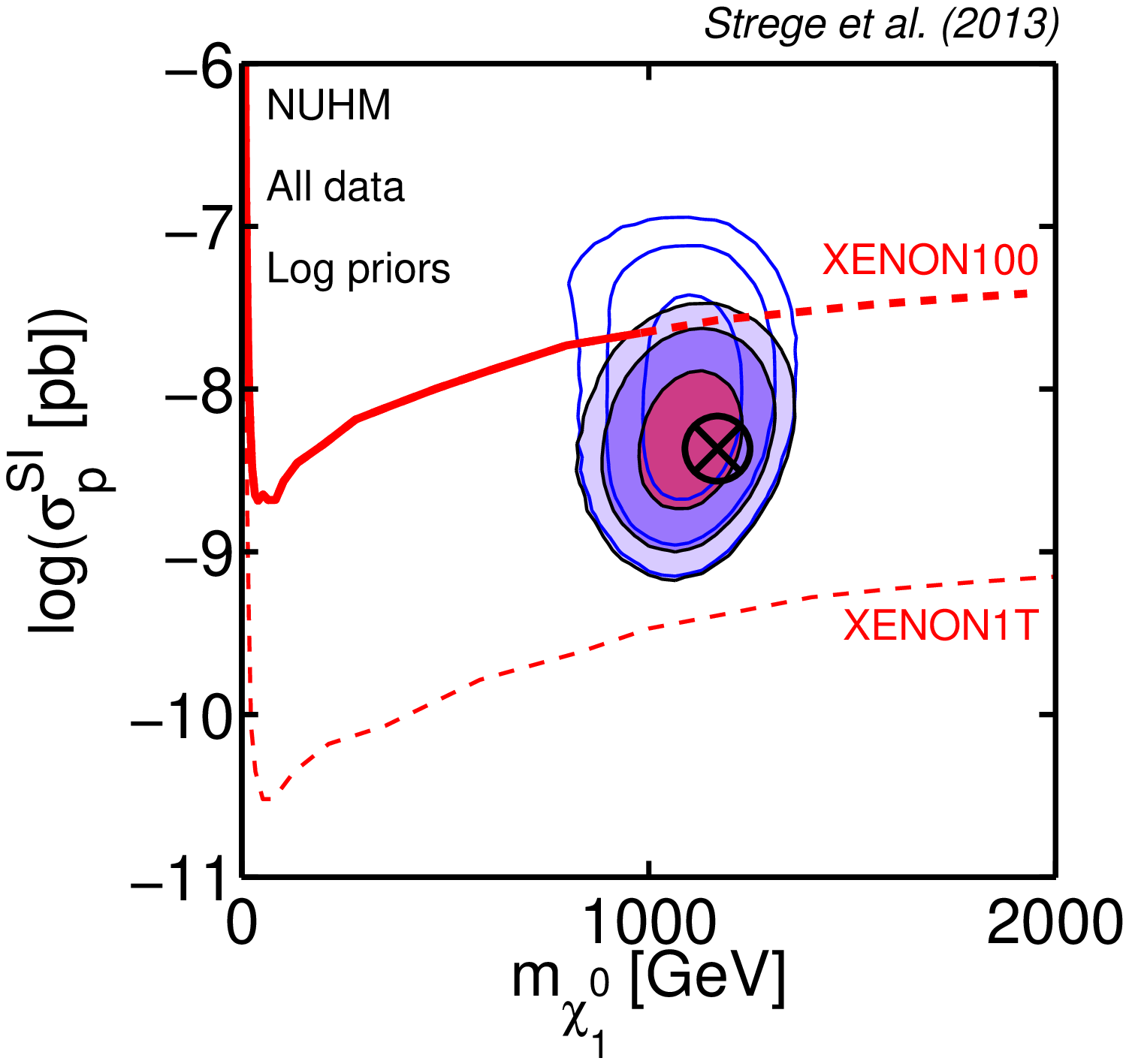}
%\expandafter\includegraphics\expandafter
%[\rowofthree]{Plots/NUHM/all/NUHM_log_pp_2D_5_sq.eps}
%\expandafter\includegraphics\expandafter
%[\rowofthree]{Plots/NUHM/all/NUHM_log_pp_2D_6_sq.eps}\\
\expandafter\includegraphics\expandafter[\rowofthree]{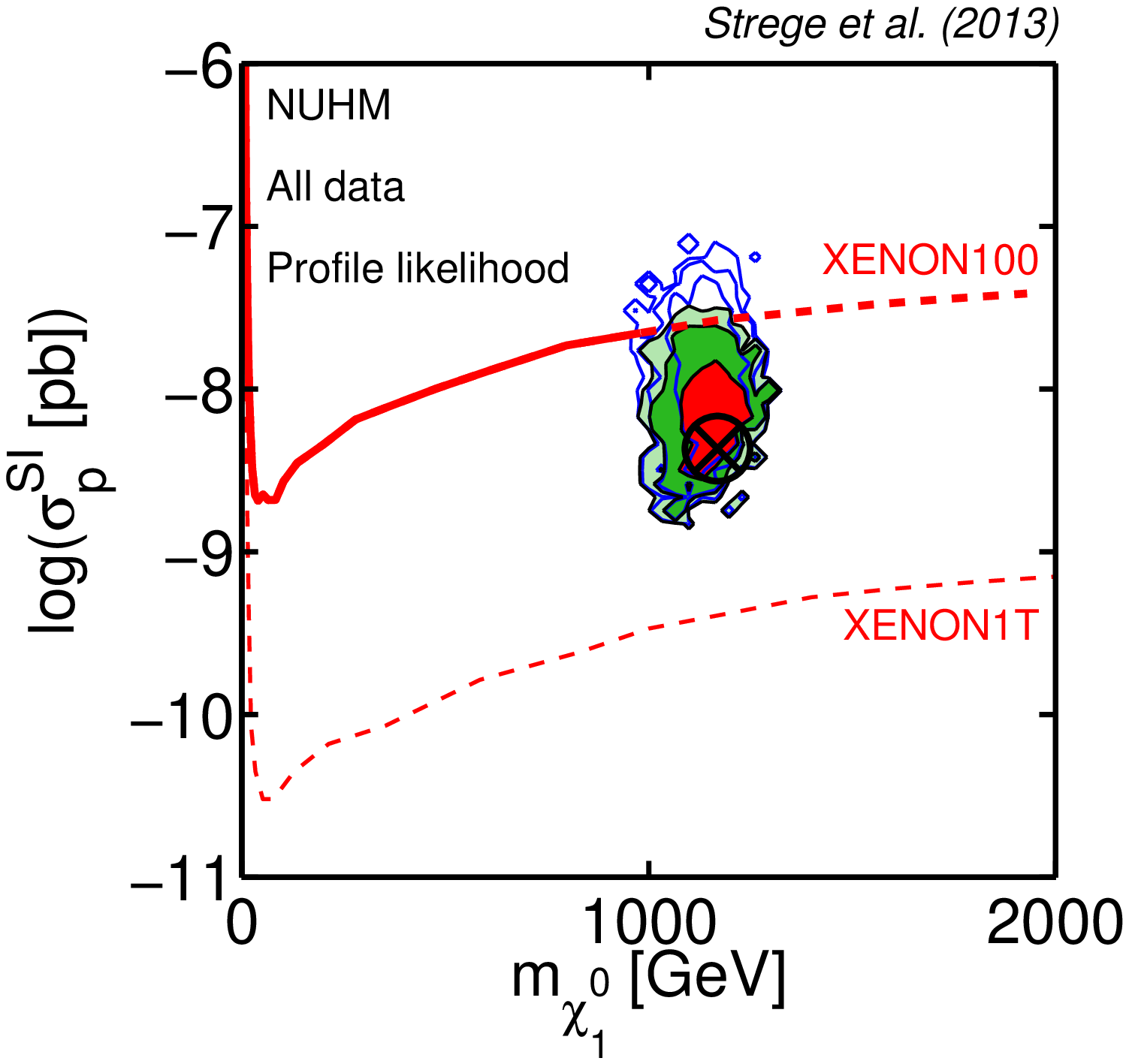} 
%\expandafter\includegraphics\expandafter
%[\rowofthree]{Plots/NUHM/all/NUHM_PL_pp_2D_5_sq.eps} 
%\expandafter\includegraphics\expandafter
%[\rowofthree]{Plots/NUHM/all/NUHM_PL_pp_2D_6_sq.eps}
\caption{\fontsize{9}{9} \selectfont  Constraints on the neutralino mass and SI scattering cross-section in the NUHM, including all available present-day data (WMAP 7-year, LHC \fiveinvfb\ SUSY null search and Higgs detection, XENON100 2012 direct detection limits). Black, filled contours depict the marginalised posterior pdf (left: flat priors; middle: log priors) and the profile likelihood (right), showing 68\%, 95\% and 99\% credible/confidence regions. The encircled black cross is the overall best-fit point. Blue/empty contours show constraints without the latest XENON100 results. The 90\% XENON100 exclusion limit (from Ref.~\cite{Aprile:2012nq}) is shown as a red/solid line. Ref.~\cite{Aprile:2012nq} only shows the limit for $m_{\neut} < 1000$ GeV; we show the extension of this limit to higher WIMP masses as a red/dashed line. We also show the expected reach of XENON1T as a red/dashed line. 
\label{NUHM2D_derived}}  
\end{figure*}

The posterior distributions for both log and flat priors strongly favour large values of the mass parameters, especially $m_{1/2}$. The LHC exclusion limit has essentially no impact on the parameter space, since values of $m_{1/2}$ above the LHC limits are already favoured by other constraints. The $m_h$ constraint plays a dominant role. As in the cMSSM, the mass of the lightest Higgs boson scales with $m_{1/2}$. As a result, values of $m_{1/2} < 2$ TeV are excluded at 99\% level. Since the mass of the lightest Higgs is not very sensitive to $m_0$, almost the entire prior range of $m_0$ is allowed, with the exception of very small values. Posterior distributions for the two different prior choices agree quite well, although contours for the posterior pdf with flat priors are shifted towards larger values of $m_0$, due to volume effects.

In the $(\tan \beta,A_0)$ plane (top central panels) only very limited constraints are placed on the parameters. The 68\% credible interval spans almost the entire prior range of $A_0$, for both the posterior pdfs with log and flat priors. Values of $\tan \beta < 40$ are favoured, but larger values $\tan \beta \approx 50$ are still allowed at 99\% level.

The Higgsino mass parameter $\mu$ is strongly constrained to values $\mu \sim 1$ TeV, independent of the choice of prior.  As the neutralino is Higgsino-like (as shown below), the WMAP relic abundance fixes its mass and hence $\mu$. In contrast, $m_A$ is almost unconstrained within the prior range, only very small values of $m_A$ are ruled out, mainly as a consequence of the $\brbsmumu$ constraint~\cite{Ellis:2006jy}. For the flat prior, the $68\%$ contour stretches to larger values of $m_A$, as expected, since the flat prior gives a large {\it a priori} statistical weight to large masses. The log scan shows a slight preference for values $m_A < 3$ TeV, but the $95\%$ region still touches the upper prior boundary of $m_A$. %\rr{It seems that  heavy binos are not favoured though mA is free to achieve the AF. I think this requirement is subdominant compared with the Higgsino scenario. }

Turning now to the profile likelihood (bottom row of Fig. \ref{NUHM2D}),  results are qualitatively similar to the Bayesian pdf, but the favoured region in the $(m_{1/2},m_0)$ plane is much more localised.  A diagonal region at large $m_{1/2} > 2$ TeV and intermediate values of $m_0 = [1,2]$ TeV is favoured. In addition, a small island at 68\% confidence survives at larger $m_0$. Results in the $(\tan \beta,A_0)$  and $(m_A,\mu)$ planes are also qualitatively similar to what has been discussed above for the posterior pdf, although slightly more localised.

The main impact of the new XENON100 limit is to push contours towards larger values of $m_0$. The regions ruled out correspond to the Focus Point region, which leads to large SI cross-sections and is therefore disfavoured by the XENON100 limit, as in the cMSSM (see Ref.~\cite{Bertone:2011nj} for a detailed discussion). The other parameters shown in Fig.~\ref{NUHM2D} are relatively insensitive to this limit.

The coordinates of the best-fit point are given in the left column of Table \ref{NUHMbf1}, the corresponding contribution to the total $\chi^2$ by the individual observables is given in Table \ref{NUHMbf2}. The best-fit point is found at large $m_{1/2}$ and intermediate $m_0$. It corresponds to a slightly negative $A_0$, and a small $\tan\beta$ value.  However, given the large extent of the 68\% confidence region in the $(A_0, \tan\beta)$ plane, there are many other values of $A_0$ (and up to $\tan\beta \lesssim 40$) that deliver a comparably good quality of fit. As was already the case for the cMSSM, the largest contribution to the overall best-fit $\chi^2$ results from the isospin asymmetry $\DeltaO$ (see below). Other experimental constraints are in good agreement with the best-fit point. The p-value corresponding to this best-fit point is $0.26$, so that from the frequentist statistical perspective this model cannot be ruled out.
\begin{table*}
\begin{center}
%\begin{tabular}{|l|ll| l l}
\begin{tabular}{| l | c | c |}
\hline
\multirow{3}{*}{NUHM} & \multirow{2}{*}{LHC 2012 +} & LHC 2012 + \\
&  \multirow{2}{*}{XENON100} & XENON100 \\
& & w/o $\delta a_\mu^{SUSY}$ \\\hline 
\multicolumn{3}{|c|}{Best-fit NUHM parameters}\\\hline
$ m_0$ [GeV] & 1524.76 & 3411.36\\
$ m_{1/2}$ [GeV] & 3836.97 & 3911.16\\
$A_0$ [GeV] & -478.54 & -3519.45\\
$\mu$ [GeV] & 1149.27 & 1132.91\\
$m_A$ [GeV] & 773.47 & 681.35\\
$ \tan\beta$ & 15.37 & 9.38\\
\hline 
\multicolumn{3}{|c|}{Best-fit nuisance parameters}\\\hline
$M_t$ [GeV] & 173.809 & 173.380\\
$m_b(m_b)^{\bar{MS}}$ [GeV] & 4.240 & 4.219\\
$[\alpha_{em}(M_Z)^{\bar{MS}}]^{-1}$ & 127.949 & 127.956\\
$\alpha_s(M_Z)^{\bar{MS}}$ & 0.118 & 0.117\\ \hline
$\rho_{\loc}$ [GeV/cm$^3$] & 0.387 & 0.390 \\
$v_{\lsr}$ [km/s] & 227.0 & 226.5 \\
$v_{\esc}$ [km/s] & 544.7 & 539.5 \\
$v_d$ [km/s] & 261.2 & 278.9 \\ \hline
$f_{Tu} \times 10^{2}$ & 2.788 & 2.819 \\
$f_{Td} \times 10^{2}$ & 3.699 & 3.985 \\
$f_{Ts}$ & 0.360 & 0.359 \\ \hline
\multicolumn{3}{|c|}{Best-fit observables}\\
\hline
$m_h$ [GeV] & 126.1 & 126.2\\
$m_\neut$ [GeV] & 1169.1 & 1153.0\\
$\delta a_\mu^{SUSY} \times 10^{10}$ & 27.92 & 0.19\\ 
$BR(\bar{B} \rightarrow X_s\gamma) \times 10^4$ & 3.59 & 3.64\\
$\DeltaO  \times 10^{2}$ & 7.45 & 7.40\\ 
$\brbsmumu \times 10^{9}$ & 2.73 & 2.71\\
$\Dstaunu \times 10^{2}$ & 5.10 & 5.10\\
$\Dsmunu \times 10^{3}$ & 5.23 & 5.24\\
$\Dmunu \times 10^{4}$ & 3.85 & 3.85\\
$\sigmaSI$ [pb] & $4.3 \times 10^{-9}$ & $3.9 \times 10^{-9}$\\
$\sigmaSD$ [pb] & $3.0 \times 10^{-7}$ & $2.6 \times 10^{-7}$\\
$\Omega_\neut h^2$ & 0.1159 & 0.1123\\
\hline
\end{tabular}
\end{center}
\caption{Best-fit model parameters (top section), nuisance parameters (central section) and derived observables (bottom section) in the NUHM. Data included in each column is as in Table~\ref{cMSSMbf1}.
 \label{NUHMbf1}}
\end{table*}

\begin{table*}
\begin{center}
\begin{tabular}{| l | c | c |}
\hline
\multirow{3}{*}{NUHM} & \multirow{2}{*}{LHC 2012 +} & LHC 2012 + \\
& \multirow{2}{*}{XENON100} & XENON100 \\
& & w/o $\delta a_\mu^{SUSY}$ \\\hline 
& \multicolumn{2}{c|}{Gaussian constraints} \\
\hline
SM nuisance (4 parameters) & 0.831 & 0.199\\
Astro nuisance (4 parameters) & 0.343 & 0.051\\
Hadronic nuisance (3 parameters) & 0.215 & 0.118\\
$M_W$ & 1.539 & 1.745\\
$\sin^2\theta_{eff}$ & 0.022 & 0.009\\
$\delta a_\mu^{SUSY} $ & 0.009 & N/A\\
$BR(\bar{B} \rightarrow X_s\gamma) $ & 0.008 & 0.053\\
$\Delta M_{B_s}$ & 0.132 & 0.132\\
$\RBtaunu$  & 1.453 & 1.405\\
$\DeltaO$  & 3.570 & 3.500\\
$\RBDtaunuBDenu $ & 0.816 & 0.807\\
$\Rl$ & 0.017 & 0.019\\
$\Dstaunu $ & 0.544 & 0.542\\
$\Dsmunu  $ & 1.472 & 1.468\\
$\Dmunu $  & 0.008 & 0.008\\
$\Omega_\neut h^2$ & 0.141 & 0.011\\
$m_h$ & 0.021 & 0.032\\
$\brbsmumu$ & 0.096 & 0.105\\ 
\hline
& \multicolumn{2}{c|}{Exclusion limits} \\\hline
XENON100 & 0.070 & 0.033\\
LHC & 0.0 & 0.0\\ 
Sparticles (LEP) & 0.0 & 0.0\\ \hline
Total $\chi^2$ & 11.31 & 10.24\\
Total Gaussian $\chi^2$ (dof) & 11.24 (9) & 10.21 (8) \\
Gaussian $\chi^2/$dof & 1.25 & 1.28 \\
p-value (Gaussian constraints only) & 0.26 & 0.25 \\
\hline
\end{tabular}
\end{center}
\caption{Breakdown of the total best-fit $\chi^2$ by observable for the NUHM. Data included in each column is as in Table~\ref{cMSSMbf2}. \label{NUHMbf2}}
\end{table*}

\subsection{Implications for direct detection and future SUSY searches} 

The favoured regions in the $(m_{\neut},\sigmaSI)$ plane are shown in Fig. \ref{NUHM2D_derived}.  Before inclusion of the new XENON100 limit, the posterior distributions (blue/empty contours) favour spin-independent cross-sections in the range $\sigmaSI = [10^{-7},10^{-9}]$ pb, and a relatively small range of neutralino masses around $m_{\neut} \sim 1$ TeV, as a consequence of the Higgsino-like character of the neutralino. As can be seen by comparing the blue and the black contours, the XENON100 2012 limit (red/solid line) rules out part of this otherwise unconstrained region.  The picture is similar in terms of the profile likelihood. We also display the expected 90\% exclusion limit from the future XENON1T direct detection experiment. XENON1T will probe the entire currently favoured NUHM parameter space, independently of the statistical perspective. Therefore, direct detection prospects for this model remain very good given all present-day experimental constraints.
%\rr{We can get large sigmas since still we can have a mixed state for not very large m12. As long as it increases the neutralino becomes pure Higgsino and the sigma drops. I guess that in this limit the lightest Higgs exchanges is the dominant channel. mH goes as mA  which is alredy large}

As in the cMSSM, the spin-dependent cross-section remains out of reach even for future multi-ton scale detectors. The favoured region spans the interval $\sigmaSD \in [10^{-5.5}, 10^{-6.5}]$ pb, with the best-fit point found at the bottom end of the range.

\begin{figure*}%[htp]
\expandafter\includegraphics\expandafter[\rowoffour]{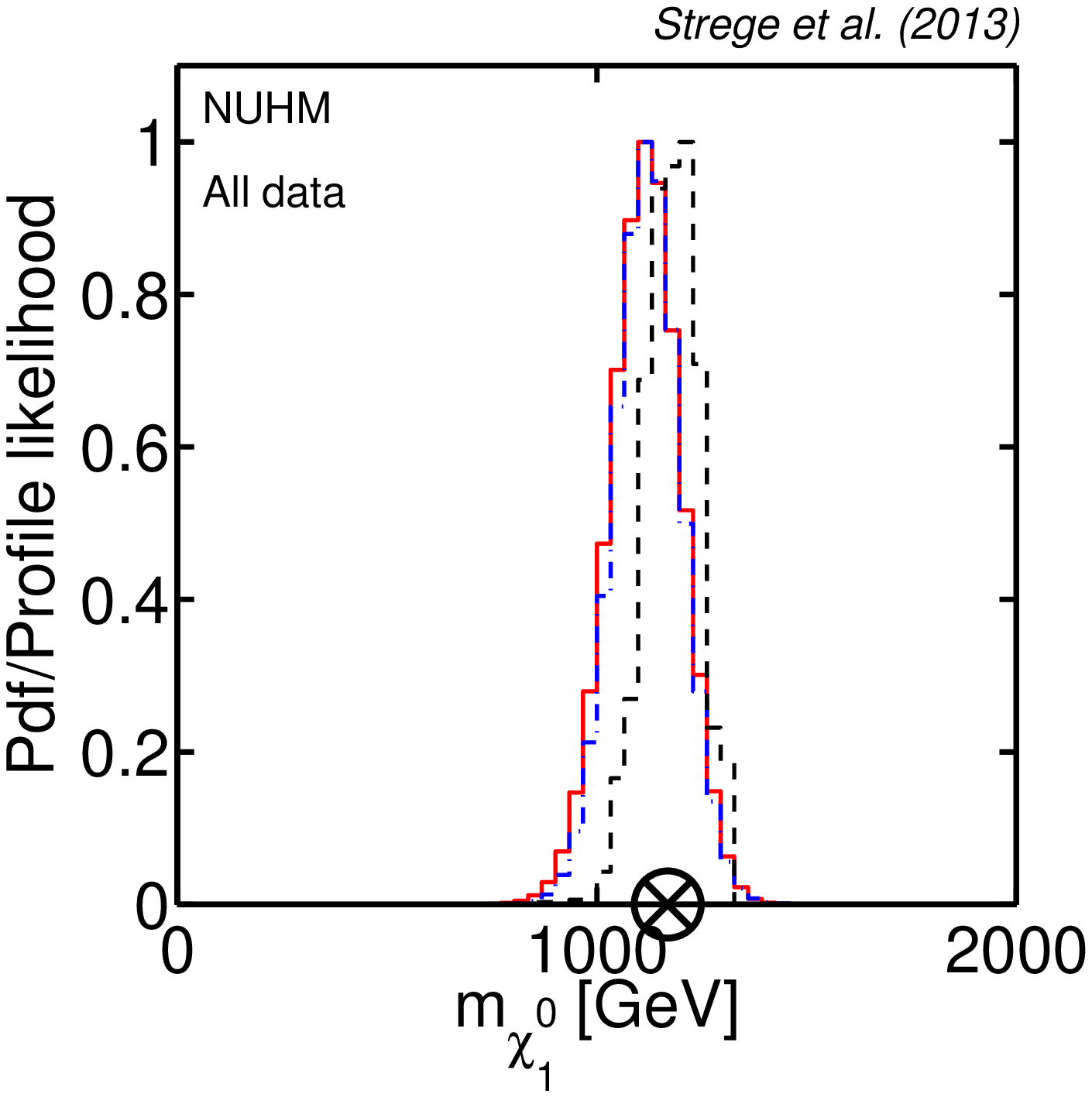}  
\expandafter\includegraphics\expandafter[\rowoffour]{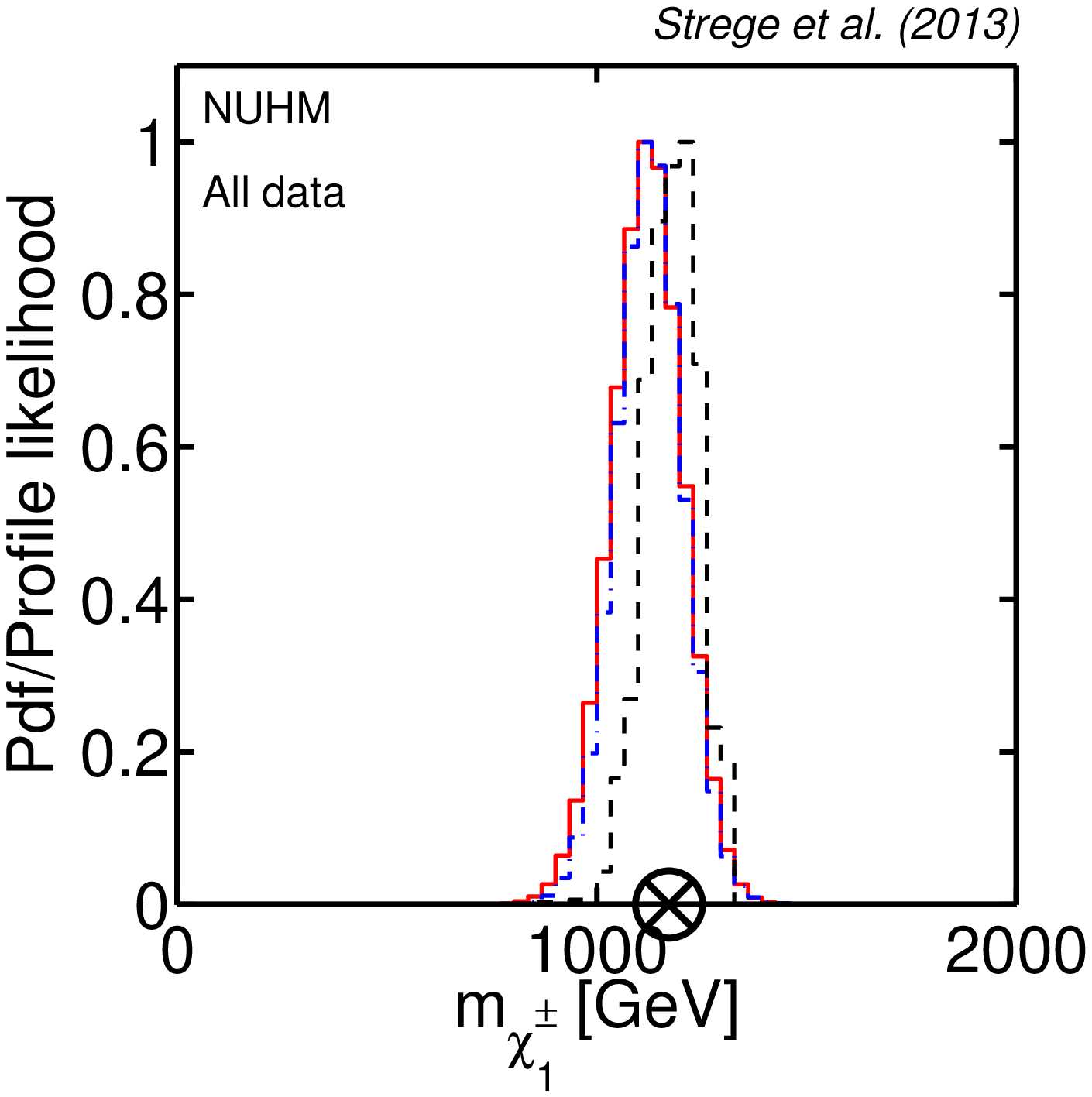}
\expandafter\includegraphics\expandafter[\rowoffour]{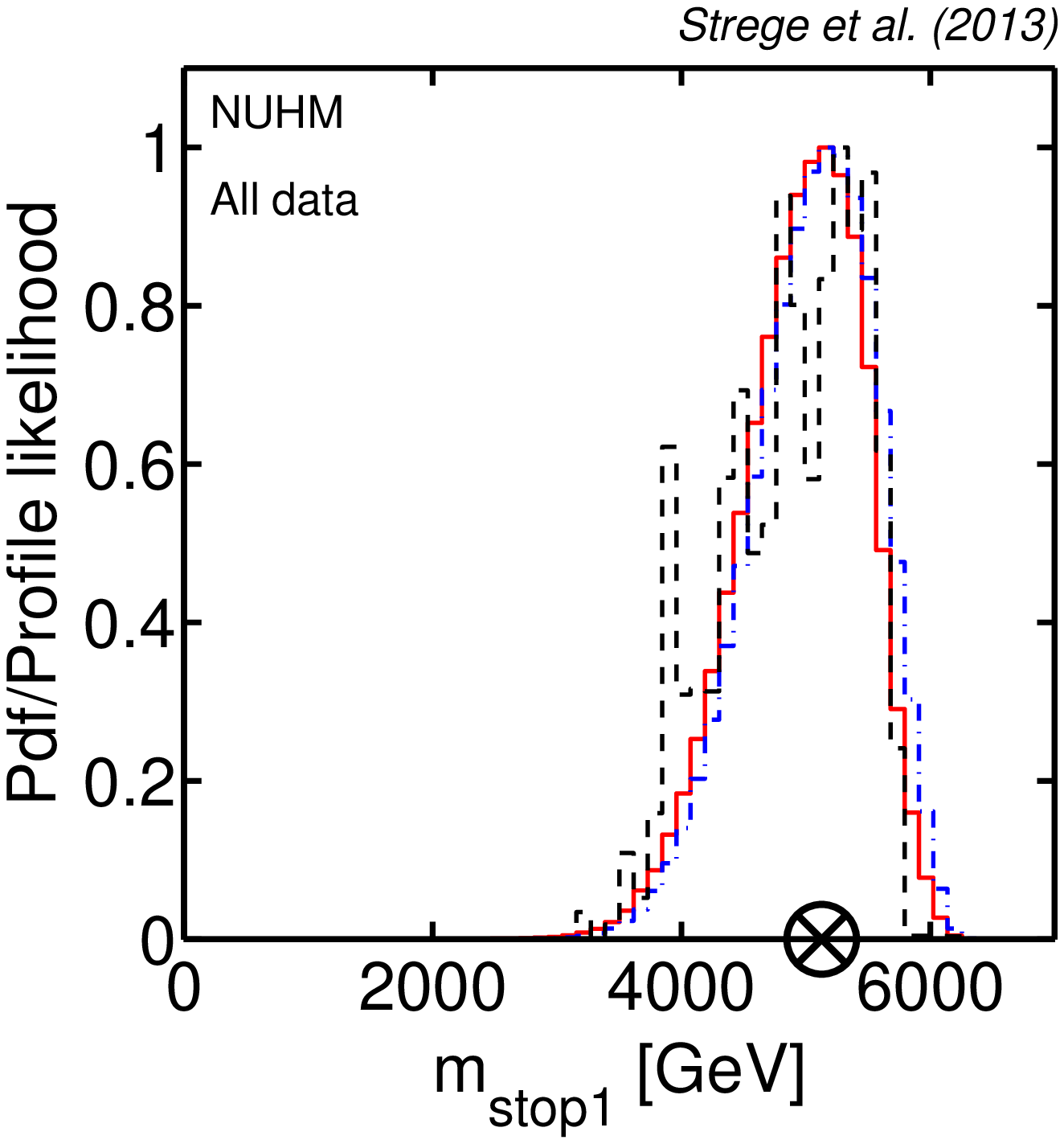}
\expandafter\includegraphics\expandafter[\rowoffour]{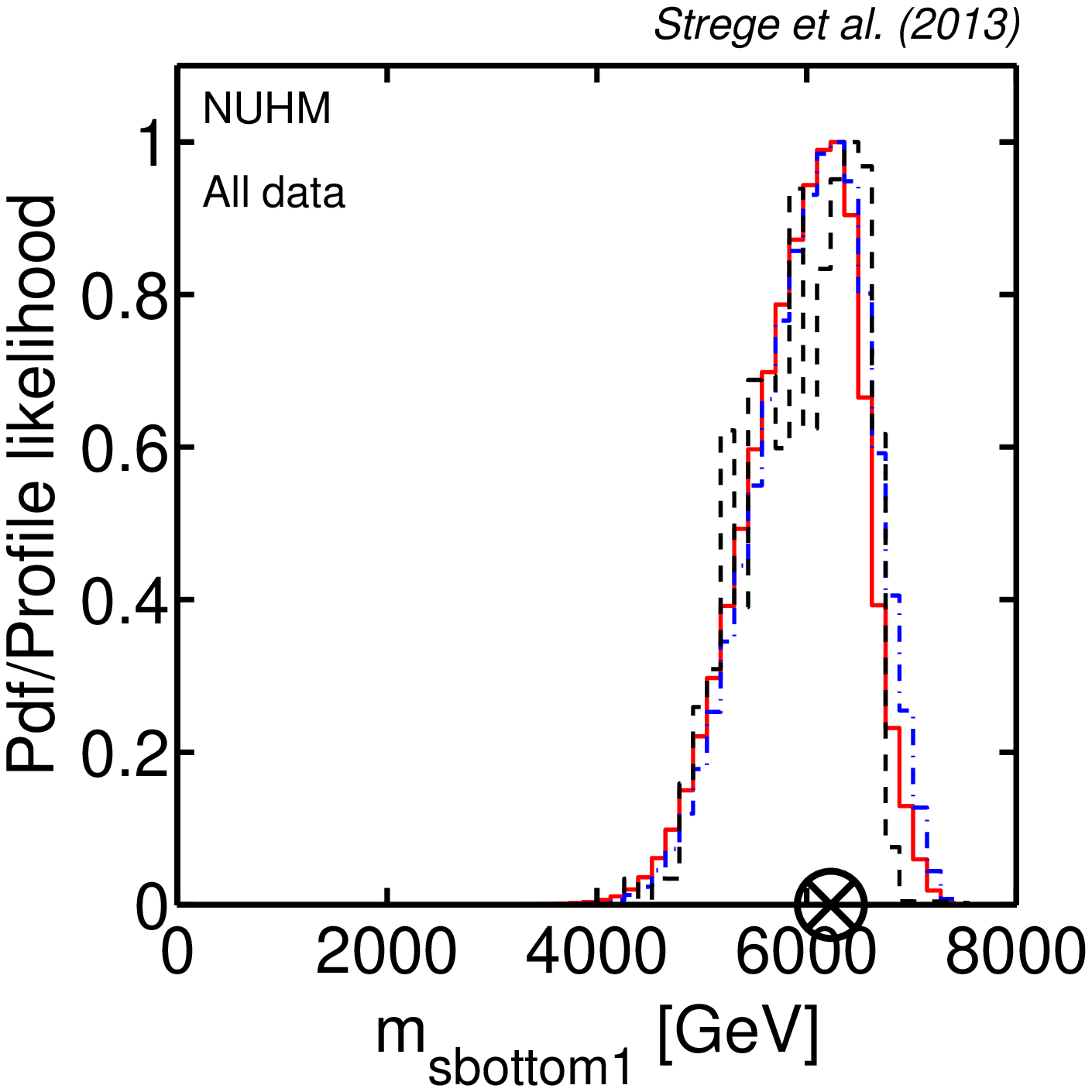}\\
\expandafter\includegraphics\expandafter[\rowoffour]{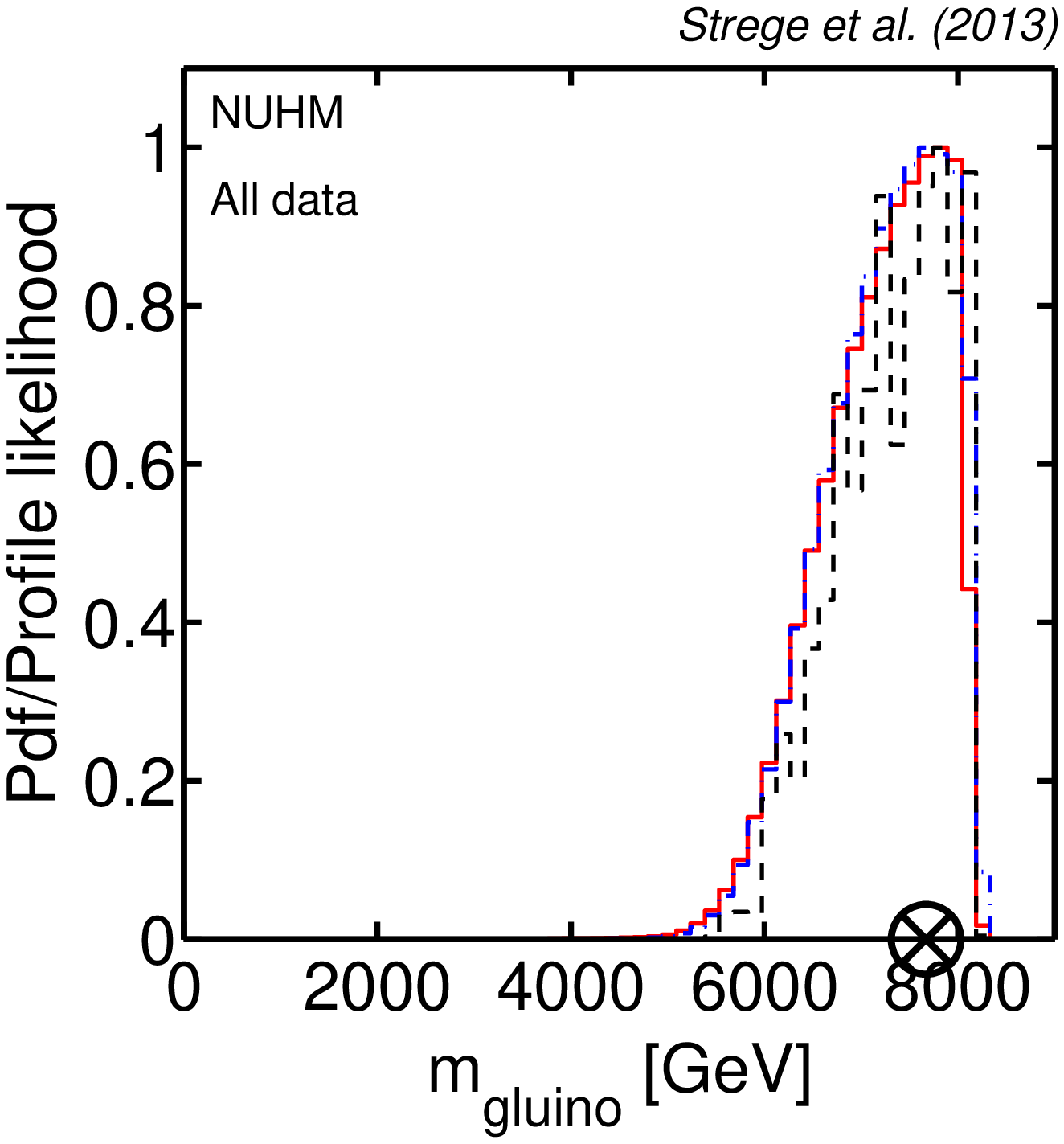} 
\expandafter\includegraphics\expandafter[\rowoffour]{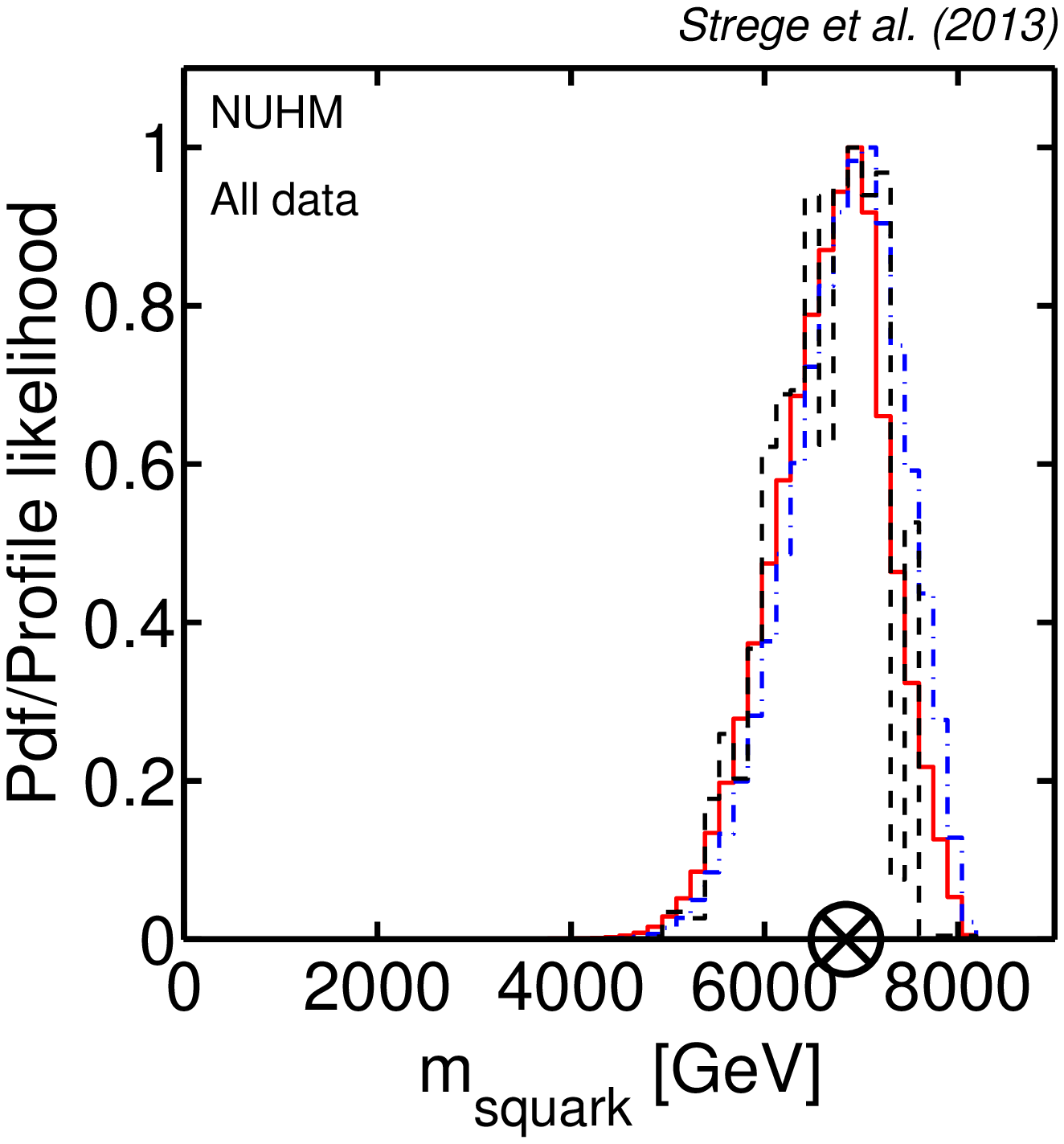}
\expandafter\includegraphics\expandafter[\rowoffour]{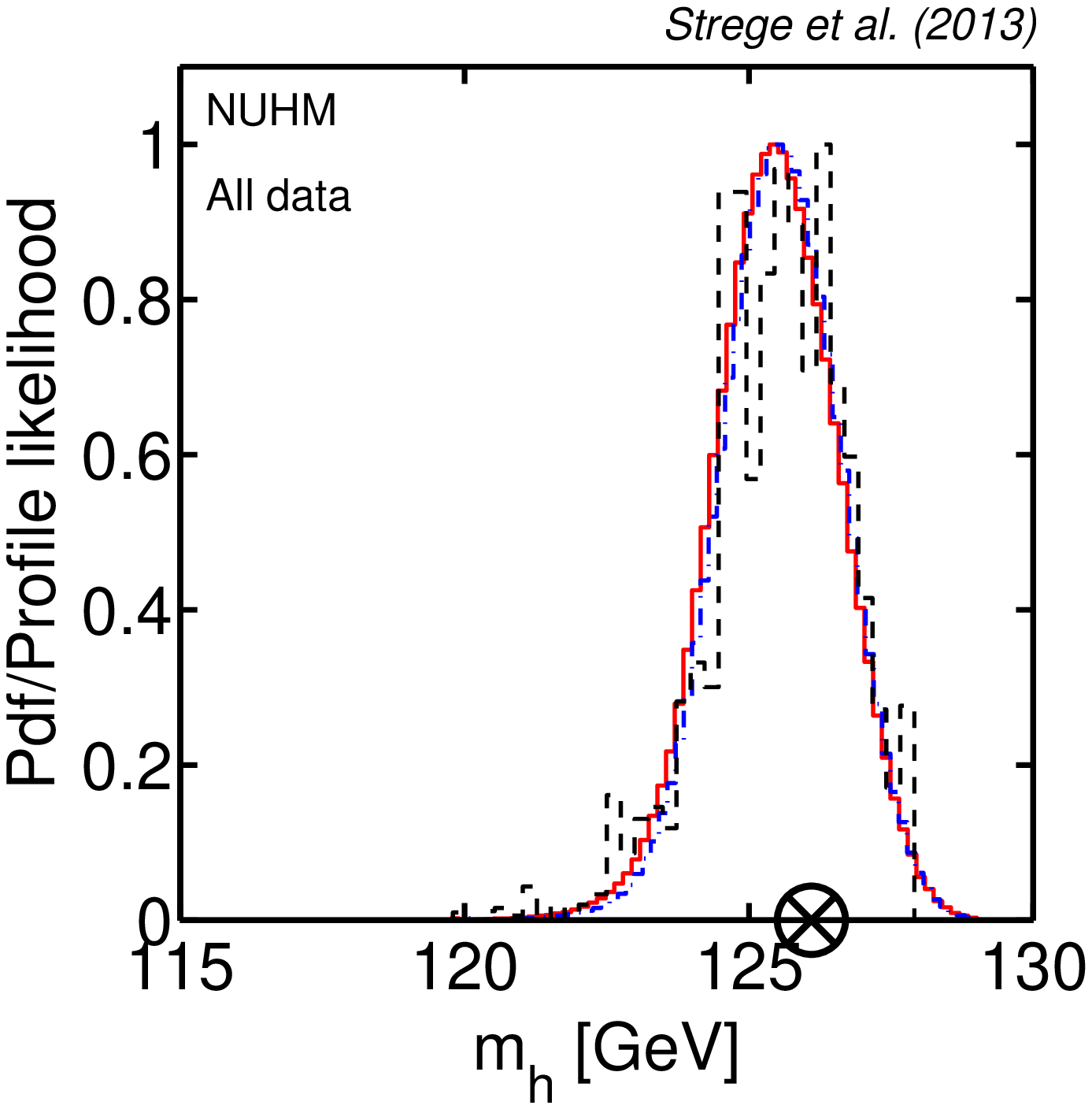}
\expandafter\includegraphics\expandafter[\rowoffour]{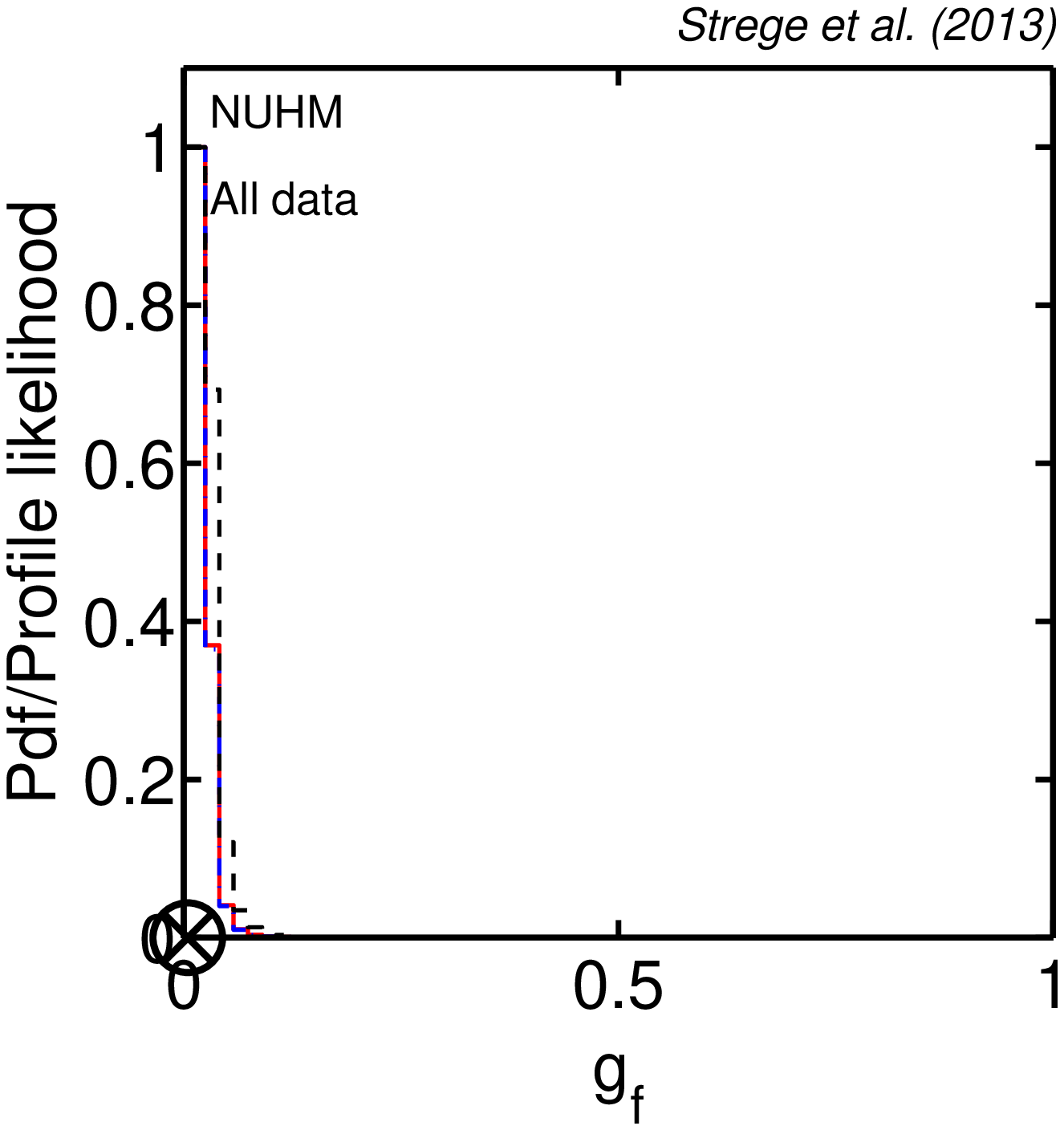} \\
%
%\expandafter\includegraphics\expandafter[\rowoffour]{Plots/NUHM_1D/NUHM_pp_1D_20_sq.eps}
\expandafter\includegraphics\expandafter[\rowoffour]{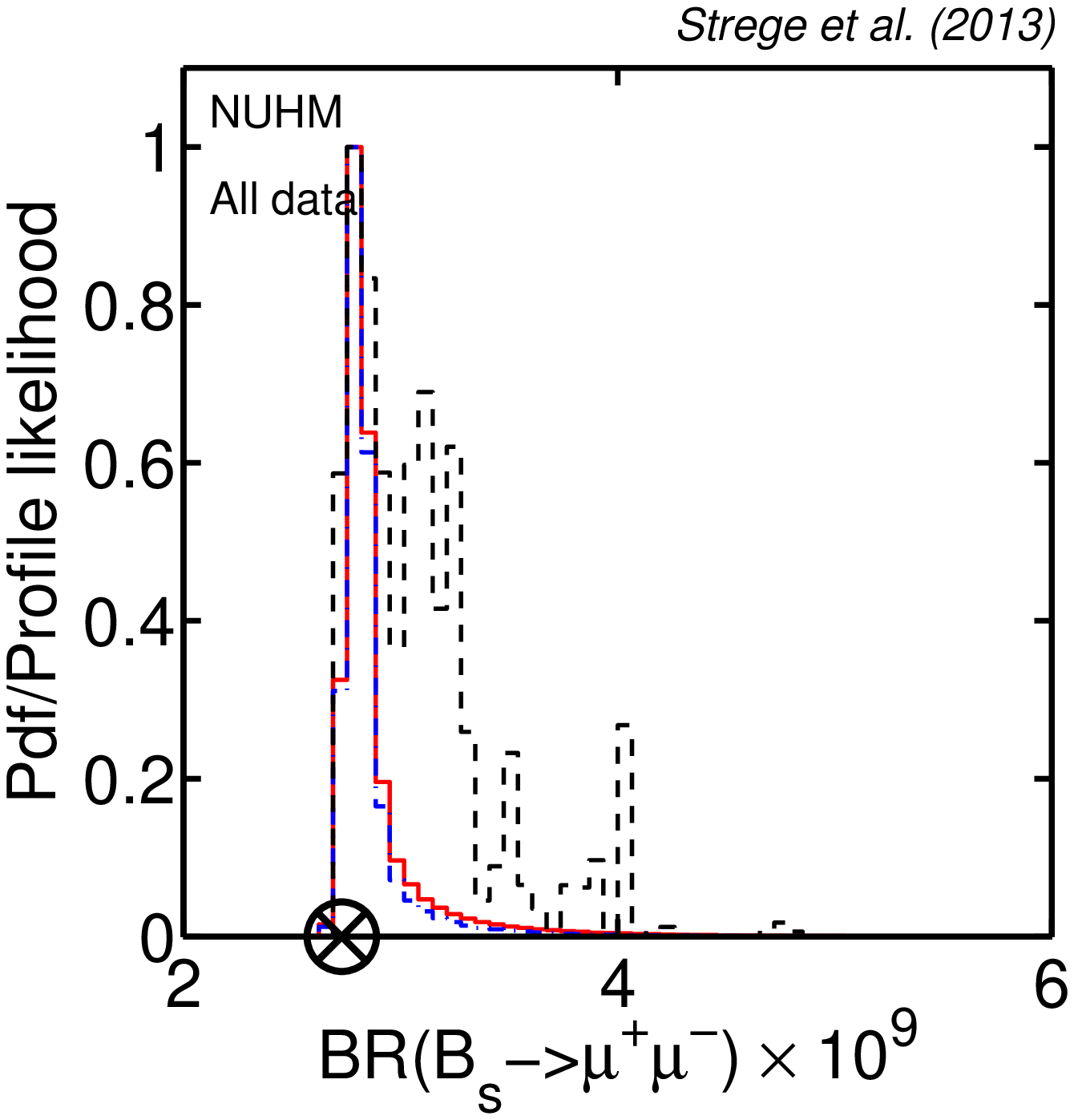}
\expandafter\includegraphics\expandafter[\rowoffour]{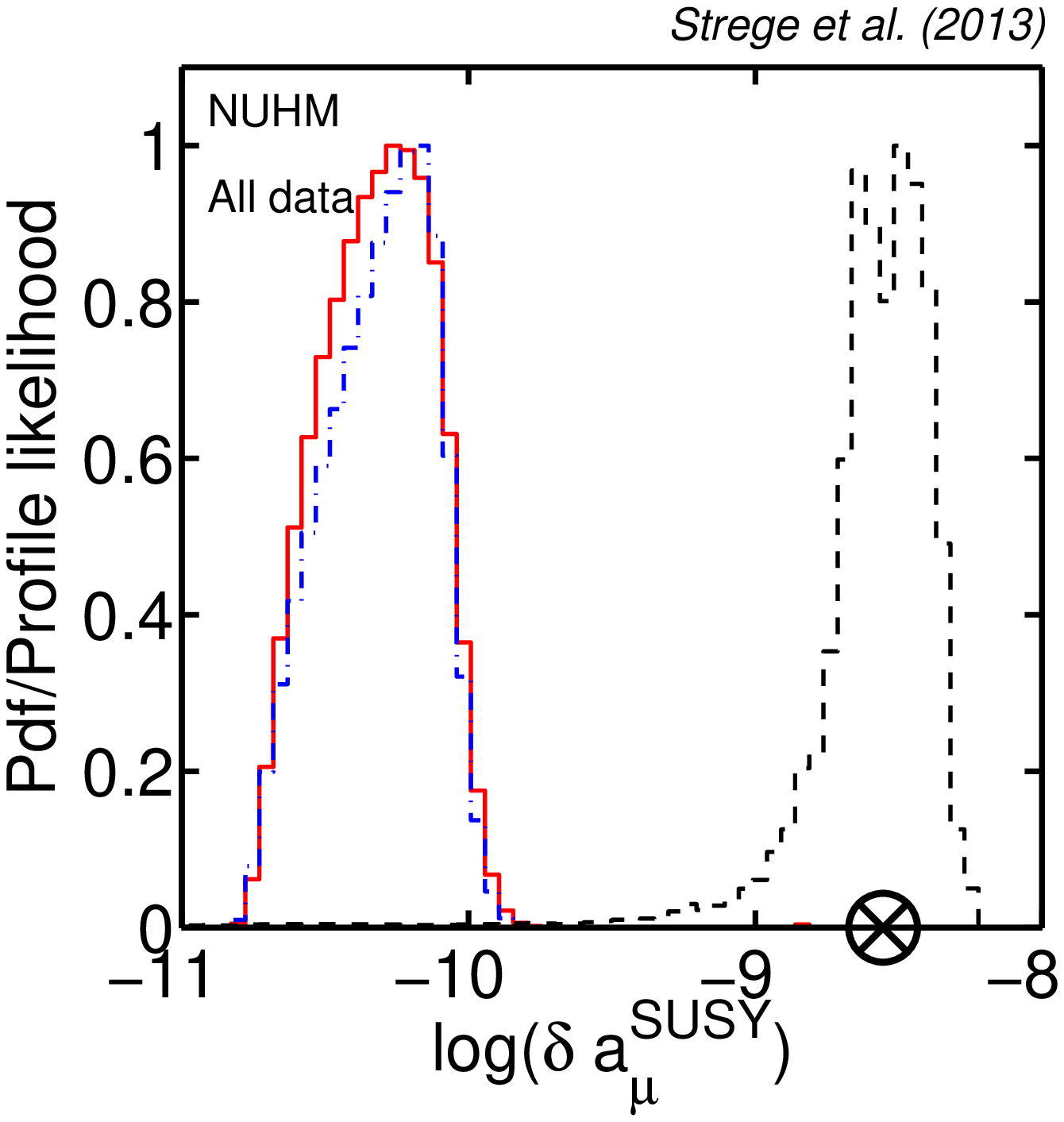} 
\expandafter\includegraphics\expandafter[\rowoffour]{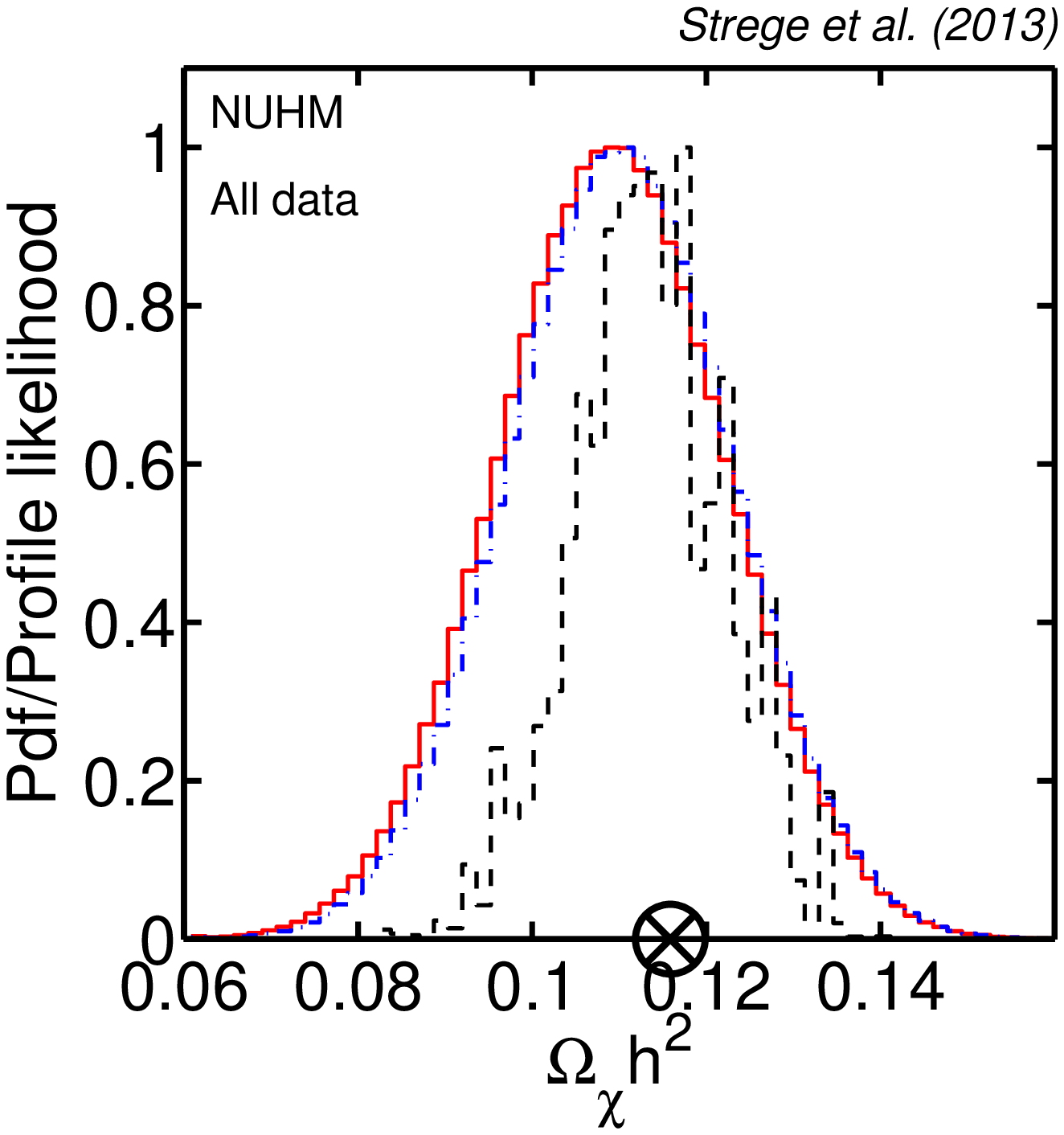}
\expandafter\includegraphics\expandafter[\rowoffour]{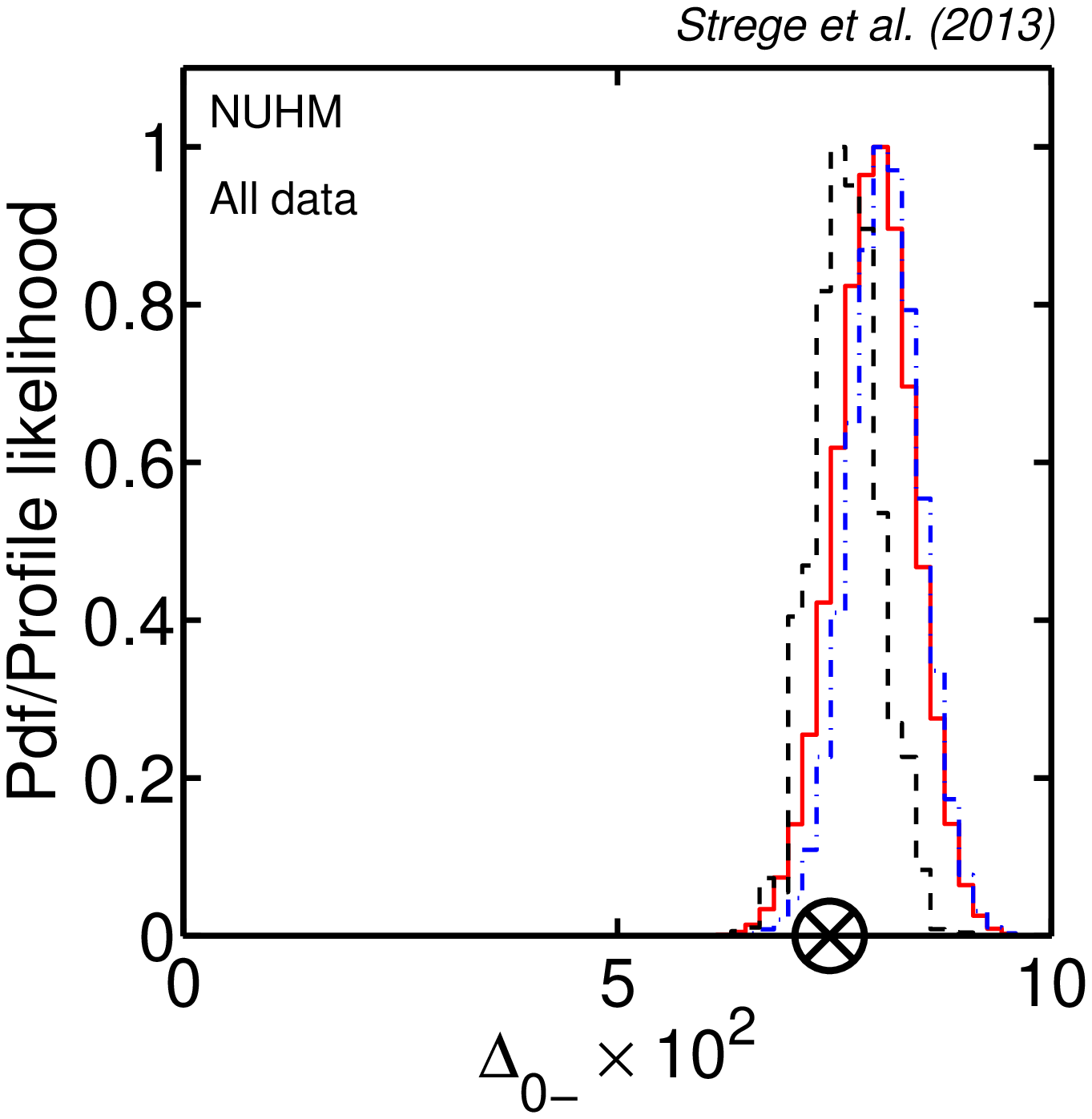} 

\caption{\fontsize{9}{9} \selectfont  1D marginal pdf for flat priors (dash-dot/blue), log priors (thick solid/red) and 1D profile likelihood (dashed/black) in the NUHM, including all current experimental constraints.  Top row, from left to right: neutralino mass, lightest chargino, stop and sbottom masses. Central row: gluino mass, average squark mass, lightest Higgs mass boson mass, gaugino fraction. Bottom row: $\brbsmumu$ branching ratio, anomalous magnetic moment of the muon, dark matter relic abundance and isospin asymmetry. The best-fit point is indicated by the encircled black cross. \label{fig:NUHM_1D}} 
\end{figure*}

In Fig.~\ref{fig:NUHM_1D}, we show the 1D posterior pdf and profile likelihood for some observables and derived quantities. The first six panels show some sparticle masses of interest. Both the lightest neutralino and lightest chargino masses are highly concentrated around 1 TeV. The favoured masses of the lightest stop and bottom are $m_{\textnormal{stop1}} \approx 5000$ GeV and $m_{\textnormal{sbottom1}} \approx 6000$ GeV, respectively. The favoured gluino and average squark masses are even larger, $m_{\textnormal{gluino}},m_{\textnormal{squark}} \approx 7000$ GeV. The favoured sparticle masses are far beyond the current reach of the LHC, and will not be accessible to the  LHC operating at 14 TeV collision energy, nor the HL-LHC upgrade. This is true for both the posterior pdf and the profile likelihood function, which are in excellent agreement. Therefore, detection prospects of the NUHM at colliders are dim, and for discovery of this model alternative search strategies, such as direct detection experiments, have to be relied on.

\begin{figure*}%[htp]
\centering
\expandafter\includegraphics\expandafter[\rowofthree]{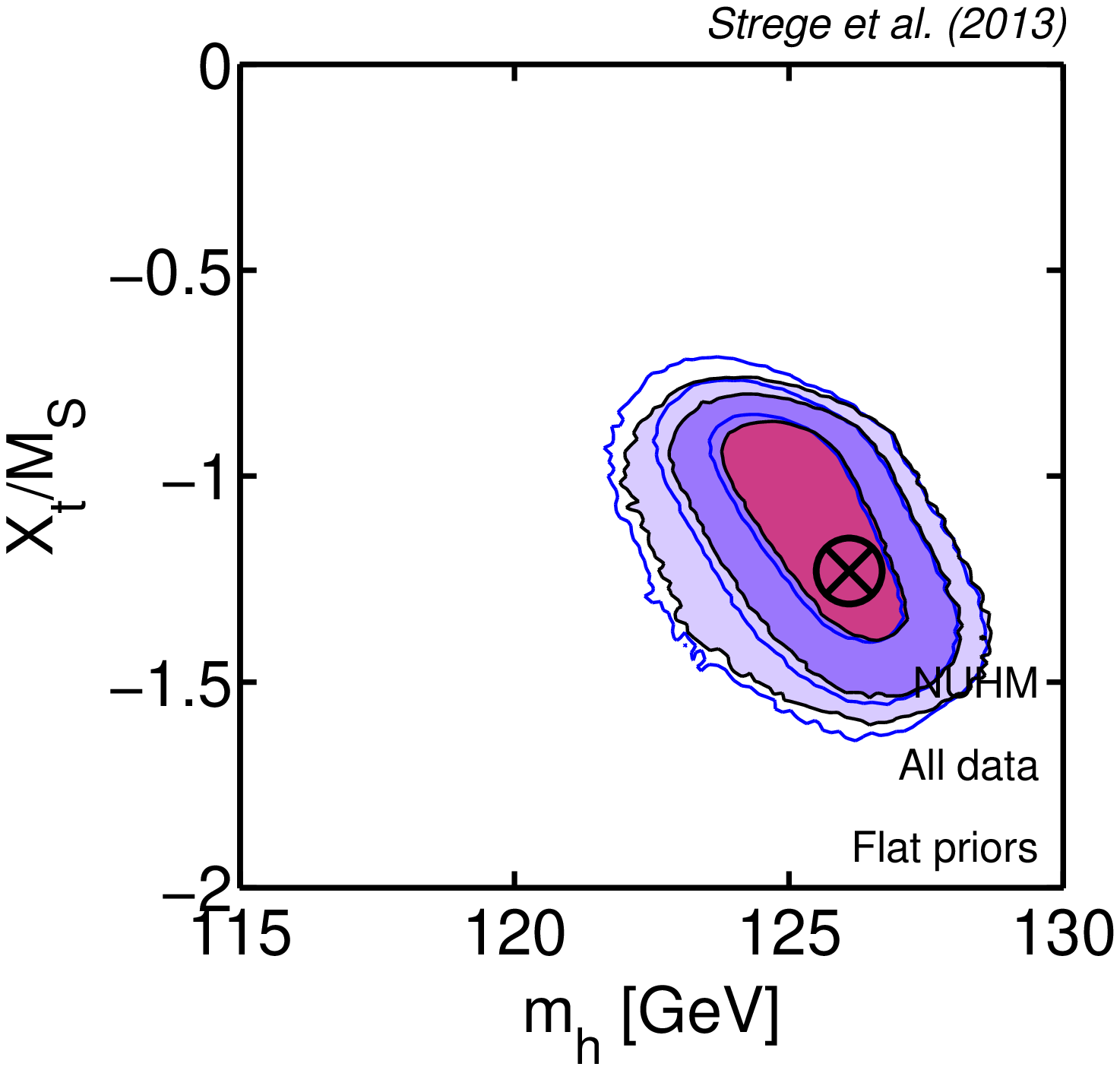}
\expandafter\includegraphics\expandafter[\rowofthree]{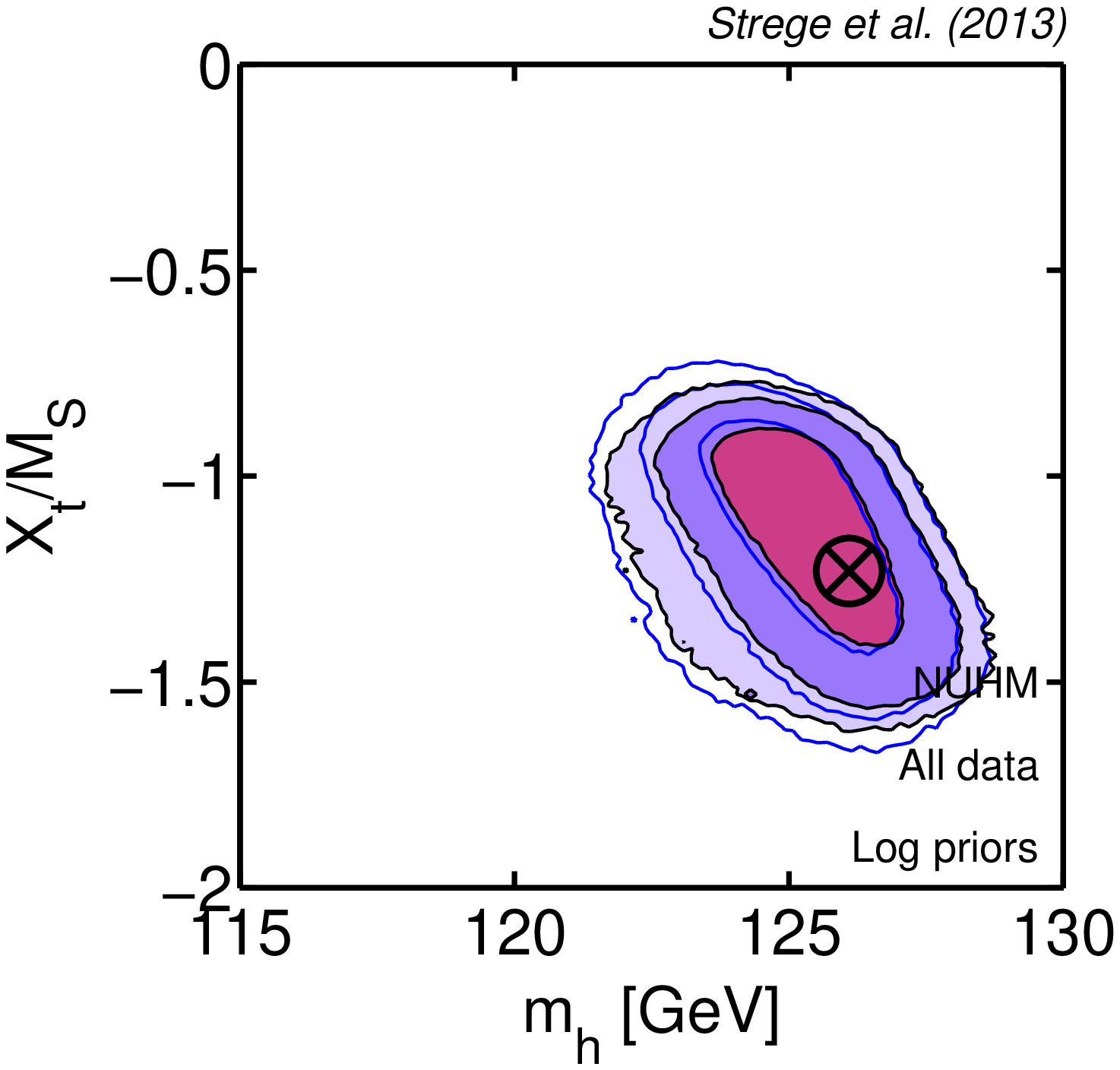}
\expandafter\includegraphics\expandafter[\rowofthree]{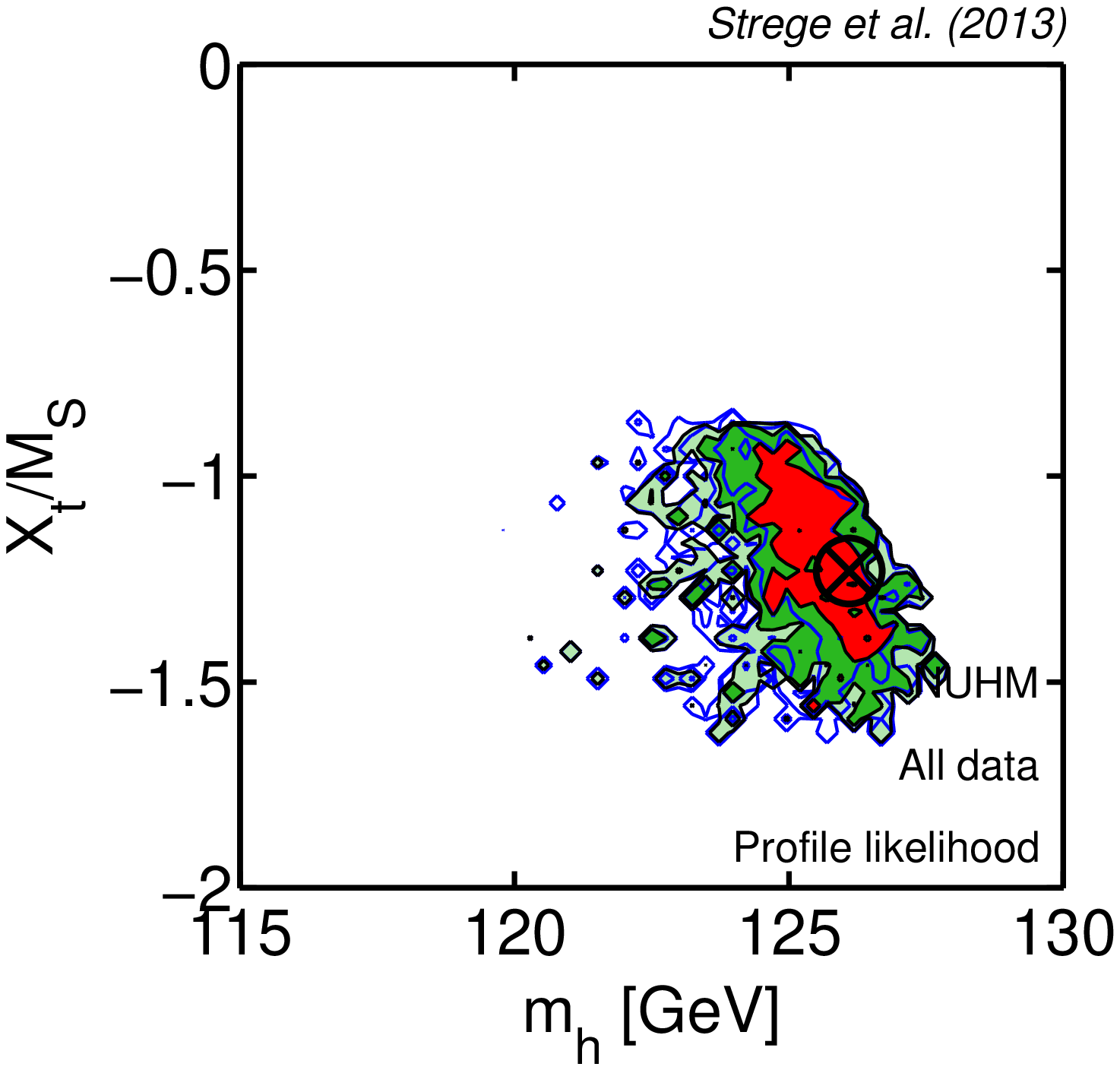} 
\caption{\fontsize{9}{9} \selectfont  Favoured regions in the $(m_h, X_t/M_S)$ plane in the NUHM, including all available present-day data. Black, filled contours depict the marginalised posterior pdf (left: flat priors; middle: log priors) and the profile likelihood (right), showing 68\%, 95\% and 99\% credible/confidence regions. The encircled black cross is the overall best-fit point. Blue/empty contours show constraints without the latest XENON100 results. Compared to the cMSSM, the maximal mixing scenario is not realised.
\label{NUHM2D_XtMS}}  
\end{figure*}

As can be seen from the 1D distributions for $m_h$, in the NUHM a Higgs mass $m_h \sim 126$ GeV can easily be realised, and both the Bayesian pdfs and the profile likelihood function peak at the experimentally measured value. The reason why $m_h \sim 126$ GeV is easily achieved, while this value is disfavoured in the cMSSM, is that in the NUHM much larger values of $m_{1/2}$ are allowed, leading to larger stop masses, and thus larger values of $m_h$. The favoured regions in NUHM parameter space in the $(X_t/M_S,m_h)$ plane are shown in Fig.~\ref{NUHM2D_XtMS}. As can be seen, moderate values of $X_t/M_S$ are favoured, and the maximal mixing scenario is not realised. By a combination of large stop masses and moderate stop mixing the Higgs mass is increased, so that $m_h \sim 126$ GeV can comfortably be achieved. Since the cMSSM is obtained from the NUHM by imposing extra universality conditions, in principle the maximal mixing scenario could be realised at low NUHM masses, similar to what was found for the cMSSM in section~\ref{sec:cMSSM_Higgs}. However, this scenario requires large fine-tuning and, due to the small stop masses in this region, can only achieve values of $m_h$ slightly lower than the experimental constraint. Therefore, this region is disfavoured with respect to the high-mass region, in which $m_h \sim 126$ GeV can easily be achieved.

The 1D distributions for the gaugino fraction $g_f$ are shown in the right-hand panel in the central row of Fig.~\ref{fig:NUHM_1D}. One of the main new features in the NUHM compared to the cMSSM is the possibility of dark matter with a large Higgsino fraction (i.e. a small gaugino fraction $g_f < 0.3$). Higgsino-like dark matter arises from the fact that in the NUHM $\mu$ is a free parameter, so that it can be adjusted to give the correct dark matter relic density required by the WMAP constraint. As can be seen in Fig.~\ref{fig:NUHM_1D}, current experimental constraints rule out the possibility of gaugino-like dark matter ($g_f \gg 0.5$) and favour regions of parameter space that correspond to a strongly Higgsino-like LSP with $g_f \leq 0.1$ at $99\%$ level. For Higgsino-like dark matter $|\mu| \approx m_{\neut}$, and a relatively large $\mu$ is required to fulfil the relic density constraint. A mass of $m_{\chi} \sim \mathcal{O}(100)$ GeV underproduces dark matter in the universe, but higher masses $m_{\neut} \approx 1$ TeV can correctly reproduce the WMAP results. Increasing $\mu$ further leads to a larger dark matter relic density, in conflict with the experimental constraint. This explains the concentration of $m_{\neut}$ around $1$ TeV.

The 1D distributions for $\brbsmumu$ are confined to a relatively small range $\brbsmumu \sim (2.5,4) \times 10^{-9}$. This ranges is comfortably within the $1 \sigma$ error range of the new LHCb measurement of $\brbsmumu$, so that this constraint has a minimal impact on the NUHM parameter space.

While for almost all other quantities the profile likelihood results are in good agreement with the Bayesian posterior pdfs, this agreement breaks down for the anomalous magnetic moment of the muon (see central left panel in the bottom row of Fig.~\ref{fig:NUHM_1D}). The profile likelihood function peaks in a region that satisfies the \gmt\ constraint, while the bulk of the 1D posterior distributions (for both choices of priors) has a strong preference for SM-like values of \gmt. This is mainly a consequence of the Higgs mass measurement and other constraints preferring large values of the mass parameters, while the \gmt\ constraint is most easily satisfied at low masses. In the high-mass regions favoured in the NUHM the vast majority of points lead to a SM-like value of \gmt\, so that the posterior pdfs strongly favour these values. The best-fit, and hence the peak of the profile likelihood, is found in a region of parameter space where the \gmt\ constraint and the other constraints are simultaneously satisfied. This requires strong fine-tuning, so that only a very small number of such points are found by our scans. This explains the relatively lower resolution observed for the profile likelihood function in Fig. \ref{NUHM2D}. 

The 1D distributions for $\Omega_{\chi}h^2$ agree well with the WMAP constraint. In contrast, similar to what was observed for the cMSSM, values of $\Delta_{0-}$ in agreement with the experimental measurement are difficult to realise in the NUHM, as can also be seen by the sizeable contribution of this observable to the best-fit $\chi^2$ value in Table~\ref{NUHMbf2}. Large negative SUSY contributions to $\Delta_{0-}$ are difficult to achieve in this model, so that SM-like values $\Delta_{0-} \sim 8 \times 10^{-2}$ are favoured. The SUSY contribution to $\Delta_{0-}$ is minimised, since the regions favoured in NUHM parameter space correspond to large $m_{1/2}$, relatively small $\tan \beta$, and there is a preference for vanishing or positive $A_0$. 

\subsection{Comparison with previous analyses}

Our results can be contrasted with an earlier study of the NUHM in Ref.~\cite{Roszkowski:2009sm}. In this study it was found that regions in NUHM parameter space corresponding to neutralino dark matter with a large gaugino fraction $g_f > 0.7$, found at low values of $m_0$ and $m_{1/2}$, were strongly favoured, while regions corresponding to Higgsino-like neutralino dark matter were less favoured. In contrast, our results show a strong preference for Higgsino-like dark matter, corresponding to large scalar and gaugino masses, while gaugino-like dark matter is ruled out at 99\% level from both the Bayesian and the profile likelihood statistical perspective.

While a large fraction of the NUHM parameter space favoured in Ref. \cite{Roszkowski:2009sm} is ruled out by the LHC exclusion limit, the constraint on the Higgs mass also plays a dominant role in our findings.  As can be seen in Fig. 2 of Ref.  \cite{Roszkowski:2009sm}, small values of $m_h \approx 116$ GeV, now ruled out by the $m_h$ constraint, were previously  favoured. In order to achieve Higgsino-like (instead of the previously favoured gaugino-like) neutralino dark matter one requires $|\mu| < M_1$, where $M_1$ is the soft mass of the bino. The bino mass scales with the gaugino mass as $M_1 \simeq 0.4 m_{1/2}$. Since $\mu \approx 1000$ GeV is required by the relic density constraint, this explains the strong preference for large  $m_{1/2} > 2$ TeV shown in Fig. \ref{NUHM2D}. As mentioned in the previous section, large values of $m_{1/2}$ lead to an increase of $m_h$, making it easier to satisfy the LHC Higgs constraint. Therefore, the constraint on $m_h$ strongly favours Higgsino-like dark matter over gaugino-like dark matter, which is predominantly found at small $m_{1/2}$ (see Ref. \cite{Roszkowski:2009sm}).  

Further differences with the study in Ref.~\cite{Roszkowski:2009sm} concern the statistical setup, as in that work the soft masses $m_{H_u}, m_{H_d}$ of the two Higgs doublets are used as input parameters, rather than $\mu, m_A$ as in this work. Given the non--linear relationship between the two sets of quantities, flat priors in one set do not correspond to flat priors in the other, hence the Bayesian posterior is affected by the Jacobian of the transformation. The profile likelihood is, in principle, prior independent. However, the present work makes use of a much higher resolution to ensure a reliable mapping of the profile likelihood function than the scanning methodology adopted in \cite{Roszkowski:2009sm}.

\subsection{Impact of the $\delta a_\mu^{SUSY}$ constraint}

 \begin{figure*}%[htp]
\centering
\expandafter\includegraphics\expandafter
[\rowofthree]{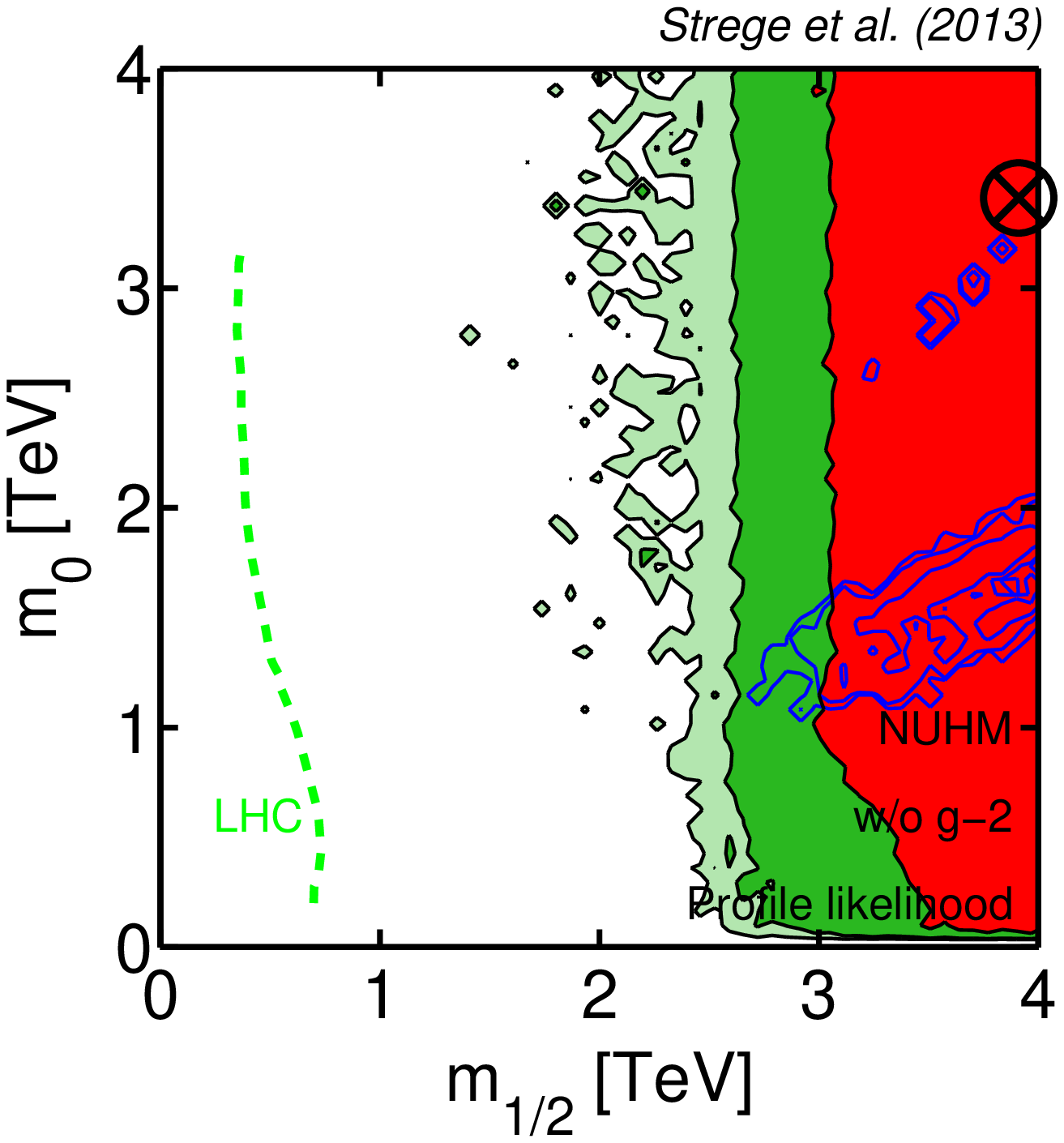} 
\expandafter\includegraphics\expandafter
[\rowofthree]{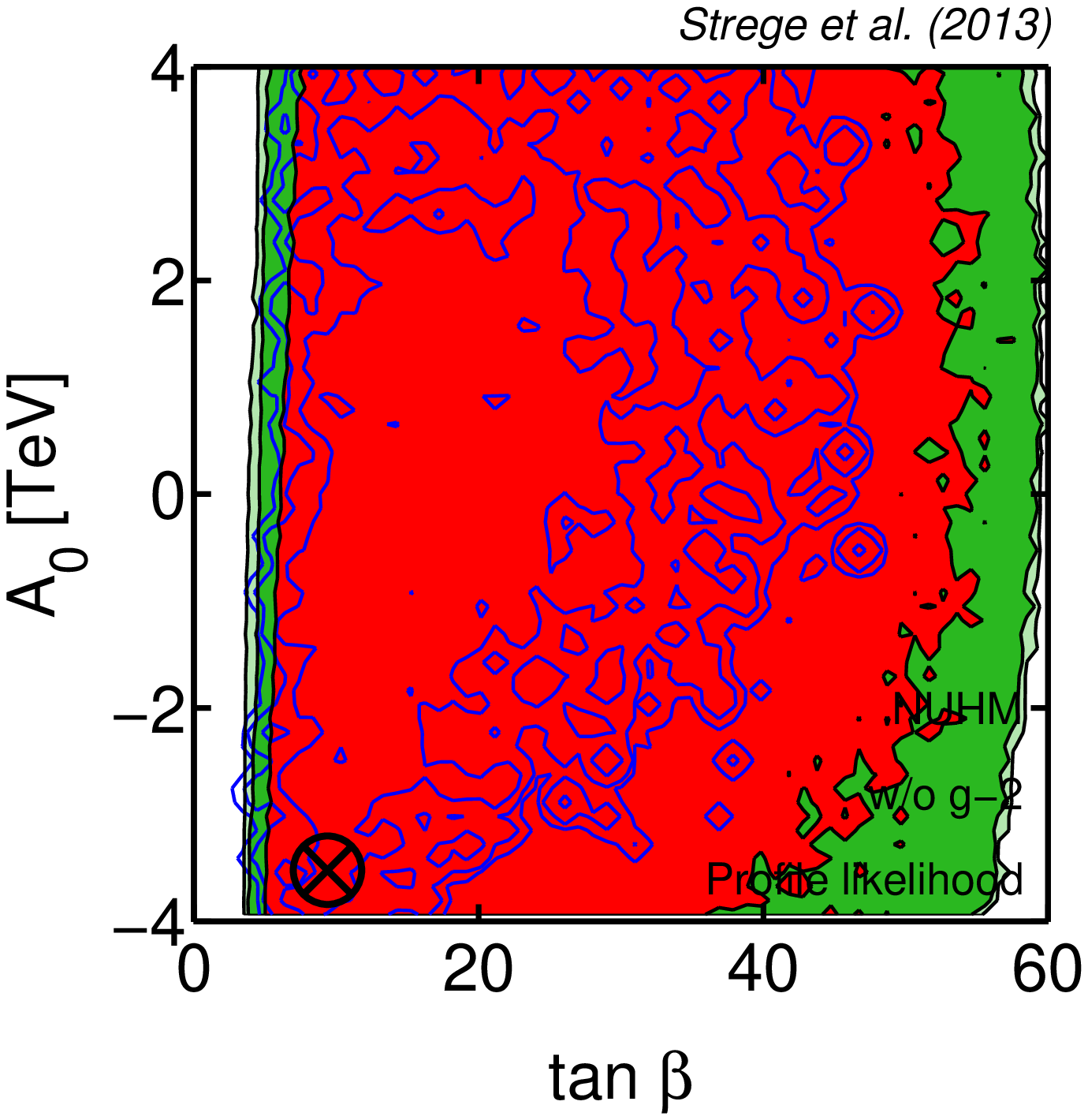} \\
\expandafter\includegraphics\expandafter
[\rowofthree]{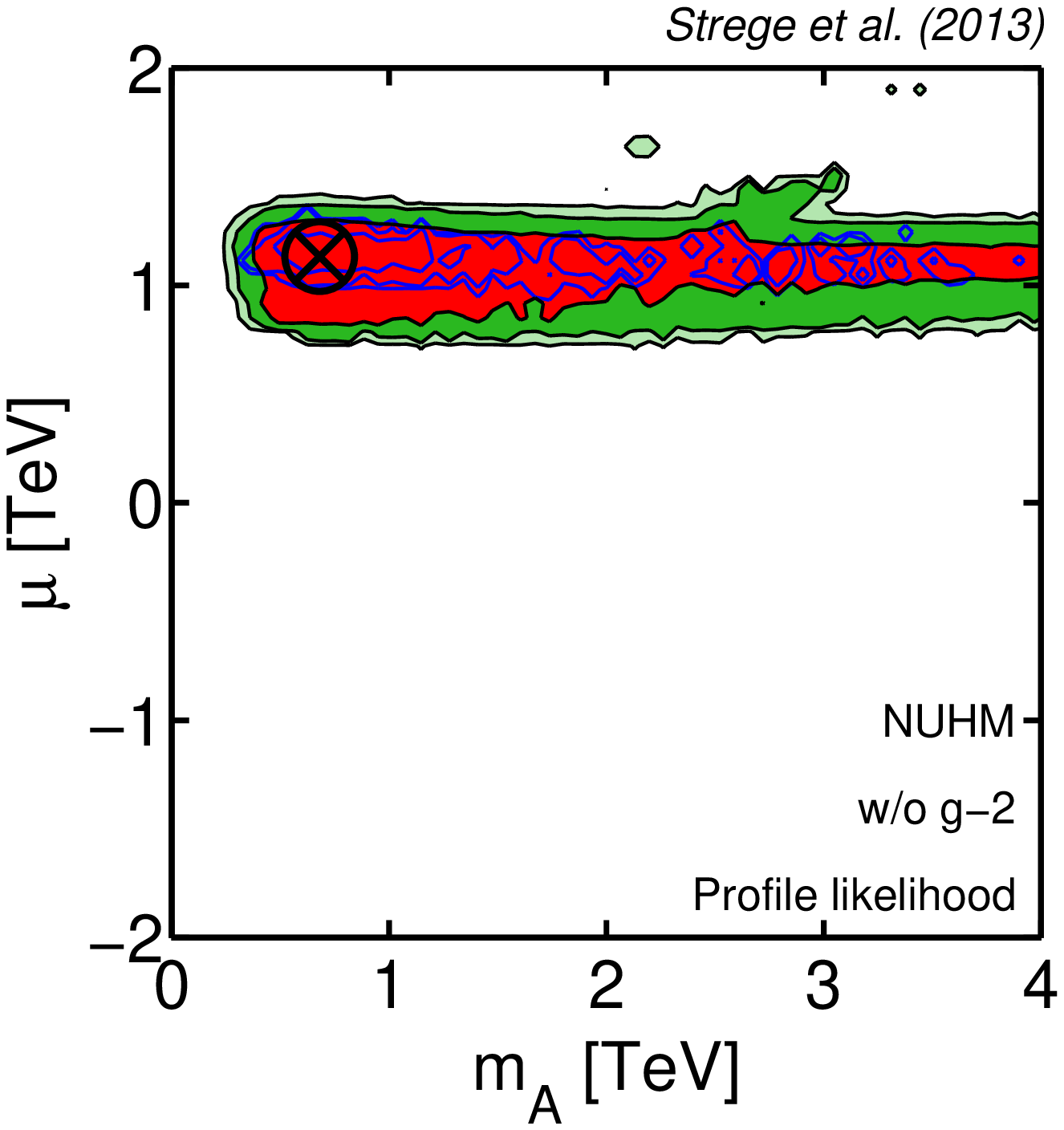}
\expandafter\includegraphics\expandafter
[\rowofthree]{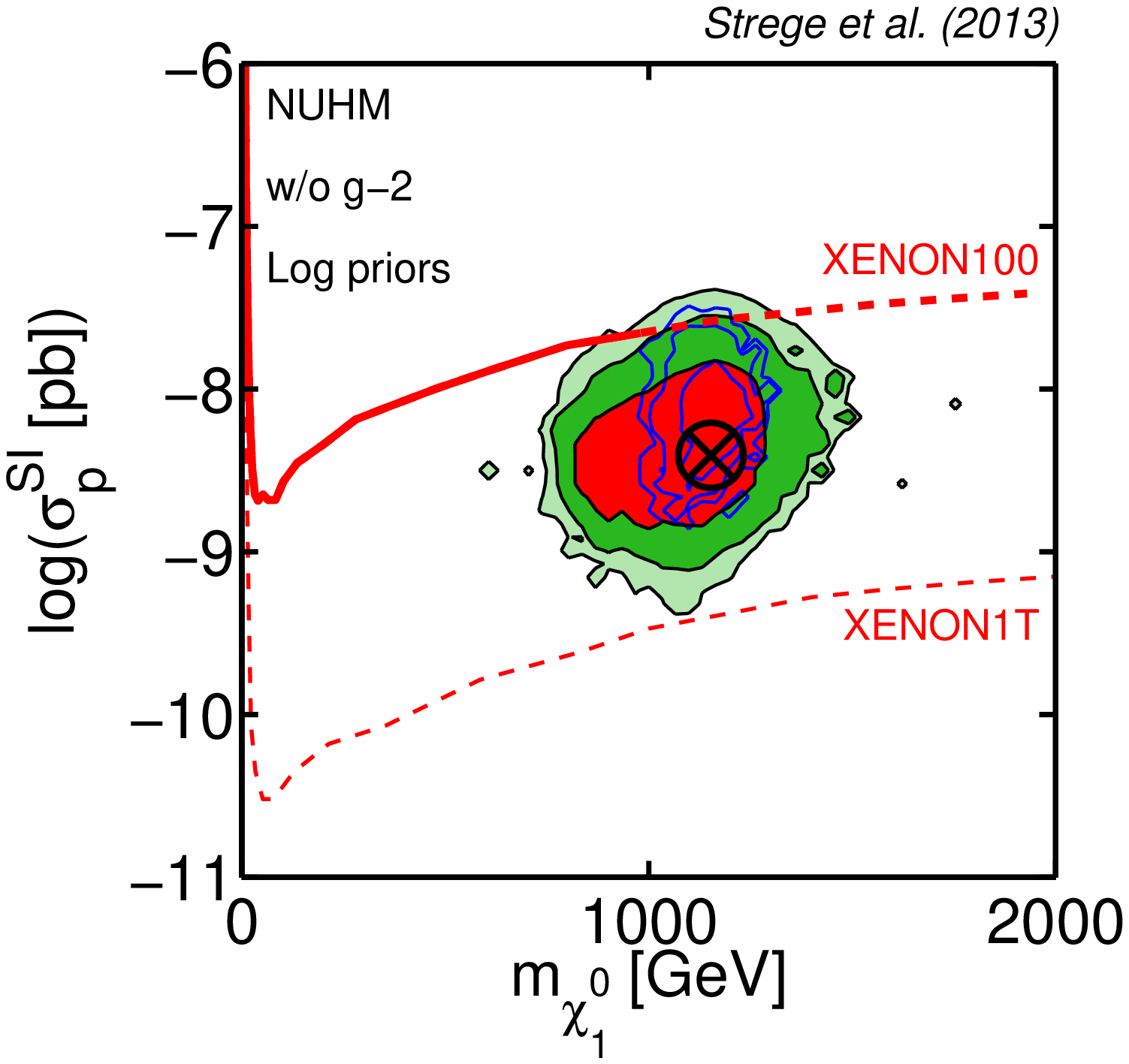} \\
\caption{\fontsize{9}{9} \selectfont  Profile likelihood (black/filled, 68\%, 95\% and 99\% CL) in the NUHM parameter space including the LHC Higgs discovery and 5.8 fb$^{-1}$ exclusion limit,  the latest XENON100 limit and all other data, {\em except the $\delta a_\mu^{SUSY}$ constraint}. The bottom-right plot shows the profile likelihood in the spin-independent scattering cross-section vs neutralino mass plane. Blue/empty contours include the $\delta a_\mu^{SUSY}$ constraint (as in Fig.~\ref{NUHM2D}), for comparison.  The posterior pdfs are identical to Fig.~\ref{NUHM2D}, up to numerical noise, and hence are not shown here. The encircled black cross shows the overall best-fit point.\label{fig:NUHM_2D_wogmt}}  
\end{figure*}

\begin{figure*}%[htp]
%\centering
\expandafter\includegraphics\expandafter
[\rowoffour]{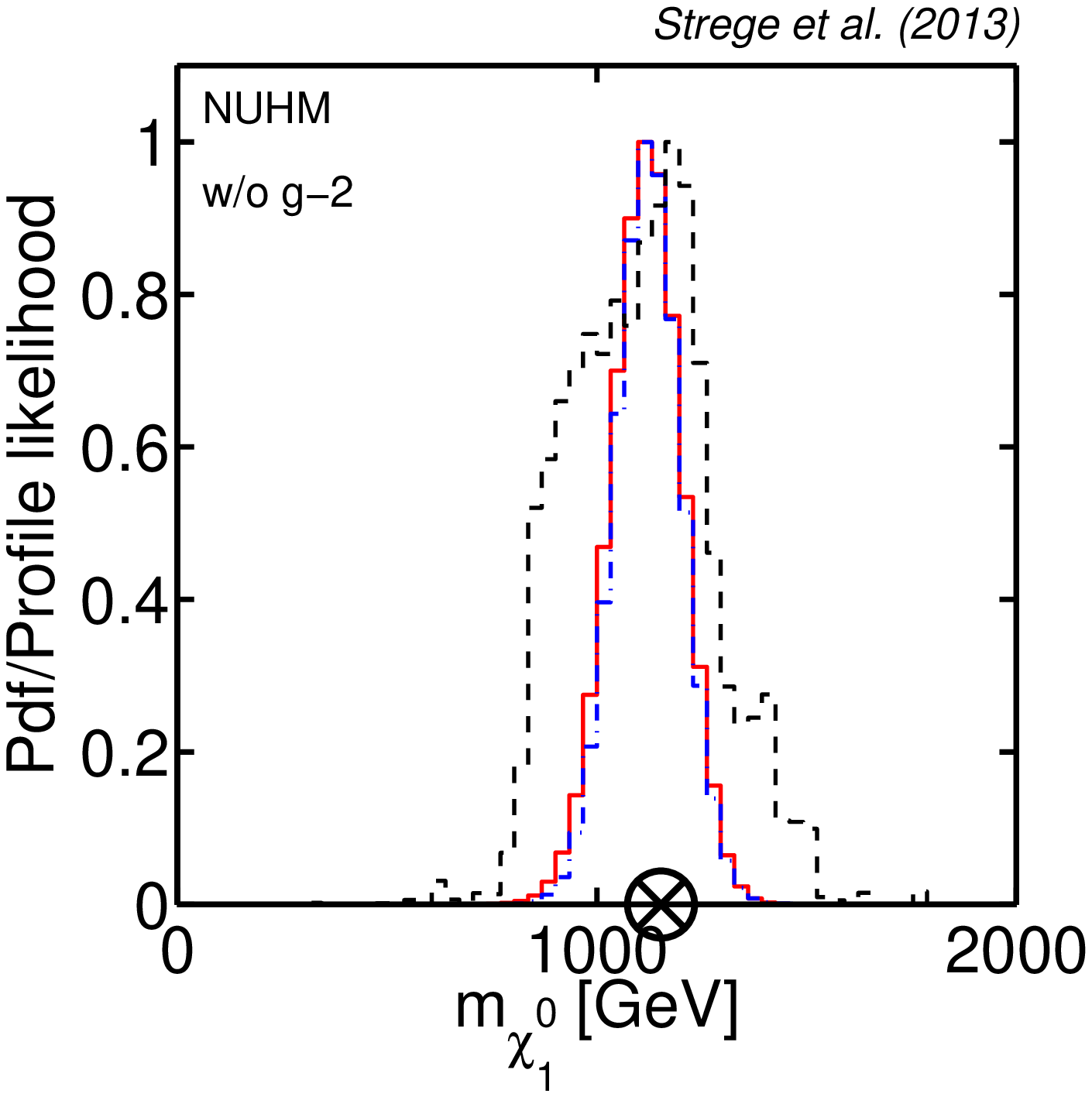}  
\expandafter\includegraphics\expandafter
[\rowoffour]{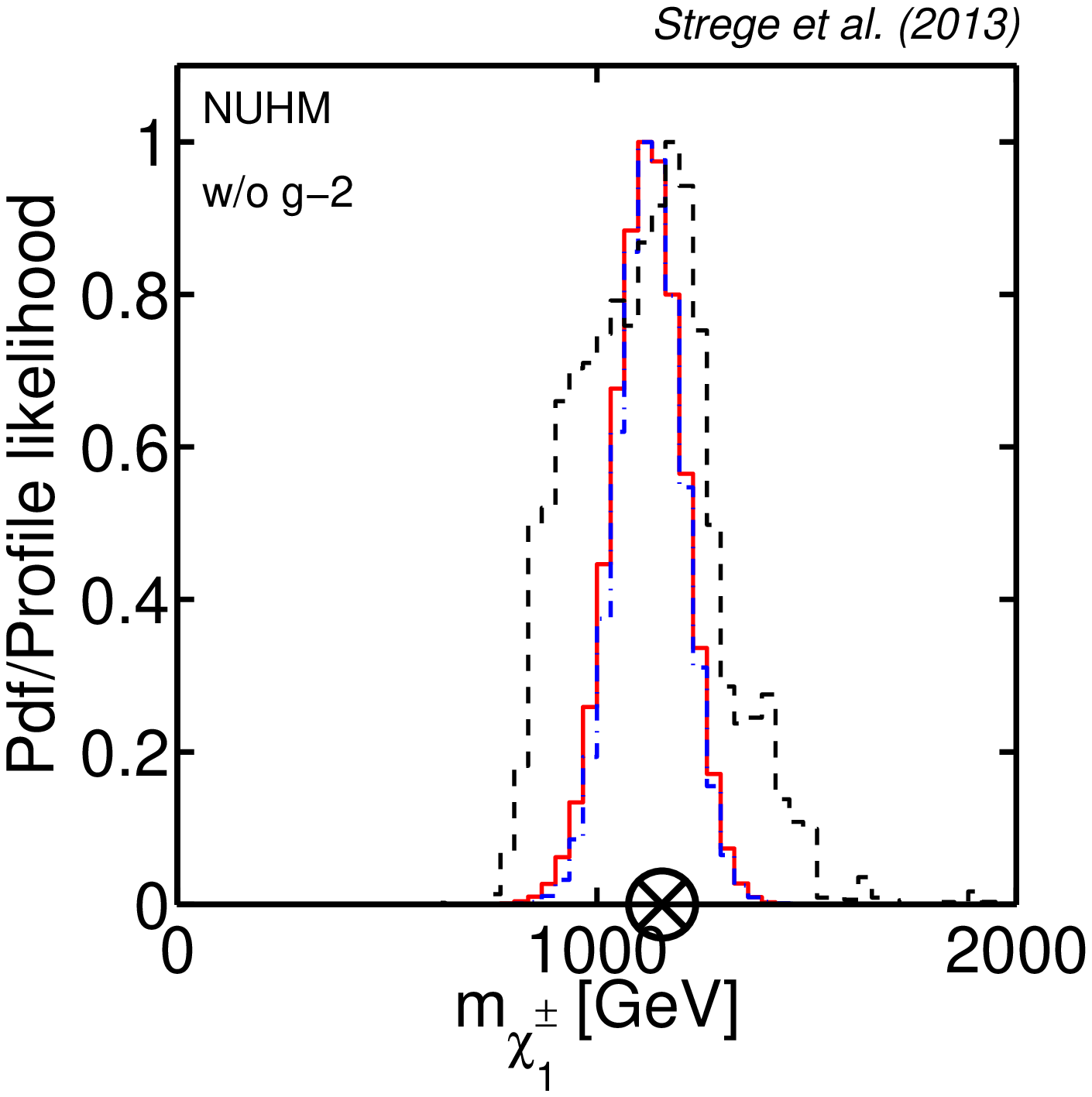}
\expandafter\includegraphics\expandafter
[\rowoffour]{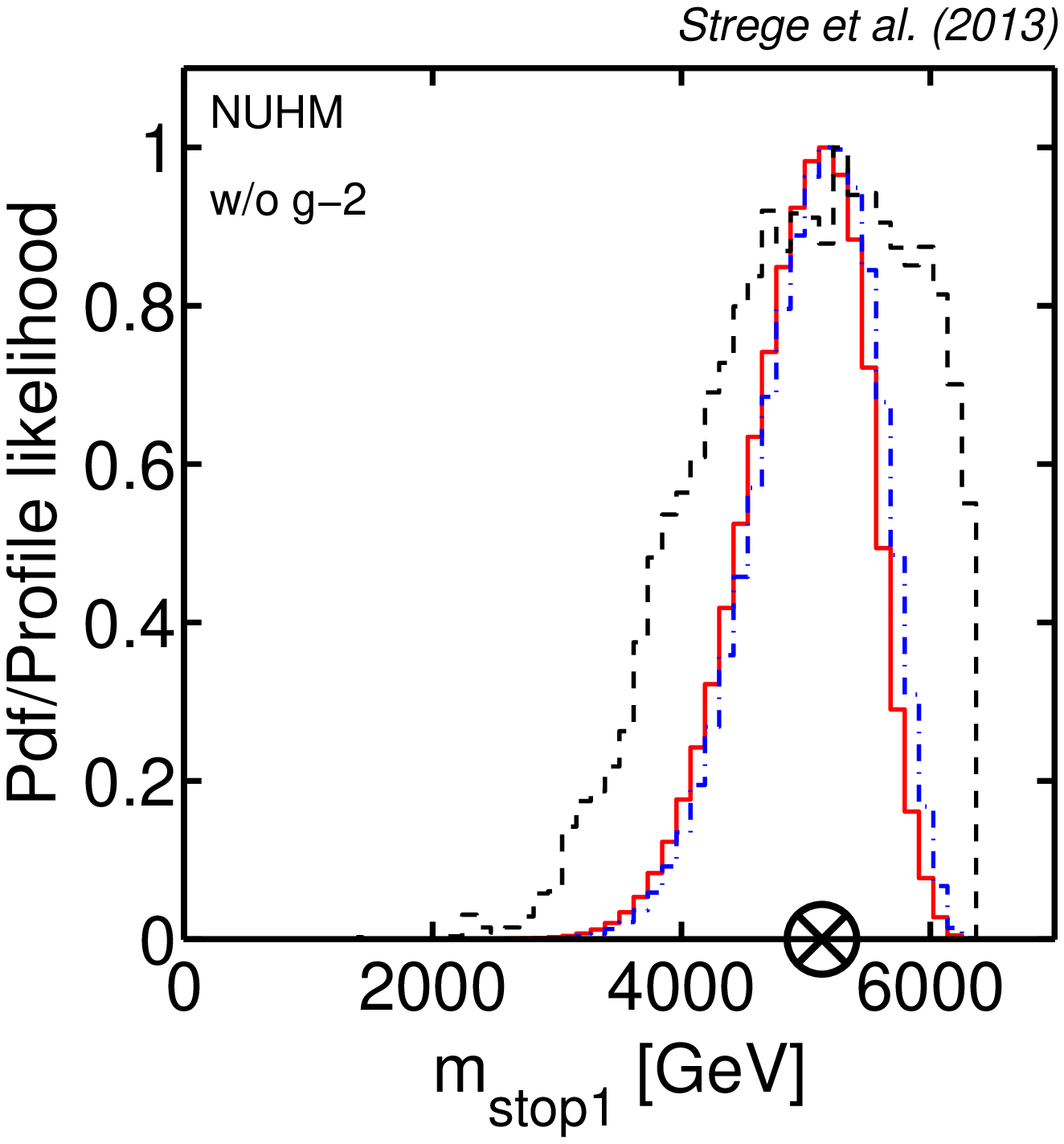}
\expandafter\includegraphics\expandafter
[\rowoffour]{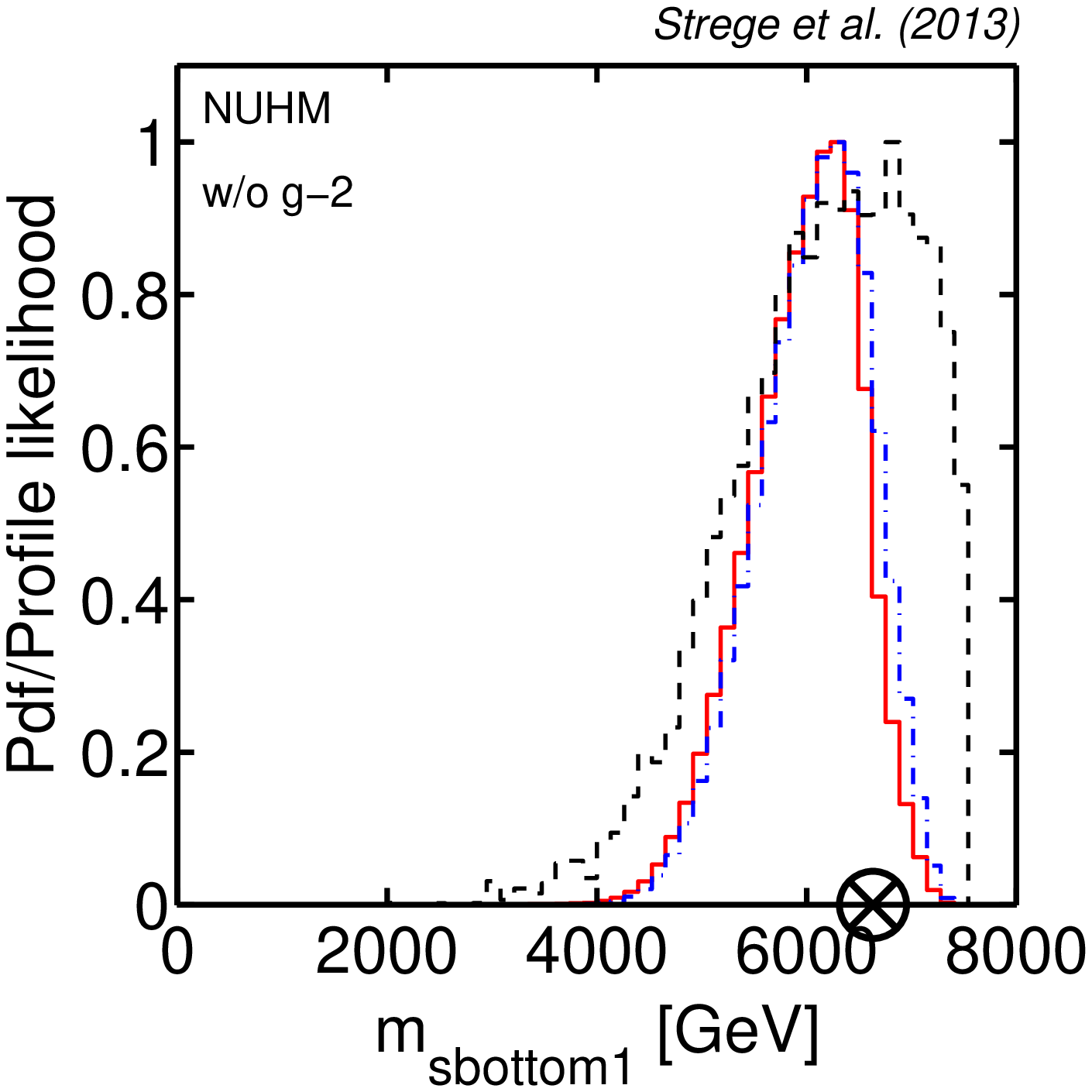}\\
\expandafter\includegraphics\expandafter
[\rowoffour]{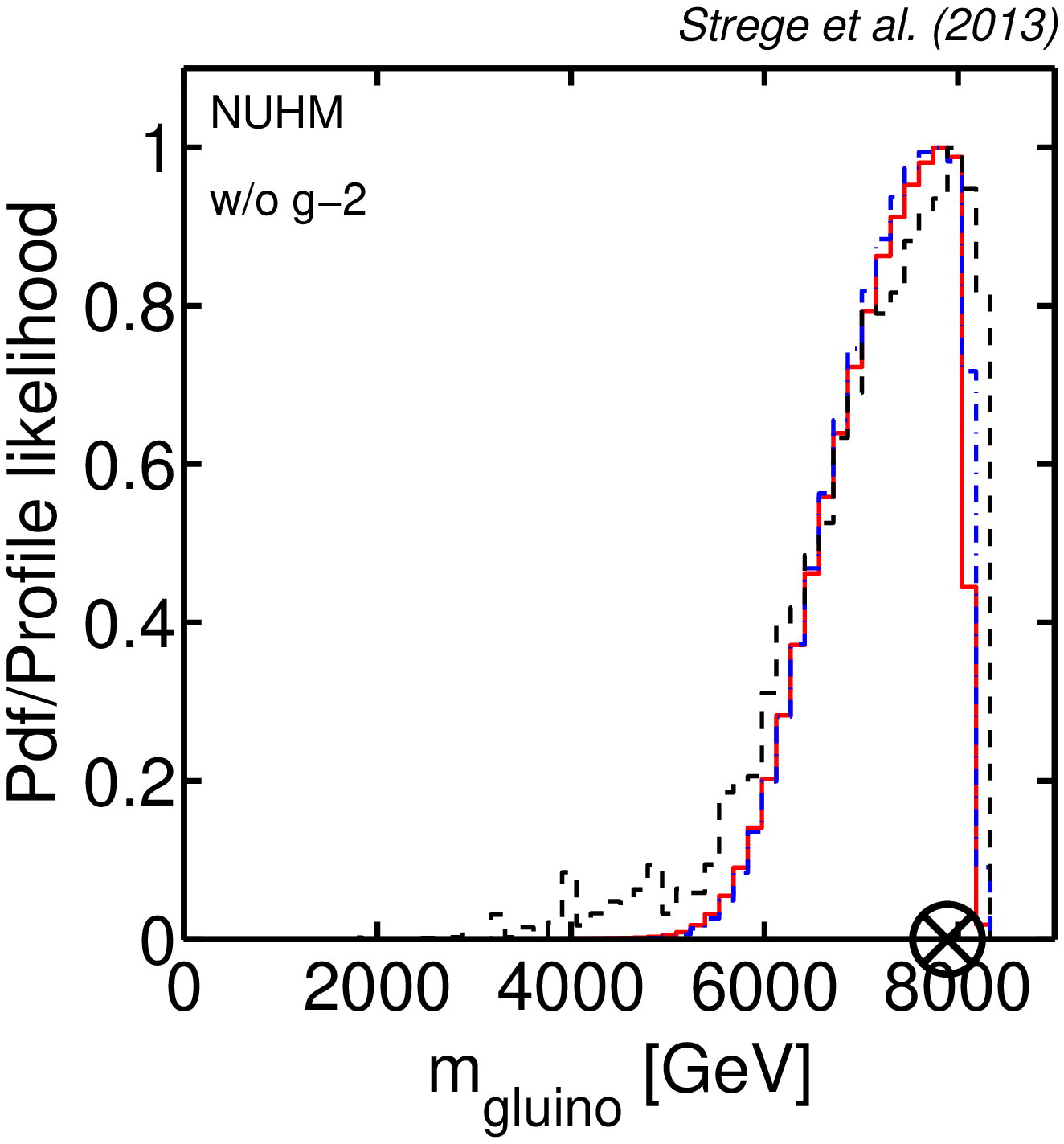}
\expandafter\includegraphics\expandafter
[\rowoffour]{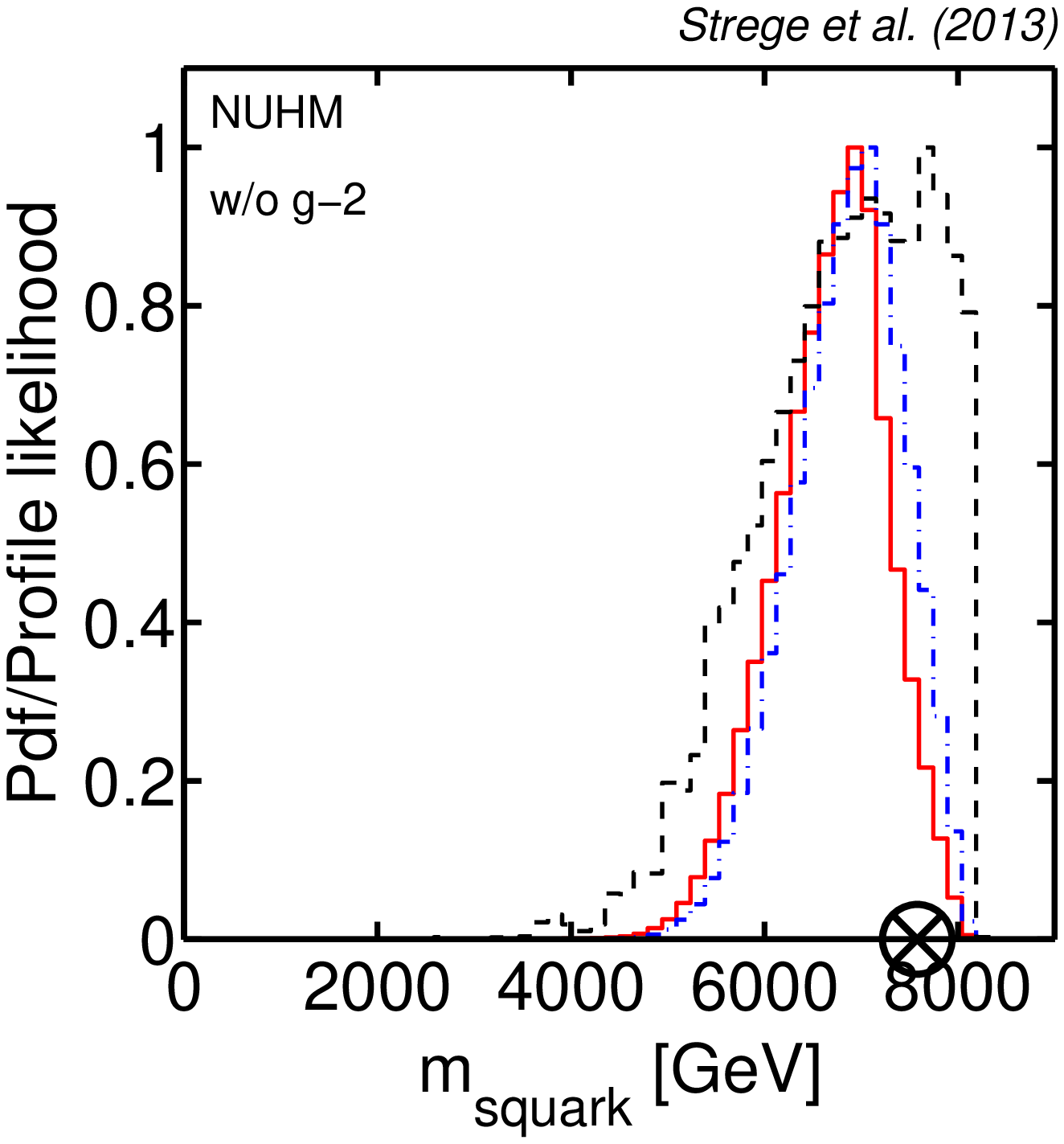} 
\expandafter\includegraphics\expandafter
[\rowoffour]{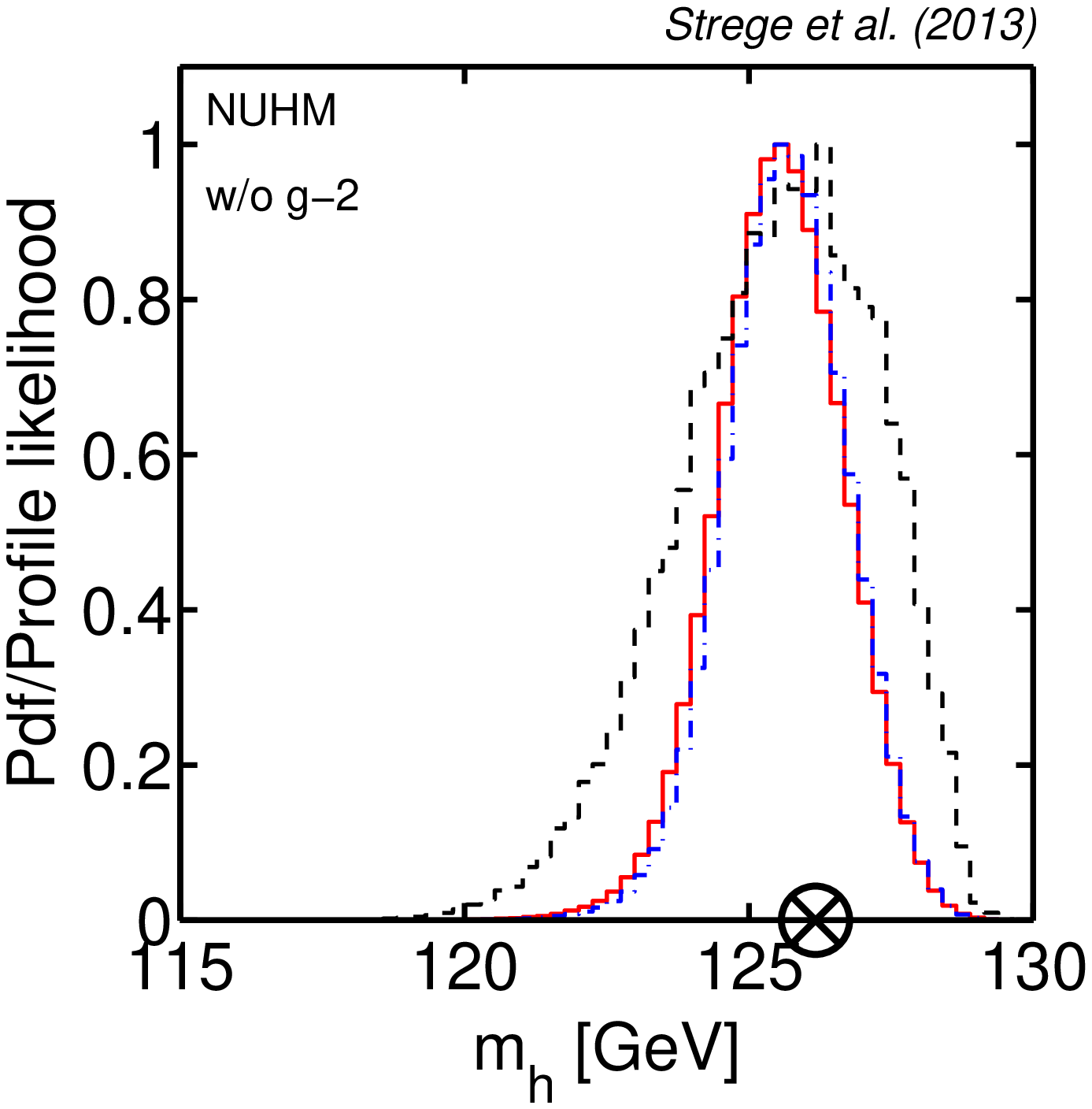}
\expandafter\includegraphics\expandafter
[\rowoffour]{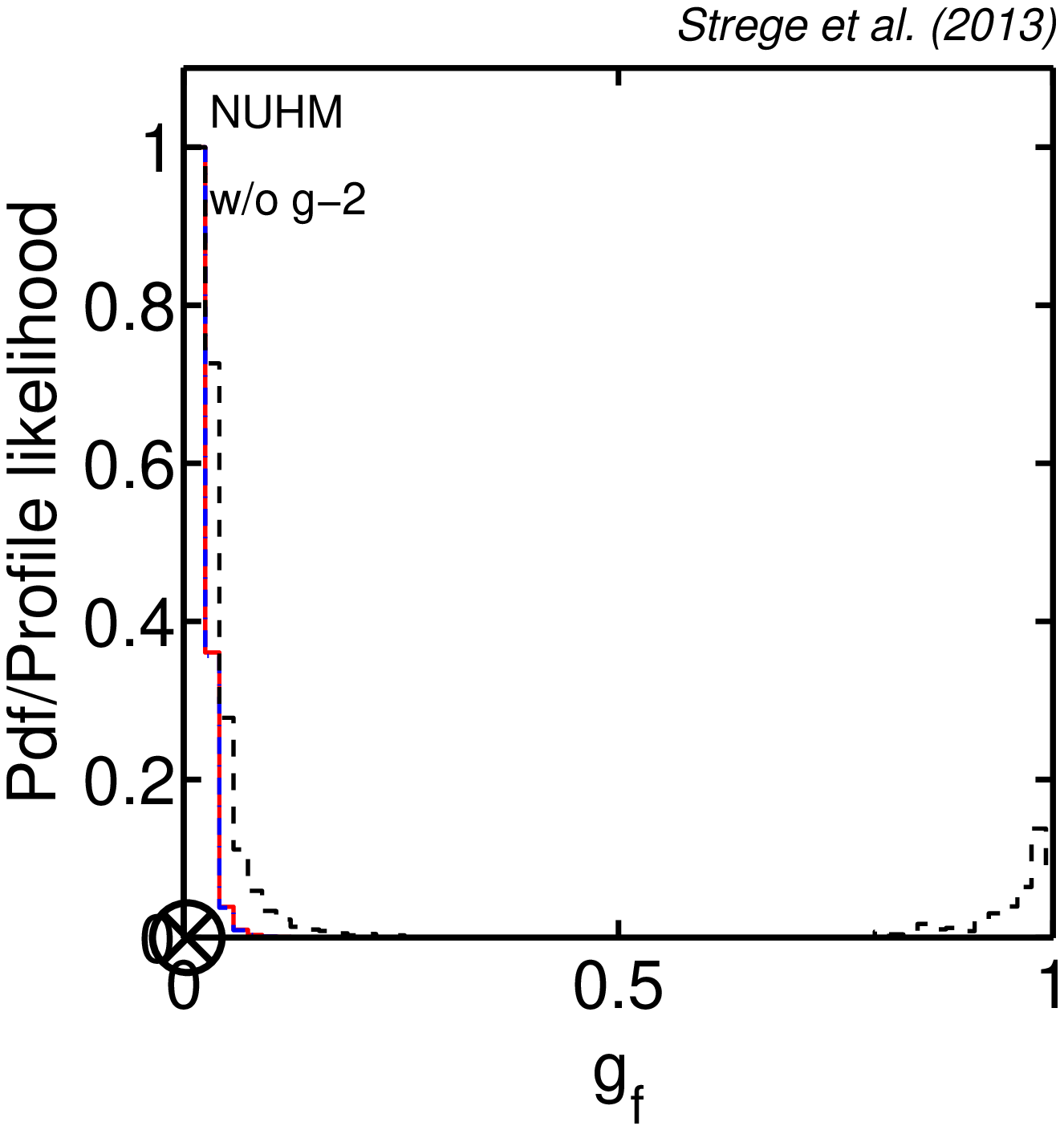} \\
\expandafter\includegraphics\expandafter
[\rowoffour]{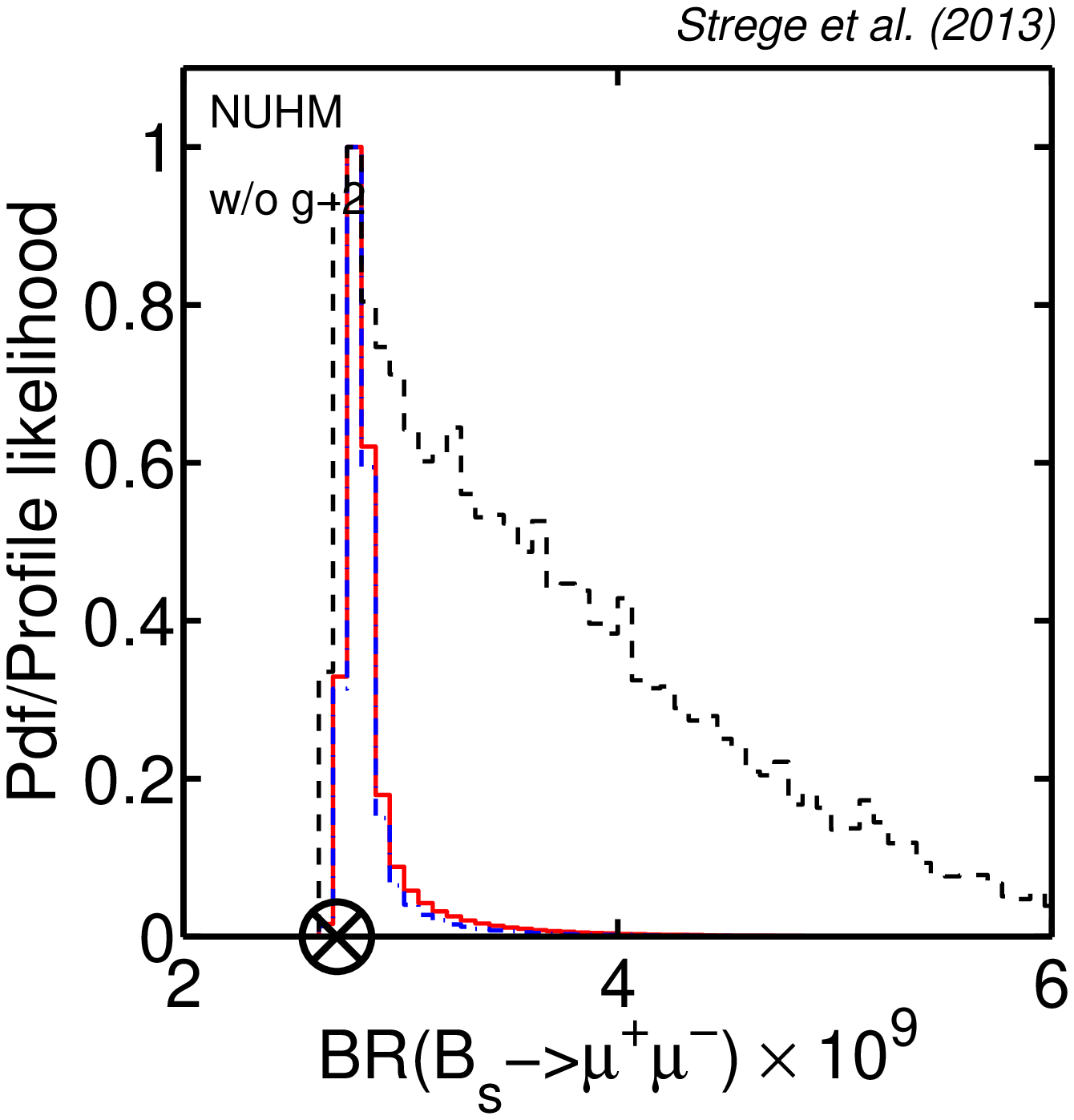}
\expandafter\includegraphics\expandafter
[\rowoffour]{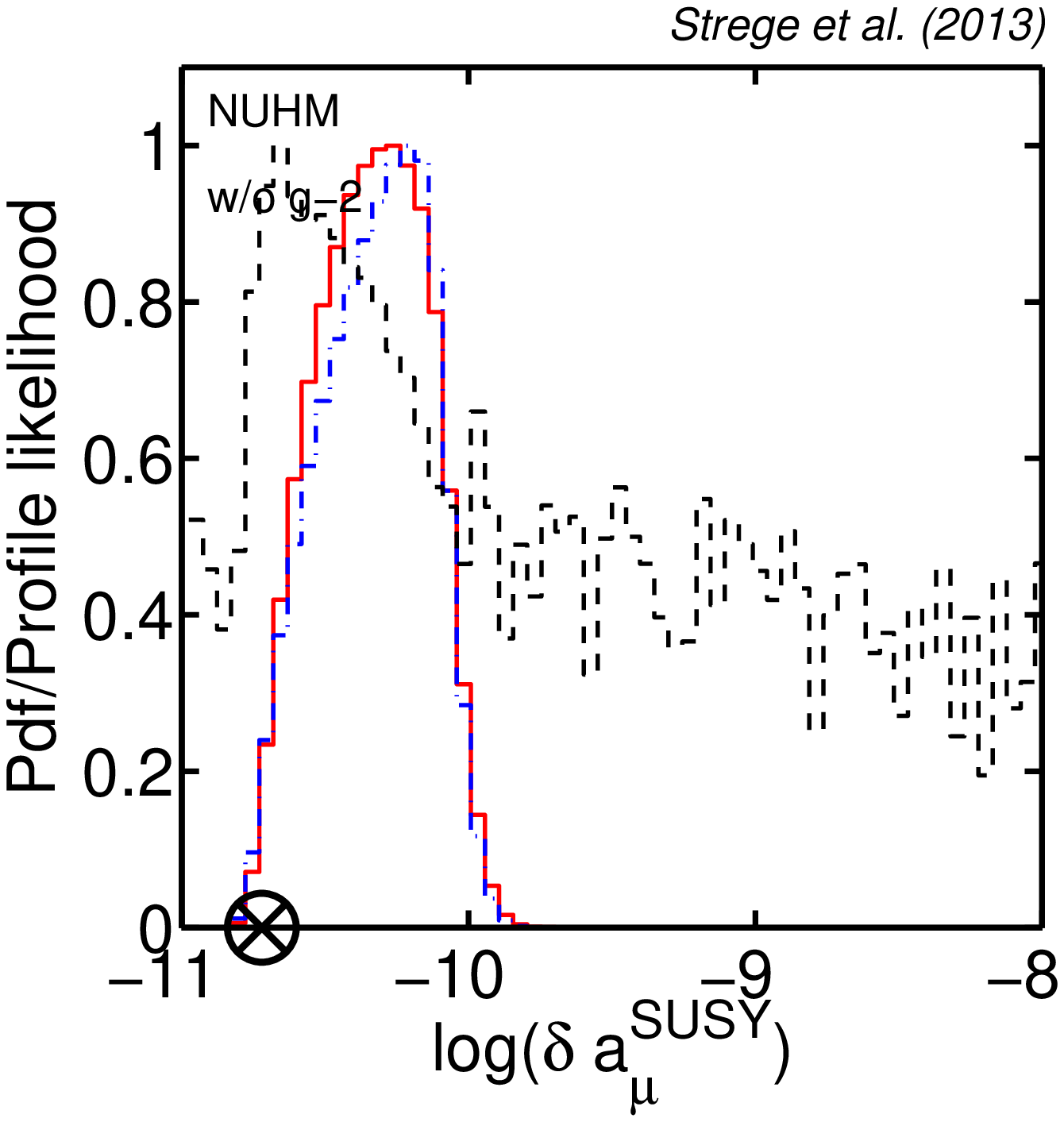}
\expandafter\includegraphics\expandafter
[\rowoffour]{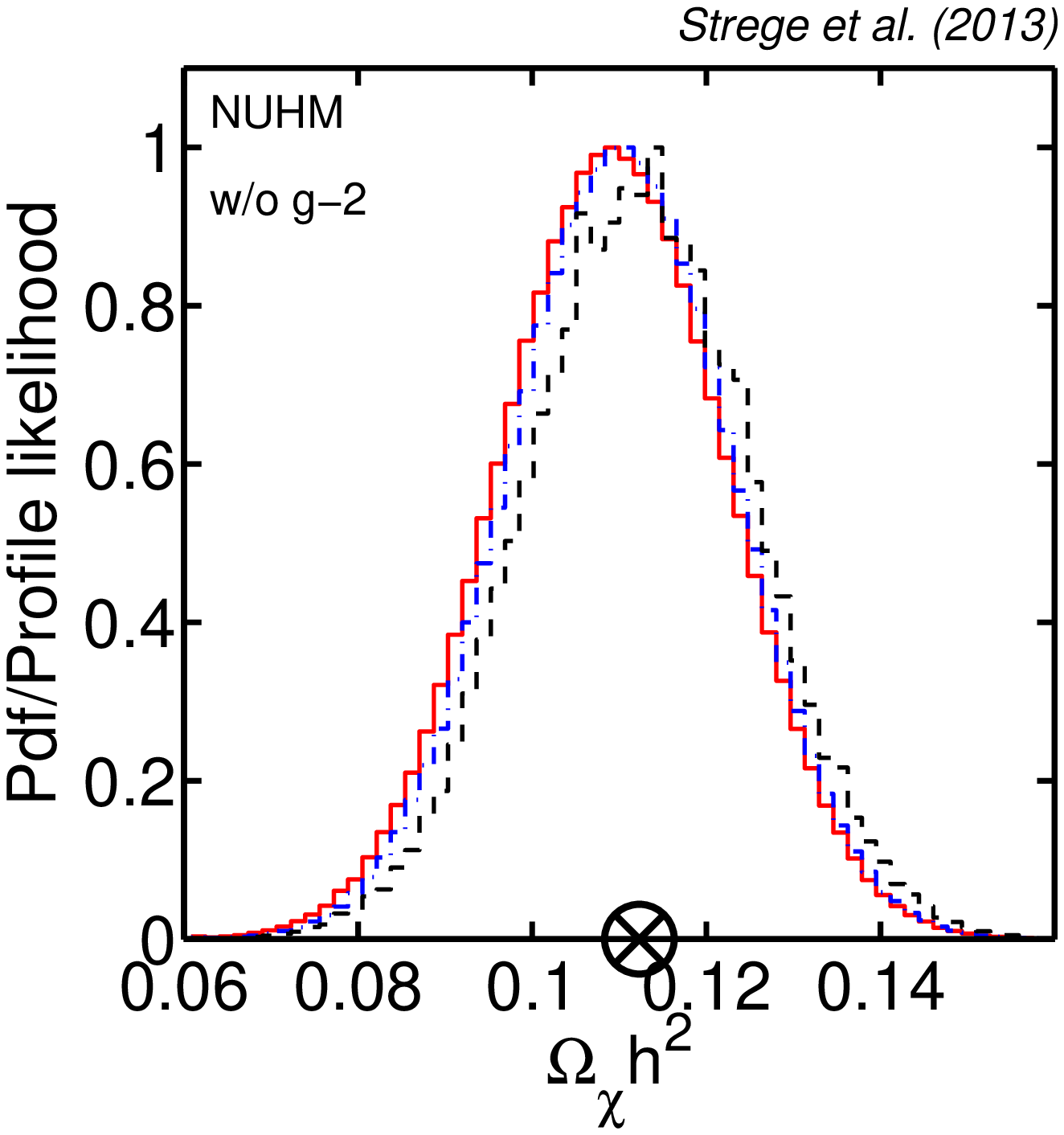}
\expandafter\includegraphics\expandafter
[\rowoffour]{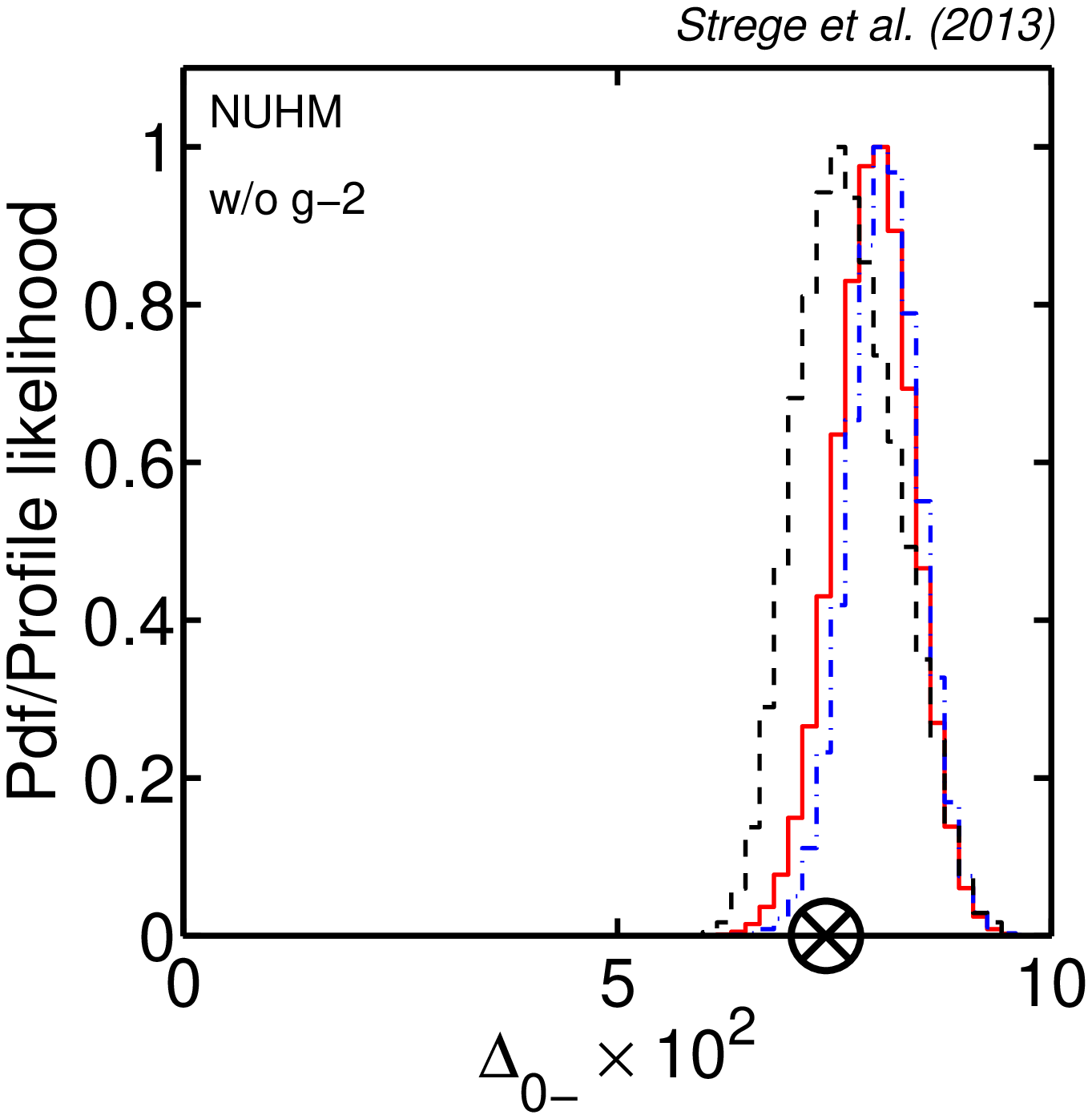} \\
%
%\expandafter\includegraphics\expandafter
%[\rowoffour]{Plots/NUHM_wog2_1D/NUHM_wog2_pp_1D_19_sq.eps} 
\caption{\fontsize{9}{9} \selectfont 1D marginal pdf for flat priors (dash-dot/blue), log priors (thick solid/red) and 1D profile likelihood (dashed/black) in the NUHM, including all current experimental constraints but {\em excluding the \gmt\ constraint}.  Quantities as in Fig.~\ref{fig:NUHM_1D}.  The best-fit point is indicated by the encircled black cross. \label{fig:NUHM_1D_wogmt}} 
\end{figure*}

Given the strong impact of the \gmt\ constraint on the profile likelihood results it is of interest to compare Fig. \ref{NUHM2D} to the results obtained when excluding the \gmt\ constraint from the analysis. We carried out a second set of scans including exactly the same constraints as before, except for the \gmt\ constraint. The resulting 2D and 1D distributions for \gmt\ are shown in Figs.~\ref{fig:NUHM_2D_wogmt} and \ref{fig:NUHM_1D_wogmt}, respectively.  The posterior pdfs are identical to what was found in Fig.~\ref{fig:NUHM_1D}, up to numerical noise (as can be verified by comparing Fig.~\ref{fig:NUHM_1D_wogmt} with Fig.~\ref{fig:NUHM_1D}), hence we do not display the 2D posterior pdf for this case.  

Dropping the \gmt\ constraint has almost no impact on the Bayesian results since the posterior pdfs including that constraint already favoured SM-like values, in tension with the experimental value, as discussed above. In contrast, the 1D profile likelihood function differs strongly from the previous results. The best-fit point is shifted to much smaller values of \gmt\ (see Fig.~\ref{fig:NUHM_1D_wogmt}), and the profile likelihood analysis now agrees much better with the Bayesian results: high values of the gaugino mass parameter $m_{1/2} > 2$ TeV are favoured at 99\% confidence level, while $m_0$ is almost unconstrained, with a small preference for $m_0 > 1$ TeV. Results in the other planes qualitatively agree with the profile likelihood analysis including the \gmt\ constraint, but the contours are more spread out, stretching to higher values of $\tan \beta$ (top right panel) and spanning a larger range of $m_{A}$ (bottom left panel).  In the direct detection plane (bottom right panel), the profile likelihood contours cover a large cross-section range, and a significantly larger range of $m_{\neut}$ than before. When excluding the \gmt\ constraint from the scan, results for the Bayesian posteriors for both choices of priors and for the profile likelihood function agree well. 

The 1D profile likelihood function for the sparticle masses (see Fig.~\ref{fig:NUHM_1D_wogmt}) are more spread out, but conclusions remain qualitatively similar to the analysis including the \gmt\ constraint. Intriguingly, a second, less prominent peak in the profile likelihood is observed for values $g_f \approx 1$. This corresponds to the SC region, where $m_\neut \approx m_{\tilde{\tau}_1}$ and the WMAP relic abundance is achieved via co-annihilation. In this case, the neutralino is bino-like, as in the cMSSM. However, this region of parameter space remains disfavoured by the Higgs mass constraint, even when the \gmt\ constraint is dropped. Hence this secondary peak in the profile likelihood for $g_f$ is much lower than that providing the overall best-fit. 

Both the 1D profile likelihood for $\brbsmumu$ and for \gmt\ are now significantly more spread out.  The profile likelihood for the relic density of dark matter is in much better agreement with the posterior pdfs. Since no more fine-tuning to satisfy the \gmt\ is required, there is a lot more freedom to adjust the parameters to satisfy other constraints. For the same reason, a small shift to smaller values of $\Delta_{0-}$, in better agreement with the experimental constraint, is observed, although qualitatively the profile likelihood for this quantity remains similar to the distribution obtained in the previous section.

Taken together, these findings suggest that the \gmt\ constraint strongly impacts on the favoured regions in the NUHM from a profile likelihood perspective, while the Bayesian pdf is much more robust with respect to this constraint. 

Details about the best-fit point found when excluding the \gmt\ constraint from the analysis are given in the right-hand column of Table~\ref{NUHMbf1} and Table~\ref{NUHMbf2}. The best-fit point is shifted towards larger $m_0$ and more negative $A_0$, and corresponds to a significantly lower value of \gmt; values of the other parameters remain similar to the best-fit point found in the analysis including the \gmt\ constraint. The largest contribution to the total best-fit $\chi^2$ still results from the constraint on the isospin asymmetry. The p-value of the best-fit is 0.25, which is almost identical to the p-value of 0.26 found when including the \gmt\ constraint. This suggests that, while the \gmt\ constraint has a strong impact on the shape of the profile likelihood function, the overall viability of the NUHM remains similar, independent of the inclusion of the \gmt\ constraint in the analysis.

%However, in contrast to the posterior distributions, the profile likelihood function is very spread out. Even when excluding the experimental constraint on \gmt from the analysis a small fraction of the samples still achieves the experimentally favoured \gmt value, although points corresponding to small \gmt are in better agreement with other experimental constraints.

% \begin{figure*}%[htp]
%\centering
%\expandafter\includegraphics\expandafter
%[\rowofthree]{Plots/NUHM_wog2/all/NUHM_PL_wog2_pp_2D_4_sq.eps}
%\expandafter\includegraphics\expandafter
%[\rowofthree]{Plots/NUHM_wog2/all/NUHM_PL_wog2_pp_2D_6_sq.eps} \\
%\caption{\fontsize{9}{9} \selectfont Left: Comparison of NUHM regions favoured in the SI scattering cross-section vs neutralino mass plane (left) and the $X_t/M_S$ vs $m_h$ plane (right) with (without) the \gmt\ constraint (blue/empty vs solid filled contours). \label{fig:DD_comparison_NUHM}}  
%\end{figure*}

\subsection{Comparison with the cMSSM}

By comparing the 2D distributions for the cMSSM in Fig. \ref{cMSSM_2D_alldata} with the NUHM results in Fig.~\ref{NUHM2D}, phenomenological differences between the NUHM and the cMSSM become apparent. 

In both models the measured value of the Higgs mass is difficult to achieve, so that this constraint plays a dominant role in determining the favoured regions of parameter space. However, the phenomenological consequences are very different: in the cMSSM the constraint on $m_h$ leads to a strong preference for the SC region at low scalar and gaugino masses, where the maximal mixing scenario is realised. In contrast, in the NUHM large values of $m_{1/2}$ are favoured, since the resulting large stop masses combined with a small amount of stop mixing can reproduce the measured value of $m_h$. Such large values of $m_{1/2}$ are disfavoured in the cMSSM by the relic density constraint. This difference is particularly pronounced when comparing the profile likelihood for the two models, which favours very large values of $m_{1/2} > 3$ TeV in the NUHM, but is constrained to $m_{1/2} \leq 1$ TeV in the cMSSM. Posterior and profile likelihood contours in the $(\tan \beta,A_0)$ plane are much more spread out in the NUHM, while in the cMSSM small favoured regions can easily be identified. 

These differences also lead to very different observational consequences for the NUHM and the cMSSM. In the NUHM a Higgsino-like neutralino with mass $m_{\neut} \sim 1$ TeV is predicted, while the cMSSM essentially always leads to gaugino-like neutralino dark matter with a mass of a few hundred GeV. In the cMSSM $m_{\neut} \sim 1$ TeV is excluded at more than $99\%$ from both the Bayesian and the frequentist statistical perspective. Also, for the reasons explained above, in the cMSSM lower values of the SI cross-sections are favoured than in the NUHM, which is especially apparent when comparing the profile likelihood functions in the $(m_{\neut},\sigmaSI)$ plane in Fig.~\ref{cMSSM_2D_alldata} and Fig.~\ref{NUHM2D_derived}. Therefore, detection prospects of the NUHM by direct detection experiments are much more promising than for the cMSSM. In contrast, the LHC operating at 14 TeV collision energy will probe the majority of the currently favoured region of cMSSM parameter space, while in the NUHM very large gaugino and squark masses are favoured, so that most of the currently favoured NUHM parameter space will not be accessible to the LHC, even after the High-Luminosity upgrade. This difference makes it possible to distinguish experimentally between these two models given a positive signal at the LHC, or in a future direct detection experiment.

%Dof:
%cMSSM: 11
%cMSSM wog2: 10
%NUHM: 9
%NUHM wog2: 8

%%%%%%%%%%%%%%%%%%%%%%%%%%%%%%%%%%%%%%%%%%%%%%%%%%%%%%%%%%%%%%%%%%%%%%%
%
% CONCLUSIONS 
%
%%%%%%%%%%%%%%%%%%%%%%%%%%%%%%%%%%%%%%%%%%%%%%%%%%%%%%%%%%%%%%%%%%%%%%%

\section{Conclusions}
\label{secconclusion}

In this paper we have presented up-to-date global fits of two models of minimal Supersymmetry, the cMSSM and the NUHM, including the latest constraints from LHC SUSY and Higgs searches, and the XENON100 direct detection experiment. We also included for the first time the new LHCb measurement of $\brbsmumu$, but found this constraint to have a negligible impact on both models. In contrast, the LHC constraint on the mass of the lightest Higgs boson has a very strong impact on the cMSSM and NUHM parameter spaces. Achieving large Higgs masses of $m_h \approx 126$ GeV in these models is difficult, and requires a significant amount of fine-tuning, either in the form of very heavy stops (and thus very heavy squarks in general), or the maximal mixing scenario.

In the cMSSM this leads to a strong preference of the profile likelihood function for the stau-coannihilation (SC) region, in which the maximal mixing scenario can be realised. In contrast, the Bayesian posterior shows a varying degree of preference for larger values of $m_{1/2}$ (in the A-funnel region), depending on the choice of priors.  Conclusions about the detection prospects for the cMSSM depend on the statistical perspective: the posterior pdfs suggest encouraging detection prospects at future direct detection experiments, and reasonably good prospects at the LHC operating at 14 TeV collision energy. In contrast, the region favoured by the profile likelihood will be challenging to explore with future direct detection experiments. However, the profile likelihood favours very small sparticle masses, and the SC region is already significantly constrained by the 5.8 fb$^{-1}$ ATLAS exclusion limit. This region -- which also contains our overall best-fit point -- will be further probed within the next months, and will fully be explored by the LHC operating at 14 TeV collision energy, leading to excellent prospects for either detecting or conclusively ruling out the cMSSM.  A goodness-of-fit test using all available present-day constraints does not allow to rule out the cMSSM at any meaningful significance level. Although this result has to be interpreted with care, due to the approximation involved in adopting an asymptotic chi-squared distribution, it appears that the cMSSM is still not ruled out by current experimental constraints. Nevertheless, the region of cMSSM parameter space surviving the increasingly tight constraints set by the combination of all data sets is shrinking. 

In contrast to the cMSSM, where dark matter is almost exclusively gaugino-like, in the NUHM Higgsino-like dark matter is strongly favoured. This leads to a strong preference for large $m_{1/2} > 3$ TeV, that are disfavoured in the cMSSM due to the dark matter relic density constraint. As a result, very large stop masses are favoured in the NUHM that, in combination with a moderate amount of stop mixing, can reproduce the measured value of $m_h$. The resulting preference for large sparticle masses renders detection prospects of the NUHM at the LHC operating at 14 TeV collision energy negative, even after the High-Luminosity upgrade~\cite{HL_LHC}. In contrast, the recent XENON100 limit was found to have a significant impact on this model, and future direct detection experiments will explore the entire currently favoured NUHM parameter space, from both the Bayesian and the profile likelihood statistical perspective.  As for the cMSSM, according to the p-value of the best-fit point the NUHM is not ruled out at any meaningful significance by combined constraints from all available present-day data sets.

The preference for Higgsino-like dark matter with $m_{\neut} \sim 1$ TeV at cross-sections easily accessible to future direct detection experiments is an important phenomenological difference between the NUHM and the cMSSM. Detection prospects at the LHC are also very different for these two models, very promising for the cMSSM, while no signal at the LHC is expected for the NUHM. These phenomenological differences make it possible to distinguish between these models using future data from the LHC and direct detection experiments. This might require the adoption of a suitable scaling Ansatz for the local dark matter density, as introduced in~\cite{Bertone:2010rv}.

While we have shown that both of these constrained models of SUSY are not ruled out yet in a quantitative manner (as our null results for the significance tests demonstrated), several highly complementary data sets now strongly impact on the cMSSM and NUHM parameter spaces and previously strongly favoured regions have been ruled out. The difficulty to simultaneously satisfy all experimental constraints in these models is becoming increasingly apparent, and strong degrees of fine-tuning are required to achieve satisfactory likelihood values. This motivates the study of more general SUSY models, such as the phenomenological MSSM (pMSSM) or the NMSSM, which are expected to be much more weakly constrained in light of current experimental data sets. The techniques used in this paper have provided a high-resolution mapping of the 15-dimensional cMSSM and 17-dimensional NUHM parameter space (including nuisance parameters), and are therefore expected to be useful in exploring less constrained models of SUSY with a richer phenomenology, such as e.g. the pMSSM.

%%%%%%%%%%%%%%%%%

%\vspace{0.5cm}
\acknowledgments  
We would like to thank Filip Moortgat, Giacomo Polesello and David van Dyk for useful discussions. C.S. is partially supported by a scholarship of the ``Studienstiftung des deutschen Volkes''. 
M.F. is supported by a Leverhulme Trust grant. The work of G.B. is supported by the ERC Starting Grant {\it WIMPs Kairos}.   R.R.dA, is supported by the Ram\'on y Cajal program of the Spanish MICINN
and also thanks the support of the MICINN under grant FPA2011-29678 and the
Invisibles European ITN project  (FP7-PEOPLE-2011-ITN, PITN-GA-2011-289442-INVISIBLES). We thank the support of the Consolider-Ingenio 2010 Programme under grant MultiDark CSD2009-00064.  The use of Imperial's High Performance Computing cluster is gratefully acknowledged. Special thanks to Simon Burbidge for help with the scans.
C.S., R.R.dA. and R.T. would like to thank the GRAPPA Institute at the University of Amsterdam for hospitality.

\end{document}